\documentclass[aps,prd, amsmath,amssymb,superscriptaddress,showkeys,nofootinbib,reprint,preprintnumbers]{revtex4-1}
\usepackage{newtxtext,newtxmath,natbib}
\usepackage{mdframed,bm}
\usepackage[T1]{fontenc}
\usepackage{ae,aecompl}
\usepackage{graphicx}	
\usepackage{amsmath}	
\usepackage{amssymb}	
\usepackage{booktabs}
\usepackage{color}
\usepackage{enumitem}
\usepackage{subfig}

\usepackage{physics}
\usepackage{cleveref}
\usepackage{comment}
\usepackage{xspace}
\usepackage{aas_macros}
\usepackage{ulem}
\usepackage{siunitx}

\usepackage{soul} 
\usepackage{amsmath}
\usepackage{amssymb}
\usepackage{xspace}
\usepackage[dvipsnames]{xcolor}

\newcommand{\vs}{vs.\xspace}

\newcommand{\COMMENT}[3]{{\textcolor{#1}{(#2: #3)}}}

\newcommand{\XX}{{\color{red}{XX}}\xspace}

\newcommand{\todo}[1]{\COMMENT{magenta}{TBD}{#1}}

\newcommand{\code}[1]{\textsc{#1}\xspace}

\newcommand{\buzzard}{\code{Buzzard v2.0}}

\newcommand{\bandvar}[2][]{%
  \ifthenelse{\isempty{#1}}{\var{#2}}{\var{#2\_#1}}%
}

\newcommand{\LCDM}{\ensuremath{\rm \Lambda CDM}\xspace}
\newcommand{\lcdm}{\LCDM}

\newcommand{\mcal}{\textsc{Metacalibration}\xspace}

\newcommand{\mcalRg}{\mbox{\boldmath $R_\gamma$}}
\newcommand{\mcalRs}{\mbox{\boldmath $R_{\rm s}$}}
\newcommand{\sompz}{\textsc{SOMPZ}\xspace}

\newcommand{\snr}{S/N\xspace}

\newcommand{\xipm}{\xi_\pm\xspace}

\usepackage[dvipsnames]{xcolor}

\begin{document}
\preprint{DES-2019-0479}
\preprint{FERMILAB-PUB-21-250-AE}
\title{Dark Energy Survey Year 3 Results: \\Cosmology from Cosmic Shear and Robustness to Data Calibration}

\author{A.~Amon}\email{amon2018@stanford.edu}
\affiliation{Kavli Institute for Particle Astrophysics \& Cosmology, P. O. Box 2450, Stanford University, Stanford, CA 94305, USA}
\author{D.~Gruen}
\affiliation{Department of Physics, Stanford University, 382 Via Pueblo Mall, Stanford, CA 94305, USA}
\affiliation{Kavli Institute for Particle Astrophysics \& Cosmology, P. O. Box 2450, Stanford University, Stanford, CA 94305, USA}
\affiliation{SLAC National Accelerator Laboratory, Menlo Park, CA 94025, USA}
\author{M.~A.~Troxel}
\affiliation{Department of Physics, Duke University Durham, NC 27708, USA}
\author{N.~MacCrann}
\affiliation{Department of Applied Mathematics and Theoretical Physics, University of Cambridge, Cambridge CB3 0WA, UK}
\author{S.~Dodelson}
\affiliation{Department of Physics, Carnegie Mellon University, Pittsburgh, Pennsylvania 15312, USA}
\author{A.~Choi}
\affiliation{Center for Cosmology and Astro-Particle Physics, The Ohio State University, Columbus, OH 43210, USA}
\author{C.~Doux}
\affiliation{Department of Physics and Astronomy, University of Pennsylvania, Philadelphia, PA 19104, USA}
\author{L.~F.~Secco}
\affiliation{Department of Physics and Astronomy, University of Pennsylvania, Philadelphia, PA 19104, USA}
\affiliation{Kavli Institute for Cosmological Physics, University of Chicago, Chicago, IL 60637, USA}
\author{S.~Samuroff}
\affiliation{Department of Physics, Carnegie Mellon University, Pittsburgh, Pennsylvania 15312, USA}
\author{E.~Krause}
\affiliation{Department of Astronomy/Steward Observatory, University of Arizona, 933 North Cherry Avenue, Tucson, AZ 85721-0065, USA}
% ...

\collaboration{DES Collaboration}

\date{\today}
\label{firstpage}

\begin{abstract}
This work, together with its companion paper, \citet*{y3-cosmicshear2}, presents the Dark Energy Survey Year 3 cosmic shear measurements and cosmological constraints based on an analysis of over 100 million source galaxies. With the data spanning 4143 deg$^2$ on the sky, divided into four redshift bins, we produce a measurement with a signal-to-noise of 40. We conduct a blind analysis in the context of the $\Lambda$CDM model and find a 3\% constraint of the clustering amplitude, $S_8\equiv \sigma_8 (\Omega_{\rm m}/0.3)^{0.5} = 0.759^{+0.025}_{-0.023}$. A \textit{\lcdm-Optimized} analysis, which safely includes smaller scale information, yields a 2\% precision measurement of $S_8= 0.772^{+0.018}_{-0.017}$ that is consistent with the fiducial case. The two low-redshift measurements are statistically consistent with the \textit{Planck} Cosmic Microwave Background result, however, both recovered $S_8$ values are lower than the high-redshift prediction by $2.3\sigma$ and $2.1\sigma$  ($p$-values of 0.02 and 0.05), respectively. The measurements are shown to be internally consistent across redshift bins, angular scales and correlation functions. The analysis is demonstrated to be robust to calibration systematics, with the $S_8$ posterior consistent when varying the choice of redshift calibration sample, the modeling of redshift uncertainty and methodology. Similarly, we find that the corrections included to account for the blending of galaxies shifts our best-fit $S_8$ by $0.5\sigma$ without incurring a substantial increase in uncertainty. We examine the limiting factors for the precision of the cosmological constraints and find observational systematics to be subdominant to the modeling of astrophysics. Specifically, we identify the uncertainties in modeling baryonic effects and intrinsic alignments as the limiting systematics.
\end{abstract}

\maketitle


\section{Introduction}
 
The current era of precision cosmology has delivered measurements of cosmological parameters at percent-level accuracy, and a standard cosmological model that fits data. This era was enabled by decades of progress at the nexus of instrumentation, observations, cosmology theory and analysis  \citep{weinberg1972,Peebles1980}. On the one hand, increasingly strong evidence for dark matter, made over many decades \citep{Bertone:2016nfn,Trimble:1987ee}, along with the discovery of dark energy and the accelerating universe in the late 1990s \citep{Perlmutter:1998np,Riess:1998cb}, paved the way for the current standard model. On the other hand, advances in our ability to collect, process, and analyze data from such diverse observations such as type Ia supernovae (SNIa),  the cosmic microwave background (CMB) anisotropies, and the distribution of galaxies and other tracers of large-scale structure (LSS), improve our ability to test theories for the accelerating universe \citep{Frieman08,Weinberg:2012es}, as well as physics at moments after the Big Bang \citep{Abazajian:2016yjj}.

While the model's success has been reinforced by agreement across a broad range of observations, the advancing precision has revealed some tensions among cosmological parameters measured by different observational probes. Most notably, there is a 3-5$\sigma$ tension in the Hubble constant, $H_0$, between low-redshift measurements made by the distance-ladder technique and those from the CMB at $z\approx1100$ (see Refs. \cite{Verde2019,DiValentino:2021izs} for a summary). Another widely discussed discrepancy, though less statistically significant, is that between constraints on the parameter $S_8\equiv \sigma_8 (\Omega_{\rm m}/0.3)^{0.5}$ --- the amplitude of matter density fluctuations, $\sigma_8$, scaled by the square root of the matter density, $\Omega_{\rm m}$. This quantity is consistently found to be 2-3$\sigma$ lower when measured in LSS data, including from the Dark Energy Survey\footnote{ https://www.darkenergysurvey.org/} (DES) \cite{Heymans13,Troxel2018,y1keypaper,hikage19, asgari20, heymans20,unwise}, than the constraint by the CMB \cite{Planck2018}. New and improved data and analysis methods are key to bring these tensions into sharp focus, in order to test whether they are attributed to new physics, or are caused by unforeseen systematic errors.

Weak gravitational lensing of large-scale structure, \textit{cosmic shear}, is a powerful method that is sensitive to both the geometry and the growth of cosmic structure in the Universe. This is the measurement of small but coherent distortions of the observed shapes of galaxies as their light passes through the intervening structure on its way to Earth. Measurements of these distortions carry information about the projected mass density to the source galaxy, and hence the amplitude, shape, and time evolution of the matter power spectrum. They are also sensitive to the geometrical ratio of distances to both the lens structures and the source galaxy (for reviews, see \citep{bartelschneider,hoekstrajain,kilbinger15}). 

Although proposed half a century ago \citep{Kristian67}, weak lensing by large-scale structure was not detected until 2000 
\citep{bacon00,kaiser00,vanwaerbeke00, wittman2000}. 
Cosmic shear has made strides since its first detection two decades ago with multiple surveys from the ground and space, owing to rapid advancements in technology, galaxy surveys and methodology \cite{Hoekstra:2002cj,Refregier:2002ux,Jarvis:2002vs,Hamana:2002yd,Brown:2002wt,Rhodes:2003wj,Heymans:2004zp,Jarvis:2005ck,Hetterscheidt:2006up,Massey:2007gh,Leauthaud:2007fb,Benjamin:2007ys,Fu:2007qq,Schrabback:2009ba,Huff:2011gq,Huff:2011aa,Lin:2011bc,Jee:2012hr,Jee:2015jta}. A significant step forward, both in quantity and quality of the data, was made with the Canada-France-Hawaii Lensing Survey (CFHTLens) \citep{Erben:2012zw,lensfit,Kilbinger:2012qz,Kitching:2014dtq, heymans06} while the current state of the art is being pursued by DES, the Hyper Suprime-Cam Subaru Strategic Program\footnote{ https://hsc.mtk.nao.ac.jp/ssp/} (HSC; \cite{HSC2017}) and the ESO Kilo-Degree Survey\footnote{http://kids.strw.leidenuniv.nl/} (KiDS; \cite{Kuijken15}). These current surveys use an increasingly sophisticated set of tools to make cosmic shear measurements over large areas on the sky and provide competitive constraints on cosmological parameters \citep{dls,svcosmicshear,hildebrandt2017kids,hikage19,Troxel2018,Troxel:2018qll,hildebrandt20,Ham20,asgari20}. 
Cosmic shear is highly complementary to probes utilizing the positions of galaxies. Of particular note is the joint analysis of cosmic shear with galaxy-galaxy lensing and galaxy clustering, commonly referred to as \textit{3$\times$2pt}, performed either within a single survey \citep{y1keypaper}, or in tandem with spectroscopic surveys \citep{heymans20}, and varying statistics \citep{joudaki_KiDS2df, vanuitert_KiDSGAMA}.
Such an analysis utilizes multiple sources of information to break degeneracies in cosmological parameters, and as importantly, nuisance parameters that describe observational and astrophysical systematic effects. The field is at the brink of a new epoch, both in terms of the advances in analysis and with upcoming surveys spanning `full-sky' to unprecedented depth. These upcoming surveys will revolutionize the field in the decades to come. They include the ground-based Vera C. Rubin Observatory's Legacy Survey of Space and Time (LSST\footnote{http://www.lsst.org/lsst}; \citep{LSST12}), and the space missions Euclid\footnote{sci.esa.int/euclid/} \citep{Laureijs2011} and the Nancy Grace Roman Space Telescope\footnote{https://roman.gsfc.nasa.gov/} \citep{Spergel15}.

While cosmic shear is a cornerstone to the future precise measurements of dark matter, dark energy, neutrino mass, and other fundamental quantities, its accurate measurement and modeling are challenging. First, cosmic shear is a percent-level effect that is statistically extracted from millions of galaxies. Second, the signal encoded in the shapes can be contaminated with a number of systematic effects (for a review, see \citep{MandelbaumRev}), such as leakage of the Point-Spread Function (PSF) that must be modeled and accounted for through robust shape measurement techniques \citep{lensfit,Huff_Mandelbaum_2017, SheldonMcal2017,y3-piff, metadetect,bfd} and validated through rigorous testing \citep{y3-shapecatalog, giblin20, mandelbaum17}.

Furthermore, interpreting the cosmic shear signal demands accurate estimation of the distribution of the galaxies' redshifts, in order to not incur a bias in cosmological constraints \citep[see, for example][]{Bernsteinhuterer10}.
Approaches to this calibration challenge may employ template-fitting or machine-learning to empirically learn the relationship between photometry and redshift based on a training sample, none of which is \textit{a priori} designed to meet the needs of, and quantify the resulting uncertainty for, weak lensing analyses. 
Existing methods are limited by incomplete information: in the case of template fitting, a complete description of the distribution of galaxy SEDs across luminosity and redshift; in the case of machine learning, a very large, perfectly representative spectroscopic training sample \citep{Newman2015}. The resulting selection biases have spurred debate on the use of complete photometric redshift or incomplete spectroscopic redshift training samples \citep{Newman2015,gruen17, hoyle,wright20, hildebrandt20, joudaki2020, hartley2020}. Regardless of the choice taken, consistency of different techniques, and the use of independent information increase confidence in the calibration \citep[see][for DES Y3]{y3-sompz,y3-sourcewz,y3-shearratio}.

Notably, the robust calibration of both the shear signals and redshift distributions necessitates understanding and mitigating the impact of blending, or crowded images, through realistic image simulations \citep{Dawson16, mandelbaum17b}. The rejection of galaxies with nearby neighbors \citep{lensfit, jarvis2016} is limited to handling only recognized blends, does not account for the increased occurrence of blending in high-density regions, and is unfeasible for deeper surveys \citep{bosch2017}. Detailed image simulation analyses are required to shed light on the resulting systematic errors \citep{mandelbaum17, samuroff17, kannawadi19}. The corresponding DES Y3 analysis has shown that the selection of galaxies based on their multi-band photometry has to be performed consistently between the data and simulations in order to understand the impact of blended sources at different redshift on both the shape measurement and on the effective redshift distribution for lensing \citep{y3-imagesims}.  

Finally astrophysical effects, such as intrinsic alignments \citep[e.g.][]{croft2000, heavens2000, troxel15} and the impact of the physics of galaxy formation on weak lensing observables \citep[e.g.][]{chisari, wechsler2018}, must be sufficiently modeled, or nulled, such that no residual impact is detected \citep{Krause17}. This is particularly challenging as the nuisance parameters incurred in intrinsic alignment modeling can absorb systematic errors in the calibrations of photometric redshift distributions \citep[e.g.][]{wright20}. Tests of internal consistency across redshift, angular scales and statistics can provide essential checks of sufficient mitigation against both theoretical and observational systematics \citep{Efstathiou18}. 

This work presents the cosmological constraints from cosmic shear measurements with the DES wide-field survey, using data taken during its first three years of observations and presents its robustness to data calibration. A companion paper to this work, \citet*{y3-cosmicshear2}, demonstrates the robustness of these cosmic shear cosmological constraints to modeling choices, in particular, to intrinsic alignments, baryonic effects, higher-order lensing effects and neutrinos. Cosmic shear is analysed using a common framework with those from galaxy-galaxy lensing (\citet{y3-gglensing}) and galaxy clustering (\citet{y3-galaxyclustering}), combined as \textit{2$\times$2pt} in  \citet*{y3-2x2ptaltlensresults, y3-2x2ptbiasmodelling, y3-2x2ptmagnification} in a joint DES Y3 \textit{3$\times$2pt} analysis presented in \citet{y3-3x2ptkp}. The measurements presented in this work are supported by a number of accompanying infrastructure papers: 
\begin{itemize}
 \item The construction and validation of the photometric `Gold' catalog of high-quality objects in DES Y3 is described in \citet{y3-gold}. New PSF modeling (\citet{y3-piff}) combined with weak lensing shape measurement based upon Ref.~\citep{SheldonMcal2017} gives a catalog of 100 million selected galaxies that are validated in \citet*{y3-shapecatalog}.
 \item Redshift calibration methodology is summarised in \citet*{y3-sompz}. This framework utilises external training data from narrow-band photometric and spectroscopic sources and DES deep observations with overlapping near-infrared data, presented in \citet*{y3-deepfields}, mapped to the wide-field with an improved image injection measurements (\citet{y3-balrog}). The full scheme incorporates new, independent methods, a two-step reweighting with self-organising maps (\citet{buchs19}), clustering redshifts (\citet*{y3-sourcewz}) derived from correlations with  the DES foreground  \textsc{redMaGiC} lens samples and small-scale shear ratios (\citet*{y3-shearratio}), to accurately constrain the redshift distributions. The latter two methods use the two DES foreground lens samples, \textsc{redMaGiC} and \textsc{Maglim} (\citet{y3-2x2maglimforecast}), in their measurements and are included at different stages in the pipeline, such that the ratios are able to constrain other nuisance parameter in the analysis. We test alternative techniques for modeling and marginalising over the uncertainty on the tomographic distributions, implementing a new tool, \textsc{Hyperrank}, (\citet{y3-hyperrank}).
\item State-of-the-art shear calibration with realistic image simulations and new methodology to account for the impact of blending on the effective redshift distribution for lensing measurements in \citet{y3-imagesims}.
\item  The general methodology, likelihood analysis and covariance used in the cosmological analyses shown in this work and \citet{y3-3x2ptkp} is presented in \citet{y3-generalmethods} and \citet{y3-covariances} and this methodology is independently validated using realistic simulations in \citet{y3-simvalidation}.
\item The statistical framework to assess the internal consistency of the DES data and measurements is presented in \citet{y3-inttensions} and the consistency with independent, external data in \citet{y3-tensions}.
\end{itemize}

This paper is structured as follows: in Sections~\ref{sec:data}, \ref{sec:calib}, and \ref{sec:imsims} the DES Y3 data and the calibration of the shear and redshift distributions are described. Section~\ref{sec:sims} provides an overview of the cosmological simulations used in validating the model and methods. Section~\ref{sec:measurement} presents the cosmic shear measurements,  covariance matrix validation and the blinding methodology and Section~\ref{sec:model}, the model. We detail the cosmological constraints in Section~\ref{sec:results}, and their internal consistency in Section~\ref{sec:IC}. Finally, in Sections~\ref{sec:robustredshift} and Section~\ref{sec:robustshear}, the robustness of the cosmological constraints to systematics in the data calibration is assessed through a range of validation tests.  More technical details of the analysis are provided in the appendices. 

\section{Dark Energy Survey Year 3 Data}\label{sec:data}

\begin{figure*}
\centering
\begin{minipage}{.5\textwidth}
  \centering
  \includegraphics[width=\textwidth]{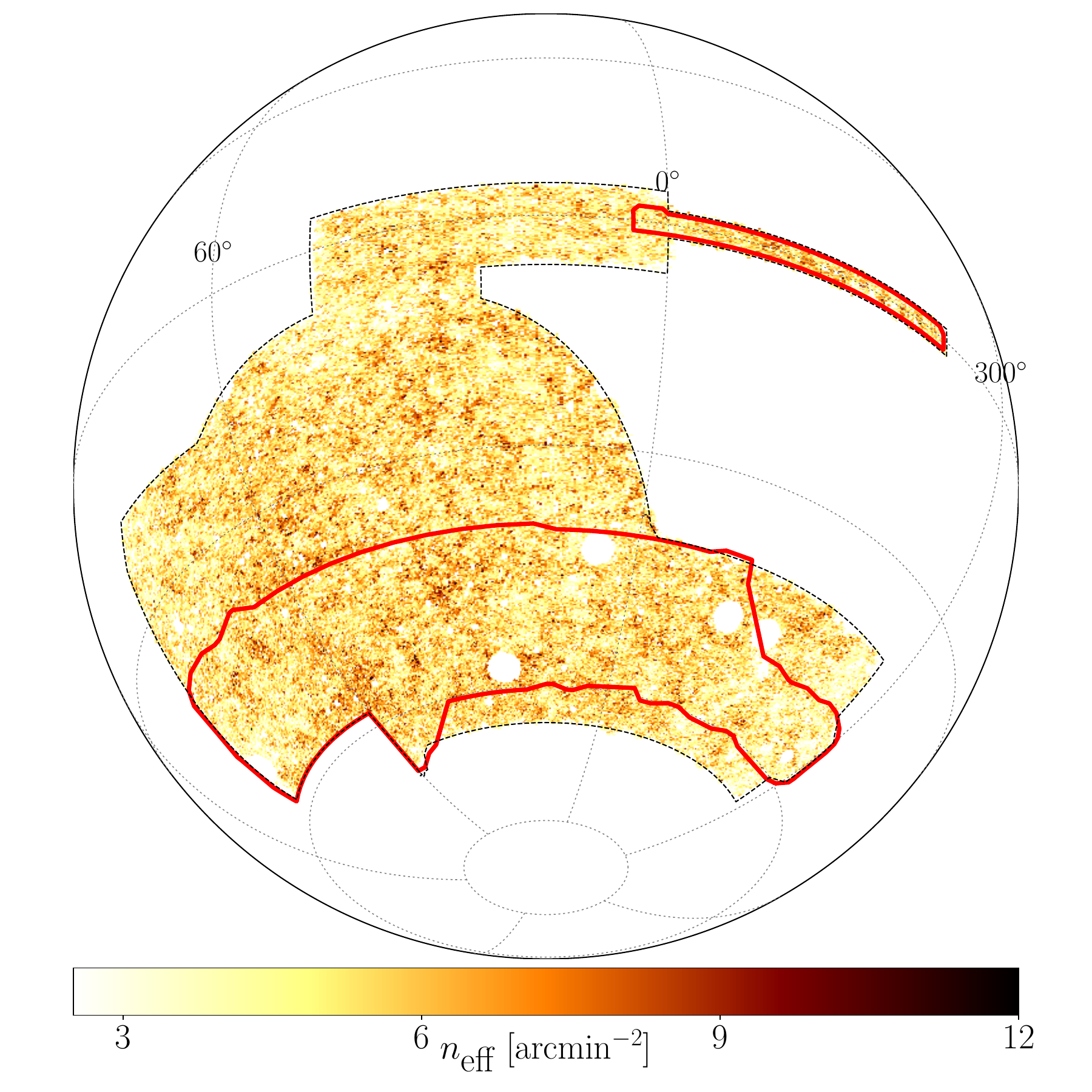}
  \label{fig:test1}
\end{minipage}%
\begin{minipage}{.5\textwidth}
  \centering
  \includegraphics[width=\textwidth]{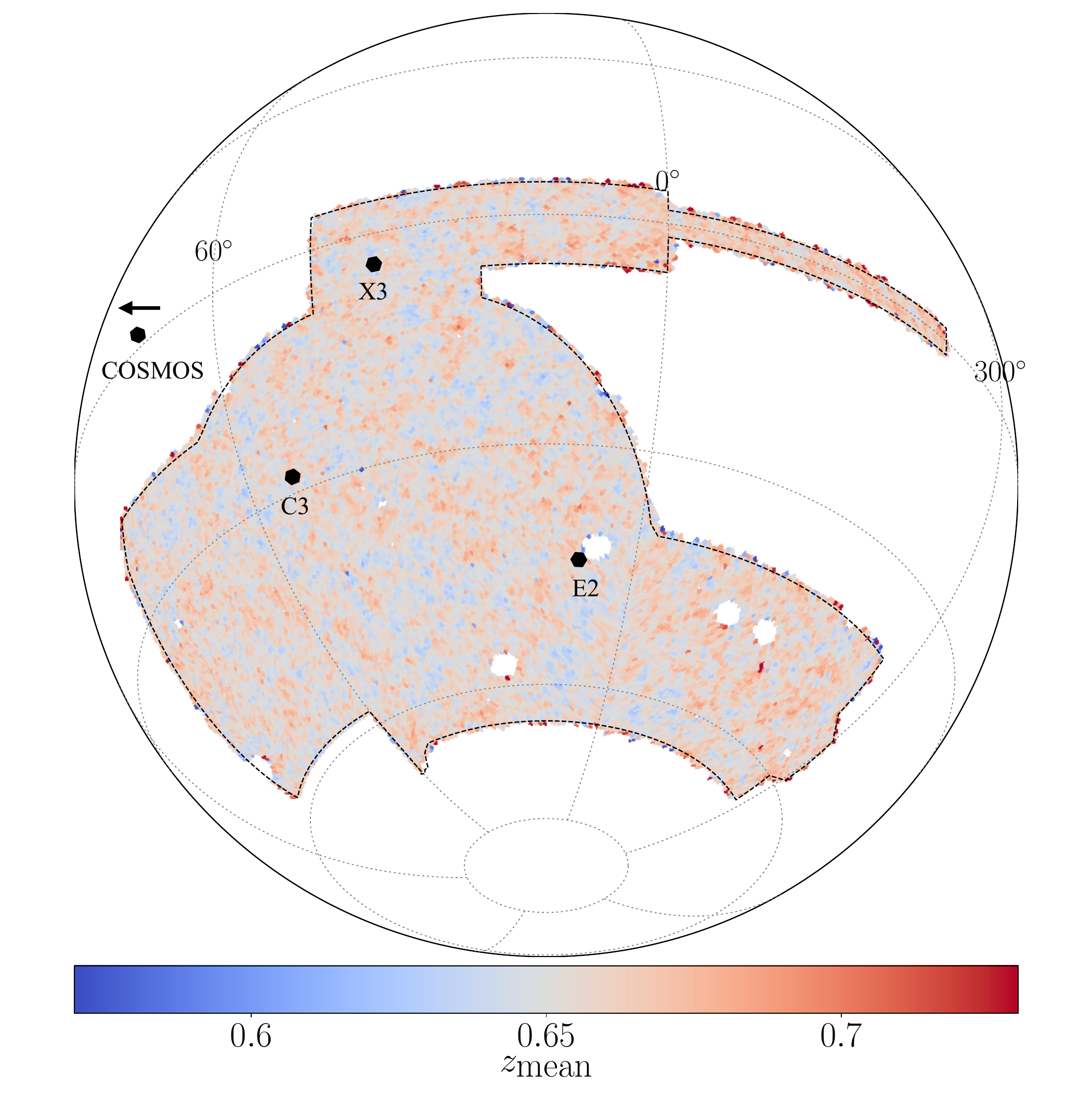}
  \label{fig:test2}
\end{minipage}
\caption{DES Y3 footprint showing the variation in the number density across the sky, as determined with the \mcal catalog (left) and the variation in mean redshift of that catalog (right). Overlaid on the left is the red outline of the Y1 footprint and on the right, the locations of the four DES Deep Fields \citep{y3-deepfields} (the fourth field, \textsc{COSMOS}, is positioned at $\sim150$deg, outside of the DES footprint, but has been rotated here to be shown on the map). The catalog spans a final effective area of 4143 deg$^2$ with an average number density of 5.59 arcmin$^{-2}$ and a mean redshift of 0.63. 
}
\label{fig:map}
\end{figure*}

The Dark Energy Survey has completed a six year observing program at the Cerro Tololo Inter-American Observatory (CTIO), Chile, using the Blanco telescope and the 570-megapixel Dark Energy Camera (DECam) \citep{flaugher15}. Ultimately, the complete survey spans 5000 square degrees of the Southern hemisphere in the $grizY$ bands with exposure times of 90 seconds in $griz$ and 45 seconds in $Y$ \citep{des-overview}. With 10 overlapping exposures in each of the $griz$ bands over the full wide-field area, the survey reaches a limiting magnitude of $i \sim 24$.

This DES Year 3 (Y3)  analysis exploits the data acquired over the first three years of observations, from August 15, 2013 to February 12, 2016, approximately 345 nights. The survey has a mean depth of four exposures that reaches an $i$-band signal-to-noise of $\sim10$ for extended objects up to $i_{AB}\sim23.0$ \citep{y3-gold}. The median recorded seeing (FWHM) in the $riz$ bands is  0.98, 0.89, 0.85 arcsec, respectively.

The DES Y3 dataset exhibits a number of improvements compared to DES Y1, including updates in brighter-fatter correction, sky-background modeling, better morphological star-galaxy classification and astrometric solutions, reduced photometric residuals compared to Gaia with photometric calibration uniformity $< 3$ mmag, the introduction of  per-object chromatic corrections, improvements to PSF modeling \citep{y3-piff} and better flagging of imaging artifacts. These improvements, as well as the production and validation of a `Gold' catalog of 390 million objects are described in detail in Ref.~\cite{y3-gold}.

\subsection{Shape measurement}\label{sec:shape}

\begin{table*}
    \caption{DES Y3 data properties per redshift bin: the unweighted number of objects passing the weak-lensing selection, the effective number density $n_{\rm eff}$ (gal/arcmin$^2$), calculated with $A_{\rm eff}= 4143$ deg$^2$, the per-component shape noise, $\sigma_e$, the mean redshift, $z_{\rm mean}$, the spread in the mean redshift of the redshift distribution realisations, $\Delta z$,  the mean shear per component, $c_{1,2}=\langle e_{1,2} \rangle$, the mean shear response, $\left<\boldsymbol{R}_\gamma \right>$, selection response, $\left<\boldsymbol{R}_{\rm s} \right>$, and the shear calibration parameter, $m$. Also shown are the variations of observational systematic uncertainties, $\Delta z_{\rm no z-blend}$ and $m_{\rm no z-blend}$, which do not account for the redshift-mixing effect due to blending, discussed in Section~\ref{sec:imsims}. }  
\label{tab:datastats}
\begin{center}
\begin{ruledtabular}
\begin{tabular}{cccccccccccccc}
Bin & no. objects & $n_{\rm eff}$ & $\sigma_e$ &  $z_{\rm mean}$  & $\langle e_1 \rangle \times 10^{4}$  & $\langle e_2 \rangle \times 10^{4}$ & $\Delta z$& $m$ & $\Delta z_{\rm no z-blend}$& $m_{\rm no z-blend}$ & $\left<\boldsymbol{R}_\gamma \right>$  & $\left<\boldsymbol{R}_{\rm s} \right>$  \tabularnewline
\hline 
Full & 100 204 026 &5.590 & 0.268 &  0.633 &  - & -  & - & - & - & - \tabularnewline
0 & 24 940 465 & 1.476 & 0.243 & 0.336 & 3.22 & 1.60  &    0.018 & -0.006 $ \pm $ 0.009 & 0.016  & -0.013 $ \pm $ 0.003 & 0.7636 & 0.0046 \tabularnewline
1 & 25 280 405 & 1.479 & 0.262 & 0.521 & 3.36 & 0.38 &  0.015 & -0.020 $ \pm $ 0.008 & 0.013  & -0.018 $ \pm $ 0.004 & 0.7182 & 0.0083\tabularnewline
2 & 24 891 859 & 1.484 &  0.259 &  0.742 & 3.77 & 0.07&  0.011 & -0.024 $ \pm $ 0.008 & 0.006  & -0.023 $ \pm $ 0.004 & 0.6887 & 0.0126 \tabularnewline
3 & 25 091 297 & 1.461 & 0.301 &  0.964  & 4.06 & -0.27 &  0.017 & -0.037 $ \pm $ 0.008 & 0.015  & -0.036 $ \pm $ 0.006 & 0.6154 & 0.0145\tabularnewline
\end{tabular}
\end{ruledtabular}
\end{center}
\end{table*}

The relation of measured galaxy shape to gravitational shear is commonly expressed for each of their two components as \citep{heymans06}
\begin{equation}
\label{eqn:shearbias}
    \epsilon^{\rm obs} = (1+m)(\epsilon^{\rm int} + \gamma) + c\, ,
\end{equation}
i.e.,~a combination of the intrinsic ellipticity of the galaxy, $\epsilon^{\rm int}$, and an additional lensing-induced shear, $\gamma$, which is the cosmologically interesting weak lensing signal. In this affine approximation, contamination of the measurement, or systematics, come in two variants. Additive bias $c$ can result, for example, from insufficient modeling of the PSF that causes a leakage into the measured galaxy ellipticity (see Section~\ref{app:add}). Multiplicative bias, $m\ne0$, can arise, for example, from noise bias, model bias, or mis-estimation of the PSF size. While the affine relation does not account for many complexities of real-world shear measurement, e.g.~the impact of measurement noise, the spatial correlation of additive bias \citep{amara07,vanuitert16,kitching21}, or the redshift-mixing effect of galaxy blending \citep{y3-imagesims}, it is nevertheless a useful starting point for shear calibration.

For the DES Y3 analysis, \mcal is used to produce the shear catalog \citep{SheldonMcal2017, Huff_Mandelbaum_2017}. This method calibrates shear statistics from the imaging data itself, without reliance on prior information about galaxy properties. In its Y3 implementation, \mcal measures the properties of a galaxy, including its  ellipticity, using a single Gaussian as a pre-seeing model that is fit to each detected object, for all the available epochs, using the $riz$-band images. The $g$-band data are excluded in the fit due to insufficient PSF modeling, as detailed in \citep{y3-piff}. During the fit, light from neighbouring objects is masked in order to reduce blending effects. The galaxy image is then deconvolved with the PSF, artificially sheared, reconvolved by a symmetrized version of the PSF and the ellipticity remeasured. Done repeatedly, this results in one unsheared and four artificially-sheared versions of the shape catalog (each component of ellipticity is sheared both positively and negatively). These are used to construct the shear response matrix, \mcalRg, of each galaxy via numerical derivatives of the ellipticity, as
 \begin{eqnarray}
\left(\mcalRg\right)_{i,j}  = \frac{e_i^{\rm{s}_{j+}}- e_i^{\rm{s}_{j-}}}{\Delta \gamma_j} \, .
\end{eqnarray}
Here $e_i^{\rm{s}_{j+}}$ is the $i$th ellipticity component measured on an image sheared positively by $\Delta \gamma_j$ in the $j$th ellipticity component.
The response matrix is a noisy quantity on a single galaxy basis, but ensemble averages, $\left<\boldsymbol{R}_\gamma \right>$, are precisely measured for large samples, and known to a sufficient degree of accuracy given the size of the DES Y3 sample. The pipeline is largely based upon that employed in the DES Y1 analysis, now with the inclusion of an inverse variance weighting for each galaxy, $w$, employed to boost the signal-to-noise of the data. The details of this implementation can be found in \citep{y3-shapecatalog}. 

The \mcal\ pipeline is designed to self-calibrate biases in the shear estimation by correcting for not only the response of the shear estimator, but also the selection biases \citep{Huff_Mandelbaum_2017}. To this end, measurements of flux, size, signal-to-noise ratio and other selection-relevant properties are made on the unsheared and sheared images. Selection response can be estimated by selecting objects based on sheared measurements and computing \mcalRs\ as 
 \begin{eqnarray}
\left<\mcalRs \right>_{i,j}  = \frac{\left< e_i\right>^{\rm{s}_{j+}}-\left< e_i\right>^{\rm{s}_{j-}}}{\Delta \gamma_j} \, .
\end{eqnarray}
Here $\left<e_i\right>^{\rm{s}_{j+}}$ represents the mean of the $i$th ellipticity component measured on images without applied shear. The average is taken over the group of galaxies selected into a given bin using the parameters extracted from positively sheared images, $\Delta \gamma_j$, in component $j$. $\left<e_i\right>^{\rm{s}_{j-}}$ is the analogous quantity for negatively sheared images. To calibrate the mean shear of the catalog, it is sufficient to consider the total response matrix per redshift bin as the sum of shear and selection response:
\begin{eqnarray}
\label{eqn:response}
\left<\boldsymbol{R} \right> =  \left<\boldsymbol{R}_\gamma \right> + \left<\boldsymbol{R_{\rm s}} \right> \, ,
\end{eqnarray}
which are quoted in Table~\ref{tab:datastats}. For the purposes of our statistics, it is an excellent approximation to correct for shear response with a scalar $R=\left[R_{11}+R_{22}\right]/2$ instead of the full response matrix \citep[see appendix~A of][]{y3-shapecatalog}. The \mcal\ response accounts for shear biases at a level of a few parts in a thousand in the absence of blends and detection biases \citep{SheldonMcal2017}.  The current \mcal\ implementation, however, does not correct for a shear-dependent detection bias \citep{metadetect} and the redshift-mixing effect of blending, which leave a multiplicative factor at the level of $m\sim2-3$\% and are calibrated using image simulations in \citep{y3-imagesims} and discussed in Section~\ref{sec:imsims}.

\subsection{The shape catalog}

Of the 390 million detected and measured objects, only those that pass a number of criteria are included in the weak lensing catalog \citep{y3-shapecatalog}. This is encoded in an index column in the catalog and encompasses selections based on a combination of galaxy and PSF properties, which are designed to reduce potential systematic biases due to blending, PSF mis-estimation and stellar contamination by imposing selections on the S/N of objects, the size, the PSF size ratio and binary stars (see Section 4.2 of Ref. \citep{y3-shapecatalog} for a complete description). In addition, magnitude selections are imposed ($i=18-23.5$ and $rz=15-26$) to exclude from the analysis galaxies for which robust redshift estimation is difficult. Finally, we mask the catalog to limit to the area that enables coherent combination with the DES lens sample (Section~\ref{sec:lensdata}), bringing the footprint's final effective area to span $A_{\rm eff}= 4143$ deg$^2$.  The resulting \mcal\ catalog yields a total of $100,204,026$ galaxies, with a weighted number density of $n_{\rm eff}=5.59$ galaxies per square arcmin, as defined by \cite{Heymans13}. Table~\ref{tab:datastats} lists the effective number density for the full catalog and the corresponding weighted ellipticity variance.  The statistical power of the \mcal\ catalog without galaxy weights is tripled compared to Y1, with weighting increasing this further by $\approx 25\%$ \cite{y3-shapecatalog}.

The spatial variation of number density over the survey footprint is shown in the left-hand panel of Figure~\ref{fig:map} in orthographic projection and equatorial coordinates, with a map that utilises \texttt{skymap}\footnote{https://github.com/kadrlica/skymap}. Overlaid are the bounds of the previous DES Y1 survey. 

The \mcal\ shape catalog has passed a library of tests aimed at identifying residual biases, detailed in \citep{y3-shapecatalog}. These have validated the measurements against systematic errors connected to PSF mis-modeling, which are negligible for the full catalog with a smaller amplitude compared to the DES Y1 analysis. In addition, the catalog was tested for robustness against the presence of spurious B-modes using two different estimators, \textsc{COSEBIs} \citep{Schneider2010} and Pseudo-$C_\ell$, which consistently resulted in a null detection (see also Appendix~\ref{App:Bmodes}). In Section~\ref{sec:robustshear}, these tests are extended to the tomographic case. Other tests included checking the dependence of the two components of the shear on a number of galaxy or survey properties, finding no significant correlations, except for a linear dependence between $\langle e_1 \rangle$ and the ratio between the galaxy size and PSF size. While the origin of this trend is unknown, its amplitude is three orders of magnitude smaller than the cosmic shear signal at all scales and thus can be safely neglected in the cosmological analysis.

\subsection{The lens sample}\label{sec:lensdata}
The DES fiducial foreground galaxy sample, \textsc{MagLim}, is divided into  six redshift bins and used in the measurement of the shear ratios (see Section~\ref{sec:SR}). These galaxies are defined by a magnitude cut that evolves linearly with their photometric redshift estimate, $z_{\rm ML}$ \citep{dnf}, as $i < az_{\rm ML} + b$, with $a = 4$ and $b = 18$, as well as a lower magnitude bound $i > 17.5$. The sample selection is optimized to prioritize brighter galaxies at low redshift and balance number density and photometric redshift accuracy in terms of cosmological constraints
obtained from galaxy clustering and galaxy-galaxy lensing \citep{y3-2x2maglimforecast}. 

An alternative lens sample is selected by \textsc{redMaGiC} \citep{RozoRM}, designed to find luminous red galaxies with precise photometric redshift estimates \citep{y3-lenswz}. It does so by selecting galaxies above some luminosity threshold based on how well they fit a red sequence template. The template is calibrated using \textsc{redMaPPer} \citep{rykoffRM} and a subset of galaxies with spectroscopic-verified redshifts. The goodness of fit threshold is chosen to maintain a desired comoving number density of galaxies. Clustering weights are assigned to \textsc{redMaGiC} galaxies to eliminate spurious correlations with observational systematics.

\section{Redshift calibration}
\label{sec:calib}

Any cosmological interpretation of weak lensing signals requires accurate knowledge of the distribution of distances to the source galaxies used in the measurement \citep{Huterer06,vanWaerbeke06}. A tomographic cosmic shear measurement requires the distribution of source galaxies into several redshift bins, each of which is characterized by a redshift distribution $n(z)$. Along any line of sight, $\bm{\theta}$, the expected observed shear, $\bm{\gamma}^{\rm obs}(\bm{\theta})$, is related to an integral of the shear experienced by sources at redshift $z$ along that line of sight, $\bm{\gamma}(\bm{\theta},z)$, as
\begin{equation}
\bm{\gamma}^{\rm obs}(\bm{\theta}) = \int \mathrm{d}z \, n(z) \, \bm{\gamma}(\bm{\theta},z) \; . 
\label{eqn:nz}
\end{equation}

The cosmic shear signal, as the angular correlation of two such $\bm{\gamma}^{\rm obs}(\bm{\theta})$ fields, is highly sensitive to biases in the estimates of $n(z)$. Such biases can result in significant shifts in the inferred cosmological parameters \citep{Bonnett16, hildebrandt20, joudaki2020}. The increased statistical precision of DES Y3 requires errors in the redshift distributions to be unprecedentedly small so as not to dominate the uncertainty budget.

In practice, photometric redshift calibration for weak lensing source galaxies relies on DES galaxies for which there exists accurate redshift information, defining the \textit{redshift sample}. To this end, the DES Y3 methodology follows the idea that galaxies can be categorised by their color in many-band photometric information, thereby tightly constrained \textit{phenotypes} \citep{Masters15, Sanchez19, buchs19}. The DES \textit{Deep Fields} \citep{y3-deepfields}, a combination of optical and near-infrared multi-band, deep photometry over a smaller area of 5.88 deg$^2$ (after masking), are used to define these phenotypes. These observations act as an intermediary between redshift sample galaxies and the wide field photometry over the DES footprint. This framework successfully reduces both the statistical and systematic uncertainty in redshift calibration \citep{y3-sompz}. 

\begin{figure}
\centering
\includegraphics[width=0.49\textwidth]{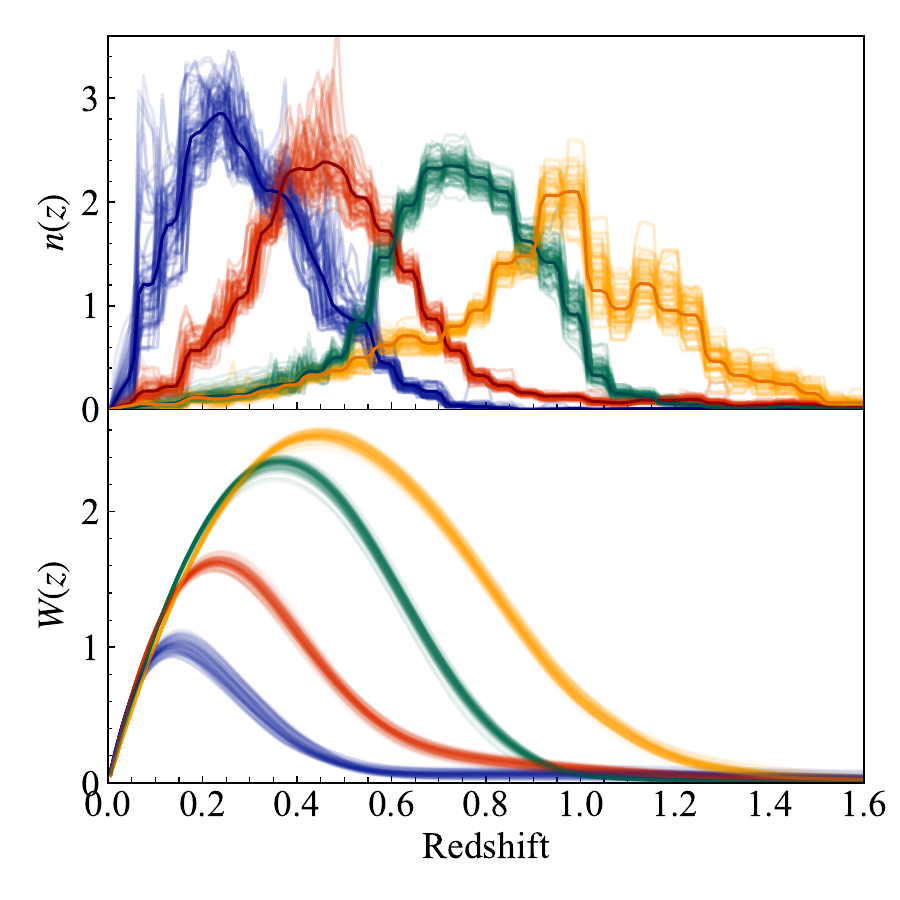}
\caption{Estimated redshift distributions for the weak lensing catalog, divided into four redshift bins (upper panel). Fainter lines indicate the ensemble of realisations, whose spread represents the uncertainty in their estimation while the darker, solid lines denote the mean of the ensemble. These are derived using the joint \textsc{SOMPZ} and \textsc{WZ} methodology, summarised in Section~\ref{sec:zmethod} and detailed in \citep{y3-sompz}. The stair-step appearance is an artifact of using a binned representation for the $n(z)$ and is immaterial to the cosmological results. In the lower panel, the lensing efficiency kernel (defined in equation~\ref{eqn:kernel}) for each of the source redshift bins, following the same color scheme, demonstrates that the DES Y3 sample is most sensitive between $z=0.1-0.7$. }
\label{fig:nzs}
\end{figure}

This section summarises the strategy and choices made for the calibration of redshift distributions, described fully in \citep{y3-sompz}. First, it summarises the primary method employed that relates redshift samples to sets of galaxies distinguishable by their wide-field photometry (Section~\ref{sec:sompz}), and how clustering cross-correlation measurements and small-scale shear ratios are folded in to validate and further constrain the proposed redshift distributions (Sections~\ref{sec:clusteringz} and \ref{sec:SR}). Next, the DES Deep Fields and redshift samples used are described in (Section~\ref{sec:zsamples}). The estimated mean redshift of the source sample is shown in the right panel of Figure~\ref{fig:map} as a function of position in the sky, showing the survey to be homogeneous  across the footprint. In Figure~\ref{fig:nzs}, the overall redshift distributions of the four bins are shown.

\subsection{Methodology}\label{sec:zmethod}

The DES Y3 weak lensing redshift methodology combines three likelihoods with complementary information \citep{y3-sompz}. The cornerstone to the method for estimating the photometric redshift distributions is a scheme based on two self-organizing maps (hereafter {\textsc{SOMPZ}}). The estimate of the redshift distributions is uncertain and this framework provides a means to generate samples of the $n(z)$ for each redshift bin that encompasses various sources of error. The clustering redshift method \citep[\textsc{WZ}]{y3-sourcewz} uses the angular cross-correlation of the weak lensing sources with galaxies of known redshift to generate realisations of $n(z)$ from the combined \textsc{SOMPZ+WZ} likelihoods. The shear ratio method, \textsc{SR}, \citep{y3-shearratio} incorporates the ratios of small-scale galaxy-galaxy lensing measurements between different source bins with the same lens bin to give an independent likelihood for each proposed redshift distribution. Ideally, weak lensing analyses approach redshift estimation with redundancy, comparing alternative methods as a means of validation. The DES Y3 cosmological analysis first assesses the consistency of these methods, demonstrated in \citep[][Figure 12]{y3-sompz} and then combines their information: we sample over an $n(z)$ generated by \textsc{SOMPZ+WZ}, with a joint likelihood from cosmic shear and \textsc{SR}.

\subsubsection{\textsc{SOMPZ}}
\label{sec:sompz}
The estimates of the redshift distribution of 100 million DES Y3 wide field galaxies rely on a \textit{redshift sample}, or external samples of 60 thousand galaxies that have spectroscopic or many-band photometric redshifts, as well as deep DECam and near-infrared photometry \citep{y3-deepfields}. To relate redshift information to the wide field galaxies, a self-organizing map formalism, \textsc{SOMPZ}, is used to compress the multi-dimensional information of both the deep and the wide field photometry \citep{Masters15,buchs19}. This is done as a two-step process that connects the three samples: the DES wide field source galaxies with $riz$ information, the DES Deep Fields that span a smaller area with $ugrizJHK_s$, and the \textit{redshift sample}. The DES Deep Fields are key to the redshift process as they contain deep eight-band photometry for 1.6 million galaxies. This information limits degeneracies of galaxy spectral energy distributions present in lower-dimensional color spaces when calibrating the weak-lensing galaxies.

A deep-field color defines a galaxy's phenotype, $c,$ and the wide field $riz$ color-magnitude is used to determine $\hat{c}$, both of which are cell identifiers for two separate self-organizing maps \citep{buchs19,y3-sompz}. Thus, the redshift distributions of wide field galaxies whose photometry best matches cell $\hat{c}$ can be written as
\begin{equation}
p(z|\hat{c}) = \sum_c p(z|c,\hat{c}) p(c|\hat{c}) \; .
\label{eqn:pzc}
\end{equation}
The choice of sample that informs the redshift of a given phenotype, $p(z|c)$, is discussed in Section~\ref{sec:zsamples}. To estimate the statistical connection between the wide-field and deep-field photometry $p(c|\hat{c})$,  the fitted light profile of each Deep Field galaxy is drawn into real DES Y3 science images multiple times in random positions with \textsc{Balrog} \citep{y3-balrog}. Processing these renderings with the DES photometry and shape measurement pipeline delivers the mapping of galaxies with noisy wide field $riz$ and successful shape measurement to the deep $ugrizJHK_s$ color space.

Four redshift bins, $\hat{b}_i$, are constructed as sets of wide field photometry SOM cells, $\hat{c}\in\hat{b}_i$, by defining redshift bin edges such that each contains an approximately equal number of source galaxies. 
Once the four redshift bins have been established, an estimate of the redshift distribution of each bin, $n_{i}(z)$, is given as 
\begin{equation}
    n_{i}(z) = p(z|\hat b_i)   \approx \sum_{\hat{c} \in \hat{b}_i} \sum_{c} \underbrace{p(z|c)}_\text{Redshift} \underbrace{p(c)}_\text{Deep} \underbrace{\frac{p(c,\hat{c})}{p(c)p(\hat{c})}}_\text{Balrog}  \underbrace{p(\hat{c})}_\text{Wide} \,.
\end{equation}
In each probability density term, contributing galaxies need to be weighted appropriately by their shear response and inverse variance weight. See \citep{y3-sompz} for more details of this procedure. 
 
An ensemble of $n(z)$ realisations is constructed to encompass four main sources of uncertainty. Specifically, as outlined in Section~\ref{sec:zsamples}, these include i) the sample variance in the color-redshift inference due to the limited Deep Field area, ii) the Deep Field zero-point error in the color information, particularly at redshifts where prominent features in the galaxy spectra transition between filters, iii) an uncertainty on the method due to the finite number of galaxies and \textsc{Balrog} realizations, and iv) the uncertainty due to limitations in the redshift sample, incorporated with an ensemble of realisations built from multiple redshift samples.

\subsubsection{Clustering redshifts}
\label{sec:clusteringz}

Clustering redshifts constrain the distances to source galaxies from their angular galaxy clustering with samples of reference galaxies within narrow redshift ranges \citep{Newman2008}. This method is based on the fact that the amplitude of this correlation function is proportional to the fraction of source galaxies in physical proximity to those reference galaxies. This information validate and refine the SOMPZ $n(z)$ with the added benefit of avoiding any reliance on the statistical colour-redshift relation and bypassing the completeness issues associated with spectroscopic survey coverage.

The tomographic SOMPZ $n(z)$ are further constrained by the measured small-scale angular cross-correlations, $w_{br}(z)$, between the source galaxies in bin $b$ and a reference population $r$ of galaxies that is known to be within a small range of redshift centered at $z$. The {\textsc{WZ}} method correlates the Y3 sources with both the redMaGiC galaxy sample (which has photo-$z$s determined with $\sigma_z\approx 0.02$) and eBOSS galaxies \citep{Ross20}, as detailed in \citep{y3-sourcewz}.  Using a linear bias model, one can predict the observed $w_{br}$ data, given assumed values for $n(z),$ the biases and lensing magnification coefficients of the source and reference populations, and several nuisance parameters, $N$, that describe potential errors in the linear-bias model. Comparison of this model to the observed $w_{br}$ measures allows the construction of a likelihood. The details of this analysis are described fully in \citet{y3-sourcewz}.

With the \textsc{SOMPZ} photometric and \textsc{WZ} clustering constraints on the redshift distribution in hand, a Hamiltonian Monte Carlo (HMC) method is used to draw samples of the 
$n(z)$ functions from the joint likelihood.
This yields \textsc{SOMPZ+WZ} samples of possible sets of $n(z)$ for the redshift bins. By marginalizing over these, one simultaneously accounts for uncertainty in the nuisance parameters.
The final set of redshift distributions are shown in  Figure~\ref{fig:nzs}. The mean, median and spread of the realisations for each source bin are quoted in Table~\ref{tab:datastats}.

\subsubsection{Shear ratios}
\label{sec:SR}
The ratios of small-scale galaxy-galaxy lensing measurements from two source bins and a shared lens sample provide geometric information \citep{JainTaylor}.  The primary dependence of shear ratios, \textsc{SR}, on distances has established them as a means of constraining and testing redshift distributions \citep{schneider2016, mandelbaum2005, hoekstra2005,Heymans2012,hildebrandt2017kids,hildebrandt20,giblin20}. When source and lens redshift bins overlap, the \textsc{SR} data also respond to intrinsic alignments, in an orthogonal direction to the cosmic-shear 2-point functions. In this analysis, \textsc{SR} are incorporated at the likelihood level to constrain the $n(z)$ and other nuisance parameters of our model, particularly those of intrinsic alignments \citep{y3-shearratio}. 

The \textsc{SR} data vector consists of nine scale-averaged lensing ratios, each constructed from the combination of tangential shear signals with a \textsc{MagLim} lens bin and two weak-lensing source bins. 
As validation, the \textsc{SR} using \textsc{RedMaGiC} are analysed and the consistency of the results are tested in Appendix~\ref{App:IAnz}. The lens data is limited to the three low-redshift \textsc{MagLim} bins, defined in \citep{y3-2x2maglimforecast}, to mitigate the impact of lens magnification. It also utilises only small angular scales, $\sim2-6$ Mpc/$h$, which correspond to a maximum angular scale of $\sim2.5-9$ arcmin depending on lens redshift, that are not used in, and therefore mostly independent of, the joint \textit{3$\times$2pt} analysis \citep[][Table 2]{y3-shearratio}. As a robustness test of the use of these scales and their impact on the analysis, the large-scale \textsc{SR} using \textsc{RedMaGiC} are incorporated in a variant analysis in Appendix~\ref{App:IAnz}.

In the inference, a Gaussian likelihood for the \textsc{SR} is assumed, using an analytical covariance matrix \citep{y3-shearratio}. The addition of the shear-ratio data to the analysis necessitates additional modeling and lens observational parameters, summarised in Table~\ref{tab:priors}. These describe the uncertainties in the redshift distribution of lens galaxies, as well as the relation between galaxies and dark matter, parametrized using a per-bin linear galaxy bias. The impact of the \textsc{SR} information, as well as the \textsc{WZ} method, on cosmological constraints is tested in Section~\ref{sec:robustredshift}.

\subsection{Deep Fields and redshift samples}
\label{sec:zsamples}

The subset of these DES Deep Field galaxies with \textsc{Balrog} \citep{y3-balrog} wide-field realizations that pass the weak-lensing selection and have external high-quality redshift information forms the \textit{redshift sample}. It is constructed from both spectroscopic and multi-band photometric redshifts as detailed in \citep[][Section 3.3]{y3-sompz}. The former consists of spectra from the following surveys: \textsc{zCOSMOS}-bright and deep \citep{lilly09}, \textsc{C3R2} \citep{Masters15,Masters17,Masters19}, \textsc{VVDS} \citep{lefevre13}, and \textsc{VIPERS} \citep{guzzo14,garilli14,scodeggio18}. The two many-band photo-$z$ catalogs used are those based on the \textsc{COSMOS} 30-band \citep{laigle16} and the \textsc{PAUS+COSMOS} 66-band catalog \citep{Alarcon20}, which adds narrow band filters from the \textsc{PAUS} to the \textsc{COSMOS} photometry. These are the underlying data for three equally-weighted redshift samples that are used in conjunction to span the uncertainty of the redshift sample selection. These prioritize either spectroscopic or high quality photometric information, where available, and are designed to be complete by using \textsc{COSMOS} 30-band photometric redshifts elsewhere.

 Spectroscopic calibration samples on their own can suffer from selection effects \citep{Newman2015,Bonnett16, gruen17, hoyle,wright20, hildebrandt20, joudaki2020, hartley2020}: at a given color or color-magnitude, the subset of galaxies with successful spectroscopic redshift measurements may not have redshifts that are representative of the full sample. This motivates our choice to never discard those galaxies in the \textsc{COSMOS} field from our redshift sample that do not have spectroscopic information, but rather to use their photometrically estimated redshifts where necessary. Conversely, even high-quality photometric catalogs suffer from biases \citep{laigle16,hildebrandt20, joudaki2020}, or missing templates or photometric outliers in photo-$z$ surveys. The maximal impact of redshift sample uncertainty on cosmic shear cosmological parameters, and therefore the robustness of this analysis to these effects, is tested by analysing DES cosmic shear with `pure' redshift samples in Section~\ref{sec:robustredshift}:
\begin{itemize}
\item \textsc{C}:  This sample includes only information from the \textsc{COSMOS} catalog and would therefore suffer from the systematic calibration biases, claimed by \citep{hildebrandt20, joudaki2020}.
\item \textsc{MB}: Even the redshift sample that is least reliant on \textsc{COSMOS} has ten percent of the redshift information derived from that sample at the faintest magnitudes. In order to test the impact of any residual calibration biases due to this subset, this `maximally-biased' (MB) sample is complete, by design, and artificially constructed to realistically alter the \textsc{COSMOS} galaxies that are not matched to spectroscopic information, by altering those redshifts with a magnitude-dependent prescription \citep{y3-sompz}. 
\end{itemize}

This approach differs to that used in the cosmic shear analysis for both DES Y1 \cite{troxel2018dark}, which focuses on \textsc{COSMOS} as the redshift sample, adjusted with clustering redshift information for the former and HSC \citep{hikage19,Ham20}, which uses the same redshift sample with a reweighting method. It is also unlike the KiDS-1000 approach \citep{asgari20}, which uses only spectroscopic information, but mitigates selection biases with 9-band information over the footprint and by removing the subset of their data that is not well-represented by the redshift sample.

\section{Shear Calibration}
\label{sec:imsims}

Accurate galaxy shape estimates are essential for cosmic shear studies. They are hindered by the fact that the majority of galaxies used in any weak lensing measurement are faint, noisy, pixellated, barely resolved due to convolution with an anisotropic PSF, or blended with neighboring galaxies. Their detection, deblending, and inclusion in the source sample is subject to biases. The \mcal\ algorithm corrects several, but not all these sources of bias, particularly not ones related to detection and (de-)blending. Our ability to calibrate biases in the shear estimation therefore relies on producing sufficiently realistic image simulations to calibrate the shear estimates with. 

The  suite of image simulations used for DES Y3 shear calibration are presented in \citep{y3-imagesims}. They are based on drawing model fits with morphological information of galaxies from HST imaging \citep{leauthaudHST} and DES Deep Field observations \citep{y3-deepfields} in the \textsc{COSMOS} field to generate mock DES Y3 observations in $riz$ bands. Objects are simulated with realistic DES observing conditions, that is, convolved with the DES PSF models of sets of $riz$ exposures overlapping a tile of the sky, with noise and masking matching the data. Image coaddition, object detection, shape measurement, assignment to redshift bins and redshift estimation are performed as in the fiducial DES pipelines. As such, the multi-band suite affords a testing ground that is well-matched to the DES observations in each redshift bin.

Shape measurement biases are taken into account approximately, as in previous work, through a multiplicative bias correction, as defined in equation~\ref{eqn:shearbias}. 
The corresponding correction is applied as an average $m$ over all the galaxies in a redshift bin. From a second rendering of simulation tiles with a shear $\Delta\gamma(z)$ applied to all galaxy images, we can estimate biases in our shear measurement. Multiplicative bias can be accounted for by replacing, in equation \ref{eqn:nz}, $n(z)$ by $n(z) \times (1+m)$. In this way, we account not only for the fraction of galaxies in an ensemble at a given redshift, but also for the sensitivity of that subset of the galaxies to the applied shear.
Thus substituting, equation~\ref{eqn:nz} reads
\begin{equation}
\Delta\gamma^{\rm obs} = \int \mathrm{d}z \, (1+m) n(z) \, \Delta\gamma(z) \, . 
\label{eqn:mnz}
\end{equation}  

Blending is, however, expected to introduce a distortion of the mean and shape of $n(z)$, in addition to its normalization. For one thing, the $m$ in equation~\ref{eqn:mnz} could well be redshift dependent, due to the changing morphology of galaxies. But a distortion of the $n(z)$ is also caused by the response of the measured shape of one galaxy to the shearing of light at a blended galaxy's different redshift, which leads to a perturbation on the $n(z)$ that should be used for lensing analyses.  Constraints on this effect are possible by using multiple different redshift-dependent $\Delta\gamma(z)$ in an image simulation. We find it to be significantly non-zero for the DES Y3 analysis choices and level of precision \citep{y3-imagesims}. A joint understanding of multiplicative shear bias and blending-related redshift calibration errors is a necessary development for future weak lensing analyses. This is especially important for deeper imaging, where blending becomes even more ubiquitous than the case of DES Y3. A comparison to simpler shear calibration simulations, in which galaxy images are sheared uniformly, or constrained to be equally separated over a 2D grid, determines that the dominant contributor to shear calibration in DES Y3 is indeed this redshift-dependent blending effect, in some regimes by factors of several. 

These effects are disentangled in Ref.~\citep{y3-imagesims}: first, uncertainties on $m$ are derived for each redshift bin using a redshift independent shear simulation $m_{\rm noz-blend}$, noted in Table~\ref{tab:datastats}. In addition, image simulations with an applied redshift-dependent shear allow for constraints on the distortion of the mean and shape of the $n(z)$.
In Section~\ref{sec:modelm}, details of how these are used to model the  calibration bias and uncertainty due to the impact of blending are given, and Section~\ref{sec:robustshear} checks the impact of redshift-mixing due to blending on our analysis, the first cosmic shear study to explicitly do so.

\section{Cosmological simulations}\label{sec:sims}
Aspects of this analysis have been validated using the \buzzard\ suite of 18 cosmological simulations, the process of which we describe in brief here. We refer the reader to the comprehensive discussion in \citep{y3-simvalidation}. 

The \buzzard\ simulations are $N$-body lightcone simulations that have been populated with galaxies using the \textsc{Addgals} algorithm \citep{addgals,DeRose2021}, endowing each galaxy with a position, velocity, spectral energy distribution, broad band photometry, half-light radius and ellipticity. Each pair of Y3 simulations is produced from a set of 3 independent $N$-body lightcones with box sizes of $[1.05,\, 2.6,\, 4.0]\, (h^{-3}\, \rm Gpc^3)$, mass resolutions of $[0.33,\, 1.6, \, 5.9] \, \times10^{11}\, h^{-1}M_{\odot}$, spanning redshift ranges in the intervals $[0.0,\, 0.32,\, 0.84, \,  2.35]$ respectively. The lightcones are run with the \textsc{L-Gadget2} $N$-body code, a memory optimized version of \textsc{Gadget2} \citep{Springel2005}, with initial conditions generated using \textsc{2LPTIC} at $z=50$. Together these produce $10,313$ deg$^2$ of unique lightcone area \citep{2LPTIC}.

The simulations are ray-traced using the spherical-harmonic transform configuration of \textsc{Calclens}, performed on an $N_{\rm side}=8192$ \textsc{HEALPix} grid \citep{Becker2013, healpix}. The lensing distortion tensor is computed at each galaxy position and is used to deflect the galaxy angular positions, shear galaxy intrinsic ellipticities (including effects of reduced shear), and magnify galaxy shapes and photometry. Convergence tests conducted on this algorithm find that resolution effects on $\xi_{\pm}$ are negligible on the scales used for this analysis \cite{DeRose2019}. 

The DES Y3 footprint mask was applied to the ray-traced simulations and each set of three $N$-body simulations yields two Y3 footprints, with 520 deg$^2$ of overlap. We apply a photometric error model to the mock wide field photometry in our simulations based on a relation
measured from \textsc{Balrog} \citep{y3-balrog} \citet[see, for more details][]{y3-simvalidation}. The DES Y3 weak lensing source selection is applied to the simulations using the PSF-convolved sizes and $i$-band \snr in order to match the non-tomographic source number density, $n_{\rm eff}=5.9$ arcmin$^{-2}$, from an earlier iteration of the \mcal source catalog. The \sompz framework is applied to divide source galaxies into redshift bins, each with a number density of $n_{\rm eff}=1.48$ arcmin$^{-2}$, and to obtain estimates of the redshift distribution of galaxies which resemble those of the data well \citep{y3-simvalidation,y3-sompz}. The shape noise of the simulations is then matched to that measured in the \mcal catalog per bin. Following the methodology in \citep{y3-sompz}, mock $n(z)$ realisations are produced, limited to uncertainties from sample variance only. These samples are found to be consistent with the true redshift distributions, and the inclusion of \textsc{WZ} and \textsc{SR} likelihoods finds consistency and reduces their variance. 

Cosmic shear two-point functions are measured in the \buzzard\ simulations without shape noise using the same pipeline as that used for the data, with \mcal responses and inverse variance weights set to 1 for all galaxies. Using the DES analysis pipeline, the best-fit model for these measurements at the true \buzzard\ cosmology, assuming the true source redshift distributions and no intrinsic alignments, reproduces the mean of the measurements made using 18 realisations of the simulation with a chi-squared value of $1.37$ per degree of freedom, for $207$ data points using the Fiducial scale selection (see Section~\ref{sec:scalecuts}) and assuming the covariance of a single simulation. Simulated likelihood analyses assuming either the true simulated source redshift distributions or calibrated photometric redshift distributions results in cosmological constraints that are unbiased. 

\section{Cosmic shear measurement}\label{sec:measurement}
 
This section presents the real-space two-point shear correlation function measurements, $\xi_{\pm}$, derived from the weak-lensing science catalog, which spans 4143 deg$^2$ sky footprint and over 100 million galaxies, divided into redshift bins according to Table~\ref{tab:datastats}. These have a signal-to-noise $S/N = 40$, computed as 
  \begin{equation}
    S/N=\frac{\xi_{\pm}^{\rm }\mathbf{C}^{-1}\xi_{\pm}^{\rm model}}{\sqrt{\xi_{\pm}^{\rm model}\mathbf{C}^{-1}\xi_{\pm}^{\rm model}}} \, ,
\end{equation}
where the covariance, $\mathbf{C}$ is defined in Section~\ref{sec:covariance} and the best-fit model to the data, $\xi_{\pm}^{\rm model}$ is presented in Section~\ref{sec:results}. After eliminating the small scales, in order to mitigate model uncertainties in the cosmology analysis, as motivated in Section~\ref{sec:scalecuts} \citep{y3-generalmethods,y3-cosmicshear2}, the measurement is found to have a $S/N = 27$. Using the \lcdm-Optimized scale selection, described in Section~\ref{sec:agg},  a $S/N = 31$ is obtained. 
It is notable that the S/N for the Y1 measurement without scale cuts (26.8) \citep{Troxel2018} is matched by the Y3 \lcdm-Optimized measurement after limiting the scales.

\subsection{Correlation function measurements}\label{sec:2ptfns}

\begin{figure*}
    \centering
    \includegraphics[width=\textwidth]{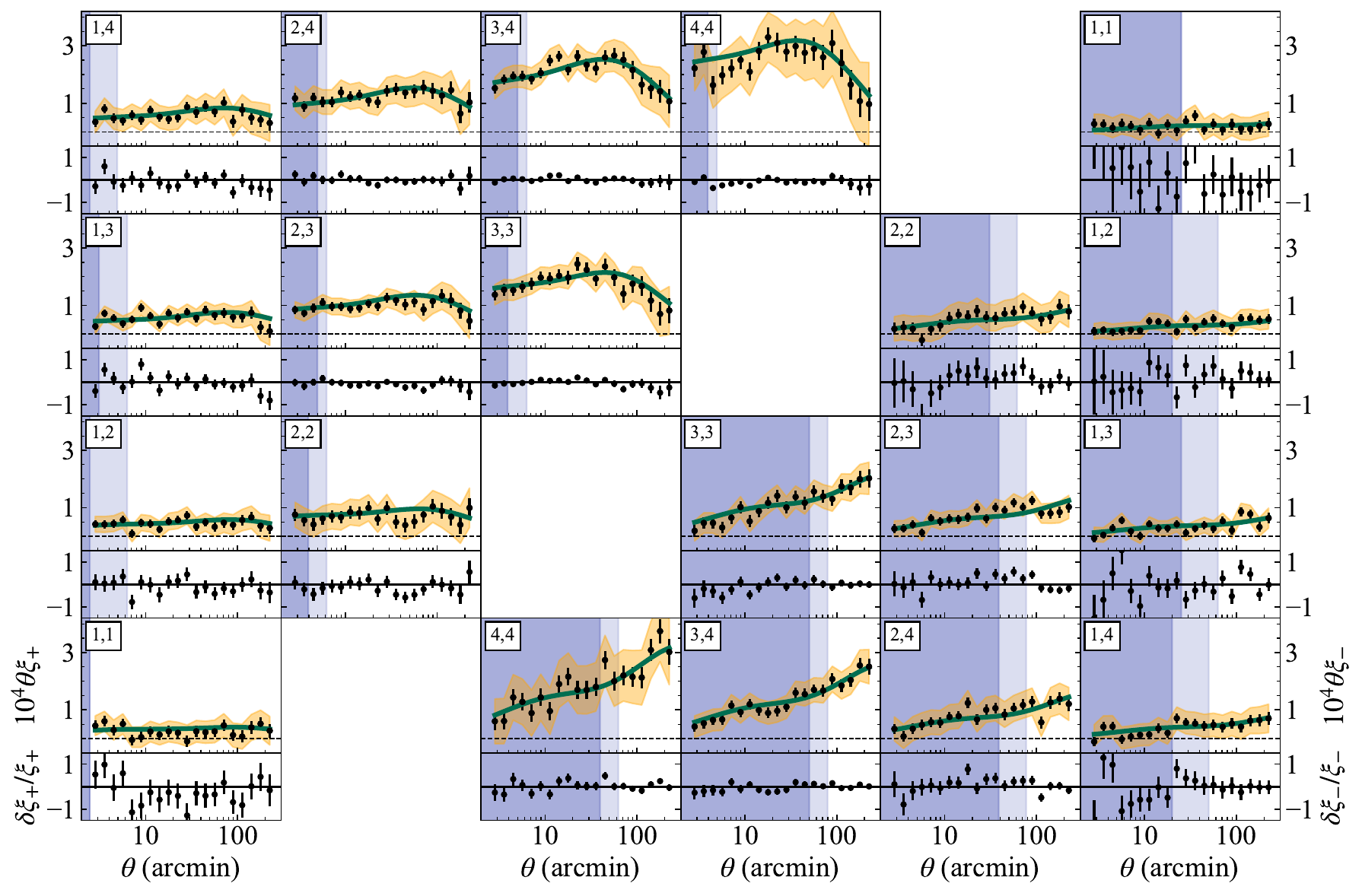}
    \caption{\label{fig:xitomo}
    Measured tomographic DES Y3 cosmic shear two-point correlation functions: $\xi_{+}(\theta)$ (left) and $\xi_{-}(\theta)$ (right), scaled by the angular separation, $\theta$, to emphasize differences relative to the best-fit model (upper panels).  The correlation functions are measured for each redshift bin pair, indicated by the label and the error bar represents the square root of the diagonal of the analytic covariance matrix. The best-fit \lcdm theoretical prediction from the cosmic shear-only tomographic analysis is denoted by a green line. Scales excluded from the analysis, due to their sensitivity to small-scale systematics, are shaded in light blue for the \textit{Fiducial} analysis and darker blue for the \textit{\lcdm-Optimized} analysis. The signal-to-noise of the  measurement is 40 using all angular scales and 27 (31) using the Fiducial (\lcdm-Optimized) scale-selection. For comparison, the yellow shaded region shows the Y1 uncertainty, with a factor of $\sim\sqrt{2}$ lower signal-to-noise.  The lower panels plot the fractional difference between the measurements and best-fit, $\delta\xi_{\pm}/\xi_{\pm}=\big(\xi_{\pm}-\xi_{\pm}^{\rm theory}\big)/\xi_{\pm}^{\rm theory}$. We find that the $\chi^2$ per effective d.o.f of the \lcdm model is $237.7/222.2 = 1.07$, and the $p$-value is 0.223. }
\end{figure*}

The estimator of the two-point correlation function or shear-shear correlation function can be written in terms of the measured radial, $\epsilon_{\rm x}$, and tangential ellipticities, $\epsilon_{\rm t}$, of a galaxy as, 
\begin{equation}
\xi^{ij}_{\pm}(\theta)= \langle\epsilon_{\rm t}\epsilon_{\rm t} \pm \epsilon_{\times}\epsilon_{\times} \rangle (\theta) \,.
\end{equation}
It is determined by averaging over all galaxy pairs $(a, b)$ separated by an angle $\theta$, for two redshift bins, $i$, and $j$, as,  
\begin{equation}
\label{eqn:2ptcorrfnestimator}
\centering
\hat{\xi}^{ij}_{\pm}(\theta) = \frac{\sum_{ab} w_a w_b[ \epsilon_{\rm t}^i \epsilon_{\rm t}^j \pm \epsilon^i_{\times}\epsilon^j_{\times}]}{\sum_{ab} w_a w_b R_a R_b} \, , 
\end{equation}
where $R$ represents a response correction from \mcal, in the for our case highly accurate scalar approximation (see Section~\ref{sec:shape}). The weighted sum utilises the per-galaxy inverse-variance weights, $w$, introduced in Section~\ref{sec:shape}, and is taken over galaxy pairs whose angular separation is within an interval $\Delta\theta$ around $\theta$. 

The tomographic DES Y3 cosmic shear data vector is shown in Figure~\ref{fig:xitomo}. It is computed via equation~\ref{eqn:2ptcorrfnestimator} using the public code TreeCorr\footnote{https://rmjarvis.github.io/TreeCorr} \citep{TreeCorr}. Twenty angular logarithmic bins are chosen to span 2.5 to 250.0 arcmin. Of these, small-scale measurements are discarded from the cosmological inference, represented as a light blue shaded region for the \textit{`Fiducial'} analysis, defined in Section~\ref{sec:scalecuts}, and a darker blue for a \textit{`\lcdm-Optimized'} analysis (see Section~\ref{sec:agg}). In the upper panel, the green line denotes the best-fit cosmological and astrophysical parameters from the tomographic analysis. The error bound indicates the square root of the diagonal of the analytic covariance matrix. For comparison, the level of uncertainty from the DES Y1 analysis is shown as the yellow shaded bands. In the lower panels, the fractional residuals, $(\xi_{\pm}-\xi_{\pm}^{\rm theory})/\xi^{\rm theory}$ demonstrate the fit of the model to the measurements. 

The data vector, $D$, comprises four auto-correlations, and six unique cross-correlations between redshift bins for each $\xi_+$ and $\xi_-$. The small angular scales of the measurements are eliminated from the analysis primarily to mitigate the impact of baryonic effects, indicated by the shaded region,  leaving 167 (60) data points for $\xi_+$ ($\xi_-$).

We have verified that an independent pipeline produces the same $\xi_{\pm}$ measurements to numerical precision. The data points shown represent the weighted mean of pair separation, but the theoretical prediction is averaged over the bin using the geometric approximation, following equation 10 in Ref.~\citep{y3-covariances}.

\subsection{Covariance Matrix}\label{sec:covariance}

To model the statistical uncertainties of our measurements of $\xi_\pm$ we assume a multi-variate Gaussian distribution for our combined data vector. The modeling of the disconnected four-point function part of the covariance matrix of that data vector (also known as the Gaussian part of the covariance) is described in \citep{y3-covariances} and includes analytic treatment of bin averaging and sky curvature. The connected four-point function part and the contribution from super-sample covariance use the public code CosmoCov\footnote{https://github.com/CosmoLike/CosmoCov} \citep{fang20b} based on the CosmoLike framework \citep{Krause17}. 

The covariance matrix, $\mathbf{C}$, is a function of both the cosmological and nuisance parameters that are required to describe the data vector, as well as the redshift distributions. In this work, following previous cosmic shear analyses, an iteratively fixed covariance matrix is used. That is, we assume a fiducial set of
input parameters for the computation of the covariance matrix used in the initial unblind analysis. Then, the covariance is recomputed
at the best-fit from the first iteration, and the final analysis is performed. This step incurs a change in cosmological parameters of much less then 1$\sigma$ \citep{y3-3x2ptkp}.

The robustness of our analysis with respect to the details of our covariance modeling is demonstrated in \citep{y3-covariances}. FLASK simulations are used to test for the impact of effective number density and shape-noise dispersions in the presence of complex survey footprints \citep{troxel2018dark}. There, it is also demonstrated that deviations from the Gaussian likelihood assumption are negligible for this analysis. If the covariance model was perfect and there were no tight priors on any of the parameters of the model for cosmic shear measurements, then the $\chi^2$ between any measurement of our combined data vector of cosmic shear correlation functions and our best-fit model to that measurement should on average be $\approx N_{\mathrm{data}} - N_{\mathrm{param}}$ = 227-28 = 199. Furthermore, in that situation the typical scatter of $\chi^2$ ($1\sigma$) is expected to be $\approx \sqrt{2(N_{\mathrm{data}} - N_{\mathrm{param}})}$. Any errors in our covariance model may cause deviations from that behaviour. We estimate the impact of numerous potential covariance errors in \citep{y3-covariances} and find that none of them significantly impact the cosmic shear part of the DES Y3 2-point function analysis --- neither with respect to $\chi^2$ nor with respect to parameter constraints.
The aforementioned behaviour of $\chi^2$ will also be altered in the presence of priors on any of the model parameters. Taking into account our Gaussian priors on multiplicative shear bias and on shifts of the redshift distributions we find in \citep{y3-covariances} that within a 68\% confidence interval our best-fit $\chi^2$ should be $\approx 220.2 \pm 20.7$.

\subsection{Mitigating observer bias}\label{sec:blind}

In order to ensure that analysis decisions are not influenced, even unconsciously, by the comparison of results to experimenters' expectations, we apply transformations to the data that are designed to obscure the cosmological results. Although the approaches differ,
a philosophy of blind analyses is necessary and accepted for all recent cosmological weak lensing analyses \citep{rcslens,svcosmicshear,hildebrandt2017kids, Troxel2018, y1keypaper, hikage19, hildebrandt20, Ham20, asgari20, heymans20}. 
For this DES Y3 analysis, these transformations, and the procedures for deciding when to remove them, make up a three-stage blinding strategy, at the levels of the shape catalog, data vector, and parameter inference. Stages, described below, are removed sequentially as a pre-determined set of criteria is fulfilled.

During stage one, galaxy shapes are transformed by altering their ellipticities, \textbf{e}, via |{\mbox{\boldmath $\eta$}}| $\equiv 2 {\rm arctanh} |\textbf{e}| \rightarrow f$|{\mbox{\boldmath $\eta$}}$ |$, with an unknown and random value 0.9 < f < 1.1. This transformation is sufficient to change the cosmology results for shear-only analyses and is similar to previous work \citep{y1shearcat, hikage19, hildebrandt20}. The transformed shear catalog is used until the non-tomographic shear validation tests \citep{y3-shapecatalog}, and the tomographic versions (see Appendix~\ref{App:additivecorr}) are passed, the three redshift methods were proved to be consistent in the $\Delta z$ parameter space \citep{y3-sompz} and similarly, Section~\ref{sec:zmethod}, the analysis choices and model pipeline fixed \citep{y3-generalmethods} and all relevant papers had completed a first round of the DES internal review process. 

As the shear catalog transformation breaks the internal consistency between shear and galaxy clustering observables, it cannot be used for a blinded combined \textit{3$\times$2pt} analysis. The second stage of blinding is implemented at the data vector  level using the method detailed in Ref.~\citep{y3-blinding}. Here, a transformation adds to each two-point shear correlation function data point a quantity equal to the difference between model predictions for that observable computed at two sets of cosmological parameter values, $\Theta_{\rm ref}$ and $\Theta_{\rm ref}+\Delta\Theta$. This has been shown to change the best-fit cosmology associated with the data vector by approximately $\Delta\Theta$ while preserving the internal consistency between components of the data vector. This blinding step is performed automatically in the measurement pipeline. $\Theta_{\rm ref}$ is chosen to be equal to the same fiducial cosmology used for our modeling tests, and the parameter shift $\Delta\Theta$ is drawn pseudo-randomly in $w$CDM parameter space, using a fixed seed to ensure that different measurements are transformed consistently.  The final stage of our blinding procedure is a simple obscuring of axes of the cosmological parameters in 1 or 2D contours, implemented as a final safeguard. 

The measurement and analysis pipeline was tested on the mock catalogues described in Section~\ref{sec:sims}, and in more detail in \citet{y3-simvalidation} and shown to recover unbiased parameters. When the analysis framework, scale selection and systematic priors were fixed, and the correlation functions measured consistently with two pipelines (see Section~\ref{sec:measurement}), the criteria for removing the third stage of blinding is met. At this point, internal consistency is assessed in terms of goodness of fit and consistency between the measurements from cosmic shear, galaxy-galaxy lensing \citep{y3-gglensing} and galaxy clustering \citep{y3-galaxyclustering}, required to meet an arbitrary $\chi^2$ criterion of $p$-value $>0.01$ with the posterior predictive distribution process (see Section~\ref{sec:IC}). Finally, parameter constraints could be revealed. Any changes to the analysis after this point are documented in Appendix~\ref{app:unblinding}. 

\section{Modeling and Analysis Choices}\label{sec:model}
This section outlines the baseline theoretical model for the cosmic shear correlation functions used in this analysis and discusses the evidence for its robustness. Systematic errors in the model are demonstrated to be subdominant to the precision of the data, in order to derive unbiased cosmological parameters \citep{y3-generalmethods}, as validated specifically for cosmic shear in Ref.~\cite{y3-cosmicshear2}. Cosmological, astrophysical, and systematic parameters are constrained for the $\Lambda$CDM model. For the case of cosmological parameters that are not well-constrained by cosmic shear, informative \textit{priors} with boundaries that widely encompass allowed values from external experiments are used, reported in Table~\ref{tab:priors}. For the massive neutrino density parameter, $\Omega_{\nu}$, we vary $\Omega_{\nu}h^2$, where $h$ is the Hubble parameter. We note that this is often fixed in other analyses \citep{Planck2018} at zero, or to the minimum mass allowed by oscillation experiments, $m_{\nu}=0.06$eV \citep{Patrignani}.

\subsection{Cosmic shear signal}
The observed angular two-point correlation for two redshift bins $i$ and $j$ is expressed in terms of the convergence power spectrum $C_{\kappa}(\ell)$ at an angular wavenumber $\ell$ as
\begin{align}
\centering
\label{eqn:2ptP}
{}& \xi_+(\theta) = \sum_{\ell} \frac{2\ell+1}{4\pi} G^\pm_\ell(\cos \theta) \qty[C_{\kappa,\textrm{EE}}^{ij}(\ell) + C_{\kappa, \textrm{BB}}^{ij}(\ell)] \, , \\
& \xi_-(\theta) = \sum_{\ell} \frac{2\ell+1}{4\pi} G^\pm_\ell(\cos \theta) \qty[C_{\kappa,\textrm{EE}}^{ij}(\ell) - C_{\kappa, \textrm{BB}}^{ij}(\ell)],
\end{align}
\noindent
where the functions $G^\pm_\ell(x)$ are computed from Legendre polynomials $P_\ell(x)$  (see e.g.  \citep{Stebbins}). Although cosmological lensing does not produce B-modes except due to  multiple-deflection effects, our baseline model does allow for a B-mode contribution from intrinsic alignments, and so we show the more general expression here. 

The 2D convergence power spectrum $C_{\kappa}$ can be related to the non-linear 3D matter power spectrum $P_{\delta}$ via the flat-sky and Limber approximations \citep{Limber53,Limber_LoVerde2008}
as, 
\begin{equation}
\centering
C_{\kappa}^{ij}(\ell) = \int_0^{\chi_{\rm H}} d\chi \,  \frac{W_i(\chi)W_j(\chi)}{\chi^2}  P_{\delta}\left(\chi,k=\frac{\ell+0.5}{z(\chi)}\right) \, ,
\label{eq:limber_gg}
\end{equation}
\noindent
where $\chi_{\rm H}$ is the horizon distance and a flat spatial geometry is assumed, so that $f_K(\chi)$, the comoving angular diameter distance, is simplified to $\chi$. The lensing efficiency kernel $W_i(\chi)$ for the redshift bin $i$ is defined as
\begin{equation}
\label{eqn:kernel}
\centering
W_i(\chi)= \frac{3H_0^2\Omega_{\rm m}}{2c^2}\frac{\chi}{a(\chi)} \int_{\chi}^{\chi_{\rm H}} d\chi' \, n_i(\chi') \frac{\chi'-\chi}{\chi'}  \, ,
\end{equation}
\noindent
where $n_i(\chi) d\chi$ is the effective number of galaxies in $d\chi$ in the $i$th redshift bin, normalised such that $\int_0^{\chi_{\rm H}} n_i(\chi) d\chi =1$.
The convergence power spectrum can be described by the amplitude of matter density fluctuations on an 8$h^{-1}$Mpc scale in linear theory, $\sigma_8$, which is related to the amplitude of the primordial scalar density perturbations, $A_{\rm s}$, and is degenerate with the matter density parameter $\Omega_{\rm m}$. That is, the power spectrum at small $k$ increases with an increase in either $\Omega_{\rm m}$ or the $\sigma_8$ normalisation. 

Higher-order contributions to the observed two-point statistics are caused by reduced shear \citep{dodelson2006,shapiro2009}, source clustering  and  magnification  \citep{Schneider2002, Schmidt2009} and the deflection of source positions \citep{seljak96, dodelson08}.
The impact of these higher order effects is verified to be negligible at the precision of this analysis, with the derivation and computation of the contributions accounted in Ref.~\citep{y3-generalmethods}. The predicted impact on cosmic shear measurements is illustrated in Figure 5 of Ref.~\citep{y3-cosmicshear2}.

\begin{table}
    \caption{Summary of cosmological, observational and astrophysical parameters and priors used in the analysis. In the case of flat priors, the prior bound to the range indicated in the `value' column while Gaussian priors are described by their mean and 1$\sigma$ width.}   
    \label{tab:priors}
\begin{center}
\begin{ruledtabular}
\begin{tabular}{ccc}
Parameter & Type & Value \tabularnewline
\hline 
\bf{Cosmological} \tabularnewline
$\Omega_{\rm m}$, Total matter density & Flat & [0.1, 0.9] \tabularnewline
 $\Omega_{\rm b}$, Baryon density & Flat & [0.03, 0.07] \tabularnewline
$10^{-9}A_{\rm s}$, Scalar spectrum amplitude  & Flat & [0.5, 5.0]\tabularnewline
$h$, Hubble parameter  & Flat & [0.55, 0.91]  \tabularnewline
$n_{\rm s}$, Spectral index  & Flat & [0.87,1.07] \tabularnewline
$\Omega_{\nu}h^2$, Neutrino mass density &  Flat & [0.00060,0.00644] \tabularnewline
*$w$, Dark energy parameter & Fixed & [-2,-1/3] \tabularnewline
\hline 
\bf{Observational} \tabularnewline
$\Delta z^1$, Source redshift 1 & Gaussian & ( 0.0, 0.018 ) \tabularnewline
$\Delta z^2$, Source redshift 2 & Gaussian  & ( 0.0, 0.015 ) \tabularnewline
$\Delta z^3$, Source redshift 3 & Gaussian  & ( 0.0, 0.011 ) \tabularnewline
$\Delta z^4$, Source redshift 4  & Gaussian  & ( 0.0, 0.017 ) \tabularnewline
$m^1$, Shear calibration 1 & Gaussian & ( -0.006, 0.009 )\tabularnewline
$m^2$, Shear calibration 2 & Gaussian  & ( -0.020, 0.008 )\tabularnewline
 $m^3$, Shear calibration 3 & Gaussian  & ( -0.024, 0.008 )\tabularnewline
$m^4$, Shear calibration 4 & Gaussian  & ( -0.037, 0.008 )\tabularnewline
\hline 
\bf{Intrinsic alignment} \tabularnewline
$a_1$, Tidal alignment amplitude        & Flat & $[-5,5]$ \tabularnewline
$a_2$, Tidal torque amplitude        & Flat & $[-5,5]$ \tabularnewline
$\eta_1$, Tidal alignment redshift index     & Flat & $[-5,5]$ \tabularnewline
$\eta_2$, Tidal torque redshift index      & Flat & $[-5,5]$ \tabularnewline
$b_{\rm ta}$, Tidal alignment bias  & Flat & $[0,2]$ \tabularnewline
\hline 
\bf{Shear-ratio} \tabularnewline
Galaxy bias, $b^{1-3}_{\rm g}$  & Flat & $[0.8,3]$ \tabularnewline
$\Delta z_l^{1}$, Lens redshift 1     & Gaussian & ( -0.009, 0.007  ) \tabularnewline
$\Delta z_l^{2}$, Lens redshift 2     & Gaussian & ( -0.035, 0.011 )  \tabularnewline
$\Delta z_l^{3}$, Lens redshift 3      & Gaussian & ( -0.005, 0.006 )  \tabularnewline
 $\sigma z_{\rm l}^{1}$, Lens redshift width 1     & Gaussian & ( 0.98,0.06 ) \tabularnewline
$\sigma z_{\rm l}^{2}$, Lens redshift width 2       & Gaussian & (1.31,0.09 )  \tabularnewline
$\sigma z_{\rm l}^{3}$, Lens redshift width 3       & Gaussian & ( 0.87,0.05 )  \tabularnewline
\end{tabular}
\end{ruledtabular}
\end{center}
\end{table}

\subsection{Intrinsic Alignments}

If galaxy orientation was truly random, any measured $\xi_\pm$ signal would be attributed to gravitational lensing. In reality, we expect \textit{intrinsic alignment}, whereby galaxy shapes are correlated with the local environments in which they formed and evolved \citep{hirata04, troxel15, kiessling15}.
One can approximate the apparent shape of a galaxy as the superposition of an intrinsic shape alignment, $\gamma_{\rm I}$, and a true shear due to lensing, $\gamma_{\rm G}$, such that the observed shape is given by $\epsilon^{\mathrm{obs}} = \epsilon + \gamma_{\rm G} + \gamma_{\rm I}$, where the galaxy shape in the absence of lensing and intrinsic alignments, $\epsilon$, is uncorrelated across the sky. 
The total harmonic-space power spectrum can be written as a sum over corresponding terms,
\begin{equation}
C_l^{\rm total} = C^{ij}_{\rm GG}(\ell) + C^{ij}_{\rm GI}(\ell) + C^{ji}_{\rm IG}(\ell) + C^{ij}_{\rm II}(\ell).
\end{equation}
The `gravitational-gravitational' term, $C^{ij}_{\rm GG}(\ell)$, corresponds to the convergence power spectrum given by equation~\ref{eq:limber_gg}. The `intrinsic-intrinsic' contribution, $C^{ij}_{\rm II}(\ell)$, arises due to correlations between the intrinsic shapes of two physically nearby galaxies, while the `gravitational-intrinsic' terms, $C^{ij}_{\rm GI/IG}(\ell)$, arise in pairs of galaxies for which some common structure affects the intrinsic shape of one of the galaxies, and the shear on the other.

The baseline intrinsic alignment model used in this work, a nonlinear perturbative prescription, the Tidal Alignment and Tidal Torquing model (TATT; \cite{blazek19}), is motivated by tests on simulated data \citep{y3-generalmethods,y3-cosmicshear2}. This model choice represents a departure from previous weak lensing analyses where the more simple nonlinear alignment (NLA) model was opted for \citep[e.g.][]{Troxel2018, hildebrandt20, Ham20, asgari20}. In brief, the TATT model allows three contributions in the gravitational tidal field which capture the `tidal alignment', linear in the tidal field, the `tidal torquing' \citep{mackey02,codis15}, quadratic in the tidal field, and the impact of source density weighting \citep{blazek15}. At fixed redshift, the TATT power spectra depends on a tidal alignment amplitude, $A_1$, a tidal  torquing  amplitude, $A_2$, and an effective source linear bias of the galaxies, $b_{\rm ta}$. The redshift evolution of $A_1$ and $A_2$ is parametrized as a power law, governed by $\eta_1$ and $\eta_2$ given by,
\begin{align}
  {}&  A_1(z) = -a_1 \bar{C}_1 \frac{\rho_{\rm crit} \Omega_{\rm m}}{D(z)} \left( \frac{1+z}{1+z_0}\right )^{\eta_1}\, , \\
   &  A_2(z) =   5 a_2 \bar{C}_1 \frac{\rho_{\rm crit} \Omega_{\rm m}}{D(z)^2} \left( \frac{1+z}{1+z_0}\right )^{\eta_2}\, , \\
  &  A_{1\delta}(z) = b_{\rm ta} A_1(z)\, ,
\end{align}
\noindent
where a normalization constant, by convention is fixed to $\bar{C}_1=5\times10^{-14} {\rm M}h^{-1}{\rm Mpc}^2$ \citep{brown02}, the pivot redshift is fixed at $z_0=0.62$ \citep{Troxel2018} and $b_{\rm ta}$ is assumed to be constant in redshift. Given the absence of informative priors, the analysis marginalises over the five intrinsic alignment parameters that govern the amplitude and redshift dependence of the signal, $(a_1, a_2, \eta_1, \eta_2, b_{\rm ta})$, with wide flat priors summarised in Table~\ref{tab:priors}. In the limit $A_2,b_{\rm ta} \rightarrow 0$, TATT reduces to the more commonly used NLA. 

\subsection{The matter power spectrum and baryonic effects}\label{sec:scalecuts}

Modeling the impact of baryonic feedback effects on the small-scale matter power spectrum is a leading systematic uncertainty in cosmic shear surveys \citep{vandaalen2014, semboloni2013effect, harnois2015baryons, chisari}. Power is surpressed by Active Galactic Nuclei (AGN) feedback processes at $k\sim10 h$/Mpc, as well as enhanced at smaller scales due to the more efficient cooling and star formation. These effects on the matter power spectrum can be modeled using, for example, empirical halo models fitted to hydrodynamic simulations \citep{mead2015,hildebrandt20, asgari20}, `baryonification' models \citep{schneider2019quantifying} and Principle Component Analysis (PCA) \citep{huang2020dark}, though these are limited by the large range of behaviours exhibited.

Instead, to mitigate any bias, this analysis adopts a gravity-only power spectrum and limits the measurements to larger angular scales \citep{y3-generalmethods, y3-cosmicshear2}. To define the maximum scale up to which plausible models may cause significant impact, synthetic cosmic shear data vectors are contaminated with baryonic effects measured from \textsc{EAGLE} \citep{mcalpine} and \textsc{OWLS-AGN} \citep{vanDaalen11} hydrodynamic simulations, according to
\begin{equation}
    P_{\delta,b}(k,z) = \frac{P_{\rm hydro}(k,z)}{P_{\rm DMO}(k,z)}P_{\delta} (k,z)
\end{equation}
where $P_{\rm hydro}(k,z)$ and $P_{\rm DMO}(k,z)$ are the matter power spectra measured from hydrodynamic and dark matter only simulations of the same suite. $P_{\delta}(k,z)$ is the non-linear matter power spectrum analytically calculated with the \textsc{Halofit} model \citep{smith2003, takahashi2012}. As detailed in Ref.~\citep{y3-generalmethods}, a threshold for the minimum set of angular scales is determined to ensure that the bias in the \textit{3$\times$2pt} analysis is below $0.3\sigma_{2D}$, for the 2D $\Omega_{\rm m} - S_8$ parameter space for \lcdm (found to be $0.09\sigma$, in the baseline analysis) and in $\Omega_{\rm m}-w$ for $w$CDM ($0.23\sigma_{2D}$), by balancing the cuts made for cosmic shear and \textit{2$\times$2pt}. With that criteria, the cosmic shear scale selection is identified where the bias incurred in the contaminated analysis compared to the baseline is $\Delta \chi^2 = 0.5$, which corresponds to a residual bias of $0.02\sigma_{2D}$ (see Figure 5 in Ref.~\citep{y3-generalmethods}). The angular bounds that satisfy this requirement are indicated by the shaded region in Figure~\ref{fig:xitomo}. Taking these into account, the data vector has 166 and 61 angular bins in  $\xi_+$ and $\xi_-$ and a total of 227 data points. 

\subsection{Modeling blending}\label{sec:modelm}

\begin{figure*}
    \centering
    \includegraphics[width=\textwidth]{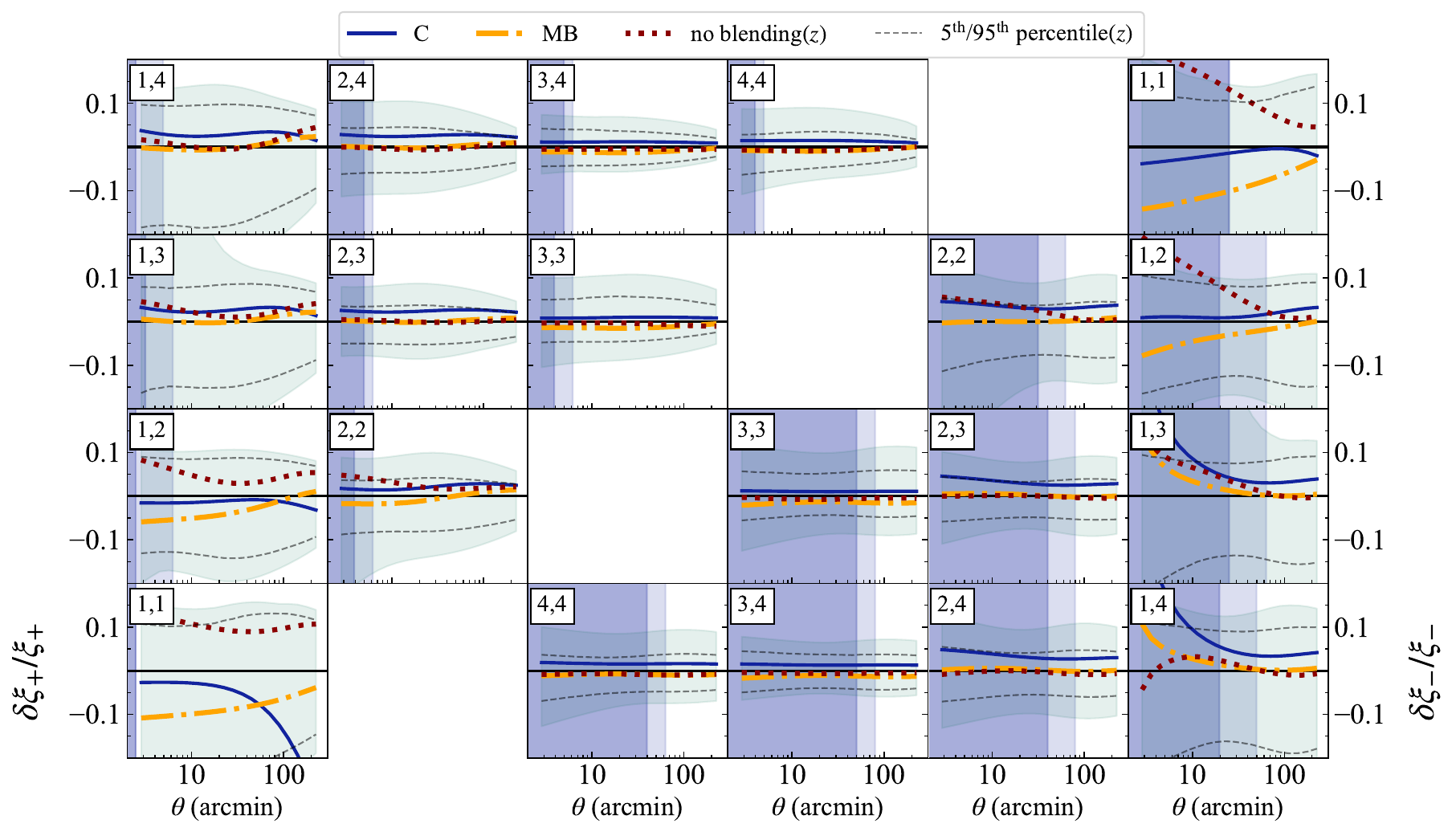}
    
    \caption{\label{fig:xitheoryredshift} Impact of choices in redshift calibration on predicted cosmic shear observables.  The  fractional difference between the fiducial simulated signal and one with an alternative analysis choice is shown, $\delta\xi_{\pm}/\xi_{\pm}$. Plotted are predicted data vectors (i) with the purely \textsc{COSMOS} \textsc{C}-redshift sample (blue solid line) (ii)  with the artificially biased spectroscopic \textsc{MB}-redshift sample (yellow dash-dotted line) (iii) without accounting for the \textit{redshift-mixing} effects of blending (red dotted line). Fiducial (\lcdm-Optimized) scale cuts are shown as (dark) blue shaded regions. The shaded green regions represent the simulated signals corresponding to the full range of \textsc{hyperrank} $n(z)$ realisations described in Section~\ref{sec:hyperrank}, and the dashed grey lines show the 5th and 95th percentiles of these simulated signals.}
\end{figure*}

The DES Y3 catalog uses the \mcal\ method to remove the largest calibration biases. However, blending is not fully accounted for in this approach, resulting in percent-level residual multiplicative bias. This is usually modeled  in the shear correlation functions by the approximation \citep{heymans06,Huterer06}
\begin{equation}
\xi^{ij}_{\rm obs} = (1+m^i)(1+m^j)\xi^{ij} \; ,
\end{equation}
where $m^i$ is the multiplicative bias for photometric redshift bin $i$, and is marginalized over with some prior usually inferred from image simulations, as in the case of this analysis. As detailed in Ref.~\citep{y3-imagesims}, shear calibration biases for a given redshift bin cannot be fully described by a single $m^i$ factor, firstly, because the $m^i$ may evolve over the redshift bin, i.e. $m^i=m^i(z)$. Moreover, in the presence of blending, the shape estimation for galaxies at one redshift may be influenced by the shear that blended galaxies at a different redshift are subject to. A general approach allowing for both of these effects is to quantify biases in the effective redshift distribution $n_{\gamma}(z)$. A mean multiplicative bias is related to the normalization of this effective redshift distribution, but biases to its shape are also expected due to redshift evolution of $m$ and blending. 

Ref.~\cite{y3-imagesims} calibrates this, assuming a model for the correct effective $n_{\gamma}^i(z)$ for a redshift bin $i$ as
\begin{equation}
    n_{\gamma}^i(z)= [1+F_i(z)]n^i_{\rm obs}(z) + G_i(z).
\end{equation}
where $F_i(z)$ and $G_i(z)$ are functions of $z$, constrained using image simulations in \citep{y3-imagesims}. They allow for multiplicative and additive deviations from the \mcal response-weighted redshift distribution, $n^i_{\rm obs}(z)$, presented in Section~\ref{sec:calib}. The term $F_i(z)$ models the impact on the effective weighting of the discrete sources used to construct $n_i(z)$, such as (possibly redshift-dependent) multiplicative bias effects. The term $G_i(z)$ captures responses to shear of light at redshifts other than that of the primary galaxy of a detection, due to blending of that primary galaxy with galaxies at different redshifts. Uncertainty in both effects are incorporated into the redshift calibration by producing samples of possible $n_{\gamma}^i(z)$. The impact of this modification to the $n^i_{\rm obs}(z)$ on the shear correlation functions is demonstrated in Figure~\ref{fig:xitheoryredshift} as `no blending(z)'. The use of the normalized $n(z)$ samples in the analysis pipeline gives shear calibration priors per bin, $m^i$, inferred from the $n_{\gamma}(z)$ samples, that are sampled. The blending-based perturbations are also reflected in widened priors on the mean redshift of each redshift bin. Both priors are listed in Table~\ref{tab:datastats}.

To demonstrate the importance of this previously uncalibrated effect, a variant analysis is performed in Section~\ref{sec:robustshear} that neglects the impact of the redshift-mixing blending.  In this test, the redshift distributions without shear-calibration correction are used with the $m_{\rm noz-blend}$ priors, listed in Table~\ref{tab:datastats} of 3-6\% width, that represent the impact from masking, detection biases and blending, drawn from \citep[][their Table 3]{y3-imagesims} and quoted. The impact of the assumption  that ignores the correlation between an individual $n(z)$ realisation and the residual multiplicative factor, $m$, is tested at the likelihood level. To do so the analysis is modified to use \textsc{Hyperrank} to sample the ensemble $n(z)$, with an $m$ associated with each realisation, labelled `Full blending treatment'.

\subsection{Redshift uncertainty}
\label{sec:modelz}

An approximation previously used for propagating photometric redshift calibration uncertainties into cosmic shear cosmological parameter constraints \citep[e.g.,][]{svcosmicshear,choideltaz,joudakicfht,Troxel2018,hikage19,asgari20, Ham20} relies on introducing a nuisance parameter, $\Delta z^i$, for the offset of the mean redshift of each source bin $i$ from its estimate,
\begin{equation}
n_i(z) \rightarrow n_i(z-\Delta z^i) \, .
\end{equation}

The prior on the $\Delta z^i$ encapsulates the statistical and systematic calibration uncertainty. However, 
uncertainties in the estimated redshift distributions are not limited to the mean redshift, but rather include, e.g., the extendedness 
of a redshift distribution's tail. The assumption of a prior purely on $\Delta z^i$ could potentially lead to a mis-estimation of the confidence intervals in cosmological parameters, or possibly biases in inferred cosmological, or particularly, intrinsic alignment parameters, especially as the statistical power of the measurement improves. Some studies in the literature have thus chosen to account for uncertainty in the redshift distributions differently than with such a prior on their mean. For instance, several recent cosmic shear studies \citep{hildebrandt2017kids, Troxel2018, hildebrandt20, joudaki2020} compare cosmological constraints achieved with different redshift calibration strategies, e.g., with different redshift samples, to gauge the uncertainty.

\subsubsection{Full shape uncertainty}
\label{sec:hyperrank}

While the ensemble variance of mean redshift can be expressed as a set of $\Delta z^i$ priors, it is more accurate to account for the full uncertainty in the shape of the $n(z)$. The DES Y3 effort has developed techniques to marginalize over the full shape uncertainty of the redshift distribution in the likelihood analysis. The set of candidate $n(z)$ samples described in Section~\ref{sec:sompz} encapsulates the full uncertainty on redshift calibration. These samples preserve correlations between redshift bins, uncertainties on higher-order moments of the $n(z)$, and non-Gaussianities in those systematic errors. The methodology for an alternative to the $\Delta z^i$ marginalization approach, \textsc{Hyperrank}, is presented in Ref.~\citep{y3-hyperrank}. Here, a single realization of the $n(z)$ of the four bins is selected in each likelihood evaluation. The sampling across a set of such realizations describes the full uncertainty and preserves the correlation of $n(z)$ variations across redshift and between bins.

To avoid low Markov chain sampling efficiency by selecting a random realization on each evaluation, \textsc{Hyperrank} constructs a mapping between the $n(z)$ ensemble, characterized by a set of descriptive parameters (in this case, the mean redshift of the three bins with the largest variance across the realizations of the ensemble) and a multidimensional grid. Coordinates of this grid are sampled in the likelihood analysis and the $n(z)$ realization mapped to the closest point is used in each step.

Prior to unblinding, in a simulated analysis, the impact of accounting for the full shape uncertainty was contrasted against marginalising over an approximate uncertainty on the mean, as well as neglecting any redshift uncertainty. This Y3 analysis was found to be insensitive to any redshift uncertainty, given the high precision of the redshift calibration priors, in tandem with the current level of statistical power of the DES Y3 data. While this is an important step forward in the methodology for future analyses, given the significant additional computing time required for \textsc{Hyperrank}, the primary analysis uses the mean of the ensemble and the uncertainty approximated as a shift in the mean, with the priors given in Table~\ref{tab:priors}. As validation, in Section~\ref{sec:robustredshift}, the robustness of this approximation is tested.

\subsubsection{Impact on cosmic shear}

Figure~\ref{fig:xitheoryredshift} illustrates the impact of the effects of sources of uncertainty in redshift calibration on the expected cosmic shear data vector. In particular, the predicted signals that use redshift distributions calibrated from the pure \textsc{COSMOS} photometric sample, \textsc{C}, (blue) and the (artificially) pure spectroscopic sample, \textsc{MB}, (yellow), defined in Section~\ref{sec:zsamples} are compared to the fiducial simulated data vector, computed with each $n(z)$ realisation. Their spread, and the uncertainty on the redshift distribution is indicated by the shaded green region with the 5th and 95th percentile denoted by the grey dashed lines. The red line indicates the impact of the redshift-mixing effect of blending on the redshift distribution (see Section~\ref{sec:modelm}). Differences are small --- well within the range of uncertainty allowed by the $n(z)$ ensemble. The highest significance of differences is seen in the lowest redshift bin, which is of relatively small importance for cosmological constraints. The size of the effects of redshift sample choice and of accounting for the impact of blending on the $n(z)$ are comparable. Due to the coherence of the impact across scales and bins, investigating the impact on cosmology is still warranted, as explored in Section~\ref{sec:robustredshift}. 

\subsection{Bayesian Inference}

For parameter inference, the likelihood, $\mathcal{L}$, of the data vector, $D$, given the model, $T$, with parameters, $\textbf{p}$, can be expressed as $\mathcal{L}(D | \textbf{p})$. The latter is assumed to be a multivariate Gaussian,
\begin{equation}
\ln{\mathcal{L}(D | \textbf{p})} = -\frac{1}{2} \sum_{ij} \Big(D_i-T_i(\textbf{p})\Big)\,\,[\mathbf{C}^{-1}]_{ij}\,\,\Big(D_j-T_j(\textbf{p})\Big) \,.
\end{equation}
$D_i$ represents the $i$th element of the data vector $\xi_{\pm}$, presented in Section~\ref{sec:measurement}, and its covariance,$\mathbf{C}$ . It initially contains 20 angular data points each over the combinations of 4 redshift bins and 2 correlation functions, which amounts to 227 data points after limiting the angular scales. The corresponding theoretical prediction for the statistics,  $T_i(\textbf{p})$, are detailed in this section.  The Bayesian \textit{posteriors} of the cosmological parameters, $\mathcal{P}(\textbf{p}|D)$, are computed as the product of the likelihood with the priors, $P(\textbf{p})$, listed in Table~\ref{tab:priors}, following Bayes' theorem:
\begin{equation}\label{eq: Bayes theorem}
\mathcal{P}(\textbf{p}|D)=\frac{P(\textbf{p})\mathcal{L}(D|\textbf{p})}{P(D)},
\end{equation}
where $P(D)$ is the \textit{evidence} of the data. 

The posterior is sampled with the \textsc{Polychord} sampler \cite{Handley15}. The analysis pipeline is built upon the framework \textsc{CosmoSIS} \citep{zuntz15} and validated through an independent implementation of the analysis pipeline in \textsc{cosmolike} \citep{Krause17,y3-generalmethods}. \textsc{CosmoSIS} is a modular cosmological parameter estimation code and in order to calculate the linear matter power spectrum $P_{\rm \delta}(k, z)$, it uses \textsc{CAMB} \citep{Howlett2012, CAMB}. Although the fiducial sampler settings (500 live points, tolerance 0.01) have been
tested to demonstrate the accuracy of the posteriors and Bayesian evidence estimates \citep[see][]{Lemos2020}, the position of the peak inferred from the posterior samples in 28-dimensional space is noisy. Hence we use the \textsc{MaxLike} minimizer after the chain has run to have a reliable estimate of the \textit{maximum a posteriori} point (MAP). 

Even in the case of the baseline framework applied to a synthetic, noiseless data vector generated from the same model, the marginalized parameter posteriors can appear biased from the input parameter values due to parameter volume or \textit{projection effects}, which occur when parameters of interest are not well-constrained by the data or are degenerate with other parameters that are prior informed (see Figure 2 in Ref.~\citep{y3-generalmethods}). Related to projection effects, prior to unblinding, the scope of noise realisations was studied. In particular, the possibility of bimodal astrophysical nuisance parameter posteriors and their impact was investigated, discussed in Appendix~\ref{app:power}. The MAP is equivalent to the best-fit (b.f.) and recovers the input parameter values. In presence of noise, the mean provides a stable single point estimate of any single parameter value so, following Section 4 of Ref.~\citep{y3-generalmethods}, we report the 1D marginalized mean and its asymmetric $\pm34\%$ confidence intervals, together with the MAP.

\subsection{An `Optimized' analysis}\label{sec:agg}
The DES Y3 approach emphasizes conservative modeling choices for robust cosmological posteriors. In addition to the \textit{Fiducial} analysis, an approach that is optimized for \lcdm is investigated. In particular, the scale cuts implemented to account for small-scale baryonic physics, are revisited to optimise the analysis post-unblinding. The Fiducial scale selection was chosen to be robust for the joint $w$CDM \textit{3$\times$2pt} analysis using the procedures described in Section~\ref{sec:scalecuts}. As a result, for cosmic shear-only in \lcdm, this resulted in a conservative choice of scale cuts, with potential biases in cosmological parameters inferred to be significantly lower than the required threshold.

We consider an \textit{`\lcdm-Optimized'} analysis, where the scale cuts are relaxed maintaining that potential biases for \textit{3$\times$2pt} satisfy the limit of $\leq 0.3\sigma_{2D}$ for the 2D $\Omega_{\rm m} - S_8$ parameter space for \lcdm. For cosmic shear in \lcdm, this scale selection is still conservative, with potential biases at most $0.14\sigma_{2D}$. This scale selection is illustrated in Figure~\ref{fig:xitomo} by the darker shaded region,
giving 184 and 89 angular bins in $\xi_+$ and $\xi_-$, and a total of 273 data points. Throughout this work, the \lcdm-Optimized analysis is presented as a robust \lcdm result alongside the Fiducial, which was used for the un-blinding and internal consistency requirements.

\section{Cosmological constraints}\label{sec:results}

\begin{figure*}
\centering
\begin{minipage}{.5\textwidth}
  \centering
  \includegraphics[width=\textwidth]{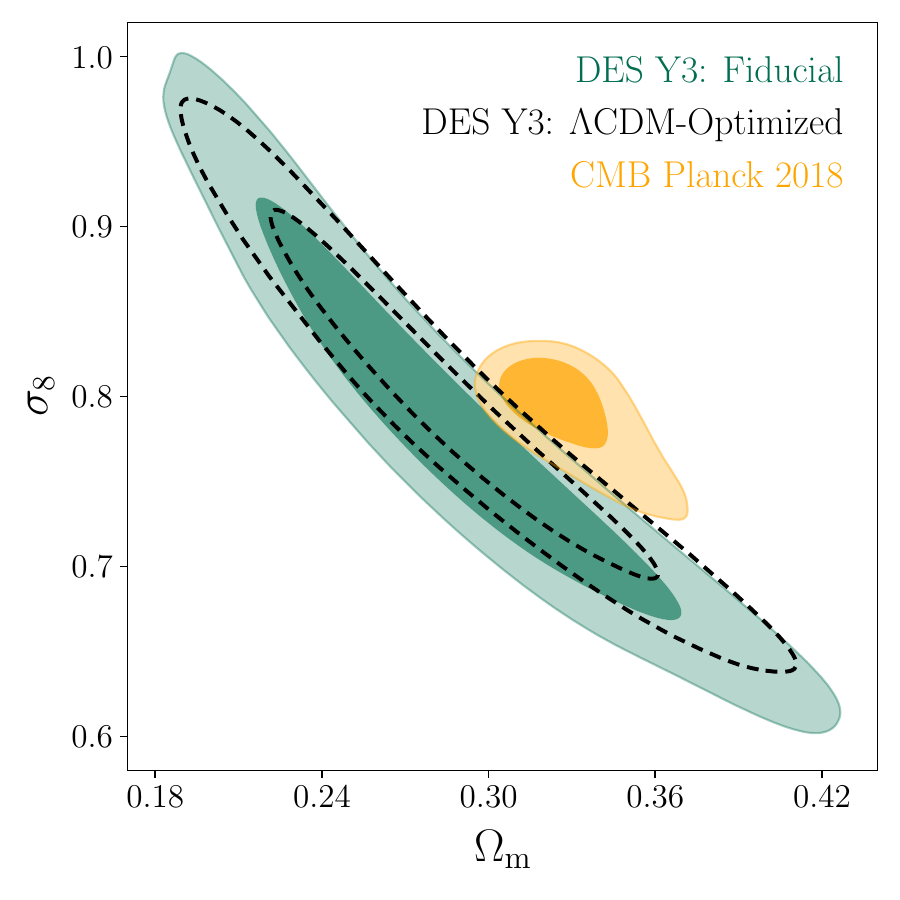}
\end{minipage}%
\begin{minipage}{.5\textwidth}
  \centering
  \includegraphics[width=\textwidth]{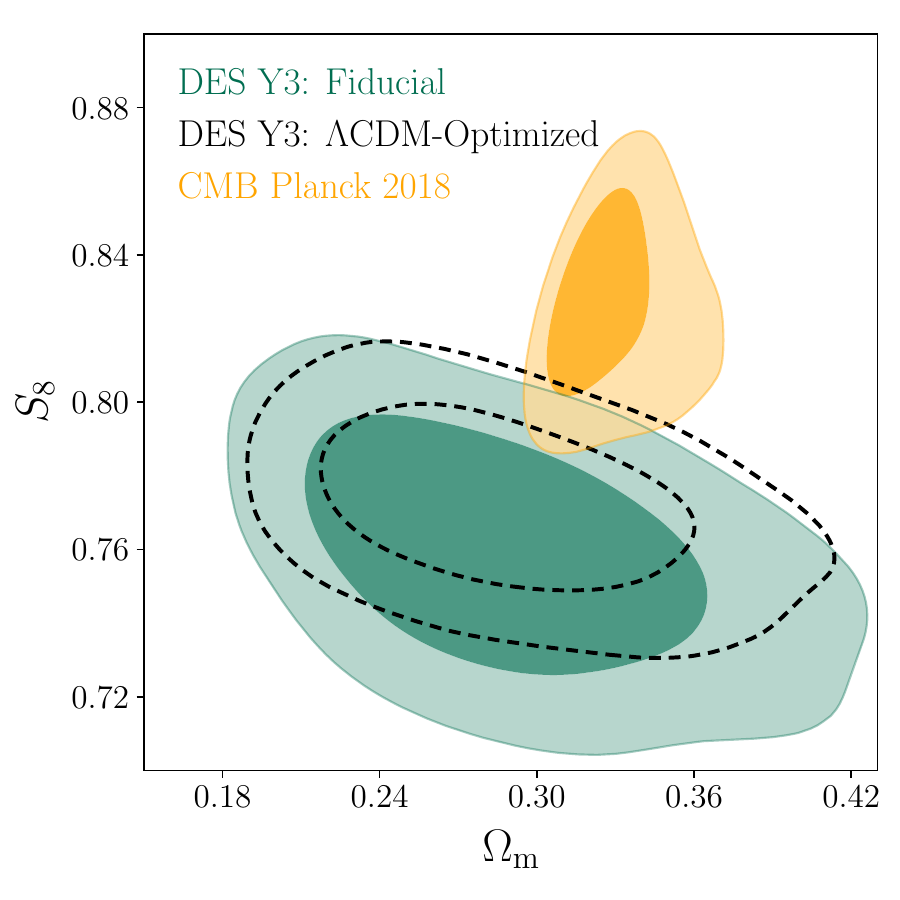}
\end{minipage}
\caption{Cosmological constraints on the clustering amplitude, $\sigma_8$, (left) and $S_8$ (right) with the matter density, $\Omega_{\rm m}$ in \lcdm. The marginalised posterior contours (inner 68\% and outer 95\% confidence levels) are shown for the \textit{Fiducial} DES Y3 analysis in green and \textit{Planck} 2018 CMB in yellow \citep{Planck2018}. The black dashed contours represent the \textit{\lcdm-Optimized} analysis, that preserves more small-scale information compared to the Fiducial analysis, as described in Section~\ref{sec:agg}. }
\label{fig:result}
\end{figure*}

In this section, we present constraints on cosmological parameters: the \textit{Fiducial} analysis is described in Section~\ref{sec:lcdmresults}
and the results of the \textit{\lcdm-Optimized} analysis (see Section~\ref{sec:agg}) in Section~\ref{sec:aggresults}. This analysis presents cosmic shear measurements with a signal-to-noise of 40. These are compared to constraints from the CMB measurements by \textit{Planck} \citep{Planck2018} for a test of the $\Lambda$CDM model over the history of cosmic evolution. In addition, the posteriors are compared to those from other weak-lensing analyses, with caution that, for this comparison, these analyses differ in many aspects and cannot be quantitatively assessed without a homogeneous framework \citep{chang2018}. In \citep{y3-3x2ptkp}, these cosmic shear results are combined with measurements of galaxy-galaxy lensing and galaxy clustering.

Although we sample over the normalization of the matter power spectrum $A_{\rm s}$, results are presented in terms of the commonly used $S_8$ parameter, defined as
$S_8 = \sigma_8(\Omega_{\rm m}/0.3)^{\alpha}$,
in terms of the matter density parameter and the amplitude of fluctuations, with $\alpha=0.5$. The constraints are quantified in terms of the 68\% confidence limit, which defines the area around the peak of the posterior within which 68\% of the probability lies as well as the figure of merit (FoM),  to compare the relative constraining power of results in 2D. The FoM is defined for two parameters, in this case $\Omega_{\rm m}$ and $S_8$, and their covariance, $\mathbf{C}_{\Omega_{\rm m},S_8}$, as \citep{huterer01, wang08}
\begin{equation}
\label{eqn:FOM}
\textrm{FoM}_{\Omega_{\rm m},S_8} = (\textrm{det}\, \mathbf{C}_{\Omega_{\rm m},S_8})^{-0.5} \,,
\end{equation} 
which is in analogy to the Dark Energy Task Force recommendation for the dark energy FoM \citep{fom}.

\subsection{Fiducial results} \label{sec:lcdmresults}

The constraints are obtained by marginalising over 6 cosmological parameters in the $\Lambda$CDM model, including a free neutrino mass density (assuming a normal mass hierarchy), and 22 systematic and astrophysical parameters, as summarised in Table~\ref{tab:priors}. In Figure~\ref{fig:result} we show the Fiducial DES Y3 cosmic shear posteriors for the \LCDM model, projected into 2D parameters $\Omega_{\rm m}, \sigma_8$, and $S_8$. These are represented by green filled contours, denoting the 68\% and 95\% confidence levels. 

The marginalized  mean (and MAP, or \textit{maximum a posteriori} point) values of $S_8$, $\Omega_{\rm m}$ and $\sigma_8$ and are found with 68\% confidence intervals to be 
\begin{align}
S_8 &{} =  0.759_{-0.025}^{+0.023} \quad (0.755)  \\ 
\Omega_{\rm m} &{} =  0.290_{-0.063}^{+0.039} \quad (0.293) \\
\sigma_8 &{} =  0.783^{+0.073}_{-0.092} \quad (0.763)\, ,
\end{align}
\noindent 
constituting a 3\% fractional uncertainty on $S_8$.

The best-fit \LCDM prediction is over-plotted on the cosmic shear two-point measurements in Figure~\ref{fig:xitomo} as a black line. A total $\chi^2$ of 237.7 is found for the Fiducial measurement with the \LCDM best-fit model. The analysis has 202 degrees of freedom (227 data points and 28 free parameters), but when accounting for the informative priors used following Ref. \citep{raverihu}, the effective dimensionality of parameter space is reduced to $\sim$5, as many parameters in the analysis are not fully constrained by the data. From that, we estimate 222 effective degrees of freedom, giving $\chi^2/$d.o.f = 237.7 / 222 = 1.07. The probability of getting a higher $\chi^2$ value can be derived assuming our data vector is drawn from a multivariate Gaussian likelihood with our assumed covariance matrix is precisely and fully characterized. This leads to $p$-value = 0.223. The FoM for the analysis is found to be 927. These constraints are summarised in Figure~\ref{fig:summary}, alongside a raft of robustness tests. The parameter constraints, goodness of fit and FoM are tabulated in Table~\ref{tab:summary}.  In both, we distinguish robustness tests that are not expected to give consistent results,  such as by neglecting to account for systematics, by an asterisk and an open symbol.

In comparison, Figure~\ref{fig:result} shows constraints from the \textit{Planck} satellite CMB temperature and polarization measurements \citep{Planck2018} in yellow.
These include the \textit{Planck} measurements of the auto power spectra of temperature $C_{\ell}^{\rm TT}$, of $E$-modes $C_{\ell}^{\rm EE}$, and their cross-power spectra $C_{\ell}^{\rm TE}$, using the `Plik' version for $\ell$ >30 in addition to the temperature and $E$-mode power spectra, $C_{\ell}^{\rm TT}$, $C_{\ell}^{\rm EE}$ measurements in the range 2< $\ell$ <29 and using no lensing, for a distinct high-redshift result. These are reanalysed using the DES Fiducial cosmological priors, primarily to allow variations in $\Omega_{\nu} h^2$, in order to assess consistency between the two measurements. For the DES Y3 analysis, we compute the Bayesian Suspiciousness \citep{Handley:2019wlz, y3-tensions}, an evidence-based method that corrects for the prior dependence of the constraints and in the full parameter-space. As in Ref.~\citep{y3-3x2ptkp}, we conclude that two data sets are statistically consistent if the $p$-value implied by our  tension metrics is greater than 0.01.

There has been significant discussion in the literature regarding the consistency of low- and high-redshift cosmological probes and specifically of cosmic shear constraints with those from measurements by the \textit{Planck} satellite CMB. We, nevertheless, find no significant evidence for disagreement with CMB from DES cosmic shear on its own, compatible at the level of $2.30 \pm 0.34\sigma$, yielding a $p$-value of $\sim0.02$. It is notable that the constraint on the value of the $S_8$ parameter is determined to be lower than that from \textit{Planck} by 2.3$\sigma$. In Appendix~\ref{app:as}, the $\Omega_{\rm m}-A_{\rm s}$ posterior is also shown, along with the other cosmological parameters. While the DES constraint in this parameter space is weak, it is interesting that there is no evidence for a tension in this parameter direction. A more detailed interpretation of consistency of DES cosmic shear with external probes is given in the companion paper \citep{y3-cosmicshear2}, and in greater detail and with higher significance, Ref.~\citep{y3-3x2ptkp},  where the cosmic shear measurements are jointly analysed with galaxy clustering and galaxy-galaxy lensing to give tighter cosmological constraints. 

\begin{figure*}
\begin{minipage}{.5\linewidth}
\centering
\subfloat{\includegraphics[width=.99\textwidth]{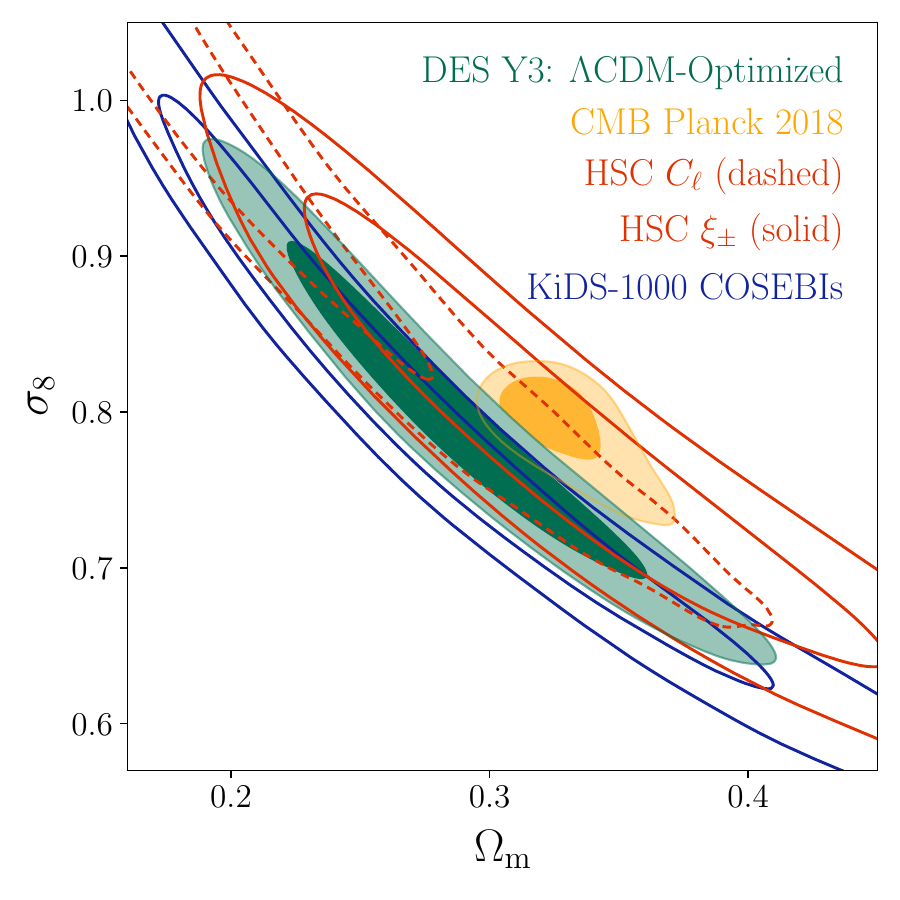}}
\end{minipage}%
\begin{minipage}{.5\linewidth}
\centering
\subfloat{\includegraphics[width=.99\textwidth]{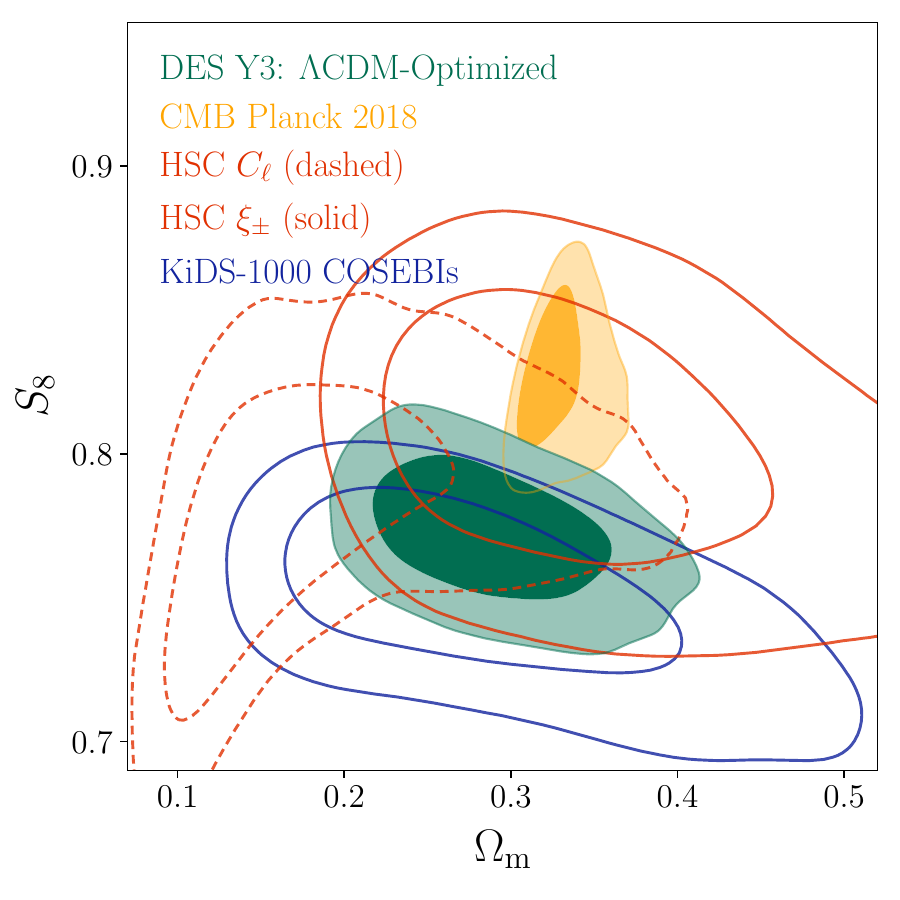}}
\end{minipage}
\centering
\subfloat{\includegraphics[width=.9\textwidth]{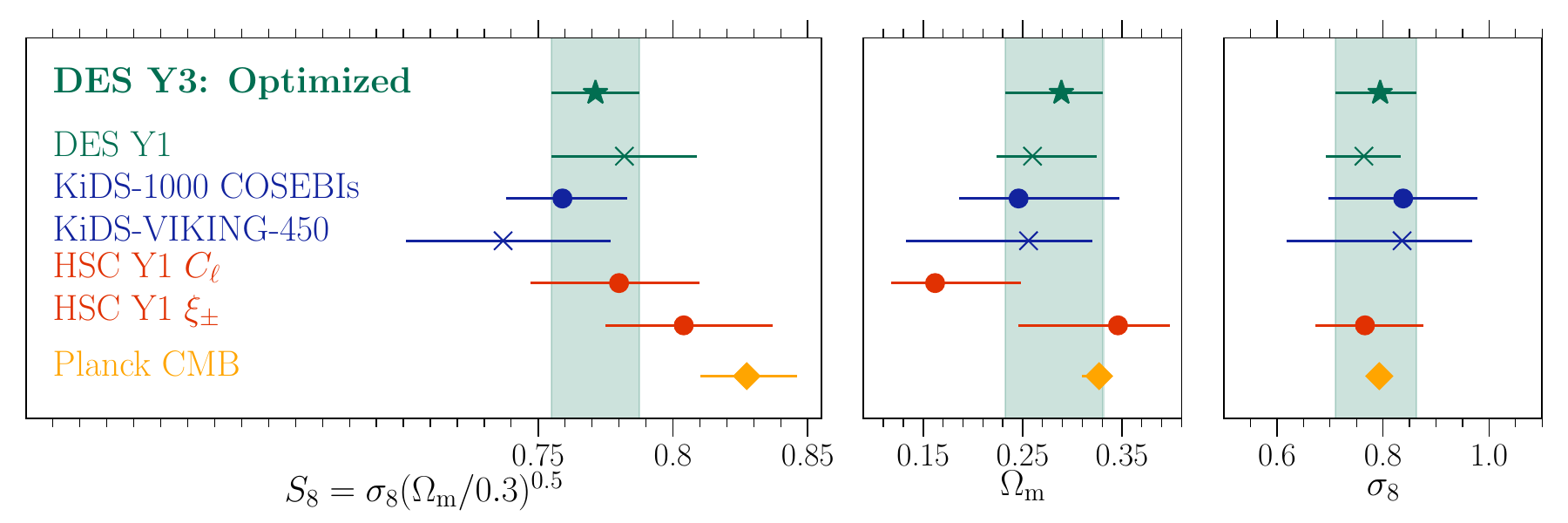}}
\caption{\label{fig:resultagg} Recent cosmic shear results: The constraints on $\Omega_{\rm m}$, $\sigma_8$ (upper left) and $S_8$ (upper right) are shown for the DES Y3 \lcdm-Optimized analysis (green) and \textit{Planck} 2018 CMB \citep{Planck2018} (with free neutrino mass density, yellow). In addition, shown are KiDS-1000 COSEBIs analysis \citep{asgari20} (blue) and the HSC $C_{\ell}$ \citep{hikage19} (red, dashed) and HSC $\xi_{\pm}$ \citep{Ham20} (red, solid) results. Lower: Summary of 1D constraints, including the previous DES and KiDS results \citep{Troxel2018,hildebrandt20}.  The mean 1D values are indicated with filled symbols and 68\% confidence limits as horizontal bars. The \lcdm-Optimized DES Y3 result is also indicated by the green shaded region. 
Note that external weak-lensing experiments employ different analysis choices, including the less general NLA intrinsic alignment model, and measure alternative statistics using different scales than DES Y3, which limits the ability to compare these results on equal footing.}
\end{figure*}

The Fiducial posteriors exhibit a bimodality intrinsic alignment parameters, that are degenerate with the $S_8$ parameter. In Appendix~\ref{app:power}, we discuss the extent of the impact of noise on cosmological analyses with an investigation of multiple noisy simulated runs.

\subsection{\lcdm-Optimized results}\label{sec:aggresults}

The `\lcdm-Optimized' analysis, introduced in Section~\ref{sec:agg}, allows for additional small-scale information to be used safely.  Figures~\ref{fig:result} and~\ref{fig:resultagg} show the cosmological constraints using the \lcdm-Optimized scales as a dotted black contour, finding substantially improved precision. Compared to the fiducial result, it is found to be consistent within 0.5$\sigma$ in the $S_8$ parameter. The 1D mean (and best-fit) values of $S_8$, $\Omega_{\rm m}$ and $\sigma_8$ and are found with 68\% confidence intervals to be 
\begin{align}
S_8  &{}=  0.772^{+0.017}_{-0.017} \quad (0.774) \\ 
\Omega_{\rm m}  &{} = 0.289^{+0.036}_{-0.056} \quad (0.279) \\
\sigma_8  &{} =  0.80^{+0.072}_{-0.076} \quad (0.802)  \, ,
\end{align}
\noindent 
with a $\sigma_{\rm Fid}/\sigma_{\rm Opt}=1.47\times$ smaller uncertainty on $S_8$, now at a level of a relative error of 2\%.

As is the case for the Fiducial analysis, in the full parameter-space, the \lcdm-Optimized DES result finds no significant evidence for disagreement with \textit{Planck} CMB data. Using the Suspiciousness statistic, we enhanced compatibility at the level of $2.0 \pm 0.4\sigma$, corresponding to a tension probability of $p$value$\sim0.05$. We note that while the inclusion of smaller scales does result in a smaller uncertainty on $S_8$, the mean does shift slightly towards higher values, such that a $2.1\sigma$-level difference with \textit{Planck} is preserved. Therefore, we observe that the moderate $\sim2.3\sigma$ tension observed in the main analysis reduces to $\sim2.0\sigma$ when the additional small-scale data is used. 
The $\chi^2$/d.o.f = 285/268.2 = 1.06 which gives a $p$-value = 0.22. The FoM for the analysis is found to be 1362, as quoted in Table~\ref{tab:summary}. 

The gain in $S_8$ constraining power and small shift can be, in part, attributed to eliminating the most extreme lobes of the Fiducial intrinsic alignment $a_1-a_2$ posterior that is degenerate with $S_8$ (see Appendix~\ref{app:power} for further discussion). In terms of $\Omega_{\rm m}$ and $\sigma_8$, the small scales now included contribute a 7\% and 3\% improvement in precision, respectively. In Section~\ref{sec:syslimit}, we assess the limiting systematics for this \lcdm-Optimized analysis.

In Figure~\ref{fig:resultagg}, the \lcdm-Optimized $S_8$, $\Omega_{\rm m}$ and $\sigma_8$  results are shown alongside the most recent \textsc{COSEBIs} cosmic shear constraints from the KiDS-1000 survey \citep{asgari20} shown in blue and the HSC-Y1 constraints from their Fourier-space analysis in red \citep{hikage19}. While plotted for illustrative purposes, we caution the reader against direct comparisons of cosmological parameters as the priors, measurement statistics, scales allowed and other analysis choices such as the intrinsic alignment modeling adopted by other surveys vary and can lead to different conclusions \citep{chang2018}. Nevertheless, it is evident that in recent years, all cosmic shear analyses find a lower value of $S_8$ than the inferred \textit{Planck} 2018 constraint under a flat \lcdm model, although with varying levels of significance. A summary of the most recent $S_8$ result from each `Stage 3' weak-lensing survey is shown in Figure~\ref{fig:resultagg}. A more quantitative comparison using a unified set of analysis choices is left to future, collaborative work.

\begin{figure*}
    \centering
    \includegraphics[width=.85\textwidth]{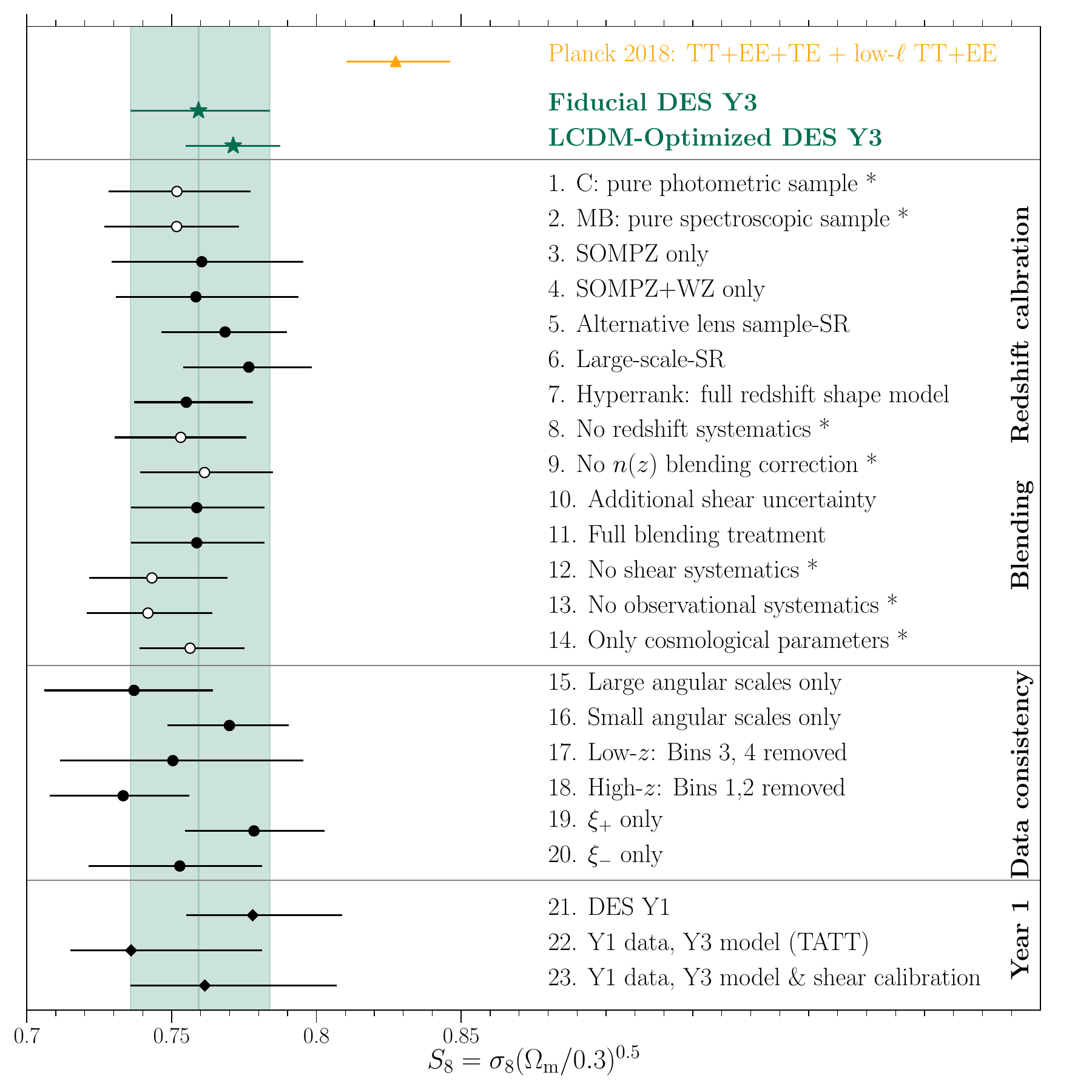}
    \caption{\label{fig:summary}
    Summary of marginalised 1D constraints on $S_8$ in \lcdm, testing various choices. The mean of the $S_8$ marginalized posterior is indicated by the symbol and 68\% confidence intervals are shown as horizontal bars. Empty symbols represent those analyses which are not necessarily expected to agree with the Fiducial result. For the Fiducial analysis, this is also represented as the green shaded region. Tests 1-14 validate the robustness against observational systematics while Tests 15-20 investigate internal consistency of the measurements, splitting the Y3 data into subsets, as well as comparing to the DES Y1 data in Tests 21-23. We distinguish variations on the baseline setup that are not necessarily required to give consistent results (e.g., by neglecting observational systematics) by an asterisk and an open symbol. We show the CMB \textit{Planck} 2018 \citep{Planck2018} (high $\ell$ TT+EE+TE + low $\ell$ TT+EE) constraint in yellow, reanalysed using DES cosmological priors, with free neutrino mass density.  The numerical parameter values are listed in Table~\ref{tab:summary}.} 
\end{figure*}
\section{Internal consistency}\label{sec:IC}

In this section, we investigate the consistency of cosmic shear cosmological results in both data space and parameter space. To do so, we analyse the data when excluding particular subsets. We test the impact of each tomographic redshift bin by excluding them one at a time from the data vector; we test the consistency between the small and large angular scale measurements; and we assess the $\xi_-$ measurements compared to the $\xi_+$ measurements. Finally, we demonstrate the consistency between the Y3 results and previous DES cosmic shear analyses.

\begin{figure*}
\centering
\begin{minipage}{.33\textwidth}
  \centering
  \includegraphics[width=.98\textwidth]{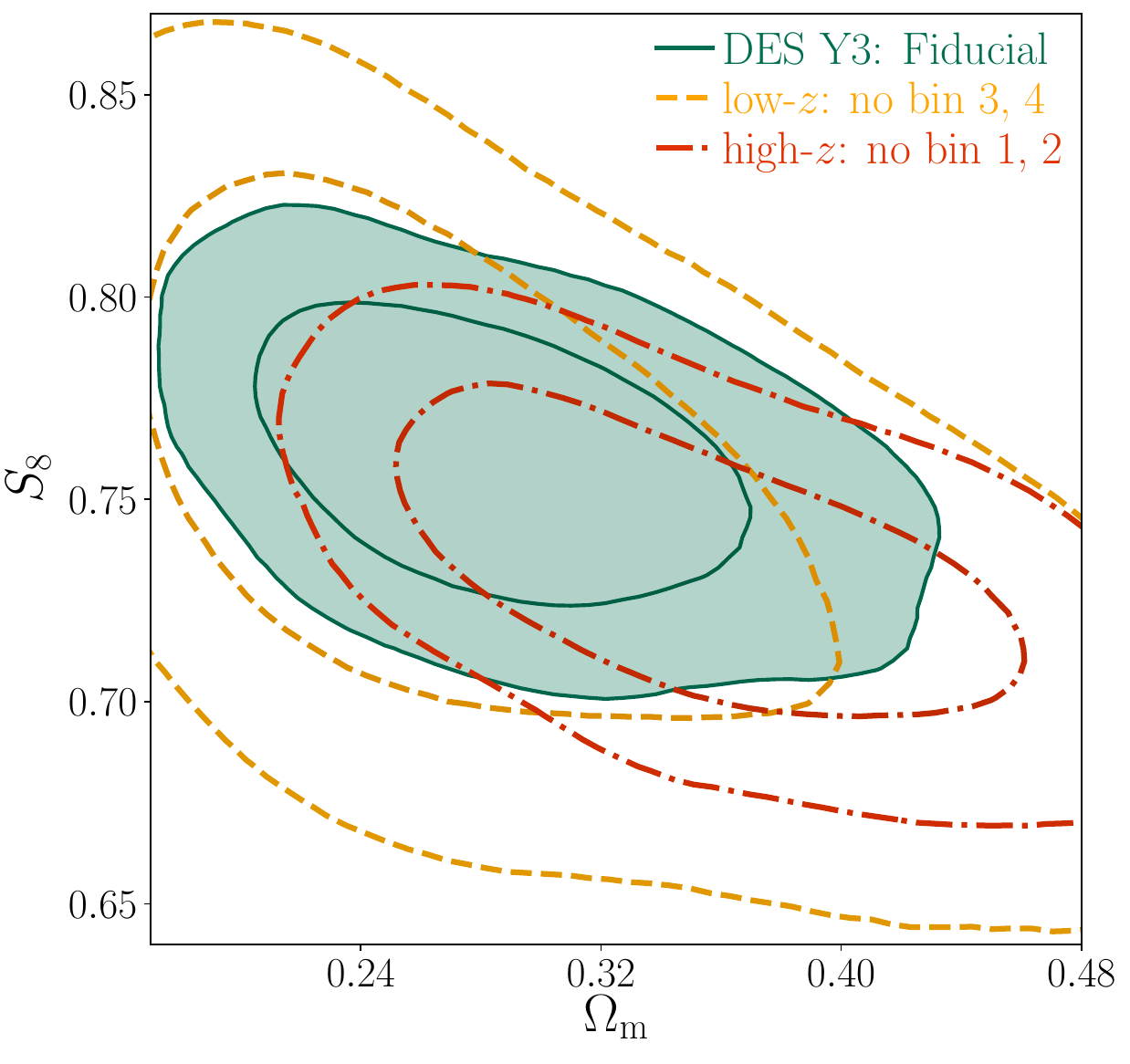}
\end{minipage}%
\begin{minipage}{.33\textwidth}
  \centering
  \includegraphics[width=.98\textwidth]{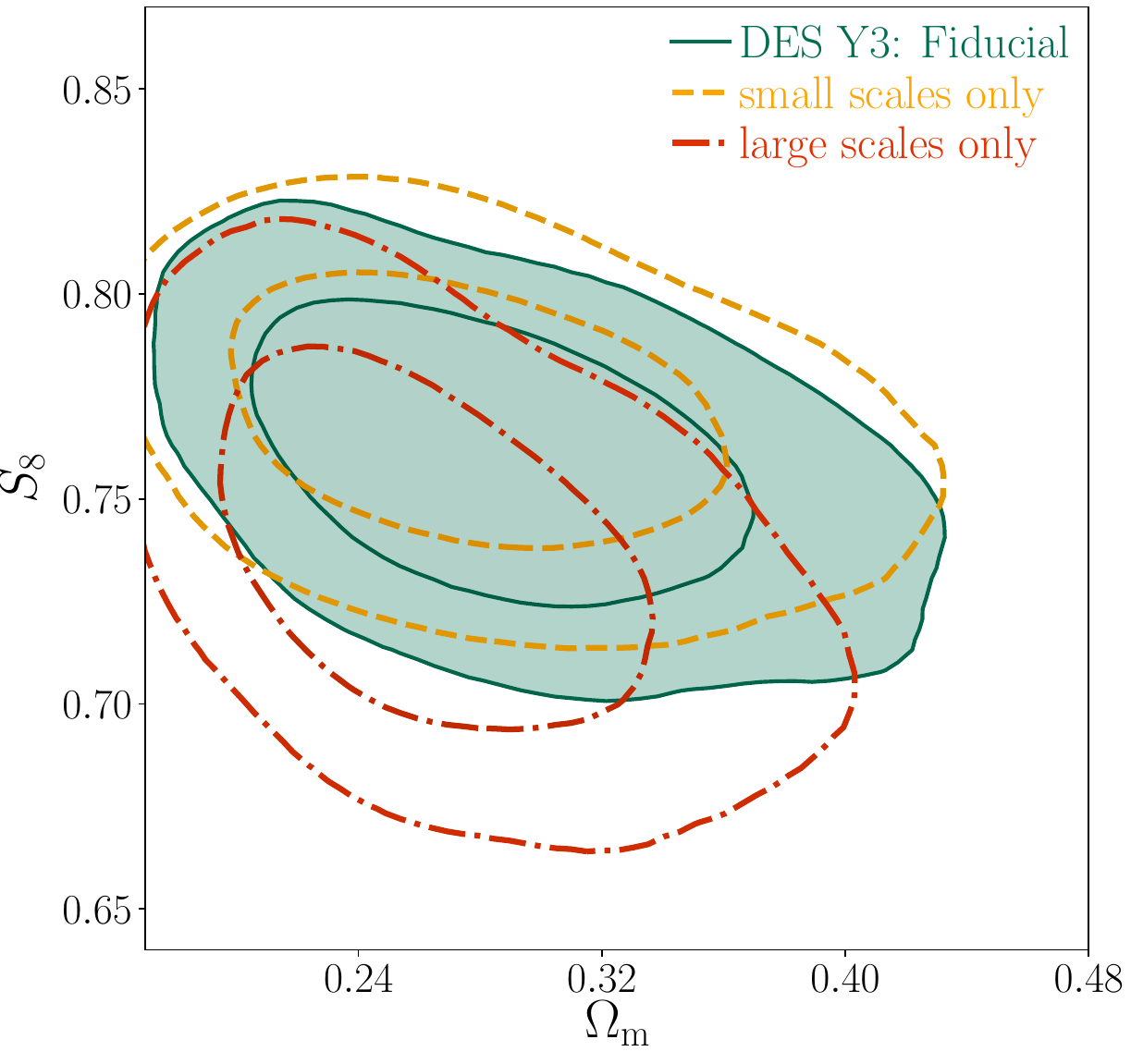}
\end{minipage}
\begin{minipage}{.33\textwidth}
  \centering
  \includegraphics[width=.98\textwidth]{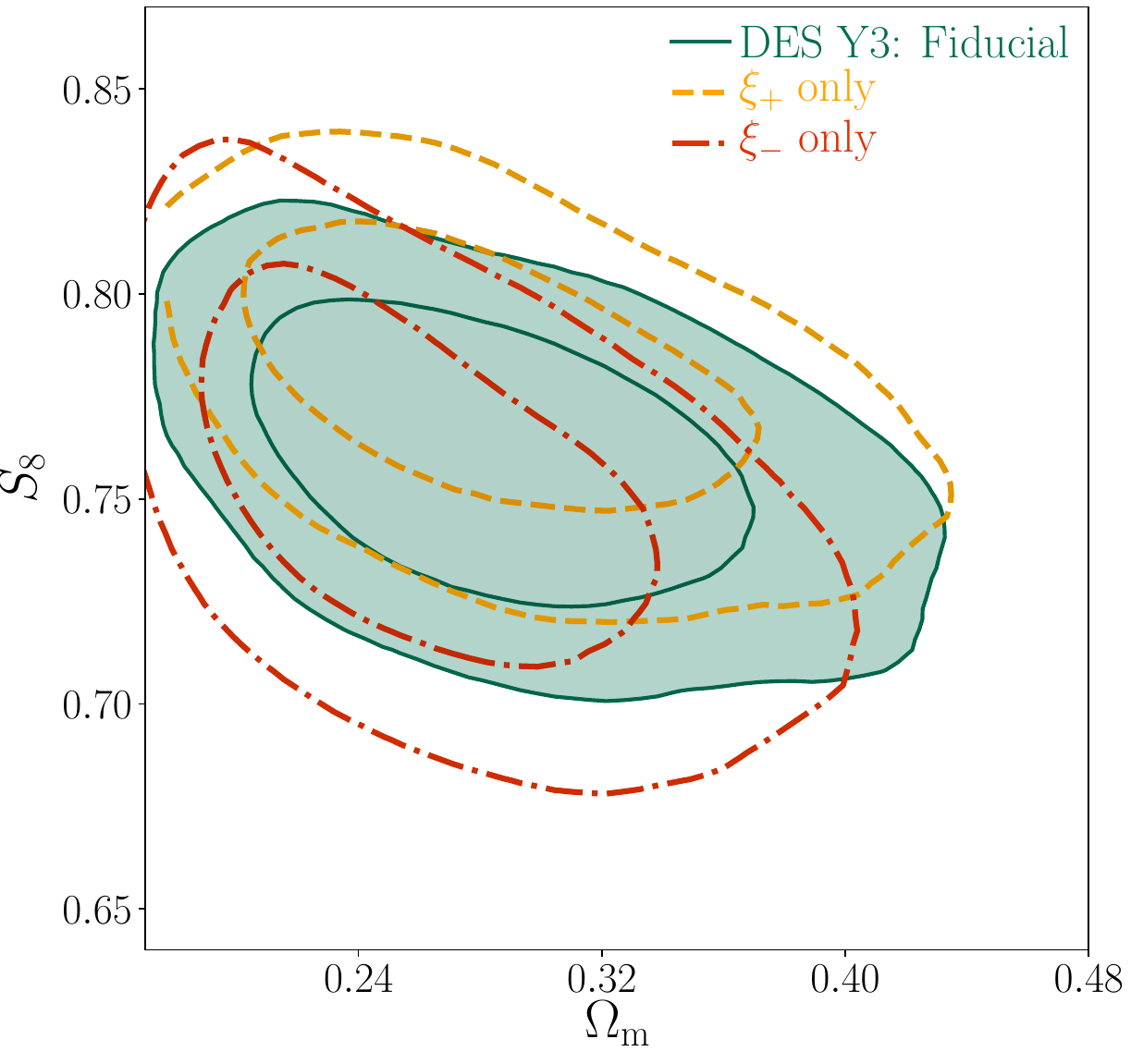}
\end{minipage}
\caption{\label{fig:consistency}Internal consistency: comparing Fiducial \lcdm constraints in the $\Omega_{\rm m}-S_8$ plane from subsets of the data. The filled green contours show the Fiducial analysis. On the left, redshift bins from low- and high-redshift are excluded to test the sensitivity of the results to redshift-dependent observational and astrophysical systematics. In the centre, subsets of the data vector are excluded to demonstrate robustness against angular scale-dependent systematics, with the red contour using $\theta > \theta_{\rm lim}^+=30$ arcmin and $\theta > \theta_{\rm lim}^-=100$ arcmin for $\xi_+$ and $\xi_-$, respectively, and the yellow contour using the small scales, divided at the same $\theta_{\rm lim}$. On the right, we compare the posteriors from $\xi_-$-only (red) and $\xi_+$-only (yellow) data. }
\end{figure*}
\subsection{Internal consistency methodology}

For tests in data space, the Posterior Predictive Distribution (PPD) methodology detailed in Ref.~\citep{y3-inttensions} is used. We consider two types of internal consistency tests: goodness of fit tests, and consistency tests. Given two subsets $\mathbf{d}$ and $\mathbf{d}^{\prime}$ of the full data vector, the distribution $P(\mathbf{d}^{\prime}_{\rm sim}|\mathbf{d}_{\rm obs})$ of simulated realizations of the latter is considered, given observations for the subset of the former, $\mathbf{d}_{\rm obs}$.  This distribution can be sampled and those samples can be compared to actual observations of $\mathbf{d}^{\prime}$, denoted $\mathbf{d}^{\prime}_{\rm obs}$. This comparison provides a meaningful way for evaluating the consistency of $\mathbf{d}^{\prime}$ with $\mathbf{d}$, given a single model.
To sample the distribution, a sample from the posterior of model parameters is first obtained given observations for $\mathbf{d}$, $P(\textbf{p}|\mathbf{d}_{\rm obs})$. Then, for each such \textbf{p}, the distribution $P(\mathbf{d}^{\prime}_{\rm sim}|\mathbf{d}_{\rm obs},\textbf{p})$ is sampled. 
Two cases are considered. In goodness-of-fit tests, $\mathbf{d}$ and $\mathbf{d}^{\prime}$ refer to independent realizations of the same subset, such that $P(\mathbf{d}^{\prime}_{\rm sim}|\mathbf{d}_{\rm obs},\textbf{p})=P(\mathbf{d}^{\prime}_{\rm sim}|\textbf{p})$ reduces to the likelihood of $\mathbf{d}^{\prime}_{\rm sim}$ at parameters $\textbf{p}$. For consistency tests, $\mathbf{d}$ and $\mathbf{d}^{\prime}$ refer to disjoint subsets of a single realization of the full data vector. In this case, $P(\mathbf{d}^{\prime}_{\rm sim}|\mathbf{d}_{\rm obs},\textbf{p})$ is the conditional likelihood, that is, the distribution of $\mathbf{d}^{\prime}_{\rm sim}$ at parameters $\textbf{p}$ given observations for $\mathbf{d}_{\rm obs}$. Here  a multivariate Gaussian likelihood is assumed, so the conditional likelihood is also multivariate Gaussian, with shifted mean and covariance accounting for the correlation between the subsets $\mathbf{d}$ and $\mathbf{d}^{\prime}$ \citep[see][for details]{y3-inttensions}. 
To perform the comparison of the results, a statistic is defined to compare PPD realizations and observations of $\mathbf{d}^{\prime}$, integrated over parameter space. For this, $\chi^2(\mathbf{d}^{\prime},\textbf{p})$ is used, defined with respect to the model at parameters $\textbf{p}$. This statistic determines the statistical significance of consistency through a $p$-value, which is given by the fraction of parameter samples $\textbf{p}$ for which $\chi^2(\mathbf{d}^{\prime}_{\rm sim},\textbf{p}) > \chi^2(\mathbf{d}^{\prime}_{\rm obs},\textbf{p})$.

As demonstrated in \citep{y3-inttensions}, this choice of test statistic may yield overly conservative $p$-values, especially in the case where parameter posteriors from $\mathbf{d}$ and $\mathbf{d}^{\prime}$ are very different. For this reason, a calibration procedure is followed as outlined there, based on repeated tests for simulated data vectors consistently drawn at the fiducial cosmology. A threshold of $\tilde{p}>0.01$ is defined as consistency, where $\tilde{p}$ is the calibrated $p$-value. Note that the calibration has little impact for the particular set of tests presented here.

An alternative to data-space consistency metrics like the PPD is to consider parameter-space consistency metrics. In parameter space, posteriors obtained with the full data vector are compared  to those obtained with subsets are shown and parameter shifts are reported. It is important to note that these shifts should be treated with caution: we do not expect identical constraints as the subsets of the data, and therefore their posterior distributions, are, on the one hand, correlated, and on the other, subject to noise.

\subsection{Tomographic redshift bins}
\label{sec:testbinning}

In this section, we investigate the impact of removing individual redshift bins from the analysis, as well as the two low and high redshift bins jointly. For the analysis with `Bin 1 removed', we remove all auto- and cross-correlations involving redshift bin 1 (that is, 1-1, 1-2, 1-3, 1-4), and for other cases correspondingly. 
This test captures potential inconsistencies across redshift bins, which may arise both due to deviations from \LCDM as well as redshift-dependent systematics. Deviations from the model, in particular from the non-linear power spectrum (e.g. the effect of baryons) and intrinsic alignments, preferentially impact the low-redshift bins, which have the lowest cosmological signal-to-noise. Thus, these consistency checks complement the tests of model robustness in the companion paper \citep{y3-cosmicshear2}. On the other hand, the sources of uncertainty in the redshift distributions described in Section~~\ref{sec:sompz}, such as the uncertainty in the photometric calibration or in the choice of redshift sample, are largest for the lowest and highest redshift bins, respectively. Furthermore, observational shear systematics, such as blending or PSF uncertainty, may have a larger impact on higher redshift bins with predominantly fainter and smaller galaxies. The PPD test, in particular, addresses the question of the consistency of observations in one redshift bin with respect to predictions given observations in other redshift bins (and our modeling of \LCDM).

Removing any one redshift bin at a time is found to produce consistent constraints within $\sim$0.3$\sigma$ in the $S_8$ parameter. 
For each of these configurations, we compare data for the redshift bin that was removed, with PPD realizations for predictions of that bin by the others. We find each subset to pass the chosen threshold, with $\tilde{p}$-values of $\tilde{p}_{1-4}=[0.357, 0.394, 0.014, 0.427]$.
These are discussed in Appendix~\ref{App:IC}.
Only bin 3 is found to yield a $\tilde{p}$-value close to our consistency threshold. It is notable that firstly, this bin has the smallest reported uncertainty on the mean redshift and secondly, when discarded from the analysis, we observe the most significant change of the intrinsic alignment $a_1-a_2$ parameter posterior, then centered on zero, as shown in Figure~\ref{fig:ICfull}.

We inspect the consistency when both redshift bins 1 and 2 are removed from the analysis (Test 16 `High-$z$'), or both bins 3 and 4 (Test 15 `Low-$z$'). 
Figure~\ref{fig:consistency} (left panel) overlays constraints derived from the Fiducial analysis (green) with those obtained when using only the two low-$z$ (yellow) or two high-$z$ (red) bins. We find consistency within 
$\sim1\sigma$, with the high-redshift variant preferring a lower value of $S_8$ by 0.8$\sigma$. When repeating PPD tests removing either half, we find that predictions of the low-redshift bins derived from high-redshift observations over-predict the data, albeit with a large uncertainty pertaining to the loss of constraining power on the intrinsic alignment parameters. We obtain $\tilde{p}$-values of 0.993 and 0.207, showing general good agreement in data space.

It is difficult to predict, either from first principles or empirically, how the intrinsic alignment contamination in DES Y3 should evolve with redshift  \citep[see e.g.][]{joachimi10}. Intrinsic alignments are known to depend significantly on luminosity and color, and therefore, redshift \citep[see the discussion in][]{Troxel2018,samuroff18}. In this analysis, a power law redshift scaling is adopted to capture this effect. In practice, therefore, the intrinsic alignment constraint requires accurate knowledge of the redshift distributions and their errors. In order to ensure that systematic errors in the data are not absorbed by the intrinsic alignment model, 
we investigate the consistency of astrophysical parameters, discussed in Appendix~\ref{App:IC} \citep{Efstathiou18}.

\subsection{Angular ranges}
In the Fiducial analysis, any scales from the $\xi_{\pm}$ data vector that have a fractional contribution from baryonic effects, as predicted by the OWLS `AGN' simulation, that exceeds $\Delta\chi^2<0.5$.  This is summarised in Section~\ref{sec:model}, and detailed in Ref.~\citep{y3-generalmethods}. The procedure removes a large number of data points at small scales, particularly in $\xi_{-}$ where the impact of baryonic physics is larger. On the other hand, large scales might be sensitive to some unaccounted for additive systematics, as described in Section~\ref{sec:modelm}.

Here, we inspect the consistency of the angular small scale and large scale contributions to the data vector, by comparing their respective cosmological constraints. As the analysis does not marginalise over baryonic feedback, and as we subtract a bias due to residual mean shear in the signal, this test ensures the robustness of the scale cuts applied. The split is chosen to be at $\theta=\SI{30}{\arcminute}$ for $\xi^+$ and $\theta=\SI{100}{\arcminute}$ for $\xi^-$, to roughly balance the constraining power between the subsets. Results are shown in the $\Omega_{\rm m}-S_8$ plane in Figure~\ref{fig:consistency}, central panel, with the Fiducial analysis as the green outline, along with results from the smaller (yellow) and larger (red) scale selection. We find consistent results with respect to the Fiducial analysis, within $\sim0.5\sigma$ in the $S_8$ parameter. The two PPD tests comparing large- and small-scale data pass, with $\tilde{p}$-values of 0.66 when comparing data at large scales with predictions from small-scale data, and $\tilde{p}$=0.083 for the opposite test, as reported in  Table~\ref{tab:IC}.

An alternative approach to account for small-scale baryonic feedback effects is to preserve the small-scale information and marginalise over any model uncertainty with \textsc{HMcode} \citep{mead2015}. To this end, an approach is investigated in \citep{y3-cosmicshear2} that uses this modeling and maintains the Fiducial scale cuts. This finds consistency within $1\sigma$, with a small preference for higher values of $S_8$ (compared to the slightly lower $S_8$ preferred by the large scales-only analysis) that is attributed to prior volume effects with unconstrained baryon nuisance parameters. Overall, given the conservative choice of scale selection and these complementary robustness tests, we deem the Fiducial constraints robust to small-scale systematics.

\begin{figure}
    \centering
-   \includegraphics[width=.47\textwidth]{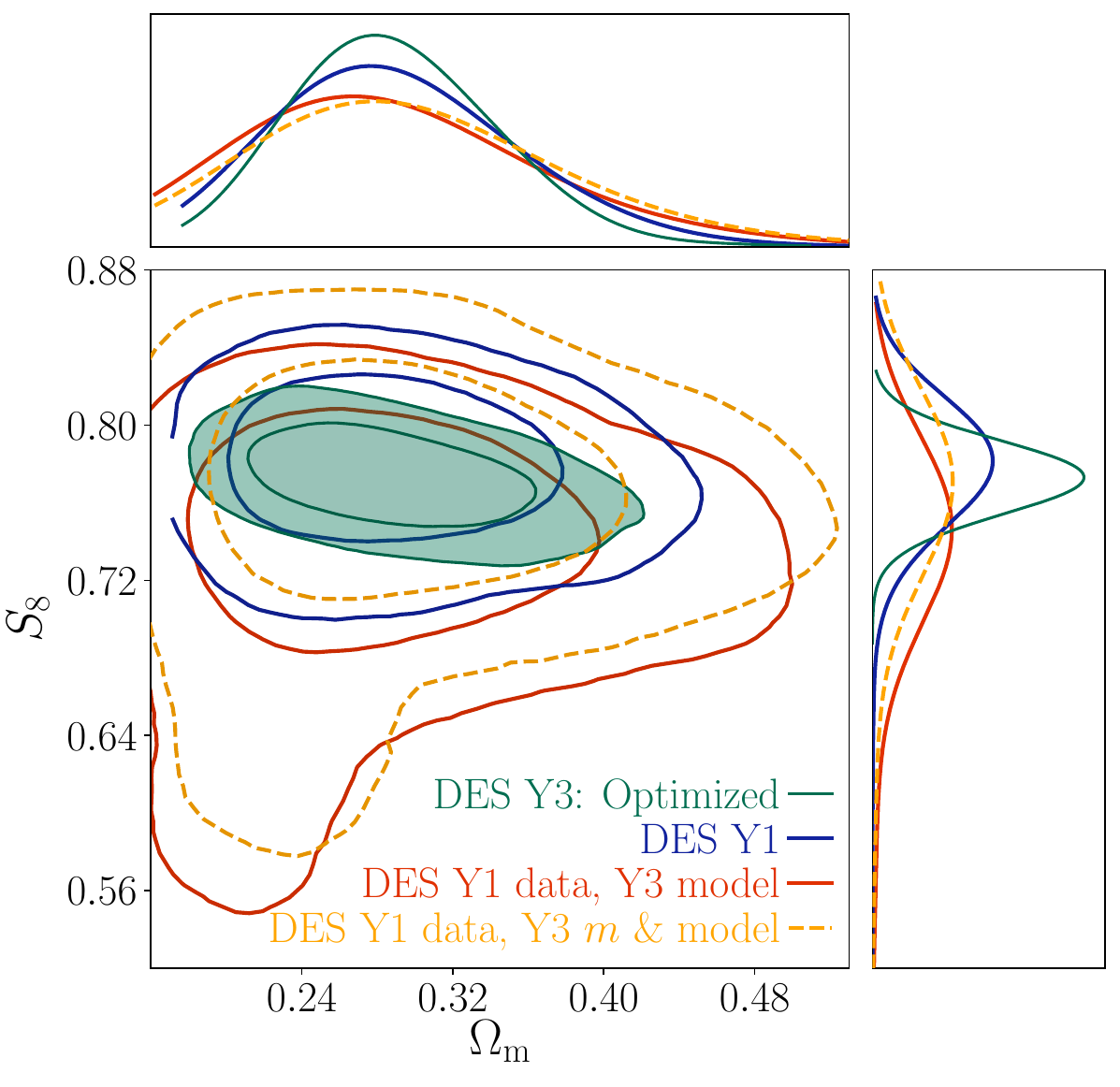}
    \caption{\label{fig:y1} Comparison to the DES Y1 posteriors: The \lcdm constraints in the $\Omega_{\rm m}-S_8$ plane for the Y3 \lcdm-Optimized (green) are compared to DES Y1, as published in Ref.~\citep{Troxel2018} (blue). For a more robust comparison, Y1 reanalysed with the Y3 modeling choices is shown in red (in particular, the intrinsic alignment model switching from NLA to TATT). Within this homogeneous modeling framework, we find the Y3 \lcdm-Optimized constraints to be consistent within $\sim0.5\sigma$ and substantially more precise than the Y1 case, with $2\times$, $1.24\times$ and $1.22\times$ smaller uncertainty on $S_8$, $\Omega_{\rm m}$ and $\sigma_8$, respectively. In yellow, the Y3 modeling choices are again used, as well as the better-informed Y3 shear calibration correction. The Y3 \lcdm-Optimized $S_8$ result, when compared to this, is $2.2\times$ more constraining and consistent with those from Y1 within 0.25$\sigma$ in $S_8$.}
\end{figure}

\subsection{$\xi_+$ versus $\xi_-$}

We test the consistency of the two components of the cosmic shear two-point function, $\xi_+$ and $\xi_-$. The motivation for this split is that $\xi_-$ is more sensitive to smaller scales than $\xi_+$, and is therefore more likely to be impacted by any unaccounted for baryonic feedback and intrinsic alignments. In addition, residual B-modes, sourced by either observational or astrophysical systematics, or higher-order shear effects, are more likely to be prominent on small-scales. Considering $\xi_+$ measurements conditioned on $\xi_-$ observations with a PPD test, we find a $\tilde{p}$-value of 0.262, and 0.451 for the opposite test. When considering the $S_8$ parameter, we find consistency within $\sim0.5\sigma$, with posteriors from these subsets of the data shown in Figure~\ref{fig:consistency} (right panel).

\begin{figure*}
\centering
\begin{minipage}{.33\textwidth}\hspace{1em}
  \centering
  \includegraphics[width=.98\textwidth]{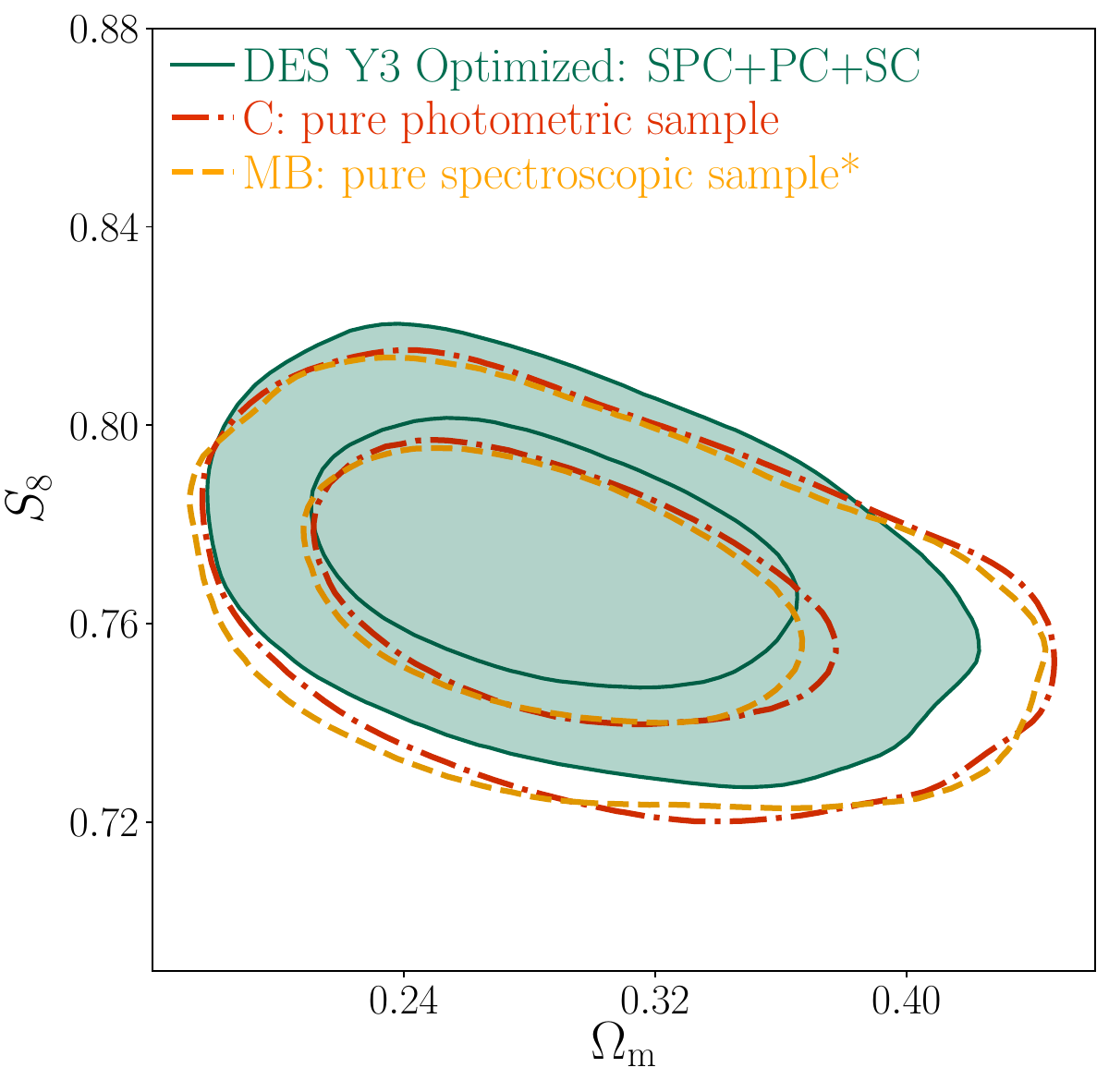}
\end{minipage}%
\begin{minipage}{.33\textwidth}\hspace{1em}
  \centering
  \includegraphics[width=.98\textwidth]{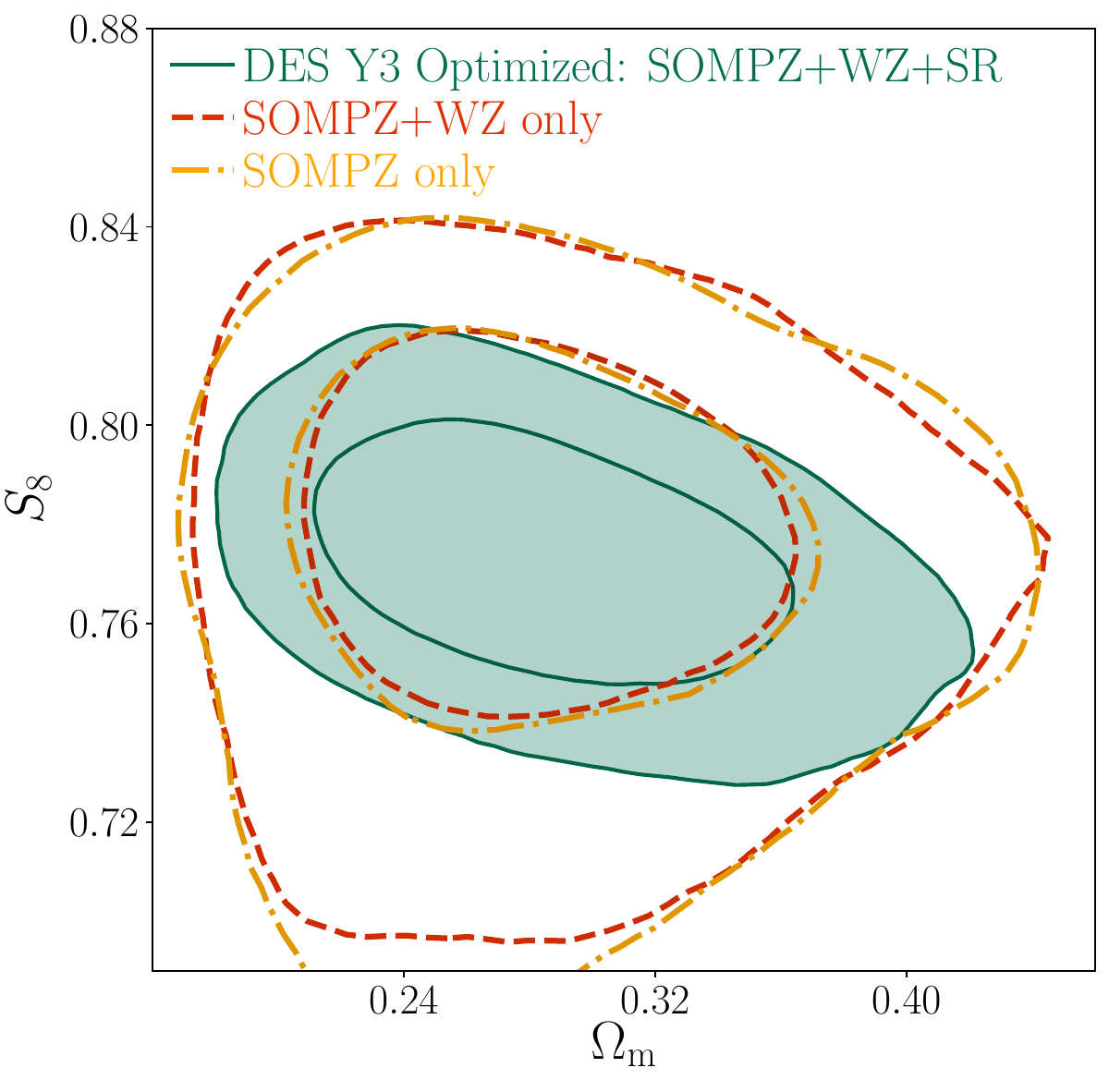}
\end{minipage}
\begin{minipage}{.33\textwidth}\hspace{1em}
  \centering
  \includegraphics[width=.98\textwidth]{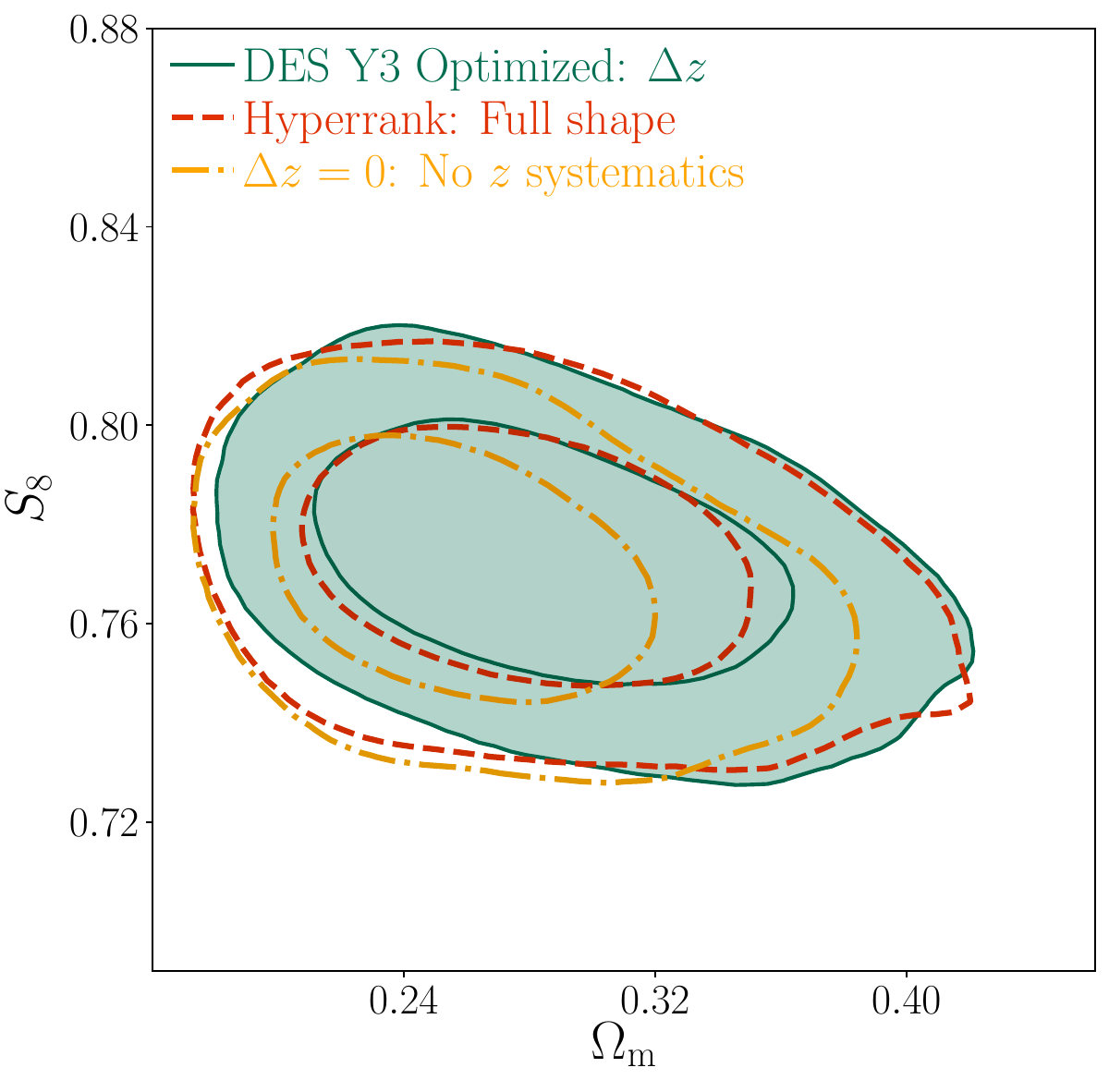}
\end{minipage}
\caption{\label{fig:zuncertainty} Robustness to redshift calibration for the \lcdm-Optimized analysis: A comparison of \lcdm constraints in the $\Omega_{\rm m}-S_8$ plane for alternative redshift choices. In all panels, the filled green contours show the optimised analysis. On the left, it is shown in comparison to `pure' redshift calibration samples, \textsc{C} (red), derived purely from \textsc{COSMOS} and \textsc{MB} (yellow), the `maximally biased' spectroscopic sample, showing the (in)sensitivity of the analysis to spectroscopic vs. photometric training samples. In the middle, the tiers of the methodology: \textsc{SOMPZ}, \textsc{WZ} and \textsc{SR}, are stepped back, showing the consistency of the likelihoods from each method. For this analysis, the \textsc{SR} information is effective, finding 25\% tighter $S_8$ constraints, with the improvement primarily attributed to the breaking of bimodal intrinsic alignment posteriors, which are degenerate with $S_8$, as well as constraining the redshift uncertainty. On the right, the fiducial modeling of the uncertainty in the redshift calibration that accounts for only shifts in the mean of each redshift bin, $\Delta z$, is compared to an analysis where the full-shape uncertainty is accounted for by varying $n(z)$ realisations with \textsc{Hyperrank} (red). A constraint that neglects any redshift calibration correction is shown in yellow.}
\end{figure*}

\subsection{Previous DES results}
In this section we investigate the consistency of the DES Y3 constraints with DES Y1 cosmic shear results from \citep{Troxel2018}. In Figure~\ref{fig:y1}, the $\Omega_{\rm m}-S_8$ constraints are overlaid. Overall, we find that our results are in agreement with the Y1 constraint to $0.5\sigma$ in $S_8$. We caution that a comparison of the two results is not straightforward, as analysis choices differ. More specifically, the Y1 analysis used the simpler NLA intrinsic alignment model, rather than TATT, the baseline choice for the Y3 analysis. Furthermore, as the two shape catalogs share a non-negligible fraction of imaging data, the correlation between DES Y3 and Y1 is difficult to quantify, which limits a real assessment of the statistical significance of any deviation. 

For a comparison that is more on equal footing, we re-analyze the DES Y1 measurements with the DES Y3 analysis pipeline, preserving the DES Y1 observational systematic choices and priors (red). This results in significantly degraded constraints and an, albeit not significantly, lower $S_8$ posterior compared to the published Y1 result (blue). This is primarily attributed to the change in the intrinsic alignment model from NLA to TATT, with a similar effect seen in Figure 15 of Ref.~\citep{Troxel2018}. Within this homogeneous modeling framework, we find the Y3 \lcdm-Optimized constraint to be consistent within $\sim0.5\sigma$ and substantially more precise than the Y1 case, with $2\times$, $1.24\times$ and $1.22\times$ smaller uncertainty on $S_8$, $\Omega_{\rm m}$ and $\sigma_8$, respectively. Furthermore, the Y1 data vector is re-analyzed with the Y3 shear calibration priors to give the shaded blue posteriors. The significant improvements in the realism of the image simulations for Y3, as well as more sophisticated understanding and modeling of the effects of blending, suggest that the Y3 calibration is likely to be more accurate for both the Y3 and the Y1 data set. These Y3 multiplicative corrections result in a shift toward higher values of $S_8$. Compared to this Y1 re-analyzed constraint, the \lcdm-Optimized Y3 cosmic shear result is consistent within $\sim0.25\sigma$ in  $S_8$, with a factor of $2.2\times$ improvement in the precision on $S_8$. 

\section{Robustness to redshift calibration}
\label{sec:robustredshift}

Determination of the true ensemble redshift distribution is critical for cosmological weak lensing analyses. To lowest order, lensing is primarily sensitive to the mean redshift and the width of the redshift distribution of each bin \citep{amara07}.

In this section we show how different choices for the redshift distribution impact the results. In particular, we demonstrate the robustness to (A) the choice of redshift sample, (B) the redshift methodology, and (C) the modeling of redshift uncertainties. These tests are shown for the higher precision \lcdm-Optimized analysis in Figure~\ref{fig:zuncertainty}. Corresponding tests for the Fiducial analysis are summarised in Table~\ref{tab:summary} and Figure~\ref{fig:summary} as Tests 1-8.

\subsection{Redshift sample}
\label{sec:testredshift}
In Section~\ref{sec:zsamples}, we summarise the choice of redshift samples that form the basis of the ensemble $n(z)$, which we consider to be complete and to span any photometric-spectrosopic differences. However, it can be argued that this framework underestimates the uncertainty as it does not span the extremities of available redshift information: neither an $n(z)$ derived solely from \textsc{COSMOS} , nor one solely from spectroscopic information are included.
In this section, we test the sensitivity of our analysis to these extremities by analysing the data with an $n(z)$ that is purely based on \textsc{COSMOS} , $C$. On the other hand, since a solely spectroscopic-based $n(z)$ cannot be directly calibrated for DES data without substantial selection biases, we use an artificially-constructed training sample. To this end, we modify the redshift sample that is least reliant on \textsc{COSMOS}  redshifts, such that the 10\% of weighted  information that does still derive from this sample is manipulated to reflect the \textsc{COSMOS} -spectroscopic biases calibrated per magnitude bin. This \textsc{MB} sample can be deemed as most different to $C$. 

As a robustness test, we substitute $n(z)$ with the `pure' redshift samples. Figure~\ref{fig:zuncertainty} (left panel) shows results in the $\Omega_{\rm m}-S_8$ plane. The two extreme choices of redshift calibration sample, used in the analysis of the red and yellow contours, show that no significant difference in cosmological parameters. Even the fractional shift compared to the baseline case in this plot are due to minor differences in processing between the samples. The figure here uses the most constraining, $\Lambda$CDM-Optimized scales. A comparison for the case of the less-constraining Fiducial analysis is shown in Tests 1 and 2 in Figure~\ref{fig:summary} and Table~\ref{tab:summary}).

The test indicates that the DES Y3 redshift methodology of using the multi-band deep information, building complete redshift samples, and accounting for an uncertainty due to the choice of redshift sample, yield cosmological constraints robust to this choice, even at the \lcdm-Optimized precision. The differences are substantially smaller than previous analyses had indicated for the \textsc{COSMOS} -spectroscopic calibration \citep{hildebrandt20, joudaki2020}. We note that this is consistent with tests done on the DES Y1 calibration of \citep{hoyle} post-unblinding that avoid selection biases, including those in \citep[][their appendix A]{joudaki2020}.

\subsection{Redshift method}
The redshift estimation for the DES Y3 analysis uses a combination of approaches. As described in Section~\ref{sec:zmethod}, the information is drawn jointly from the flux/color self-organising map-based method, \textsc{SOMPZ}, the clustering redshift method, \textsc{WZ}, and the shear-ratio method, \textsc{SR}. First, it is shown in \citep{y3-sompz} that the posteriors on $\langle z\rangle$ from these methods are consistent, before their information is combined at different points in the analysis pipeline, with the \textsc{SR} incorporated at the point of evaluation of cosmological likelihoods. For robustness, we test the impact of including each of these methods on the cosmological parameter level separately. 

The central panel of Figure~\ref{fig:zuncertainty} shows the \lcdm-Optimized analysis using the $n(z)$ created by \textsc{SOMPZ} (yellow) and that informed by \textsc{SOMPZ+WZ} jointly (red), compared to the \lcdm-Optimized result, which combines \textsc{SOMPZ+WZ+SR}. We find that the $S_8$ constraints do not shift when each method is included and are consistent, and thus that our results are robust to these variants in the methodology. The analogous test for the Fiducial analysis is shown in Tests 3 and 4 in Figure~\ref{fig:summary} and Table~\ref{tab:summary}).

In support of that, we show the posteriors on the $\Delta z$ systematic parameters for these variant analyses in Appendix~\ref{App:IAnz}. The \textsc{WZ} analysis requires marginalizing over flexible models of the redshift evolution of clustering bias of the weak lensing sources, which are largely degenerate with $\langle z \rangle$ \citep{y3-sourcewz, vandenbusch2020}. As a result, the \textsc{WZ} primarily constrains the shape of the redshift distribution, rather than the mean, and has relatively little impact on cosmological posteriors. For this analysis, the \textsc{SR} information is effective, finding 25\% tighter constraints with the improvement primarily attributed to the breaking of degeneracies in the TATT intrinsic alignment model posteriors, which are bimodal (see Appendix~\ref{app:power}), rather than a substantially tightened posterior on redshift distributions \cite{y3-shearratio}. Indeed, as discussed in that appendix, when a less-conservative model for intrinsic alignments is used, our cosmological posteriors with and without the inclusion of \textsc{SR} remain consistent. In that test, while the $a_1$ posterior is significantly tighter with the inclusion of \textsc{SR}, as that parameter is de-correlated with $S_8$, the impact on the $S_8$ constraint is negligible. In addition, Appendix~\ref{App:IAnz} demonstrates the robustness of the posteriors to choices in the \textsc{SR} method. More specifically, it compares to an analysis that uses the alternative lens galaxy sample, as well as one that uses \textsc{SR} measured from large scales only.

\subsection{Modeling redshift uncertainty}
For the Fiducial and \lcdm-Optimized analyses, the uncertainty in the redshift estimation is incorporated as an uncorrelated shift in the mean redshift of each bin. The latter is computed as the spread spanned by the ensemble of $n(z)$ realisations that result from the DES redshift calibration and image simulation studies and is modeled as a Gaussian prior on $\Delta z$. As discussed in Section~\ref{sec:modelz}, while the DES Y3 methodology is capable of sampling the full realisations in the likelihood analysis with \textsc{Hyperrank}, thereby capturing variations in the shape of the $n(z)$ as well as correlations between redshift bins, we do not include that more accurate framework in the baseline analysis. As this modeling choice was made based on a weighing of the impact of this approximation on posteriors in a simulated analysis, versus increased run-time,  we test the robustness of the decision here. We analyse the data using the full ensemble of realisations and \textsc{Hyperrank}.  To assess the overall impact of the redshift uncertainty on the cosmological constraints, we perform an analysis ignoring these nuisance parameters, using only the mean of the $n(z)$ realisations and setting $\Delta z=0$ for all bins. 

Figure~\ref{fig:zuncertainty} (right-hand panel) demonstrates  consistency in the  $\Omega_{\rm m}-S_8$ plane for the \lcdm-Optimized case: modeling only the mean of the redshift distribution (green) sufficiently captures the effect of redshift bias uncertainty, as analysed with \textsc{Hyperrank} (yellow). The full-shape constraint is only marginally degraded, and is unbiased, illustrating that uncertainties in the shape of the redshift distribution are sub-dominant for cosmic shear at the current statistical precision. Furthermore, it shows that in this plane the $n(z)$ are calibrated to sufficient accuracy and precision, as cosmological constraints are not significantly impacted ($\sim0.3\sigma$) by not marginalizing over redshift calibration (red). The analogous test for the Fiducial analysis is shown in Tests 7,8 in Figure~\ref{fig:summary} and Table~\ref{tab:summary}. In addition, we demonstrate the validity of the claim of consistency when considering the intrinsic alignment parameters and the effective redshift parameter constraints in Appendix~\ref{App:IAnz}. 

\section{Robustness to blending and shear calibration}\label{sec:robustshear} 

The amplitude of the cosmic shear signal and thus the inferred $S_8$ parameter depends directly on the multiplicative shear calibration. Accurate shear calibration relies on highly realistic image simulations that sufficiently match the properties of the data. Ref.~\citep{y3-imagesims} finds blending to be the dominant contribution to the mean multiplicative bias of Y3 shape catalogs, at approximately $-2$\%, and finds that the magnitude of this correction increases with redshift. We infer that in the presence of object blending, a systematic that is more prominent in deeper data, it is important that multi-band simulations allow for the redshift analysis applied to the data to be repeated on simulations in order to capture the coupled effects on both shear and photometric redshift calibration. This cosmic shear analysis is the first to account for the effect of blending jointly on shear and redshift calibration.  In this section we illustrate how these advancements in the methodology impact the cosmology constraints.

\begin{figure}
    \centering
    \includegraphics[width=0.48\textwidth]{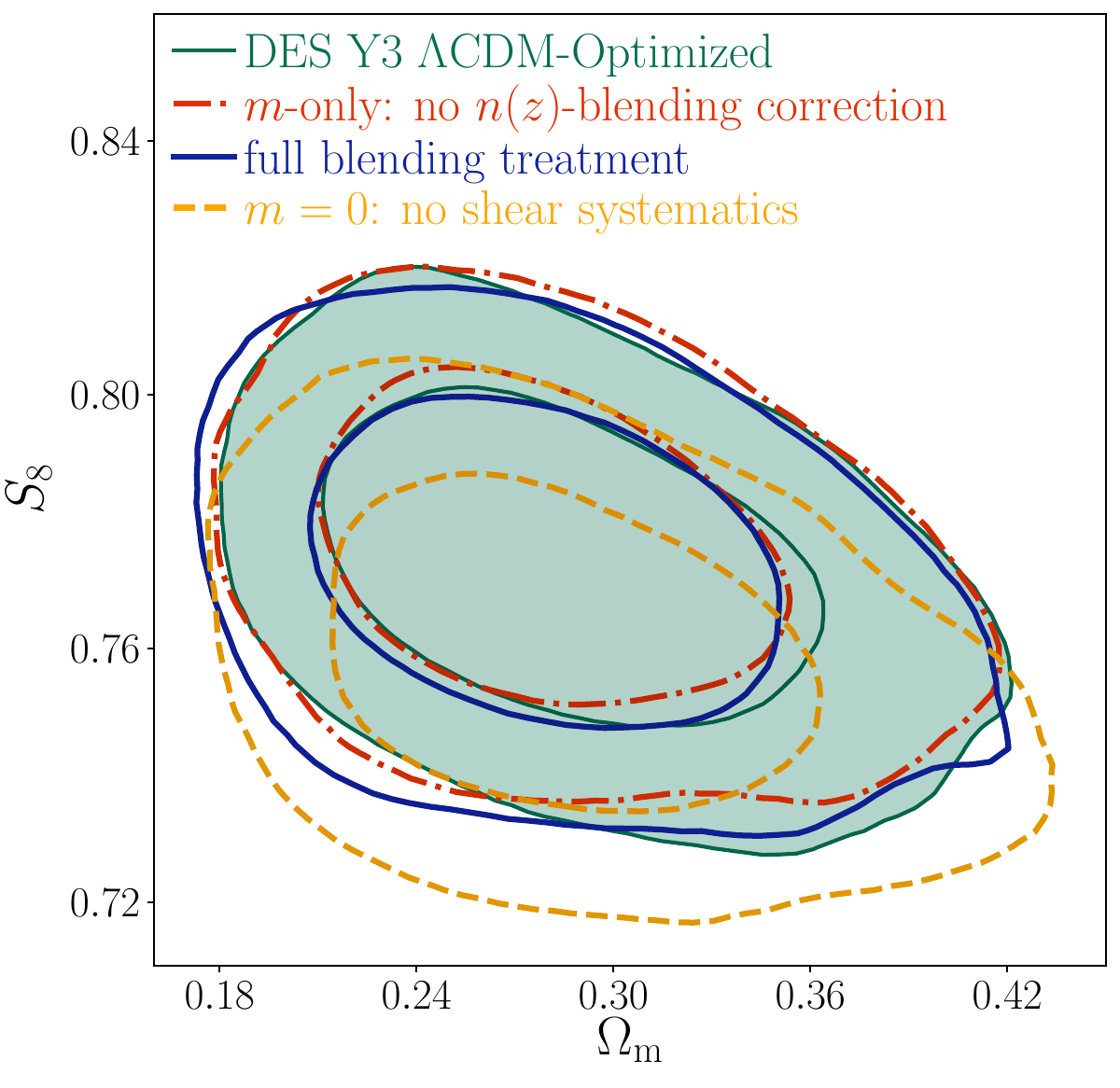}
    \caption{\label{fig:blending} Robustness to blending: A comparison of \lcdm constraints in the $S_8-\Omega_{\rm m}$ plane for varying complexity in shear calibration modeling. The shaded green contours show the \lcdm-Optimized analysis. Also shown are variants that neglect the redshift-dependent impact of blending (red), that ignores any shear calibration systematics (yellow), as well as one that uses the full blending treatment, including correlations between shear and full-shape redshift calibration, analysed with \textsc{Hyperrank} (blue).} 
\end{figure}

Figure~\ref{fig:blending} shows a variant of the \lcdm-Optimized analysis that ignores the redshift-mixing blending impact, `m-only: no n(z)-blending correction' (yellow), by neglecting to account for the correction on the ensemble and mean $n(z)$. This analysis uses the uncorrected $n(z)$ and the priors derived for each redshift bin from the redshift-independent constant shear simulation, $\Delta z_{\rm no z-blend}$ and $m_{\rm no z-blend}$, quoted in Table~\ref{tab:datastats}. Comparing this variant to the \lcdm-Optimized analysis (green) in the $\Omega_{\rm m}-S_8$ plane, we find consistency. The redshift-mixing effect is subdominant for the DES Y3 cosmic shear cosmology. On the other hand, neglecting to account for shear calibration uncertainty entirely (red) incurs a significant bias ($0.5\sigma$) toward a low $S_8$.

Uncertainties on shear and redshift calibration are correlated due to the limited volume of image simulations that impact both. The baseline analysis ignores any such correlations. In order to test the robustness of this assumption, an analysis is performed that uses \textsc{Hyperrank} to sample over $n_\gamma(z)$ realisations, which naturally include both multiplicative bias-type and $n(z)$-type biases, labelled `full blending treatment'. The cosmological posteriors from this variant, Test 11, are compared to those in Figure~\ref{fig:blending}, and are shown to be consistent. 

Finally, we test the impact on the cosmological analysis of an additional, unaccounted for shear calibration uncertainty. While the image simulations used to inform these choices are well-matched to the data, they do not account for some effects such as clustering with undetected sources, which can contribute additional blending effects. We assume this to take the form of a 1\% effect that is fully correlated across redshift bins. This variant is found to produce results that are totally consistent with the Fiducial analysis within, indicated by Test 10 in Figure~\ref{fig:summary}.

\section{What limits lensing cosmological precision?}\label{sec:syslimit}

The cosmic-shear measurements are known to be altered from simple theoretical predictions by various systematic effects. The importance of work to mitigate systematics is two-fold: first, these must be calibrated accurately as to not bias cosmological parameters, but second, they must be sufficiently controlled such that they do not dominate the error budget, thereby limiting precision. Of interest for this and future experiments are the questions: how much has the cosmological accuracy been degraded by treatment of systematic effects?  Which effects dominate the uncertainty?  How much could lower-noise measurements improve cosmological accuracy in the presence of these systematic effects?

\begin{figure}
    \centering
    \includegraphics[width=0.49\textwidth]{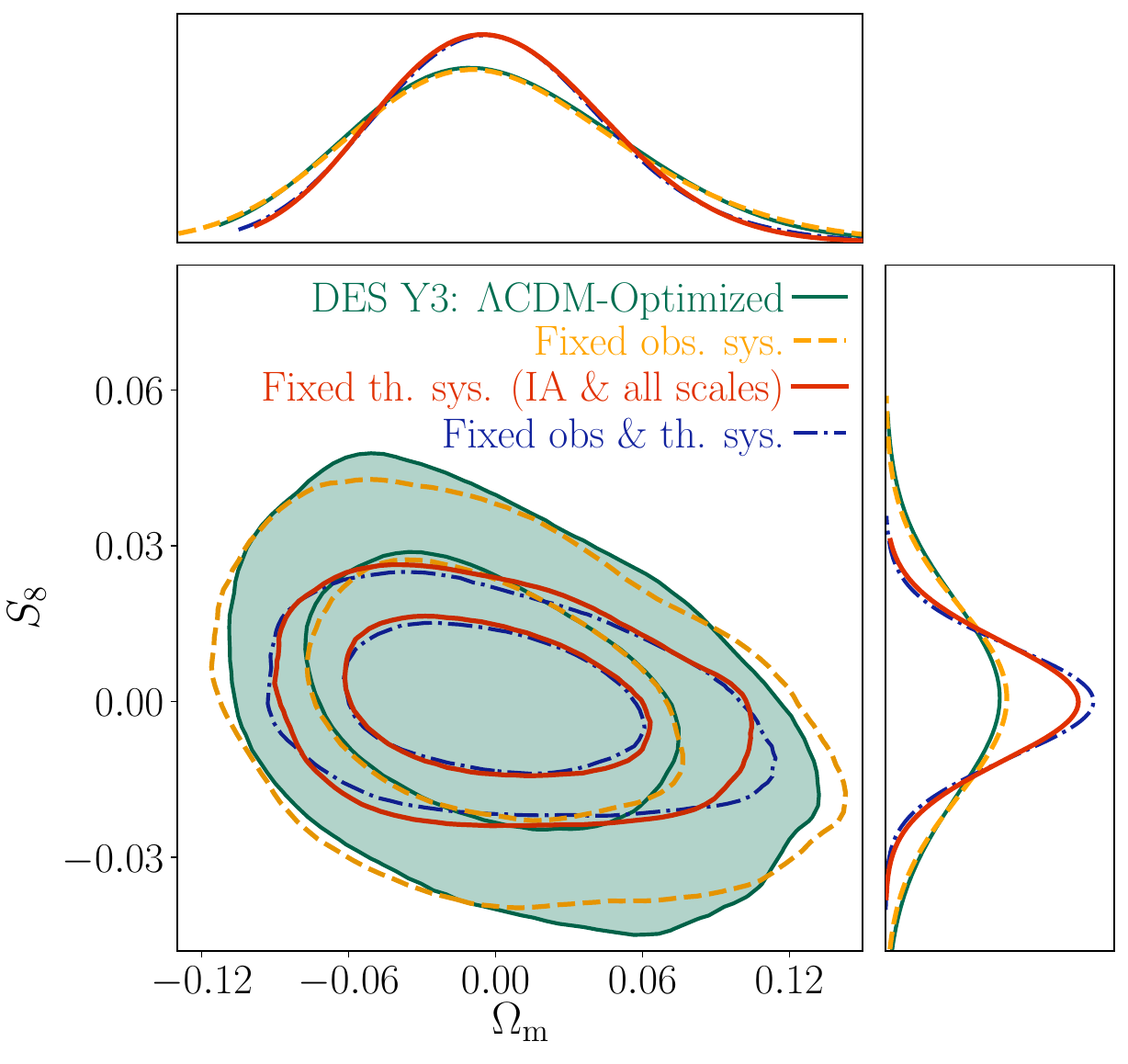}
    \caption{\label{fig:syslim} Systematics limiting cosmology: A comparison of the size of \lcdm-Optimized constraints in filled green contours to those where observational systematics, both shear and redshift priors, are fixed (yellow, dashed) and theoretical systematics are `switched off', by fixing intrinsic alignment parameters and including using the full scale-set that is limited to mitigate baryonic effects (red, solid). The analysis is not limited by observational systematics, but we find an improvement in of $\times$1.8 in $S_8$ uncertainty when fixing theoretical systematics. The constraint fixing all systematics (and therefore also using the full scale-set) is shown in blue. The mean is individually subtracted from every constraint for ease of comparing the constraining power. Both 68\% and 95\% confidence levels are shown. }
\end{figure}

Our baseline analysis marginalizes over 19 systematic parameters in addition to the 6 cosmological parameters. Of course,  this marginalization is necessary --- ignoring systematic effects, as is done in the Tests 13 and 14 of Figure~\ref{fig:summary}, will produce biased inferences.  In this investigation, we are not interested in the incurred biases. As such, we plot posterior distributions centered at zero to understand how `shutting off' the effects changes the size of posterior uncertainties in the $S_8-\Omega_{\rm m}$ plane, as shown in Figure~\ref{fig:syslim}.

We divide the systematic effects into two classes. First, we consider \textit{observational} systematics, namely the redshift-distribution and shear-measurement calibrations investigated in detail in this paper.   The uncertainties in these calibrations are propagated by marginalizing over relevant nuisance parameters, which expands the posterior distribution on cosmological parameters of interest---we will focus here on $S_8$.  The impact of observational systematics on the fiducial $S_8$ constraint can be illustrated by comparing to the case where the relevant nuisance parameters are fixed. We will refer to this case as `Fixed observational systematics'.

The second class is astrophysical or \textit{theoretical systematics}, which arise from our incomplete knowledge of non-linear processes in the Universe. This includes intrinsic alignments, baryonic physics, and growth of structure in the nonlinear regime, which are investigated in the companion paper \citep{y3-cosmicshear2}. The impact of marginalising over the intrinsic alignment model on the posterior uncertainty can be tested by fixing its parameters.  The baryonic/non-linear effects are ameliorated in the analysis by imposing scale cuts on the data vector to retain only elements for which the model is believed to be secure.  In other words, the intrinsic precision of the data is degraded in order to maintain the accuracy of the model.  To investigate the loss of constraining power from this, we produce posteriors that use the full $\xi_\pm$ data vector.  We have already seen that moving from Fiducial to \lcdm-Optimized scale selections reduces uncertainties on $S_8$ by $\approx1.5\times.$  We will refer to the combination of fixing intrinsic alignment parameters and expanding the $\xi_\pm$ data vector to all measured scales as `Fixed theory systematics.'

From the starting point of the \lcdm-Optimized analysis, we find that fixing observational systematics (yellow, dashed) produces no significant improvement in the $S_8$ posterior uncertainty.  Thus, the current uncertainties in shear and redshift calibration are small enough that they do not degrade the results and we are not limited by observational systematics.  On the other hand, fixing only the theoretical systematics (red) yields a substantially narrower $S_8$ constraint, by roughly a factor $\sigma_{\rm opt}/\sigma_{\rm fix th.}=1.8$. Therefore, the theoretical uncertainties are costing about 2/3 of the Y3 power (in terms of variance) on the cosmological parameters, with intrinsic alignment uncertainty contributing slightly more than baryon-driven scale cuts in limiting the cosmological precision. This suggests that future decreases in measurement noise in $\xi_\pm$ may not lead to concomitant decreases in cosmological uncertainties.  The precise balance will depend upon 
how many more modes have to be sacrificed to keep modelling accuracy below measurement noise, and on how well the intrinsic alignment models can self-calibrate from data. We note that this is a different scenario to the KiDS-1000 analysis, which is found to be statistics-limited 
\citep{asgari20}. That analysis uses the simpler and more constraining NLA intrinsic alignment model, COSEBIs measurements, including small angular scales, and marginalises over non-linear evolution.

Fixing both observational and theoretical systematic errors leads to the blue posteriors in Figure~\ref{fig:syslim}.  A small improvement is seen over the fixed-theory result ($\times1.9$ versus $\times1.8$ improvement, compared to the \lcdm-Optimized case), i.e. the current state of knowledge on observational errors would suffice for a DES Y3 cosmic-shear measurement with negligible theory errors.
The work on DES observational systematics, presented in a series of papers \citep{y3-balrog,y3-deepfields,y3-sompz,y3-sourcewz,y3-shearratio,y3-imagesims}, has been successful in avoiding a loss of constraining power in cosmic shear.
This does not imply that innovation is not required for a similar statement to hold for any future cosmic shear analyses. In particular, in DES Y3, uncertainties related to, for example, the redshift calibration sample \citep{y3-sompz}, or blending \citep{y3-imagesims} increased with redshift bin, highlighting that deeper data will require innovation for observational systematics to stay sub-dominant. For the case of lower measurement noise, forthcoming with DES Y6 and Rubin Observatory LSST, redshift and shear calibration may become significant contributors and require improvement.

More crucial to extracting maximal information out of weak-lensing data is an improved modeling of astrophysical effects.  Progress on this systematic is substantially inhibited by the large variation demonstrated in the different hydro-simulations \citep{chisari} for the degree in which baryon feedback impacts the matter power spectrum. While there are several options for attempting to model these small-scale effects \citep{mead2015, huang2020dark,asgari_baryons}, their current level of uncertainty demands more refined hydrodynamic simulations, tailored by complementary observations such as the Sunyaev-Zel'dovich effect. It is noteworthy, however, that pushing to smaller scales will also require more accurate modeling of the PSF on these scales and accurate modeling of higher-order lensing corrections \citep{Maccrann2017}. Important too, may be the need for a principled, yet not overly conservative, modeling of intrinsic alignments.  An intrinsic alignment model choice investigation, similar to performed \textit{a posteriori} in Ref.~\citep{y3-cosmicshear2}, which finds that simpler, less-conservative models provide a sufficient fit to the DES Y3 data, could be useful if done before the data has been unblinded. This is important, as the more widely-used NLA model gives significantly reduced uncertainty on $S_8$, ($\sigma_{\rm NLA}/\sigma_{\rm TATT}=0.85$) and the amplitude of intrinsic alignment measured in Ref.~\citep{y3-3x2ptkp}, is relatively smaller than forecasted (see discussion in Section VII B. of Ref.~\citep{y3-cosmicshear2}).  Of course, these two theoretical systematics are connected -- advancements that allow for harnessing the small-scale information in the analysis will in turn allow for self-calibration of nuisance model parameters. The degradation in constraints when marginalizing over a model for intrinsic alignments motivates combined \textit{3$\times$2pt} analyses \citep{y3-3x2ptkp}. 

\begin{table*}
    \caption{\label{tab:summary}Summary of constraints on the posterior mean value of and $68\%$ confidence bounds on $S_8$, $\Omega_{\rm m}$ and $\sigma_8$ in $\Lambda$CDM, as well as the Maximum Posterior $S_8$ value (denoted $\hat{S}_8$). The $\chi^2$ per degree of freedom (dof) and FoM (defined in equation~\ref{eqn:FOM}) for the $S_8-\Omega_{\rm m}$ plane are also shown. We distinguish variations on the Fiducial model that are not required to give consistent results (for example, by neglecting observational systematics) by an asterisk and an open symbol. A visual summary of the $S_8$ constraints can be seen in Figure~\ref{fig:summary}.}
    \renewcommand{\arraystretch}{1.4}
    \setlength{\tabcolsep}{8pt}

\begin{tabular}{lcccccc}
\hline
\hline
\ Data & $S_8$ & $\hat{S}_8$ &$\Omega_{\mathrm{m}}$ & $\sigma_8$ & FoM$_{S_8,\Omega_{\mathrm{m}}}$ & $\chi^2/\mathrm{dof}$ \\ [0.1cm]
\hline
\textbf{Fiducial} & $0.759_{-0.023}^{+0.025}$  & 0.755 & $0.290^{+0.039}_{-0.063}$ &  $0.783^{+0.073}_{-0.092}$  & 927 & $237.7/222=1.07$\\

\textbf{$\Lambda$CDM-Optimized} & $  0.772^{+0.018}_{-0.017} $  & 0.774 & $  0.289^{+0.036}_{-0.056} $  & $  0.795^{+0.072}_{-0.076} $  & 1398 & $285.0/268=1.06$
  \\ 
$\Lambda$CDM-Optimized, NLA, fixed neutrino mass & $0.788^{+0.017}_{-0.016}$ & $0.775$ & $0.279^{+0.036}_{-0.053}$ & $0.825^{+0.078}_{-0.79}$ &  1501 & $288.1/270=1.07$ \\

{}  &&&& \vspace{0.02cm}  \\
\hline
C$:$ pure photometric sample * & $  0.752^{+0.025}_{-0.024} $  & 0.766  & $  0.303^{+0.043}_{-0.074} $  & $  0.760^{+0.080}_{-0.102} $  & 831 & -
  \\ 
 MB$: $pure spectroscopic sample * & $  0.752^{+0.021}_{-0.025} $ & 0.729 & $  0.298^{+0.042}_{-0.063} $  & $  0.763^{+0.075}_{-0.091} $  & 968 & -
  \\ 
 SOMPZ only & $  0.760^{+0.035}_{-0.031} $ & 0.739 & $  0.292^{+0.043}_{-0.062} $  & $  0.780^{+0.079}_{-0.089} $  & 542 & -
  \\ 
 SOMPZ$+$WZ only & $  0.758^{+0.035}_{-0.028} $  & 0.730 & $  0.298^{+0.044}_{-0.068} $  & $  0.772^{+0.079}_{-0.092} $  & 516 & -
  \\ 
 Alternative lens sample-SR & $  0.768^{+0.021}_{-0.022} $  & 0.757 & $  0.296^{+0.041}_{-0.060} $  & $  0.783^{+0.079}_{-0.091} $  & 1055 & -
  \\ 
 Large-scale-SR & $  0.777^{+0.022}_{-0.023} $ & 0.780 & $  0.331^{+0.043}_{-0.076} $  & $  0.750^{+0.083}_{-0.088} $  & 860 & -
  \\ 
 Hyperrank$:$ full redshift shape model & $  0.755^{+0.023}_{-0.018} $  & 0.744 & $  0.287^{+0.038}_{-0.055} $  & $  0.780^{+0.067}_{-0.082} $  & 1032 & -
  \\ 
 No redshift systematics * & $  0.753^{+0.023}_{-0.023} $ & 0.752 & $  0.271^{+0.032}_{-0.054} $  & $  0.801^{+0.081}_{-0.078} $  & 1111 & -
  \\ 
 No n$(z)$ blending correction & $  0.761^{+0.024}_{-0.022} $  & 0.753 & $  0.283^{+0.036}_{-0.052} $  & $  0.791^{+0.067}_{-0.082} $  & 1091 & -
  \\ 
 Additional shear uncertainty & $  0.759^{+0.023}_{-0.023} $ & 0.758 & $  0.300^{+0.044}_{-0.057} $  & $  0.767^{+0.071}_{-0.088} $  & 971 & -
  \\ 
 No shear systematics * & $  0.743^{+0.026}_{-0.022} $ & 0.7566 & $  0.292^{+0.035}_{-0.059} $  & $  0.764^{+0.073}_{-0.079} $  & 864 & -
  \\ 
{} &&&& \vspace{0.02cm} \\
\hline
No observational systematics * & $  0.742^{+0.022}_{-0.021} $  & 0.740 & $  0.272^{+0.036}_{-0.052} $  & $  0.788^{+0.071}_{-0.086} $  & 1104 & -
  \\ 
 Only cosmological parameters * & $  0.756^{+0.019}_{-0.017} $  & 0.753 & $  0.282^{+0.038}_{-0.054} $  & $  0.789^{+0.070}_{-0.082} $  & 1403 & -
  \\ 

{\textit{}} &&&& \vspace{0.02cm} \\
 \hline
High-$z$: Bins 1,2 removed & $  0.733^{+0.023}_{-0.025} $ & 0.737 & $  0.355^{+0.050}_{-0.080} $  & $  0.683^{+0.069}_{-0.090} $  & 784 & - 
  \\ 
 Low-$z$: Bins 3,4 removed & $  0.750^{+0.045}_{-0.039} $ & 0.806 & $  0.289^{+0.034}_{-0.099} $ &  $  0.787^{+0.144}_{-0.122} $  & 355 & - 
  \\ 
 $\xi_-$ only & $  0.753^{+0.028}_{-0.032} $  & 0.747 & $  0.265^{+0.031}_{-0.061} $  & $  0.812^{+0.090}_{-0.098} $  & 829 & - 
  \\ 
 $\xi_+$ only & $  0.779^{+0.022}_{-0.020} $  & 0.777 & $  0.288^{+0.039}_{-0.062} $  & $  0.805^{+0.081}_{-0.093} $  & 1008 & - 
  \\ 
 Small angular scales & $  0.770^{+0.021}_{-0.021} $  & 0.763 & $  0.283^{+0.039}_{-0.061} $  & $  0.803^{+0.082}_{-0.086} $  & 934 & - 
  \\ 
  
  Large angular scales & $  0.737^{+0.027}_{-0.031} $  & 0.734 & $  0.268^{+0.033}_{-0.054} $  & $  0.790^{+0.079}_{-0.089} $  & 845 & - \\
{} &&&& \vspace{0.02cm} \\
\hline
DES Y1 & $0.780^{+0.027}_{-0.021}$ & - & $0.319^{+0.044}_{-0.062}$ & $0.764_{-0.072}^{+0.069}$ & 625 & $227/211 = 1.08$ \\
KiDS-1000 & $0.759^{+0.024}_{-0.021}$ & -  & $0.246^{+0.101}_{-0.060}$ & $0.838^{+0.140}_{-0.141}$ & 650 & $85.5/70.5=1.21$\\
HSC Y1 $C_\ell$ &  $0.780^{+0.030}_{-0.033}$ & - & $0.162^{+0.086}_{-0.044}$ & - & 461 & $45.4/53 = 0.86$ \\
HSC Y1 $\xi_\pm$ & $0.804^{+0.032}_{-0.029}$ & - & $0.346^{+0.052}_{-0.100}$ & $0.766^{+0.110}_{-0.093}$ &  402 & $162.3/167=0.97$ \\
Planck 2018 TT + TE + EE + lowE & $0.827^{+0.019}_{-0.017}$ & - & $0.327^{+0.008}_{-0.017}$  & $0.793^{+0.024}_{-0.009}$ &  3938 & - \\
  \\ 
\hline
\hline
\end{tabular}

\end{table*}

\section{Conclusions}
This paper, and its companion \citep{y3-cosmicshear2}, present the cosmic shear analysis of the Dark Energy Survey Year 3 (DES Y3), which spans $\sim5000$ deg$^2$ of the southern sky and contains over 100 million galaxy shapes. We present cosmic shear measurements with a signal-to-noise of 40. We constrain cosmological parameters in \lcdm, while
also varying the neutrino mass, and find a 3\% fractional uncertainty on $S_8$, with $S_8=0.759_{-0.023}^{+0.025}$ (0.755 best-fit) at 68\% confidence limits. A \textit{\lcdm-Optimized} analysis, which includes more small-scale information while still passing the requirements for robustness against baryonic effects, finds a 2\% precision constraint of $S_8=0.772^{+0.018}_{-0.017}$. This is consistent with the \textit{Fiducial} result and roughly 1.5$\times$ more constraining. The goodness of fit to the data is acceptable, with a $p$-value of 0.223 in \lcdm. In $w$CDM, the cosmic shear constraint on the dark energy equation of state, $w$, is found to be prior-dominated. 

The DES Y3 cosmic shear more than triples the survey area and number of galaxies with respect to previous lensing analyses \citep[][]{Troxel2018, asgari20,Ham20}. The  posteriors in the $S_8-\Omega_{\rm m}$ plane qualitatively agree with previous cosmic shear results from KiDS-1000 \citep{asgari20} and HSC \citep{Ham20, hikage19}, which, favor a lower value of $S_8$ than the most recent CMB measurements from \textit{Planck}  \citep{Planck2018}. Considering the full parameter space, the DES Y3 shear cosmological constraints are statistically consistent with those from \textit{Planck}, with a 2.3$\sigma$ difference  (and a $p$-value of 0.02) for the Fiducial case and 2.1$\sigma$ for the \textit{\lcdm-Optimized} analysis (and a $p$-value of 0.05). While this tension is not statistically significant, our results continue a trend of weak lensing data that are in agreement with, but display a lower value of $S_8$ than the CMB. 

Beyond substantial gains in the signal to noise of the measurements, this analysis incorporated major updates in weak-lensing methodology. Therefore, these cosmological constraints are build upon innovations in data calibration, specifically the shear and redshift calibration. Throughout this paper, we examine the rigour of analysis choices related to aspects in the data, made while blind. The robustness of theoretical modelling choices is investigated in Ref.~\citep{y3-cosmicshear2}. The main conclusions of this work are:
\begin{itemize}

\item We have demonstrated the results to be robust to photometric redshift calibration, with variant analysis exhibiting consistency at the $\sim0.5\sigma$-level. The cosmological , astrophysical and observational cosmic shear posteriors are insensitive to even extreme variations of the redshift sample, and when using either of three independent redshift methods, that together, provide more precise constraints on the redshift distributions.
We demonstrate that the DES Y3 analysis is robust to the impact of uncertainty on the shape of the redshift distributions, such as a fluctuation localized in the high-redshift tail.  

 \item  We detect, model, and calibrate the redshift-dependent impact of blending on weak lensing cosmology, with state-of-the-art multi-band image simulations described in \citep{y3-imagesims}. As the first cosmic shear analysis to account for such an effect, we find it to be an important correction. For our galaxy sample and shape measurement techniques, the image simulation-based calibration results in a shift toward higher values of $S_8$ by $0.5\sigma$. We demonstrate that the cosmology is insensitive to an additional, unaccounted for shear uncertainty.

 \item We evaluate the internal consistency of the cosmic shear measurements using a PPD method \citep{y3-inttensions}. We find all $p$-values to pass our threshold and at the analysis level, we find the $S_8$ posteriors to be consistent across small and large scales, low and high redshift and between two-point shear statistics to $\sim1\sigma$, which serves as a useful validation of some scale-dependent, or redshift-dependent unaccounted for uncertainty. Given the high-dimensionality of the analysis, we test the plausibility that inconsistencies in the intrinsic alignment parameters may absorb unaccounted-for systematics and found these to be stable removing subsets of the redshift bins.

\item We show that the level of PSF contamination in the analysis is subdominant, and that the tomographic B-mode measurements are consistent with zero. 

\end{itemize}

The companion paper contains analyses that yield the following conclusions about our results \citep{y3-cosmicshear2}:
\begin{itemize}
  \item We account for the intrinsic alignment of galaxies with a model that includes the tidal alignment and tidal torquing (quadratic) alignment. We explore different choices of intrinsic alignment parameterizations, including the NLA model, and find the cosmological parameters to be robust to within 1$\sigma$. 
  \item The matter power spectrum is modeled as dark matter-only, but we a select angular scales conservatively from hydrodynamical simulations to mitigate baryonic effects. We test for the impact of residual effects by considering models that account for non-linear physics and find the results are stable within $\sim0.5\sigma$. 
  \item We demonstrate that our posteriors are stable when inference is carried out at fixed neutrino mass, and demonstrate that higher-order shear contributions are negligible.
\end{itemize}

Finally, we investigate the limiting systematics for DES Y3 cosmic shear analysis. We demonstrate that the observational systematics are calibrated to sufficient precision, such that their uncertainties do not limit the analysis. However, we find the $S_8$ constraint to be significantly limited by systematics in the theoretical modeling, that is, due to the uncertainty in modeling of small-scale baryonic effects, as well as intrinsic alignments, which costs about two thirds of the Y3 cosmic shear power. The limitation in the former is reduced in the joint analysis with galaxy clustering and galaxy-galaxy lensing, which can be combined to give significantly more precise cosmological constraints \citep{y3-3x2ptkp}. On the other hand, improvements on the latter demand advancements in hydro-dynamical simulations and external complementary observations that probe astrophysical effects.

Looking ahead, we anticipate substantially improved cosmic shear power owing to observations with roughly twice the effective integrated exposure time per galaxy, and thus an abundance of fainter galaxies usable, in the Year 6 data. New Deep Field observations in fields with key external multi-band photometry and redshift information will enable robust redshift calibration at the enhanced depth. Despite the fact that the data calibration methodology for DES Y3 is substantially ahead of the requirements set by the ever-growing statistical power, future lensing data sets will require advances in redshift calibration and the accounting for blending, which are disproportionately more difficult for the fainter and more blended galaxy populations. Collaborative joint analyses with the complementary KiDS and HSC surveys would allow for an exchange of methodological experience. This would provide a particularly promising platform for further development in cosmic shear techniques and as a training ground for the imminent and challenging data-rich era of the surveys to be performed by Rubin Observatory, Euclid, and the Roman Space Telescope.

\section*{Acknowledgements}
This research manuscript made use of Astropy \cite{astropy:2013,astropy:2018}, GetDist \cite{getdist},ChainConsumer\footnote{samreay.github.io/ChainConsumer} \citep{Hinton16} and Matplotlib \cite{matplotlib}, and has been prepared using NASA's Astrophysics Data System Bibliographic Services.
Funding for the DES Projects has been provided by the U.S. Department of Energy, the U.S. National Science Foundation, the Ministry of Science and Education of Spain, 
the Science and Technology Facilities Council of the United Kingdom, the Higher Education Funding Council for England, the National Center for Supercomputing 
Applications at the University of Illinois at Urbana-Champaign, the Kavli Institute of Cosmological Physics at the University of Chicago, 
the Center for Cosmology and Astro-Particle Physics at the Ohio State University,
the Mitchell Institute for Fundamental Physics and Astronomy at Texas A\&M University, Financiadora de Estudos e Projetos, 
Funda{\c c}{\~a}o Carlos Chagas Filho de Amparo {\`a} Pesquisa do Estado do Rio de Janeiro, Conselho Nacional de Desenvolvimento Cient{\'i}fico e Tecnol{\'o}gico and 
the Minist{\'e}rio da Ci{\^e}ncia, Tecnologia e Inova{\c c}{\~a}o, the Deutsche Forschungsgemeinschaft and the Collaborating Institutions in the Dark Energy Survey. 

The Collaborating Institutions are Argonne National Laboratory, the University of California at Santa Cruz, the University of Cambridge, Centro de Investigaciones Energ{\'e}ticas, 
Medioambientales y Tecnol{\'o}gicas-Madrid, the University of Chicago, University College London, the DES-Brazil Consortium, the University of Edinburgh, 
the Eidgen{\"o}ssische Technische Hochschule (ETH) Z{\"u}rich, 
Fermi National Accelerator Laboratory, the University of Illinois at Urbana-Champaign, the Institut de Ci{\`e}ncies de l'Espai (IEEC/CSIC), 
the Institut de F{\'i}sica d'Altes Energies, Lawrence Berkeley National Laboratory, the Ludwig-Maximilians Universit{\"a}t M{\"u}nchen and the associated Excellence Cluster Universe, 
the University of Michigan, NFS's NOIRLab, the University of Nottingham, The Ohio State University, the University of Pennsylvania, the University of Portsmouth, 
SLAC National Accelerator Laboratory, Stanford University, the University of Sussex, Texas A\&M University, and the OzDES Membership Consortium.

Based in part on observations at Cerro Tololo Inter-American Observatory at NSF's NOIRLab (NOIRLab Prop. ID 2012B-0001; PI: J. Frieman), which is managed by the Association of Universities for Research in Astronomy (AURA) under a cooperative agreement with the National Science Foundation.

The DES data management system is supported by the National Science Foundation under Grant Numbers AST-1138766 and AST-1536171.
The DES participants from Spanish institutions are partially supported by MICINN under grants ESP2017-89838, PGC2018-094773, PGC2018-102021, SEV-2016-0588, SEV-2016-0597, and MDM-2015-0509, some of which include ERDF funds from the European Union. IFAE is partially funded by the CERCA program of the Generalitat de Catalunya.
Research leading to these results has received funding from the European Research
Council under the European Union's Seventh Framework Program (FP7/2007-2013) including ERC grant agreements 240672, 291329, and 306478.
We  acknowledge support from the Brazilian Instituto Nacional de Ci\^encia
e Tecnologia (INCT) do e-Universo (CNPq grant 465376/2014-2).

This manuscript has been authored by Fermi Research Alliance, LLC under Contract No. DE-AC02-07CH11359 with the U.S. Department of Energy, Office of Science, Office of High Energy Physics.

\bibliography{y3kp,references}

\begin{thebibliography}{219}%
\makeatletter
\providecommand \@ifxundefined [1]{%
 \@ifx{#1\undefined}
}%
\providecommand \@ifnum [1]{%
 \ifnum #1\expandafter \@firstoftwo
 \else \expandafter \@secondoftwo
 \fi
}%
\providecommand \@ifx [1]{%
 \ifx #1\expandafter \@firstoftwo
 \else \expandafter \@secondoftwo
 \fi
}%
\providecommand \natexlab [1]{#1}%
\providecommand \enquote  [1]{``#1''}%
\providecommand \bibnamefont  [1]{#1}%
\providecommand \bibfnamefont [1]{#1}%
\providecommand \citenamefont [1]{#1}%
\providecommand \href@noop [0]{\@secondoftwo}%
\providecommand \href [0]{\begingroup \@sanitize@url \@href}%
\providecommand \@href[1]{\@@startlink{#1}\@@href}%
\providecommand \@@href[1]{\endgroup#1\@@endlink}%
\providecommand \@sanitize@url [0]{\catcode `\\12\catcode `\$12\catcode
  `\&12\catcode `\#12\catcode `\^12\catcode `\_12\catcode `\%12\relax}%
\providecommand \@@startlink[1]{}%
\providecommand \@@endlink[0]{}%
\providecommand \url  [0]{\begingroup\@sanitize@url \@url }%
\providecommand \@url [1]{\endgroup\@href {#1}{\urlprefix }}%
\providecommand \urlprefix  [0]{URL }%
\providecommand \Eprint [0]{\href }%
\providecommand \doibase [0]{http://dx.doi.org/}%
\providecommand \selectlanguage [0]{\@gobble}%
\providecommand \bibinfo  [0]{\@secondoftwo}%
\providecommand \bibfield  [0]{\@secondoftwo}%
\providecommand \translation [1]{[#1]}%
\providecommand \BibitemOpen [0]{}%
\providecommand \bibitemStop [0]{}%
\providecommand \bibitemNoStop [0]{.\EOS\space}%
\providecommand \EOS [0]{\spacefactor3000\relax}%
\providecommand \BibitemShut  [1]{\csname bibitem#1\endcsname}%
\let\auto@bib@innerbib\@empty
\bibitem [{\citenamefont {{Secco}}\ \emph {et~al.}(2021)\citenamefont
  {{Secco}}, \citenamefont {{Samuroff}} \emph {et~al.}}]{y3-cosmicshear2}%
  \BibitemOpen
  \bibfield  {author} {\bibinfo {author} {\bibfnamefont {L.~F.}\ \bibnamefont
  {{Secco}}}, \bibinfo {author} {\bibfnamefont {S.}~\bibnamefont {{Samuroff}}},
   \emph {et~al.},\ }\href@noop {} {\bibfield  {journal} {\bibinfo  {journal}
  {To be submitted to PRD}\ } (\bibinfo {year} {2021})}\BibitemShut {NoStop}%
\bibitem [{\citenamefont {{Weinberg}}(1972)}]{weinberg1972}%
  \BibitemOpen
  \bibfield  {author} {\bibinfo {author} {\bibfnamefont {S.}~\bibnamefont
  {{Weinberg}}},\ }\href@noop {} {\emph {\bibinfo {title} {{Gravitation and
  Cosmology}}}}\ (\bibinfo {year} {1972})\BibitemShut {NoStop}%
\bibitem [{\citenamefont {{Peebles}}(1980)}]{Peebles1980}%
  \BibitemOpen
  \bibfield  {author} {\bibinfo {author} {\bibfnamefont {P.~J.~E.}\
  \bibnamefont {{Peebles}}},\ }\href@noop {} {\emph {\bibinfo {title} {{The
  large-scale structure of the universe}}}}\ (\bibinfo {year}
  {1980})\BibitemShut {NoStop}%
\bibitem [{\citenamefont {Bertone}\ and\ \citenamefont
  {Hooper}(2018)}]{Bertone:2016nfn}%
  \BibitemOpen
  \bibfield  {author} {\bibinfo {author} {\bibfnamefont {G.}~\bibnamefont
  {Bertone}}\ and\ \bibinfo {author} {\bibfnamefont {D.}~\bibnamefont
  {Hooper}},\ }\href {\doibase 10.1103/RevModPhys.90.045002} {\bibfield
  {journal} {\bibinfo  {journal} {Rev. Mod. Phys.}\ }\textbf {\bibinfo {volume}
  {90}},\ \bibinfo {pages} {045002} (\bibinfo {year} {2018})},\ \Eprint
  {http://arxiv.org/abs/1605.04909} {arXiv:1605.04909 [astro-ph.CO]}
  \BibitemShut {NoStop}%
\bibitem [{\citenamefont {Trimble}(1987)}]{Trimble:1987ee}%
  \BibitemOpen
  \bibfield  {author} {\bibinfo {author} {\bibfnamefont {V.}~\bibnamefont
  {Trimble}},\ }\href {\doibase 10.1146/annurev.aa.25.090187.002233} {\bibfield
   {journal} {\bibinfo  {journal} {Ann. Rev. Astron. Astrophys.}\ }\textbf
  {\bibinfo {volume} {25}},\ \bibinfo {pages} {425} (\bibinfo {year}
  {1987})}\BibitemShut {NoStop}%
\bibitem [{\citenamefont {Perlmutter}\ \emph {et~al.}(1999)\citenamefont
  {Perlmutter} \emph {et~al.}}]{Perlmutter:1998np}%
  \BibitemOpen
  \bibfield  {author} {\bibinfo {author} {\bibfnamefont {S.}~\bibnamefont
  {Perlmutter}} \emph {et~al.} (\bibinfo {collaboration} {Supernova Cosmology
  Project}),\ }\href {\doibase 10.1086/307221} {\bibfield  {journal} {\bibinfo
  {journal} {Astrophys. J.}\ }\textbf {\bibinfo {volume} {517}},\ \bibinfo
  {pages} {565} (\bibinfo {year} {1999})},\ \Eprint
  {http://arxiv.org/abs/astro-ph/9812133} {arXiv:astro-ph/9812133} \BibitemShut
  {NoStop}%
\bibitem [{\citenamefont {Riess}\ \emph {et~al.}(1998)\citenamefont {Riess}
  \emph {et~al.}}]{Riess:1998cb}%
  \BibitemOpen
  \bibfield  {author} {\bibinfo {author} {\bibfnamefont {A.~G.}\ \bibnamefont
  {Riess}} \emph {et~al.} (\bibinfo {collaboration} {Supernova Search Team}),\
  }\href {\doibase 10.1086/300499} {\bibfield  {journal} {\bibinfo  {journal}
  {Astron. J.}\ }\textbf {\bibinfo {volume} {116}},\ \bibinfo {pages} {1009}
  (\bibinfo {year} {1998})},\ \Eprint {http://arxiv.org/abs/astro-ph/9805201}
  {arXiv:astro-ph/9805201} \BibitemShut {NoStop}%
\bibitem [{\citenamefont {{Frieman}}\ \emph {et~al.}(2008)\citenamefont
  {{Frieman}}, \citenamefont {{Turner}},\ and\ \citenamefont
  {{Huterer}}}]{Frieman08}%
  \BibitemOpen
  \bibfield  {author} {\bibinfo {author} {\bibfnamefont {J.~A.}\ \bibnamefont
  {{Frieman}}}, \bibinfo {author} {\bibfnamefont {M.~S.}\ \bibnamefont
  {{Turner}}}, \ and\ \bibinfo {author} {\bibfnamefont {D.}~\bibnamefont
  {{Huterer}}},\ }\href {\doibase 10.1146/annurev.astro.46.060407.145243}
  {\bibfield  {journal} {\bibinfo  {journal} {\araa}\ }\textbf {\bibinfo
  {volume} {46}},\ \bibinfo {pages} {385} (\bibinfo {year} {2008})},\ \Eprint
  {http://arxiv.org/abs/0803.0982} {arXiv:0803.0982 [astro-ph]} \BibitemShut
  {NoStop}%
\bibitem [{\citenamefont {Weinberg}\ \emph {et~al.}(2013)\citenamefont
  {Weinberg}, \citenamefont {Mortonson}, \citenamefont {Eisenstein},
  \citenamefont {Hirata}, \citenamefont {Riess},\ and\ \citenamefont
  {Rozo}}]{Weinberg:2012es}%
  \BibitemOpen
  \bibfield  {author} {\bibinfo {author} {\bibfnamefont {D.~H.}\ \bibnamefont
  {Weinberg}}, \bibinfo {author} {\bibfnamefont {M.~J.}\ \bibnamefont
  {Mortonson}}, \bibinfo {author} {\bibfnamefont {D.~J.}\ \bibnamefont
  {Eisenstein}}, \bibinfo {author} {\bibfnamefont {C.}~\bibnamefont {Hirata}},
  \bibinfo {author} {\bibfnamefont {A.~G.}\ \bibnamefont {Riess}}, \ and\
  \bibinfo {author} {\bibfnamefont {E.}~\bibnamefont {Rozo}},\ }\href {\doibase
  10.1016/j.physrep.2013.05.001} {\bibfield  {journal} {\bibinfo  {journal}
  {Phys. Rept.}\ }\textbf {\bibinfo {volume} {530}},\ \bibinfo {pages} {87}
  (\bibinfo {year} {2013})},\ \Eprint {http://arxiv.org/abs/1201.2434}
  {arXiv:1201.2434 [astro-ph.CO]} \BibitemShut {NoStop}%
\bibitem [{\citenamefont {Abazajian}\ \emph {et~al.}(2016)\citenamefont
  {Abazajian} \emph {et~al.}}]{Abazajian:2016yjj}%
  \BibitemOpen
  \bibfield  {author} {\bibinfo {author} {\bibfnamefont {K.~N.}\ \bibnamefont
  {Abazajian}} \emph {et~al.} (\bibinfo {collaboration} {CMB-S4}),\ }\href@noop
  {} {\  (\bibinfo {year} {2016})},\ \Eprint {http://arxiv.org/abs/1610.02743}
  {arXiv:1610.02743 [astro-ph.CO]} \BibitemShut {NoStop}%
\bibitem [{\citenamefont {{Verde}}\ \emph {et~al.}(2019)\citenamefont
  {{Verde}}, \citenamefont {{Treu}},\ and\ \citenamefont
  {{Riess}}}]{Verde2019}%
  \BibitemOpen
  \bibfield  {author} {\bibinfo {author} {\bibfnamefont {L.}~\bibnamefont
  {{Verde}}}, \bibinfo {author} {\bibfnamefont {T.}~\bibnamefont {{Treu}}}, \
  and\ \bibinfo {author} {\bibfnamefont {A.~G.}\ \bibnamefont {{Riess}}},\
  }\href {\doibase 10.1038/s41550-019-0902-0} {\bibfield  {journal} {\bibinfo
  {journal} {Nature Astronomy}\ }\textbf {\bibinfo {volume} {3}},\ \bibinfo
  {pages} {891} (\bibinfo {year} {2019})},\ \Eprint
  {http://arxiv.org/abs/1907.10625} {arXiv:1907.10625 [astro-ph.CO]}
  \BibitemShut {NoStop}%
\bibitem [{\citenamefont {Di~Valentino}\ \emph {et~al.}(2021)\citenamefont
  {Di~Valentino}, \citenamefont {Mena}, \citenamefont {Pan}, \citenamefont
  {Visinelli}, \citenamefont {Yang}, \citenamefont {Melchiorri}, \citenamefont
  {Mota}, \citenamefont {Riess},\ and\ \citenamefont
  {Silk}}]{DiValentino:2021izs}%
  \BibitemOpen
  \bibfield  {author} {\bibinfo {author} {\bibfnamefont {E.}~\bibnamefont
  {Di~Valentino}}, \bibinfo {author} {\bibfnamefont {O.}~\bibnamefont {Mena}},
  \bibinfo {author} {\bibfnamefont {S.}~\bibnamefont {Pan}}, \bibinfo {author}
  {\bibfnamefont {L.}~\bibnamefont {Visinelli}}, \bibinfo {author}
  {\bibfnamefont {W.}~\bibnamefont {Yang}}, \bibinfo {author} {\bibfnamefont
  {A.}~\bibnamefont {Melchiorri}}, \bibinfo {author} {\bibfnamefont {D.~F.}\
  \bibnamefont {Mota}}, \bibinfo {author} {\bibfnamefont {A.~G.}\ \bibnamefont
  {Riess}}, \ and\ \bibinfo {author} {\bibfnamefont {J.}~\bibnamefont {Silk}},\
  }\href@noop {} {\  (\bibinfo {year} {2021})},\ \Eprint
  {http://arxiv.org/abs/2103.01183} {arXiv:2103.01183 [astro-ph.CO]}
  \BibitemShut {NoStop}%
\bibitem [{\citenamefont {{Heymans}}\ and\ \citenamefont
  {{Grocutt}}(2013)}]{Heymans13}%
  \BibitemOpen
  \bibfield  {author} {\bibinfo {author} {\bibfnamefont {C.}~\bibnamefont
  {{Heymans}}}\ and\ \bibinfo {author} {\bibfnamefont {E.}~\bibnamefont
  {{Grocutt}}},\ }\href {\doibase 10.1093/mnras/stt601} {\bibfield  {journal}
  {\bibinfo  {journal} {\mnras}\ }\textbf {\bibinfo {volume} {432}},\ \bibinfo
  {pages} {2433} (\bibinfo {year} {2013})},\ \Eprint
  {http://arxiv.org/abs/1303.1808} {arXiv:1303.1808} \BibitemShut {NoStop}%
\bibitem [{\citenamefont {{Troxel}}\ \emph {et~al.}(2018)\citenamefont
  {{Troxel}}, \citenamefont {{MacCrann}}, \citenamefont {{Zuntz}},
  \citenamefont {{Eifler}}, \citenamefont {{Krause}}, \citenamefont
  {{Dodelson}}, \citenamefont {{Gruen}}, \citenamefont {{Blazek}},
  \citenamefont {{Friedrich}}, \citenamefont {{Samuroff}},\ and\ \citenamefont
  {{DES Collaboration}}}]{Troxel2018}%
  \BibitemOpen
  \bibfield  {author} {\bibinfo {author} {\bibfnamefont {M.~A.}\ \bibnamefont
  {{Troxel}}}, \bibinfo {author} {\bibfnamefont {N.}~\bibnamefont
  {{MacCrann}}}, \bibinfo {author} {\bibfnamefont {J.}~\bibnamefont {{Zuntz}}},
  \bibinfo {author} {\bibfnamefont {T.~F.}\ \bibnamefont {{Eifler}}}, \bibinfo
  {author} {\bibfnamefont {E.}~\bibnamefont {{Krause}}}, \bibinfo {author}
  {\bibfnamefont {S.}~\bibnamefont {{Dodelson}}}, \bibinfo {author}
  {\bibfnamefont {D.}~\bibnamefont {{Gruen}}}, \bibinfo {author} {\bibfnamefont
  {J.}~\bibnamefont {{Blazek}}}, \bibinfo {author} {\bibfnamefont
  {O.}~\bibnamefont {{Friedrich}}}, \bibinfo {author} {\bibfnamefont
  {S.}~\bibnamefont {{Samuroff}}}, \ and\ \bibinfo {author} {\bibnamefont {{DES
  Collaboration}}},\ }\href {\doibase 10.1103/PhysRevD.98.043528} {\bibfield
  {journal} {\bibinfo  {journal} {\prd}\ }\textbf {\bibinfo {volume} {98}},\
  \bibinfo {eid} {043528} (\bibinfo {year} {2018})},\ \Eprint
  {http://arxiv.org/abs/1708.01538} {arXiv:1708.01538 [astro-ph.CO]}
  \BibitemShut {NoStop}%
\bibitem [{\citenamefont {{Dark Energy Survey
  Collaboration}}(2018)}]{y1keypaper}%
  \BibitemOpen
  \bibfield  {author} {\bibinfo {author} {\bibnamefont {{Dark Energy Survey
  Collaboration}}},\ }\href {\doibase 10.1103/PhysRevD.98.043526} {\bibfield
  {journal} {\bibinfo  {journal} {\prd}\ }\textbf {\bibinfo {volume} {98}},\
  \bibinfo {eid} {043526} (\bibinfo {year} {2018})},\ \Eprint
  {http://arxiv.org/abs/1708.01530} {arXiv:1708.01530 [astro-ph.CO]}
  \BibitemShut {NoStop}%
\bibitem [{\citenamefont {{Hikage}}\ \emph {et~al.}(2019)\citenamefont
  {{Hikage}}, \citenamefont {{Oguri}}, \citenamefont {{Hamana}}, \citenamefont
  {{More}}, \citenamefont {{Mandelbaum}}, \citenamefont {{Takada}},
  \citenamefont {{K{\"o}hlinger}}, \citenamefont {{Miyatake}},\ and\
  \citenamefont {{Nishizawa}}}]{hikage19}%
  \BibitemOpen
  \bibfield  {author} {\bibinfo {author} {\bibfnamefont {C.}~\bibnamefont
  {{Hikage}}}, \bibinfo {author} {\bibfnamefont {M.}~\bibnamefont {{Oguri}}},
  \bibinfo {author} {\bibfnamefont {T.}~\bibnamefont {{Hamana}}}, \bibinfo
  {author} {\bibfnamefont {S.}~\bibnamefont {{More}}}, \bibinfo {author}
  {\bibfnamefont {R.}~\bibnamefont {{Mandelbaum}}}, \bibinfo {author}
  {\bibfnamefont {M.}~\bibnamefont {{Takada}}}, \bibinfo {author}
  {\bibfnamefont {F.}~\bibnamefont {{K{\"o}hlinger}}}, \bibinfo {author}
  {\bibfnamefont {H.}~\bibnamefont {{Miyatake}}}, \ and\ \bibinfo {author}
  {\bibfnamefont {A.~J.}\ \bibnamefont {{Nishizawa}}},\ }\href {\doibase
  10.1093/pasj/psz010} {\bibfield  {journal} {\bibinfo  {journal} {\pasj}\
  }\textbf {\bibinfo {volume} {71}},\ \bibinfo {eid} {43} (\bibinfo {year}
  {2019})},\ \Eprint {http://arxiv.org/abs/1809.09148} {arXiv:1809.09148
  [astro-ph.CO]} \BibitemShut {NoStop}%
\bibitem [{\citenamefont {{Asgari}}\ \emph {et~al.}(2021)\citenamefont
  {{Asgari}}, \citenamefont {{Lin}}, \citenamefont {{Joachimi}}, \citenamefont
  {{Giblin}}, \citenamefont {{Heymans}},\ and\ \citenamefont
  {{Hildebrandt}}}]{asgari20}%
  \BibitemOpen
  \bibfield  {author} {\bibinfo {author} {\bibfnamefont {M.}~\bibnamefont
  {{Asgari}}}, \bibinfo {author} {\bibfnamefont {C.-A.}\ \bibnamefont {{Lin}}},
  \bibinfo {author} {\bibfnamefont {B.}~\bibnamefont {{Joachimi}}}, \bibinfo
  {author} {\bibfnamefont {B.}~\bibnamefont {{Giblin}}}, \bibinfo {author}
  {\bibfnamefont {C.}~\bibnamefont {{Heymans}}}, \ and\ \bibinfo {author}
  {\bibfnamefont {H.}~\bibnamefont {{Hildebrandt}}},\ }\href {\doibase
  10.1051/0004-6361/202039070} {\bibfield  {journal} {\bibinfo  {journal}
  {\aap}\ }\textbf {\bibinfo {volume} {645}},\ \bibinfo {eid} {A104} (\bibinfo
  {year} {2021})},\ \Eprint {http://arxiv.org/abs/2007.15633} {arXiv:2007.15633
  [astro-ph.CO]} \BibitemShut {NoStop}%
\bibitem [{\citenamefont {{Heymans}}\ \emph {et~al.}(2021)\citenamefont
  {{Heymans}}, \citenamefont {{Tr{\"o}ster}}, \citenamefont {{Asgari}},
  \citenamefont {{Blake}}, \citenamefont {{Hildebrandt}}, \citenamefont
  {{Joachimi}},\ and\ \citenamefont {{Kuijken}}}]{heymans20}%
  \BibitemOpen
  \bibfield  {author} {\bibinfo {author} {\bibfnamefont {C.}~\bibnamefont
  {{Heymans}}}, \bibinfo {author} {\bibfnamefont {T.}~\bibnamefont
  {{Tr{\"o}ster}}}, \bibinfo {author} {\bibfnamefont {M.}~\bibnamefont
  {{Asgari}}}, \bibinfo {author} {\bibfnamefont {C.}~\bibnamefont {{Blake}}},
  \bibinfo {author} {\bibfnamefont {H.}~\bibnamefont {{Hildebrandt}}}, \bibinfo
  {author} {\bibfnamefont {B.}~\bibnamefont {{Joachimi}}}, \ and\ \bibinfo
  {author} {\bibfnamefont {K.}~\bibnamefont {{Kuijken}}},\ }\href {\doibase
  10.1051/0004-6361/202039063} {\bibfield  {journal} {\bibinfo  {journal}
  {\aap}\ }\textbf {\bibinfo {volume} {646}},\ \bibinfo {eid} {A140} (\bibinfo
  {year} {2021})},\ \Eprint {http://arxiv.org/abs/2007.15632} {arXiv:2007.15632
  [astro-ph.CO]} \BibitemShut {NoStop}%
\bibitem [{\citenamefont {{Krolewski}}\ \emph {et~al.}(2021)\citenamefont
  {{Krolewski}}, \citenamefont {{Ferraro}},\ and\ \citenamefont
  {{White}}}]{unwise}%
  \BibitemOpen
  \bibfield  {author} {\bibinfo {author} {\bibfnamefont {A.}~\bibnamefont
  {{Krolewski}}}, \bibinfo {author} {\bibfnamefont {S.}~\bibnamefont
  {{Ferraro}}}, \ and\ \bibinfo {author} {\bibfnamefont {M.}~\bibnamefont
  {{White}}},\ }\href@noop {} {\bibfield  {journal} {\bibinfo  {journal} {arXiv
  e-prints}\ ,\ \bibinfo {eid} {arXiv:2105.03421}} (\bibinfo {year} {2021})},\
  \Eprint {http://arxiv.org/abs/2105.03421} {arXiv:2105.03421 [astro-ph.CO]}
  \BibitemShut {NoStop}%
\bibitem [{\citenamefont {{Planck Collaboration}}(2018)}]{Planck2018}%
  \BibitemOpen
  \bibfield  {author} {\bibinfo {author} {\bibnamefont {{Planck
  Collaboration}}},\ }\href@noop {} {\bibfield  {journal} {\bibinfo  {journal}
  {arXiv e-prints}\ ,\ \bibinfo {eid} {arXiv:1807.06210}} (\bibinfo {year}
  {2018})},\ \Eprint {http://arxiv.org/abs/1807.06210} {arXiv:1807.06210
  [astro-ph.CO]} \BibitemShut {NoStop}%
\bibitem [{\citenamefont {{Bartelmann}}\ and\ \citenamefont
  {{Schneider}}(2001)}]{bartelschneider}%
  \BibitemOpen
  \bibfield  {author} {\bibinfo {author} {\bibfnamefont {M.}~\bibnamefont
  {{Bartelmann}}}\ and\ \bibinfo {author} {\bibfnamefont {P.}~\bibnamefont
  {{Schneider}}},\ }\href {\doibase 10.1016/S0370-1573(00)00082-X} {\bibfield
  {journal} {\bibinfo  {journal} {\physrep}\ }\textbf {\bibinfo {volume}
  {340}},\ \bibinfo {pages} {291} (\bibinfo {year} {2001})},\ \Eprint
  {http://arxiv.org/abs/astro-ph/9912508} {arXiv:astro-ph/9912508 [astro-ph]}
  \BibitemShut {NoStop}%
\bibitem [{\citenamefont {{Hoekstra}}\ and\ \citenamefont
  {{Jain}}(2008)}]{hoekstrajain}%
  \BibitemOpen
  \bibfield  {author} {\bibinfo {author} {\bibfnamefont {H.}~\bibnamefont
  {{Hoekstra}}}\ and\ \bibinfo {author} {\bibfnamefont {B.}~\bibnamefont
  {{Jain}}},\ }\href {\doibase 10.1146/annurev.nucl.58.110707.171151}
  {\bibfield  {journal} {\bibinfo  {journal} {Annual Review of Nuclear and
  Particle Science}\ }\textbf {\bibinfo {volume} {58}},\ \bibinfo {pages} {99}
  (\bibinfo {year} {2008})},\ \Eprint {http://arxiv.org/abs/0805.0139}
  {arXiv:0805.0139 [astro-ph]} \BibitemShut {NoStop}%
\bibitem [{\citenamefont {{Kilbinger}}(2015)}]{kilbinger15}%
  \BibitemOpen
  \bibfield  {author} {\bibinfo {author} {\bibfnamefont {M.}~\bibnamefont
  {{Kilbinger}}},\ }\href {\doibase 10.1088/0034-4885/78/8/086901} {\bibfield
  {journal} {\bibinfo  {journal} {Reports on Progress in Physics}\ }\textbf
  {\bibinfo {volume} {78}},\ \bibinfo {eid} {086901} (\bibinfo {year}
  {2015})},\ \Eprint {http://arxiv.org/abs/1411.0115} {arXiv:1411.0115
  [astro-ph.CO]} \BibitemShut {NoStop}%
\bibitem [{\citenamefont {{Kristian}}(1967)}]{Kristian67}%
  \BibitemOpen
  \bibfield  {author} {\bibinfo {author} {\bibfnamefont {J.}~\bibnamefont
  {{Kristian}}},\ }\href {\doibase 10.1086/149078} {\bibfield  {journal}
  {\bibinfo  {journal} {\apj}\ }\textbf {\bibinfo {volume} {147}},\ \bibinfo
  {pages} {864} (\bibinfo {year} {1967})}\BibitemShut {NoStop}%
\bibitem [{\citenamefont {{Bacon}}\ \emph {et~al.}(2000)\citenamefont
  {{Bacon}}, \citenamefont {{Refregier}},\ and\ \citenamefont
  {{Ellis}}}]{bacon00}%
  \BibitemOpen
  \bibfield  {author} {\bibinfo {author} {\bibfnamefont {D.~J.}\ \bibnamefont
  {{Bacon}}}, \bibinfo {author} {\bibfnamefont {A.~R.}\ \bibnamefont
  {{Refregier}}}, \ and\ \bibinfo {author} {\bibfnamefont {R.~S.}\ \bibnamefont
  {{Ellis}}},\ }\href@noop {} {\bibfield  {journal} {\bibinfo  {journal}
  {\mnras}\ }\textbf {\bibinfo {volume} {318}},\ \bibinfo {pages} {625}
  (\bibinfo {year} {2000})},\ \Eprint {http://arxiv.org/abs/astro-ph/0003008}
  {astro-ph/0003008} \BibitemShut {NoStop}%
\bibitem [{\citenamefont {{Kaiser}}\ \emph {et~al.}(2000)\citenamefont
  {{Kaiser}}, \citenamefont {{Wilson}},\ and\ \citenamefont
  {{Luppino}}}]{kaiser00}%
  \BibitemOpen
  \bibfield  {author} {\bibinfo {author} {\bibfnamefont {N.}~\bibnamefont
  {{Kaiser}}}, \bibinfo {author} {\bibfnamefont {G.}~\bibnamefont {{Wilson}}},
  \ and\ \bibinfo {author} {\bibfnamefont {G.~A.}\ \bibnamefont {{Luppino}}},\
  }\href@noop {} {\bibfield  {journal} {\bibinfo  {journal}
  {arXiv:astro-ph/0003338}\ } (\bibinfo {year} {2000})},\ \Eprint
  {http://arxiv.org/abs/astro-ph/0003338} {astro-ph/0003338} \BibitemShut
  {NoStop}%
\bibitem [{\citenamefont {{Van Waerbeke}}\ \emph {et~al.}(2000)\citenamefont
  {{Van Waerbeke}}, \citenamefont {{Mellier}}, \citenamefont {{Erben}},
  \citenamefont {{Cuillandre}}, \citenamefont {{Bernardeau}}, \citenamefont
  {{Maoli}}, \citenamefont {{Bertin}}, \citenamefont {{McCracken}},
  \citenamefont {{Le F{\`e}vre}}, \citenamefont {{Fort}}, \citenamefont
  {{Dantel-Fort}}, \citenamefont {{Jain}},\ and\ \citenamefont
  {{Schneider}}}]{vanwaerbeke00}%
  \BibitemOpen
  \bibfield  {author} {\bibinfo {author} {\bibfnamefont {L.}~\bibnamefont {{Van
  Waerbeke}}}, \bibinfo {author} {\bibfnamefont {Y.}~\bibnamefont {{Mellier}}},
  \bibinfo {author} {\bibfnamefont {T.}~\bibnamefont {{Erben}}}, \bibinfo
  {author} {\bibfnamefont {J.~C.}\ \bibnamefont {{Cuillandre}}}, \bibinfo
  {author} {\bibfnamefont {F.}~\bibnamefont {{Bernardeau}}}, \bibinfo {author}
  {\bibfnamefont {R.}~\bibnamefont {{Maoli}}}, \bibinfo {author} {\bibfnamefont
  {E.}~\bibnamefont {{Bertin}}}, \bibinfo {author} {\bibfnamefont {H.~J.}\
  \bibnamefont {{McCracken}}}, \bibinfo {author} {\bibfnamefont
  {O.}~\bibnamefont {{Le F{\`e}vre}}}, \bibinfo {author} {\bibfnamefont
  {B.}~\bibnamefont {{Fort}}}, \bibinfo {author} {\bibfnamefont
  {M.}~\bibnamefont {{Dantel-Fort}}}, \bibinfo {author} {\bibfnamefont
  {B.}~\bibnamefont {{Jain}}}, \ and\ \bibinfo {author} {\bibfnamefont
  {P.}~\bibnamefont {{Schneider}}},\ }\href@noop {} {\bibfield  {journal}
  {\bibinfo  {journal} {\aap}\ }\textbf {\bibinfo {volume} {358}},\ \bibinfo
  {pages} {30} (\bibinfo {year} {2000})},\ \Eprint
  {http://arxiv.org/abs/astro-ph/0002500} {astro-ph/0002500} \BibitemShut
  {NoStop}%
\bibitem [{\citenamefont {{Wittman}}\ \emph {et~al.}(2000)\citenamefont
  {{Wittman}}, \citenamefont {{Tyson}}, \citenamefont {{Kirkman}},
  \citenamefont {{Dell'Antonio}},\ and\ \citenamefont
  {{Bernstein}}}]{wittman2000}%
  \BibitemOpen
  \bibfield  {author} {\bibinfo {author} {\bibfnamefont {D.~M.}\ \bibnamefont
  {{Wittman}}}, \bibinfo {author} {\bibfnamefont {J.~A.}\ \bibnamefont
  {{Tyson}}}, \bibinfo {author} {\bibfnamefont {D.}~\bibnamefont {{Kirkman}}},
  \bibinfo {author} {\bibfnamefont {I.}~\bibnamefont {{Dell'Antonio}}}, \ and\
  \bibinfo {author} {\bibfnamefont {G.}~\bibnamefont {{Bernstein}}},\ }\href
  {\doibase 10.1038/35012001} {\bibfield  {journal} {\bibinfo  {journal}
  {\nat}\ }\textbf {\bibinfo {volume} {405}},\ \bibinfo {pages} {143} (\bibinfo
  {year} {2000})},\ \Eprint {http://arxiv.org/abs/astro-ph/0003014}
  {arXiv:astro-ph/0003014 [astro-ph]} \BibitemShut {NoStop}%
\bibitem [{\citenamefont {Hoekstra}\ \emph {et~al.}(2002)\citenamefont
  {Hoekstra}, \citenamefont {Yee}, \citenamefont {Gladders}, \citenamefont
  {Barrientos}, \citenamefont {Hall},\ and\ \citenamefont
  {Infante}}]{Hoekstra:2002cj}%
  \BibitemOpen
  \bibfield  {author} {\bibinfo {author} {\bibfnamefont {H.}~\bibnamefont
  {Hoekstra}}, \bibinfo {author} {\bibfnamefont {H.~K.~C.}\ \bibnamefont
  {Yee}}, \bibinfo {author} {\bibfnamefont {M.~D.}\ \bibnamefont {Gladders}},
  \bibinfo {author} {\bibfnamefont {L.~F.}\ \bibnamefont {Barrientos}},
  \bibinfo {author} {\bibfnamefont {P.~B.}\ \bibnamefont {Hall}}, \ and\
  \bibinfo {author} {\bibfnamefont {L.}~\bibnamefont {Infante}},\ }\href
  {\doibase 10.1086/340298} {\bibfield  {journal} {\bibinfo  {journal}
  {Astrophys. J.}\ }\textbf {\bibinfo {volume} {572}},\ \bibinfo {pages} {55}
  (\bibinfo {year} {2002})},\ \Eprint {http://arxiv.org/abs/astro-ph/0202285}
  {arXiv:astro-ph/0202285} \BibitemShut {NoStop}%
\bibitem [{\citenamefont {Refregier}\ \emph {et~al.}(2002)\citenamefont
  {Refregier}, \citenamefont {Rhodes},\ and\ \citenamefont
  {Groth}}]{Refregier:2002ux}%
  \BibitemOpen
  \bibfield  {author} {\bibinfo {author} {\bibfnamefont {A.}~\bibnamefont
  {Refregier}}, \bibinfo {author} {\bibfnamefont {J.}~\bibnamefont {Rhodes}}, \
  and\ \bibinfo {author} {\bibfnamefont {E.~J.}\ \bibnamefont {Groth}},\ }\href
  {\doibase 10.1086/341666} {\bibfield  {journal} {\bibinfo  {journal}
  {Astrophys. J. Lett.}\ }\textbf {\bibinfo {volume} {572}},\ \bibinfo {pages}
  {L131} (\bibinfo {year} {2002})},\ \Eprint
  {http://arxiv.org/abs/astro-ph/0203131} {arXiv:astro-ph/0203131} \BibitemShut
  {NoStop}%
\bibitem [{\citenamefont {Jarvis}\ \emph {et~al.}(2003)\citenamefont {Jarvis},
  \citenamefont {Bernstein}, \citenamefont {Fischer}, \citenamefont {Smith},
  \citenamefont {Jain}, \citenamefont {Tyson},\ and\ \citenamefont
  {Wittman}}]{Jarvis:2002vs}%
  \BibitemOpen
  \bibfield  {author} {\bibinfo {author} {\bibfnamefont {M.}~\bibnamefont
  {Jarvis}}, \bibinfo {author} {\bibfnamefont {G.~M.}\ \bibnamefont
  {Bernstein}}, \bibinfo {author} {\bibfnamefont {P.}~\bibnamefont {Fischer}},
  \bibinfo {author} {\bibfnamefont {D.}~\bibnamefont {Smith}}, \bibinfo
  {author} {\bibfnamefont {B.}~\bibnamefont {Jain}}, \bibinfo {author}
  {\bibfnamefont {J.~A.}\ \bibnamefont {Tyson}}, \ and\ \bibinfo {author}
  {\bibfnamefont {D.}~\bibnamefont {Wittman}},\ }\href {\doibase
  10.1086/367799} {\bibfield  {journal} {\bibinfo  {journal} {Astron. J.}\
  }\textbf {\bibinfo {volume} {125}},\ \bibinfo {pages} {1014} (\bibinfo {year}
  {2003})},\ \Eprint {http://arxiv.org/abs/astro-ph/0210604}
  {arXiv:astro-ph/0210604} \BibitemShut {NoStop}%
\bibitem [{\citenamefont {Hamana}\ \emph {et~al.}(2003)\citenamefont {Hamana}
  \emph {et~al.}}]{Hamana:2002yd}%
  \BibitemOpen
  \bibfield  {author} {\bibinfo {author} {\bibfnamefont {T.}~\bibnamefont
  {Hamana}} \emph {et~al.},\ }\href {\doibase 10.1086/378348} {\bibfield
  {journal} {\bibinfo  {journal} {Astrophys. J.}\ }\textbf {\bibinfo {volume}
  {597}},\ \bibinfo {pages} {98} (\bibinfo {year} {2003})},\ \Eprint
  {http://arxiv.org/abs/astro-ph/0210450} {arXiv:astro-ph/0210450} \BibitemShut
  {NoStop}%
\bibitem [{\citenamefont {Brown}\ \emph {et~al.}(2003)\citenamefont {Brown},
  \citenamefont {Taylor}, \citenamefont {Bacon}, \citenamefont {Gray},
  \citenamefont {Dye}, \citenamefont {Meisenheimer},\ and\ \citenamefont
  {Wolf}}]{Brown:2002wt}%
  \BibitemOpen
  \bibfield  {author} {\bibinfo {author} {\bibfnamefont {M.~L.}\ \bibnamefont
  {Brown}}, \bibinfo {author} {\bibfnamefont {A.~N.}\ \bibnamefont {Taylor}},
  \bibinfo {author} {\bibfnamefont {D.~J.}\ \bibnamefont {Bacon}}, \bibinfo
  {author} {\bibfnamefont {M.~E.}\ \bibnamefont {Gray}}, \bibinfo {author}
  {\bibfnamefont {S.}~\bibnamefont {Dye}}, \bibinfo {author} {\bibfnamefont
  {K.}~\bibnamefont {Meisenheimer}}, \ and\ \bibinfo {author} {\bibfnamefont
  {C.}~\bibnamefont {Wolf}},\ }\href {\doibase
  10.1046/j.1365-8711.2003.06237.x} {\bibfield  {journal} {\bibinfo  {journal}
  {Mon. Not. Roy. Astron. Soc.}\ }\textbf {\bibinfo {volume} {341}},\ \bibinfo
  {pages} {100} (\bibinfo {year} {2003})},\ \Eprint
  {http://arxiv.org/abs/astro-ph/0210213} {arXiv:astro-ph/0210213} \BibitemShut
  {NoStop}%
\bibitem [{\citenamefont {Rhodes}\ \emph {et~al.}(2004)\citenamefont {Rhodes},
  \citenamefont {Refregier}, \citenamefont {Collins}, \citenamefont {Gardner},
  \citenamefont {Groth},\ and\ \citenamefont {Hill}}]{Rhodes:2003wj}%
  \BibitemOpen
  \bibfield  {author} {\bibinfo {author} {\bibfnamefont {J.}~\bibnamefont
  {Rhodes}}, \bibinfo {author} {\bibfnamefont {A.}~\bibnamefont {Refregier}},
  \bibinfo {author} {\bibfnamefont {N.~R.}\ \bibnamefont {Collins}}, \bibinfo
  {author} {\bibfnamefont {J.~P.}\ \bibnamefont {Gardner}}, \bibinfo {author}
  {\bibfnamefont {E.~J.}\ \bibnamefont {Groth}}, \ and\ \bibinfo {author}
  {\bibfnamefont {R.~S.}\ \bibnamefont {Hill}},\ }\href {\doibase
  10.1086/382181} {\bibfield  {journal} {\bibinfo  {journal} {Astrophys. J.}\
  }\textbf {\bibinfo {volume} {605}},\ \bibinfo {pages} {29} (\bibinfo {year}
  {2004})},\ \Eprint {http://arxiv.org/abs/astro-ph/0312283}
  {arXiv:astro-ph/0312283} \BibitemShut {NoStop}%
\bibitem [{\citenamefont {Heymans}\ \emph {et~al.}(2005)\citenamefont {Heymans}
  \emph {et~al.}}]{Heymans:2004zp}%
  \BibitemOpen
  \bibfield  {author} {\bibinfo {author} {\bibfnamefont {C.}~\bibnamefont
  {Heymans}} \emph {et~al.},\ }\href {\doibase
  10.1111/j.1365-2966.2005.09152.x} {\bibfield  {journal} {\bibinfo  {journal}
  {Mon. Not. Roy. Astron. Soc.}\ }\textbf {\bibinfo {volume} {361}},\ \bibinfo
  {pages} {160} (\bibinfo {year} {2005})},\ \Eprint
  {http://arxiv.org/abs/astro-ph/0411324} {arXiv:astro-ph/0411324} \BibitemShut
  {NoStop}%
\bibitem [{\citenamefont {Jarvis}\ \emph {et~al.}(2006)\citenamefont {Jarvis},
  \citenamefont {Jain}, \citenamefont {Bernstein},\ and\ \citenamefont
  {Dolney}}]{Jarvis:2005ck}%
  \BibitemOpen
  \bibfield  {author} {\bibinfo {author} {\bibfnamefont {M.}~\bibnamefont
  {Jarvis}}, \bibinfo {author} {\bibfnamefont {B.}~\bibnamefont {Jain}},
  \bibinfo {author} {\bibfnamefont {G.}~\bibnamefont {Bernstein}}, \ and\
  \bibinfo {author} {\bibfnamefont {D.}~\bibnamefont {Dolney}},\ }\href
  {\doibase 10.1086/503418} {\bibfield  {journal} {\bibinfo  {journal}
  {Astrophys. J.}\ }\textbf {\bibinfo {volume} {644}},\ \bibinfo {pages} {71}
  (\bibinfo {year} {2006})},\ \Eprint {http://arxiv.org/abs/astro-ph/0502243}
  {arXiv:astro-ph/0502243} \BibitemShut {NoStop}%
\bibitem [{\citenamefont {Hetterscheidt}\ \emph {et~al.}(2007)\citenamefont
  {Hetterscheidt}, \citenamefont {Simon}, \citenamefont {Schirmer},
  \citenamefont {Hildebrandt}, \citenamefont {Schrabback}, \citenamefont
  {Erben},\ and\ \citenamefont {Schneider}}]{Hetterscheidt:2006up}%
  \BibitemOpen
  \bibfield  {author} {\bibinfo {author} {\bibfnamefont {M.}~\bibnamefont
  {Hetterscheidt}}, \bibinfo {author} {\bibfnamefont {P.}~\bibnamefont
  {Simon}}, \bibinfo {author} {\bibfnamefont {M.}~\bibnamefont {Schirmer}},
  \bibinfo {author} {\bibfnamefont {H.}~\bibnamefont {Hildebrandt}}, \bibinfo
  {author} {\bibfnamefont {T.}~\bibnamefont {Schrabback}}, \bibinfo {author}
  {\bibfnamefont {T.}~\bibnamefont {Erben}}, \ and\ \bibinfo {author}
  {\bibfnamefont {P.}~\bibnamefont {Schneider}},\ }\href {\doibase
  10.1051/0004-6361:20065885} {\bibfield  {journal} {\bibinfo  {journal}
  {Astron. Astrophys.}\ }\textbf {\bibinfo {volume} {468}},\ \bibinfo {pages}
  {859} (\bibinfo {year} {2007})},\ \Eprint
  {http://arxiv.org/abs/astro-ph/0606571} {arXiv:astro-ph/0606571} \BibitemShut
  {NoStop}%
\bibitem [{\citenamefont {Massey}\ \emph {et~al.}(2007)\citenamefont {Massey}
  \emph {et~al.}}]{Massey:2007gh}%
  \BibitemOpen
  \bibfield  {author} {\bibinfo {author} {\bibfnamefont {R.}~\bibnamefont
  {Massey}} \emph {et~al.},\ }\href {\doibase 10.1086/516599} {\bibfield
  {journal} {\bibinfo  {journal} {Astrophys. J. Suppl.}\ }\textbf {\bibinfo
  {volume} {172}},\ \bibinfo {pages} {239} (\bibinfo {year} {2007})},\ \Eprint
  {http://arxiv.org/abs/astro-ph/0701480} {arXiv:astro-ph/0701480} \BibitemShut
  {NoStop}%
\bibitem [{\citenamefont {Leauthaud}\ \emph {et~al.}(2007)\citenamefont
  {Leauthaud} \emph {et~al.}}]{Leauthaud:2007fb}%
  \BibitemOpen
  \bibfield  {author} {\bibinfo {author} {\bibfnamefont {A.}~\bibnamefont
  {Leauthaud}} \emph {et~al.},\ }\href {\doibase 10.1086/516598} {\bibfield
  {journal} {\bibinfo  {journal} {Astrophys. J. Suppl.}\ }\textbf {\bibinfo
  {volume} {172}},\ \bibinfo {pages} {219} (\bibinfo {year} {2007})},\ \Eprint
  {http://arxiv.org/abs/astro-ph/0702359} {arXiv:astro-ph/0702359} \BibitemShut
  {NoStop}%
\bibitem [{\citenamefont {Benjamin}\ \emph {et~al.}(2007)\citenamefont
  {Benjamin}, \citenamefont {Heymans}, \citenamefont {Semboloni}, \citenamefont
  {Van~Waerbeke}, \citenamefont {Hoekstra}, \citenamefont {Erben},
  \citenamefont {Gladders}, \citenamefont {Hetterscheidt}, \citenamefont
  {Mellier},\ and\ \citenamefont {Yee}}]{Benjamin:2007ys}%
  \BibitemOpen
  \bibfield  {author} {\bibinfo {author} {\bibfnamefont {J.}~\bibnamefont
  {Benjamin}}, \bibinfo {author} {\bibfnamefont {C.}~\bibnamefont {Heymans}},
  \bibinfo {author} {\bibfnamefont {E.}~\bibnamefont {Semboloni}}, \bibinfo
  {author} {\bibfnamefont {L.}~\bibnamefont {Van~Waerbeke}}, \bibinfo {author}
  {\bibfnamefont {H.}~\bibnamefont {Hoekstra}}, \bibinfo {author}
  {\bibfnamefont {T.}~\bibnamefont {Erben}}, \bibinfo {author} {\bibfnamefont
  {M.~D.}\ \bibnamefont {Gladders}}, \bibinfo {author} {\bibfnamefont
  {M.}~\bibnamefont {Hetterscheidt}}, \bibinfo {author} {\bibfnamefont
  {Y.}~\bibnamefont {Mellier}}, \ and\ \bibinfo {author} {\bibfnamefont
  {H.~K.~C.}\ \bibnamefont {Yee}},\ }\href {\doibase
  10.1111/j.1365-2966.2007.12202.x} {\bibfield  {journal} {\bibinfo  {journal}
  {Mon. Not. Roy. Astron. Soc.}\ }\textbf {\bibinfo {volume} {381}},\ \bibinfo
  {pages} {702} (\bibinfo {year} {2007})},\ \Eprint
  {http://arxiv.org/abs/astro-ph/0703570} {arXiv:astro-ph/0703570} \BibitemShut
  {NoStop}%
\bibitem [{\citenamefont {Fu}\ \emph {et~al.}(2008)\citenamefont {Fu} \emph
  {et~al.}}]{Fu:2007qq}%
  \BibitemOpen
  \bibfield  {author} {\bibinfo {author} {\bibfnamefont {L.}~\bibnamefont {Fu}}
  \emph {et~al.},\ }\href {\doibase 10.1051/0004-6361:20078522} {\bibfield
  {journal} {\bibinfo  {journal} {Astron. Astrophys.}\ }\textbf {\bibinfo
  {volume} {479}},\ \bibinfo {pages} {9} (\bibinfo {year} {2008})},\ \Eprint
  {http://arxiv.org/abs/0712.0884} {arXiv:0712.0884 [astro-ph]} \BibitemShut
  {NoStop}%
\bibitem [{\citenamefont {Schrabback}\ \emph {et~al.}(2010)\citenamefont
  {Schrabback} \emph {et~al.}}]{Schrabback:2009ba}%
  \BibitemOpen
  \bibfield  {author} {\bibinfo {author} {\bibfnamefont {T.}~\bibnamefont
  {Schrabback}} \emph {et~al.},\ }\href {\doibase 10.1051/0004-6361/200913577}
  {\bibfield  {journal} {\bibinfo  {journal} {Astron. Astrophys.}\ }\textbf
  {\bibinfo {volume} {516}},\ \bibinfo {pages} {A63} (\bibinfo {year}
  {2010})},\ \Eprint {http://arxiv.org/abs/0911.0053} {arXiv:0911.0053
  [astro-ph.CO]} \BibitemShut {NoStop}%
\bibitem [{\citenamefont {Huff}\ \emph
  {et~al.}(2014{\natexlab{a}})\citenamefont {Huff}, \citenamefont {Hirata},
  \citenamefont {Mandelbaum}, \citenamefont {Schlegel}, \citenamefont
  {Seljak},\ and\ \citenamefont {Lupton}}]{Huff:2011gq}%
  \BibitemOpen
  \bibfield  {author} {\bibinfo {author} {\bibfnamefont {E.~M.}\ \bibnamefont
  {Huff}}, \bibinfo {author} {\bibfnamefont {C.~M.}\ \bibnamefont {Hirata}},
  \bibinfo {author} {\bibfnamefont {R.}~\bibnamefont {Mandelbaum}}, \bibinfo
  {author} {\bibfnamefont {D.}~\bibnamefont {Schlegel}}, \bibinfo {author}
  {\bibfnamefont {U.}~\bibnamefont {Seljak}}, \ and\ \bibinfo {author}
  {\bibfnamefont {R.~H.}\ \bibnamefont {Lupton}},\ }\href {\doibase
  10.1093/mnras/stu144} {\bibfield  {journal} {\bibinfo  {journal} {Mon. Not.
  Roy. Astron. Soc.}\ }\textbf {\bibinfo {volume} {440}},\ \bibinfo {pages}
  {1296} (\bibinfo {year} {2014}{\natexlab{a}})},\ \Eprint
  {http://arxiv.org/abs/1111.6958} {arXiv:1111.6958 [astro-ph.CO]} \BibitemShut
  {NoStop}%
\bibitem [{\citenamefont {Huff}\ \emph
  {et~al.}(2014{\natexlab{b}})\citenamefont {Huff}, \citenamefont {Eifler},
  \citenamefont {Hirata}, \citenamefont {Mandelbaum}, \citenamefont
  {Schlegel},\ and\ \citenamefont {Seljak}}]{Huff:2011aa}%
  \BibitemOpen
  \bibfield  {author} {\bibinfo {author} {\bibfnamefont {E.~M.}\ \bibnamefont
  {Huff}}, \bibinfo {author} {\bibfnamefont {T.}~\bibnamefont {Eifler}},
  \bibinfo {author} {\bibfnamefont {C.~M.}\ \bibnamefont {Hirata}}, \bibinfo
  {author} {\bibfnamefont {R.}~\bibnamefont {Mandelbaum}}, \bibinfo {author}
  {\bibfnamefont {D.}~\bibnamefont {Schlegel}}, \ and\ \bibinfo {author}
  {\bibfnamefont {U.}~\bibnamefont {Seljak}},\ }\href {\doibase
  10.1093/mnras/stu145} {\bibfield  {journal} {\bibinfo  {journal} {Mon. Not.
  Roy. Astron. Soc.}\ }\textbf {\bibinfo {volume} {440}},\ \bibinfo {pages}
  {1322} (\bibinfo {year} {2014}{\natexlab{b}})},\ \Eprint
  {http://arxiv.org/abs/1112.3143} {arXiv:1112.3143 [astro-ph.CO]} \BibitemShut
  {NoStop}%
\bibitem [{\citenamefont {Lin}\ \emph {et~al.}(2012)\citenamefont {Lin},
  \citenamefont {Dodelson}, \citenamefont {Seo}, \citenamefont {Soares-Santos},
  \citenamefont {Annis}, \citenamefont {Hao}, \citenamefont {Johnston},
  \citenamefont {Kubo}, \citenamefont {Reis},\ and\ \citenamefont
  {Simet}}]{Lin:2011bc}%
  \BibitemOpen
  \bibfield  {author} {\bibinfo {author} {\bibfnamefont {H.}~\bibnamefont
  {Lin}}, \bibinfo {author} {\bibfnamefont {S.}~\bibnamefont {Dodelson}},
  \bibinfo {author} {\bibfnamefont {H.-J.}\ \bibnamefont {Seo}}, \bibinfo
  {author} {\bibfnamefont {M.}~\bibnamefont {Soares-Santos}}, \bibinfo {author}
  {\bibfnamefont {J.}~\bibnamefont {Annis}}, \bibinfo {author} {\bibfnamefont
  {J.}~\bibnamefont {Hao}}, \bibinfo {author} {\bibfnamefont {D.}~\bibnamefont
  {Johnston}}, \bibinfo {author} {\bibfnamefont {J.~M.}\ \bibnamefont {Kubo}},
  \bibinfo {author} {\bibfnamefont {R.~R.~R.}\ \bibnamefont {Reis}}, \ and\
  \bibinfo {author} {\bibfnamefont {M.}~\bibnamefont {Simet}} (\bibinfo
  {collaboration} {SDSS}),\ }\href {\doibase 10.1088/0004-637X/761/1/15}
  {\bibfield  {journal} {\bibinfo  {journal} {Astrophys. J.}\ }\textbf
  {\bibinfo {volume} {761}},\ \bibinfo {pages} {15} (\bibinfo {year} {2012})},\
  \Eprint {http://arxiv.org/abs/1111.6622} {arXiv:1111.6622 [astro-ph.CO]}
  \BibitemShut {NoStop}%
\bibitem [{\citenamefont {Jee}\ \emph {et~al.}(2013{\natexlab{a}})\citenamefont
  {Jee}, \citenamefont {Tyson}, \citenamefont {Schneider}, \citenamefont
  {Wittman}, \citenamefont {Schmidt},\ and\ \citenamefont
  {Hilbert}}]{Jee:2012hr}%
  \BibitemOpen
  \bibfield  {author} {\bibinfo {author} {\bibfnamefont {M.~J.}\ \bibnamefont
  {Jee}}, \bibinfo {author} {\bibfnamefont {J.~A.}\ \bibnamefont {Tyson}},
  \bibinfo {author} {\bibfnamefont {M.~D.}\ \bibnamefont {Schneider}}, \bibinfo
  {author} {\bibfnamefont {D.}~\bibnamefont {Wittman}}, \bibinfo {author}
  {\bibfnamefont {S.}~\bibnamefont {Schmidt}}, \ and\ \bibinfo {author}
  {\bibfnamefont {S.}~\bibnamefont {Hilbert}},\ }\href {\doibase
  10.1088/0004-637X/765/1/74} {\bibfield  {journal} {\bibinfo  {journal}
  {Astrophys. J.}\ }\textbf {\bibinfo {volume} {765}},\ \bibinfo {pages} {74}
  (\bibinfo {year} {2013}{\natexlab{a}})},\ \Eprint
  {http://arxiv.org/abs/1210.2732} {arXiv:1210.2732 [astro-ph.CO]} \BibitemShut
  {NoStop}%
\bibitem [{\citenamefont {Jee}\ \emph {et~al.}(2016)\citenamefont {Jee},
  \citenamefont {Tyson}, \citenamefont {Hilbert}, \citenamefont {Schneider},
  \citenamefont {Schmidt},\ and\ \citenamefont {Wittman}}]{Jee:2015jta}%
  \BibitemOpen
  \bibfield  {author} {\bibinfo {author} {\bibfnamefont {M.~J.}\ \bibnamefont
  {Jee}}, \bibinfo {author} {\bibfnamefont {J.~A.}\ \bibnamefont {Tyson}},
  \bibinfo {author} {\bibfnamefont {S.}~\bibnamefont {Hilbert}}, \bibinfo
  {author} {\bibfnamefont {M.~D.}\ \bibnamefont {Schneider}}, \bibinfo {author}
  {\bibfnamefont {S.}~\bibnamefont {Schmidt}}, \ and\ \bibinfo {author}
  {\bibfnamefont {D.}~\bibnamefont {Wittman}},\ }\href {\doibase
  10.3847/0004-637X/824/2/77} {\bibfield  {journal} {\bibinfo  {journal}
  {Astrophys. J.}\ }\textbf {\bibinfo {volume} {824}},\ \bibinfo {pages} {77}
  (\bibinfo {year} {2016})},\ \Eprint {http://arxiv.org/abs/1510.03962}
  {arXiv:1510.03962 [astro-ph.CO]} \BibitemShut {NoStop}%
\bibitem [{\citenamefont {Erben}\ \emph {et~al.}(2013)\citenamefont {Erben}
  \emph {et~al.}}]{Erben:2012zw}%
  \BibitemOpen
  \bibfield  {author} {\bibinfo {author} {\bibfnamefont {T.}~\bibnamefont
  {Erben}} \emph {et~al.},\ }\href {\doibase 10.1093/mnras/stt928} {\bibfield
  {journal} {\bibinfo  {journal} {Mon. Not. Roy. Astron. Soc.}\ }\textbf
  {\bibinfo {volume} {433}},\ \bibinfo {pages} {2545} (\bibinfo {year}
  {2013})},\ \Eprint {http://arxiv.org/abs/1210.8156} {arXiv:1210.8156
  [astro-ph.CO]} \BibitemShut {NoStop}%
\bibitem [{\citenamefont {{Miller}}\ \emph {et~al.}(2013)\citenamefont
  {{Miller}}, \citenamefont {{Heymans}}, \citenamefont {{Kitching}},
  \citenamefont {{van Waerbeke}},\ and\ \citenamefont {{Erben}}}]{lensfit}%
  \BibitemOpen
  \bibfield  {author} {\bibinfo {author} {\bibfnamefont {L.}~\bibnamefont
  {{Miller}}}, \bibinfo {author} {\bibfnamefont {C.}~\bibnamefont {{Heymans}}},
  \bibinfo {author} {\bibfnamefont {T.~D.}\ \bibnamefont {{Kitching}}},
  \bibinfo {author} {\bibfnamefont {L.}~\bibnamefont {{van Waerbeke}}}, \ and\
  \bibinfo {author} {\bibfnamefont {T.}~\bibnamefont {{Erben}}},\ }\href
  {\doibase 10.1093/mnras/sts454} {\bibfield  {journal} {\bibinfo  {journal}
  {\mnras}\ }\textbf {\bibinfo {volume} {429}},\ \bibinfo {pages} {2858}
  (\bibinfo {year} {2013})},\ \Eprint {http://arxiv.org/abs/1210.8201}
  {arXiv:1210.8201 [astro-ph.CO]} \BibitemShut {NoStop}%
\bibitem [{\citenamefont {Kilbinger}\ \emph {et~al.}(2013)\citenamefont
  {Kilbinger} \emph {et~al.}}]{Kilbinger:2012qz}%
  \BibitemOpen
  \bibfield  {author} {\bibinfo {author} {\bibfnamefont {M.}~\bibnamefont
  {Kilbinger}} \emph {et~al.},\ }\href {\doibase 10.1093/mnras/stt041}
  {\bibfield  {journal} {\bibinfo  {journal} {Mon. Not. Roy. Astron. Soc.}\
  }\textbf {\bibinfo {volume} {430}},\ \bibinfo {pages} {2200} (\bibinfo {year}
  {2013})},\ \Eprint {http://arxiv.org/abs/1212.3338} {arXiv:1212.3338
  [astro-ph.CO]} \BibitemShut {NoStop}%
\bibitem [{\citenamefont {Kitching}\ \emph {et~al.}(2014)\citenamefont
  {Kitching} \emph {et~al.}}]{Kitching:2014dtq}%
  \BibitemOpen
  \bibfield  {author} {\bibinfo {author} {\bibfnamefont {T.~D.}\ \bibnamefont
  {Kitching}} \emph {et~al.} (\bibinfo {collaboration} {CFHTLenS}),\ }\href
  {\doibase 10.1093/mnras/stu934} {\bibfield  {journal} {\bibinfo  {journal}
  {Mon. Not. Roy. Astron. Soc.}\ }\textbf {\bibinfo {volume} {442}},\ \bibinfo
  {pages} {1326} (\bibinfo {year} {2014})},\ \Eprint
  {http://arxiv.org/abs/1401.6842} {arXiv:1401.6842 [astro-ph.CO]} \BibitemShut
  {NoStop}%
\bibitem [{\citenamefont {{Heymans}}\ and\ \citenamefont {{Van
  Waerbeke}}(2006)}]{heymans06}%
  \BibitemOpen
  \bibfield  {author} {\bibinfo {author} {\bibfnamefont {C.}~\bibnamefont
  {{Heymans}}}\ and\ \bibinfo {author} {\bibfnamefont {L.}~\bibnamefont {{Van
  Waerbeke}}},\ }\href {\doibase 10.1111/j.1365-2966.2006.10198.x} {\bibfield
  {journal} {\bibinfo  {journal} {\mnras}\ }\textbf {\bibinfo {volume} {368}},\
  \bibinfo {pages} {1323} (\bibinfo {year} {2006})},\ \Eprint
  {http://arxiv.org/abs/astro-ph/0506112} {arXiv:astro-ph/0506112 [astro-ph]}
  \BibitemShut {NoStop}%
\bibitem [{\citenamefont {{Aihara}}\ and\ \citenamefont {{HSC
  Collaboration}}(2018)}]{HSC2017}%
  \BibitemOpen
  \bibfield  {author} {\bibinfo {author} {\bibfnamefont {H.}~\bibnamefont
  {{Aihara}}}\ and\ \bibinfo {author} {\bibnamefont {{HSC Collaboration}}},\
  }\href {\doibase 10.1093/pasj/psx066} {\bibfield  {journal} {\bibinfo
  {journal} {\pasj}\ }\textbf {\bibinfo {volume} {70}},\ \bibinfo {eid} {S4}
  (\bibinfo {year} {2018})},\ \Eprint {http://arxiv.org/abs/1704.05858}
  {arXiv:1704.05858 [astro-ph.IM]} \BibitemShut {NoStop}%
\bibitem [{\citenamefont {{Kuijken}}\ \emph {et~al.}(2015)\citenamefont
  {{Kuijken}}, \citenamefont {{Heymans}}, \citenamefont {{Hildebrandt}},
  \citenamefont {{Nakajima}}, \citenamefont {{Erben}}, \citenamefont {{de
  Jong}}, \citenamefont {{Viola}}, \citenamefont {{Choi}}, \citenamefont
  {{Hoekstra}}, \citenamefont {{Miller}},\ and\ \citenamefont {{van
  Uitert}}}]{Kuijken15}%
  \BibitemOpen
  \bibfield  {author} {\bibinfo {author} {\bibfnamefont {K.}~\bibnamefont
  {{Kuijken}}}, \bibinfo {author} {\bibfnamefont {C.}~\bibnamefont
  {{Heymans}}}, \bibinfo {author} {\bibfnamefont {H.}~\bibnamefont
  {{Hildebrandt}}}, \bibinfo {author} {\bibfnamefont {R.}~\bibnamefont
  {{Nakajima}}}, \bibinfo {author} {\bibfnamefont {T.}~\bibnamefont {{Erben}}},
  \bibinfo {author} {\bibfnamefont {J.~T.~A.}\ \bibnamefont {{de Jong}}},
  \bibinfo {author} {\bibfnamefont {M.}~\bibnamefont {{Viola}}}, \bibinfo
  {author} {\bibfnamefont {A.}~\bibnamefont {{Choi}}}, \bibinfo {author}
  {\bibfnamefont {H.}~\bibnamefont {{Hoekstra}}}, \bibinfo {author}
  {\bibfnamefont {L.}~\bibnamefont {{Miller}}}, \ and\ \bibinfo {author}
  {\bibfnamefont {E.}~\bibnamefont {{van Uitert}}},\ }\href {\doibase
  10.1093/mnras/stv2140} {\bibfield  {journal} {\bibinfo  {journal} {\mnras}\
  }\textbf {\bibinfo {volume} {454}},\ \bibinfo {pages} {3500} (\bibinfo {year}
  {2015})},\ \Eprint {http://arxiv.org/abs/1507.00738} {arXiv:1507.00738}
  \BibitemShut {NoStop}%
\bibitem [{\citenamefont {Jee}\ \emph {et~al.}(2013{\natexlab{b}})\citenamefont
  {Jee}, \citenamefont {Tyson}, \citenamefont {Schneider}, \citenamefont
  {Wittman}, \citenamefont {Schmidt},\ and\ \citenamefont {Hilbert}}]{dls}%
  \BibitemOpen
  \bibfield  {author} {\bibinfo {author} {\bibfnamefont {M.~J.}\ \bibnamefont
  {Jee}}, \bibinfo {author} {\bibfnamefont {J.~A.}\ \bibnamefont {Tyson}},
  \bibinfo {author} {\bibfnamefont {M.~D.}\ \bibnamefont {Schneider}}, \bibinfo
  {author} {\bibfnamefont {D.}~\bibnamefont {Wittman}}, \bibinfo {author}
  {\bibfnamefont {S.}~\bibnamefont {Schmidt}}, \ and\ \bibinfo {author}
  {\bibfnamefont {S.}~\bibnamefont {Hilbert}},\ }\href {\doibase
  10.1088/0004-637x/765/1/74} {\bibfield  {journal} {\bibinfo  {journal} {The
  Astrophysical Journal}\ }\textbf {\bibinfo {volume} {765}},\ \bibinfo {pages}
  {74} (\bibinfo {year} {2013}{\natexlab{b}})}\BibitemShut {NoStop}%
\bibitem [{\citenamefont {{Dark Energy Survey
  Collaboration}}(2016{\natexlab{a}})}]{svcosmicshear}%
  \BibitemOpen
  \bibfield  {author} {\bibinfo {author} {\bibnamefont {{Dark Energy Survey
  Collaboration}}},\ }\href {\doibase 10.1103/PhysRevD.94.022001} {\bibfield
  {journal} {\bibinfo  {journal} {\prd}\ }\textbf {\bibinfo {volume} {94}},\
  \bibinfo {eid} {022001} (\bibinfo {year} {2016}{\natexlab{a}})},\ \Eprint
  {http://arxiv.org/abs/1507.05552} {arXiv:1507.05552 [astro-ph.CO]}
  \BibitemShut {NoStop}%
\bibitem [{\citenamefont {Hildebrandt}\ \emph {et~al.}(2017)\citenamefont
  {Hildebrandt}, \citenamefont {Viola}, \citenamefont {Heymans}, \citenamefont
  {Joudaki}, \citenamefont {Kuijken}, \citenamefont {Blake}, \citenamefont
  {Erben}, \citenamefont {Joachimi}, \citenamefont {Klaes}, \citenamefont
  {Miller} \emph {et~al.}}]{hildebrandt2017kids}%
  \BibitemOpen
  \bibfield  {author} {\bibinfo {author} {\bibfnamefont {H.}~\bibnamefont
  {Hildebrandt}}, \bibinfo {author} {\bibfnamefont {M.}~\bibnamefont {Viola}},
  \bibinfo {author} {\bibfnamefont {C.}~\bibnamefont {Heymans}}, \bibinfo
  {author} {\bibfnamefont {S.}~\bibnamefont {Joudaki}}, \bibinfo {author}
  {\bibfnamefont {K.}~\bibnamefont {Kuijken}}, \bibinfo {author} {\bibfnamefont
  {C.}~\bibnamefont {Blake}}, \bibinfo {author} {\bibfnamefont
  {T.}~\bibnamefont {Erben}}, \bibinfo {author} {\bibfnamefont
  {B.}~\bibnamefont {Joachimi}}, \bibinfo {author} {\bibfnamefont
  {D.}~\bibnamefont {Klaes}}, \bibinfo {author} {\bibfnamefont {L.~t.}\
  \bibnamefont {Miller}},  \emph {et~al.},\ }\href@noop {} {\bibfield
  {journal} {\bibinfo  {journal} {Monthly Notices of the Royal Astronomical
  Society}\ }\textbf {\bibinfo {volume} {465}},\ \bibinfo {pages} {1454}
  (\bibinfo {year} {2017})}\BibitemShut {NoStop}%
\bibitem [{\citenamefont {Troxel}\ \emph
  {et~al.}(2018{\natexlab{a}})\citenamefont {Troxel} \emph
  {et~al.}}]{Troxel:2018qll}%
  \BibitemOpen
  \bibfield  {author} {\bibinfo {author} {\bibfnamefont {M.~A.}\ \bibnamefont
  {Troxel}} \emph {et~al.} (\bibinfo {collaboration} {DES}),\ }\href {\doibase
  10.1093/mnras/sty1889} {\bibfield  {journal} {\bibinfo  {journal} {Mon. Not.
  Roy. Astron. Soc.}\ }\textbf {\bibinfo {volume} {479}},\ \bibinfo {pages}
  {4998} (\bibinfo {year} {2018}{\natexlab{a}})},\ \Eprint
  {http://arxiv.org/abs/1804.10663} {arXiv:1804.10663 [astro-ph.CO]}
  \BibitemShut {NoStop}%
\bibitem [{\citenamefont {{Hildebrandt}}\ \emph {et~al.}(2020)\citenamefont
  {{Hildebrandt}}, \citenamefont {{K{\"o}hlinger}}, \citenamefont {{van den
  Busch}}, \citenamefont {{Joachimi}},\ and\ \citenamefont
  {{Heymans}}}]{hildebrandt20}%
  \BibitemOpen
  \bibfield  {author} {\bibinfo {author} {\bibfnamefont {H.}~\bibnamefont
  {{Hildebrandt}}}, \bibinfo {author} {\bibfnamefont {F.}~\bibnamefont
  {{K{\"o}hlinger}}}, \bibinfo {author} {\bibfnamefont {J.~L.}\ \bibnamefont
  {{van den Busch}}}, \bibinfo {author} {\bibfnamefont {B.}~\bibnamefont
  {{Joachimi}}}, \ and\ \bibinfo {author} {\bibfnamefont {C.}~\bibnamefont
  {{Heymans}}},\ }\href {\doibase 10.1051/0004-6361/201834878} {\bibfield
  {journal} {\bibinfo  {journal} {\aap}\ }\textbf {\bibinfo {volume} {633}},\
  \bibinfo {eid} {A69} (\bibinfo {year} {2020})},\ \Eprint
  {http://arxiv.org/abs/1812.06076} {arXiv:1812.06076 [astro-ph.CO]}
  \BibitemShut {NoStop}%
\bibitem [{\citenamefont {{Hamana}}\ \emph {et~al.}(2020)\citenamefont
  {{Hamana}}, \citenamefont {{Shirasaki}}, \citenamefont {{Miyazaki}},
  \citenamefont {{Hikage}}, \citenamefont {{Oguri}},\ and\ \citenamefont
  {{More}}}]{Ham20}%
  \BibitemOpen
  \bibfield  {author} {\bibinfo {author} {\bibfnamefont {T.}~\bibnamefont
  {{Hamana}}}, \bibinfo {author} {\bibfnamefont {M.}~\bibnamefont
  {{Shirasaki}}}, \bibinfo {author} {\bibfnamefont {S.}~\bibnamefont
  {{Miyazaki}}}, \bibinfo {author} {\bibfnamefont {C.}~\bibnamefont
  {{Hikage}}}, \bibinfo {author} {\bibfnamefont {M.}~\bibnamefont {{Oguri}}}, \
  and\ \bibinfo {author} {\bibfnamefont {S.}~\bibnamefont {{More}}},\ }\href
  {\doibase 10.1093/pasj/psz138} {\bibfield  {journal} {\bibinfo  {journal}
  {\pasj}\ }\textbf {\bibinfo {volume} {72}},\ \bibinfo {eid} {16} (\bibinfo
  {year} {2020})},\ \Eprint {http://arxiv.org/abs/1906.06041} {arXiv:1906.06041
  [astro-ph.CO]} \BibitemShut {NoStop}%
\bibitem [{\citenamefont {{Joudaki}}\ \emph {et~al.}(2018)\citenamefont
  {{Joudaki}}, \citenamefont {{Blake}}, \citenamefont {{Johnson}},
  \citenamefont {{Amon}}, \citenamefont {{Asgari}},\ and\ \citenamefont
  {{Choi}}}]{joudaki_KiDS2df}%
  \BibitemOpen
  \bibfield  {author} {\bibinfo {author} {\bibfnamefont {S.}~\bibnamefont
  {{Joudaki}}}, \bibinfo {author} {\bibfnamefont {C.}~\bibnamefont {{Blake}}},
  \bibinfo {author} {\bibfnamefont {A.}~\bibnamefont {{Johnson}}}, \bibinfo
  {author} {\bibfnamefont {A.}~\bibnamefont {{Amon}}}, \bibinfo {author}
  {\bibfnamefont {M.}~\bibnamefont {{Asgari}}}, \ and\ \bibinfo {author}
  {\bibfnamefont {A.}~\bibnamefont {{Choi}}},\ }\href {\doibase
  10.1093/mnras/stx2820} {\bibfield  {journal} {\bibinfo  {journal} {\mnras}\
  }\textbf {\bibinfo {volume} {474}},\ \bibinfo {pages} {4894} (\bibinfo {year}
  {2018})},\ \Eprint {http://arxiv.org/abs/1707.06627} {arXiv:1707.06627
  [astro-ph.CO]} \BibitemShut {NoStop}%
\bibitem [{\citenamefont {{van Uitert}}\ \emph {et~al.}(2018)\citenamefont
  {{van Uitert}}, \citenamefont {{Joachimi}}, \citenamefont {{Joudaki}},
  \citenamefont {{Amon}},\ and\ \citenamefont
  {{Heymans}}}]{vanuitert_KiDSGAMA}%
  \BibitemOpen
  \bibfield  {author} {\bibinfo {author} {\bibfnamefont {E.}~\bibnamefont {{van
  Uitert}}}, \bibinfo {author} {\bibfnamefont {B.}~\bibnamefont {{Joachimi}}},
  \bibinfo {author} {\bibfnamefont {S.}~\bibnamefont {{Joudaki}}}, \bibinfo
  {author} {\bibfnamefont {A.}~\bibnamefont {{Amon}}}, \ and\ \bibinfo {author}
  {\bibfnamefont {C.}~\bibnamefont {{Heymans}}},\ }\href {\doibase
  10.1093/mnras/sty551} {\bibfield  {journal} {\bibinfo  {journal} {\mnras}\
  }\textbf {\bibinfo {volume} {476}},\ \bibinfo {pages} {4662} (\bibinfo {year}
  {2018})},\ \Eprint {http://arxiv.org/abs/1706.05004} {arXiv:1706.05004
  [astro-ph.CO]} \BibitemShut {NoStop}%
\bibitem [{\citenamefont {{LSST Dark Energy Science
  Collaboration}}(2012)}]{LSST12}%
  \BibitemOpen
  \bibfield  {author} {\bibinfo {author} {\bibnamefont {{LSST Dark Energy
  Science Collaboration}}},\ }\href@noop {} {\bibfield  {journal} {\bibinfo
  {journal} {arXiv e-prints}\ ,\ \bibinfo {eid} {arXiv:1211.0310}} (\bibinfo
  {year} {2012})},\ \Eprint {http://arxiv.org/abs/1211.0310} {arXiv:1211.0310
  [astro-ph.CO]} \BibitemShut {NoStop}%
\bibitem [{\citenamefont {{Laureijs}}\ \emph {et~al.}(2011)\citenamefont
  {{Laureijs}}, \citenamefont {{Amiaux}}, \citenamefont {{Arduini}},
  \citenamefont {{Augu{\`e}res}}, \citenamefont {{Brinchmann}}, \citenamefont
  {{Cole}}, \citenamefont {{Cropper}}, \citenamefont {{Dabin}}, \citenamefont
  {{Duvet}}, \citenamefont {{Ealet}},\ and\ \citenamefont
  {et~al.}}]{Laureijs2011}%
  \BibitemOpen
  \bibfield  {author} {\bibinfo {author} {\bibfnamefont {R.}~\bibnamefont
  {{Laureijs}}}, \bibinfo {author} {\bibfnamefont {J.}~\bibnamefont
  {{Amiaux}}}, \bibinfo {author} {\bibfnamefont {S.}~\bibnamefont {{Arduini}}},
  \bibinfo {author} {\bibfnamefont {J.~.}\ \bibnamefont {{Augu{\`e}res}}},
  \bibinfo {author} {\bibfnamefont {J.}~\bibnamefont {{Brinchmann}}}, \bibinfo
  {author} {\bibfnamefont {R.}~\bibnamefont {{Cole}}}, \bibinfo {author}
  {\bibfnamefont {M.}~\bibnamefont {{Cropper}}}, \bibinfo {author}
  {\bibfnamefont {C.}~\bibnamefont {{Dabin}}}, \bibinfo {author} {\bibfnamefont
  {L.}~\bibnamefont {{Duvet}}}, \bibinfo {author} {\bibfnamefont
  {A.}~\bibnamefont {{Ealet}}}, \ and\ \bibinfo {author} {\bibnamefont
  {et~al.}},\ }\href@noop {} {\bibfield  {journal} {\bibinfo  {journal} {ArXiv
  e-prints}\ } (\bibinfo {year} {2011})},\ \Eprint
  {http://arxiv.org/abs/1110.3193} {arXiv:1110.3193 [astro-ph.CO]} \BibitemShut
  {NoStop}%
\bibitem [{\citenamefont {{Spergel}}(2015)}]{Spergel15}%
  \BibitemOpen
  \bibfield  {author} {\bibinfo {author} {\bibfnamefont {D.}~\bibnamefont
  {{Spergel}}},\ }\href@noop {} {\bibfield  {journal} {\bibinfo  {journal}
  {ArXiv e-prints}\ } (\bibinfo {year} {2015})},\ \Eprint
  {http://arxiv.org/abs/1503.03757} {arXiv:1503.03757 [astro-ph.IM]}
  \BibitemShut {NoStop}%
\bibitem [{\citenamefont {{Mandelbaum}}(2018)}]{MandelbaumRev}%
  \BibitemOpen
  \bibfield  {author} {\bibinfo {author} {\bibfnamefont {R.}~\bibnamefont
  {{Mandelbaum}}},\ }\href {\doibase 10.1146/annurev-astro-081817-051928}
  {\bibfield  {journal} {\bibinfo  {journal} {\araa}\ }\textbf {\bibinfo
  {volume} {56}},\ \bibinfo {pages} {393} (\bibinfo {year} {2018})},\ \Eprint
  {http://arxiv.org/abs/1710.03235} {arXiv:1710.03235 [astro-ph.CO]}
  \BibitemShut {NoStop}%
\bibitem [{\citenamefont {{Huff}}\ and\ \citenamefont
  {{Mandelbaum}}(2017)}]{Huff_Mandelbaum_2017}%
  \BibitemOpen
  \bibfield  {author} {\bibinfo {author} {\bibfnamefont {E.}~\bibnamefont
  {{Huff}}}\ and\ \bibinfo {author} {\bibfnamefont {R.}~\bibnamefont
  {{Mandelbaum}}},\ }\href@noop {} {\bibfield  {journal} {\bibinfo  {journal}
  {arXiv e-prints}\ ,\ \bibinfo {eid} {arXiv:1702.02600}} (\bibinfo {year}
  {2017})},\ \Eprint {http://arxiv.org/abs/1702.02600} {arXiv:1702.02600
  [astro-ph.CO]} \BibitemShut {NoStop}%
\bibitem [{\citenamefont {{Sheldon}}\ and\ \citenamefont
  {{Huff}}(2017)}]{SheldonMcal2017}%
  \BibitemOpen
  \bibfield  {author} {\bibinfo {author} {\bibfnamefont {E.~S.}\ \bibnamefont
  {{Sheldon}}}\ and\ \bibinfo {author} {\bibfnamefont {E.~M.}\ \bibnamefont
  {{Huff}}},\ }\href {\doibase 10.3847/1538-4357/aa704b} {\bibfield  {journal}
  {\bibinfo  {journal} {\apj}\ }\textbf {\bibinfo {volume} {841}},\ \bibinfo
  {eid} {24} (\bibinfo {year} {2017})},\ \Eprint
  {http://arxiv.org/abs/1702.02601} {arXiv:1702.02601} \BibitemShut {NoStop}%
\bibitem [{\citenamefont {Jarvis}\ \emph {et~al.}(2021)\citenamefont {Jarvis}
  \emph {et~al.}}]{y3-piff}%
  \BibitemOpen
  \bibfield  {author} {\bibinfo {author} {\bibfnamefont {M.}~\bibnamefont
  {Jarvis}} \emph {et~al.} (\bibinfo {collaboration} {DES}),\ }\href {\doibase
  10.1093/mnras/staa3679} {\bibfield  {journal} {\bibinfo  {journal} {Mon. Not.
  Roy. Astron. Soc.}\ }\textbf {\bibinfo {volume} {501}},\ \bibinfo {pages}
  {1282} (\bibinfo {year} {2021})},\ \Eprint {http://arxiv.org/abs/2011.03409}
  {arXiv:2011.03409 [astro-ph.IM]} \BibitemShut {NoStop}%
\bibitem [{\citenamefont {{Sheldon}}\ \emph {et~al.}(2020)\citenamefont
  {{Sheldon}}, \citenamefont {{Becker}}, \citenamefont {{MacCrann}},\ and\
  \citenamefont {{Jarvis}}}]{metadetect}%
  \BibitemOpen
  \bibfield  {author} {\bibinfo {author} {\bibfnamefont {E.~S.}\ \bibnamefont
  {{Sheldon}}}, \bibinfo {author} {\bibfnamefont {M.~R.}\ \bibnamefont
  {{Becker}}}, \bibinfo {author} {\bibfnamefont {N.}~\bibnamefont
  {{MacCrann}}}, \ and\ \bibinfo {author} {\bibfnamefont {M.}~\bibnamefont
  {{Jarvis}}},\ }\href {\doibase 10.3847/1538-4357/abb595} {\bibfield
  {journal} {\bibinfo  {journal} {\apj}\ }\textbf {\bibinfo {volume} {902}},\
  \bibinfo {eid} {138} (\bibinfo {year} {2020})},\ \Eprint
  {http://arxiv.org/abs/1911.02505} {arXiv:1911.02505 [astro-ph.CO]}
  \BibitemShut {NoStop}%
\bibitem [{\citenamefont {Bernstein}\ and\ \citenamefont
  {Armstrong}(2014)}]{bfd}%
  \BibitemOpen
  \bibfield  {author} {\bibinfo {author} {\bibfnamefont {G.~M.}\ \bibnamefont
  {Bernstein}}\ and\ \bibinfo {author} {\bibfnamefont {R.}~\bibnamefont
  {Armstrong}},\ }\href {\doibase 10.1093/mnras/stt2326} {\bibfield  {journal}
  {\bibinfo  {journal} {Monthly Notices of the Royal Astronomical Society}\
  }\textbf {\bibinfo {volume} {438}},\ \bibinfo {pages} {1880–1893} (\bibinfo
  {year} {2014})}\BibitemShut {NoStop}%
\bibitem [{\citenamefont {Gatti}\ \emph {et~al.}(2021)\citenamefont {Gatti},
  \citenamefont {Sheldon} \emph {et~al.}}]{y3-shapecatalog}%
  \BibitemOpen
  \bibfield  {author} {\bibinfo {author} {\bibfnamefont {M.}~\bibnamefont
  {Gatti}}, \bibinfo {author} {\bibfnamefont {E.}~\bibnamefont {Sheldon}},
  \emph {et~al.} (\bibinfo {collaboration} {DES}),\ }\href {\doibase
  10.1093/mnras/stab918} {\bibfield  {journal} {\bibinfo  {journal} {Mon. Not.
  Roy. Astron. Soc.}\ }\textbf {\bibinfo {volume} {504}},\ \bibinfo {pages}
  {4312} (\bibinfo {year} {2021})},\ \Eprint {http://arxiv.org/abs/2011.03408}
  {arXiv:2011.03408 [astro-ph.CO]} \BibitemShut {NoStop}%
\bibitem [{\citenamefont {{Giblin}}\ \emph {et~al.}(2021)\citenamefont
  {{Giblin}}, \citenamefont {{Heymans}}, \citenamefont {{Asgari}},\ and\
  \citenamefont {{Hildebrandt}}}]{giblin20}%
  \BibitemOpen
  \bibfield  {author} {\bibinfo {author} {\bibfnamefont {B.}~\bibnamefont
  {{Giblin}}}, \bibinfo {author} {\bibfnamefont {C.}~\bibnamefont {{Heymans}}},
  \bibinfo {author} {\bibfnamefont {M.}~\bibnamefont {{Asgari}}}, \ and\
  \bibinfo {author} {\bibnamefont {{Hildebrandt}}},\ }\href {\doibase
  10.1051/0004-6361/202038850} {\bibfield  {journal} {\bibinfo  {journal}
  {\aap}\ }\textbf {\bibinfo {volume} {645}},\ \bibinfo {eid} {A105} (\bibinfo
  {year} {2021})},\ \Eprint {http://arxiv.org/abs/2007.01845} {arXiv:2007.01845
  [astro-ph.CO]} \BibitemShut {NoStop}%
\bibitem [{\citenamefont {{Mandelbaum}}\ \emph {et~al.}(2018)\citenamefont
  {{Mandelbaum}}, \citenamefont {{Miyatake}}, \citenamefont {{Hamana}},
  \citenamefont {{Oguri}},\ and\ \citenamefont {{Simet}}}]{mandelbaum17}%
  \BibitemOpen
  \bibfield  {author} {\bibinfo {author} {\bibfnamefont {R.}~\bibnamefont
  {{Mandelbaum}}}, \bibinfo {author} {\bibfnamefont {H.}~\bibnamefont
  {{Miyatake}}}, \bibinfo {author} {\bibfnamefont {T.}~\bibnamefont
  {{Hamana}}}, \bibinfo {author} {\bibfnamefont {M.}~\bibnamefont {{Oguri}}}, \
  and\ \bibinfo {author} {\bibfnamefont {M.}~\bibnamefont {{Simet}}},\ }\href
  {\doibase 10.1093/pasj/psx130} {\bibfield  {journal} {\bibinfo  {journal}
  {\pasj}\ }\textbf {\bibinfo {volume} {70}},\ \bibinfo {eid} {S25} (\bibinfo
  {year} {2018})},\ \Eprint {http://arxiv.org/abs/1705.06745} {arXiv:1705.06745
  [astro-ph.CO]} \BibitemShut {NoStop}%
\bibitem [{\citenamefont {{Bernstein}}\ and\ \citenamefont
  {{Huterer}}(2010)}]{Bernsteinhuterer10}%
  \BibitemOpen
  \bibfield  {author} {\bibinfo {author} {\bibfnamefont {G.}~\bibnamefont
  {{Bernstein}}}\ and\ \bibinfo {author} {\bibfnamefont {D.}~\bibnamefont
  {{Huterer}}},\ }\href {\doibase 10.1111/j.1365-2966.2009.15748.x} {\bibfield
  {journal} {\bibinfo  {journal} {\mnras}\ }\textbf {\bibinfo {volume} {401}},\
  \bibinfo {pages} {1399} (\bibinfo {year} {2010})},\ \Eprint
  {http://arxiv.org/abs/0902.2782} {arXiv:0902.2782 [astro-ph.CO]} \BibitemShut
  {NoStop}%
\bibitem [{\citenamefont {{Newman}}(2015)}]{Newman2015}%
  \BibitemOpen
  \bibfield  {author} {\bibinfo {author} {\bibfnamefont {J.~A.}\ \bibnamefont
  {{Newman}}},\ }\href {\doibase 10.1016/j.astropartphys.2014.06.007}
  {\bibfield  {journal} {\bibinfo  {journal} {Astroparticle Physics}\ }\textbf
  {\bibinfo {volume} {63}},\ \bibinfo {pages} {81} (\bibinfo {year} {2015})},\
  \Eprint {http://arxiv.org/abs/1309.5384} {arXiv:1309.5384 [astro-ph.CO]}
  \BibitemShut {NoStop}%
\bibitem [{\citenamefont {{Gruen}}\ and\ \citenamefont
  {{Brimioulle}}(2017)}]{gruen17}%
  \BibitemOpen
  \bibfield  {author} {\bibinfo {author} {\bibfnamefont {D.}~\bibnamefont
  {{Gruen}}}\ and\ \bibinfo {author} {\bibfnamefont {F.}~\bibnamefont
  {{Brimioulle}}},\ }\href {\doibase 10.1093/mnras/stx471} {\bibfield
  {journal} {\bibinfo  {journal} {\mnras}\ }\textbf {\bibinfo {volume} {468}},\
  \bibinfo {pages} {769} (\bibinfo {year} {2017})},\ \Eprint
  {http://arxiv.org/abs/1610.01160} {arXiv:1610.01160 [astro-ph.CO]}
  \BibitemShut {NoStop}%
\bibitem [{\citenamefont {{Hoyle}}\ \emph {et~al.}(2018)\citenamefont
  {{Hoyle}}, \citenamefont {{Gruen}}, \citenamefont {{Bernstein}},
  \citenamefont {{Rau}},\ and\ \citenamefont {{DES Collaboration}}}]{hoyle}%
  \BibitemOpen
  \bibfield  {author} {\bibinfo {author} {\bibfnamefont {B.}~\bibnamefont
  {{Hoyle}}}, \bibinfo {author} {\bibfnamefont {D.}~\bibnamefont {{Gruen}}},
  \bibinfo {author} {\bibfnamefont {G.~M.}\ \bibnamefont {{Bernstein}}},
  \bibinfo {author} {\bibfnamefont {M.~M.}\ \bibnamefont {{Rau}}}, \ and\
  \bibinfo {author} {\bibnamefont {{DES Collaboration}}},\ }\href {\doibase
  10.1093/mnras/sty957} {\bibfield  {journal} {\bibinfo  {journal} {\mnras}\
  }\textbf {\bibinfo {volume} {478}},\ \bibinfo {pages} {592} (\bibinfo {year}
  {2018})},\ \Eprint {http://arxiv.org/abs/1708.01532} {arXiv:1708.01532
  [astro-ph.CO]} \BibitemShut {NoStop}%
\bibitem [{\citenamefont {{Wright}}\ \emph {et~al.}(2020)\citenamefont
  {{Wright}}, \citenamefont {{Hildebrandt}}, \citenamefont {{van den Busch}},
  \citenamefont {{Heymans}}, \citenamefont {{Joachimi}}, \citenamefont
  {{Kannawadi}},\ and\ \citenamefont {{Kuijken}}}]{wright20}%
  \BibitemOpen
  \bibfield  {author} {\bibinfo {author} {\bibfnamefont {A.~H.}\ \bibnamefont
  {{Wright}}}, \bibinfo {author} {\bibfnamefont {H.}~\bibnamefont
  {{Hildebrandt}}}, \bibinfo {author} {\bibfnamefont {J.~L.}\ \bibnamefont
  {{van den Busch}}}, \bibinfo {author} {\bibfnamefont {C.}~\bibnamefont
  {{Heymans}}}, \bibinfo {author} {\bibfnamefont {B.}~\bibnamefont
  {{Joachimi}}}, \bibinfo {author} {\bibfnamefont {A.}~\bibnamefont
  {{Kannawadi}}}, \ and\ \bibinfo {author} {\bibfnamefont {K.}~\bibnamefont
  {{Kuijken}}},\ }\href@noop {} {\bibfield  {journal} {\bibinfo  {journal}
  {arXiv e-prints}\ ,\ \bibinfo {eid} {arXiv:2005.04207}} (\bibinfo {year}
  {2020})},\ \Eprint {http://arxiv.org/abs/2005.04207} {arXiv:2005.04207
  [astro-ph.CO]} \BibitemShut {NoStop}%
\bibitem [{\citenamefont {{Joudaki}}\ and\ \citenamefont
  {{Hildebrandt}}(2020)}]{joudaki2020}%
  \BibitemOpen
  \bibfield  {author} {\bibinfo {author} {\bibfnamefont {S.}~\bibnamefont
  {{Joudaki}}}\ and\ \bibinfo {author} {\bibfnamefont {H.}~\bibnamefont
  {{Hildebrandt}}},\ }\href {\doibase 10.1051/0004-6361/201936154} {\bibfield
  {journal} {\bibinfo  {journal} {\aap}\ }\textbf {\bibinfo {volume} {638}},\
  \bibinfo {eid} {L1} (\bibinfo {year} {2020})},\ \Eprint
  {http://arxiv.org/abs/1906.09262} {arXiv:1906.09262 [astro-ph.CO]}
  \BibitemShut {NoStop}%
\bibitem [{\citenamefont {{Hartley}}\ \emph {et~al.}(2020)\citenamefont
  {{Hartley}}, \citenamefont {{Chang}},\ and\ \citenamefont {{DES
  Collaboration}}}]{hartley2020}%
  \BibitemOpen
  \bibfield  {author} {\bibinfo {author} {\bibfnamefont {W.~G.}\ \bibnamefont
  {{Hartley}}}, \bibinfo {author} {\bibfnamefont {C.}~\bibnamefont {{Chang}}},
  \ and\ \bibinfo {author} {\bibnamefont {{DES Collaboration}}},\ }\href
  {\doibase 10.1093/mnras/staa1812} {\bibfield  {journal} {\bibinfo  {journal}
  {\mnras}\ }\textbf {\bibinfo {volume} {496}},\ \bibinfo {pages} {4769}
  (\bibinfo {year} {2020})},\ \Eprint {http://arxiv.org/abs/2003.10454}
  {arXiv:2003.10454 [astro-ph.GA]} \BibitemShut {NoStop}%
\bibitem [{\citenamefont {Myles}\ \emph {et~al.}(2020)\citenamefont {Myles},
  \citenamefont {Alarcon} \emph {et~al.}}]{y3-sompz}%
  \BibitemOpen
  \bibfield  {author} {\bibinfo {author} {\bibfnamefont {J.}~\bibnamefont
  {Myles}}, \bibinfo {author} {\bibfnamefont {A.}~\bibnamefont {Alarcon}},
  \emph {et~al.} (\bibinfo {collaboration} {DES}),\ }\href@noop {} {\bibfield
  {journal} {\bibinfo  {journal} {Submitted to MNRAS}\ } (\bibinfo {year}
  {2020})},\ \Eprint {http://arxiv.org/abs/2012.08566} {arXiv:2012.08566
  [astro-ph.CO]} \BibitemShut {NoStop}%
\bibitem [{\citenamefont {Gatti}\ \emph {et~al.}(2020)\citenamefont {Gatti},
  \citenamefont {Giannini} \emph {et~al.}}]{y3-sourcewz}%
  \BibitemOpen
  \bibfield  {author} {\bibinfo {author} {\bibfnamefont {M.}~\bibnamefont
  {Gatti}}, \bibinfo {author} {\bibfnamefont {G.}~\bibnamefont {Giannini}},
  \emph {et~al.} (\bibinfo {collaboration} {DES}),\ }\href@noop {} {\bibfield
  {journal} {\bibinfo  {journal} {Submitted to MNRAS}\ } (\bibinfo {year}
  {2020})},\ \Eprint {http://arxiv.org/abs/2012.08569} {arXiv:2012.08569
  [astro-ph.CO]} \BibitemShut {NoStop}%
\bibitem [{\citenamefont {{S\'anchez}}\ \emph {et~al.}(2021)\citenamefont
  {{S\'anchez}}, \citenamefont {{Prat}} \emph {et~al.}}]{y3-shearratio}%
  \BibitemOpen
  \bibfield  {author} {\bibinfo {author} {\bibfnamefont {C.}~\bibnamefont
  {{S\'anchez}}}, \bibinfo {author} {\bibfnamefont {J.}~\bibnamefont {{Prat}}},
   \emph {et~al.},\ }\href@noop {} {\bibfield  {journal} {\bibinfo  {journal}
  {To be submitted to PRD}\ } (\bibinfo {year} {2021})}\BibitemShut {NoStop}%
\bibitem [{\citenamefont {{Dawson}}\ \emph {et~al.}(2016)\citenamefont
  {{Dawson}}, \citenamefont {{Schneider}}, \citenamefont {{Tyson}},\ and\
  \citenamefont {{Jee}}}]{Dawson16}%
  \BibitemOpen
  \bibfield  {author} {\bibinfo {author} {\bibfnamefont {W.~A.}\ \bibnamefont
  {{Dawson}}}, \bibinfo {author} {\bibfnamefont {M.~D.}\ \bibnamefont
  {{Schneider}}}, \bibinfo {author} {\bibfnamefont {J.~A.}\ \bibnamefont
  {{Tyson}}}, \ and\ \bibinfo {author} {\bibfnamefont {M.~J.}\ \bibnamefont
  {{Jee}}},\ }\href {\doibase 10.3847/0004-637X/816/1/11} {\bibfield  {journal}
  {\bibinfo  {journal} {\apj}\ }\textbf {\bibinfo {volume} {816}},\ \bibinfo
  {eid} {11} (\bibinfo {year} {2016})},\ \Eprint
  {http://arxiv.org/abs/1406.1506} {arXiv:1406.1506 [astro-ph.CO]} \BibitemShut
  {NoStop}%
\bibitem [{\citenamefont {{Mandelbaum}}\ and\ \citenamefont
  {{Lanusse}}(2018)}]{mandelbaum17b}%
  \BibitemOpen
  \bibfield  {author} {\bibinfo {author} {\bibfnamefont {R.}~\bibnamefont
  {{Mandelbaum}}}\ and\ \bibinfo {author} {\bibfnamefont {F.}~\bibnamefont
  {{Lanusse}}},\ }\href {\doibase 10.1093/mnras/sty2420} {\bibfield  {journal}
  {\bibinfo  {journal} {\mnras}\ }\textbf {\bibinfo {volume} {481}},\ \bibinfo
  {pages} {3170} (\bibinfo {year} {2018})},\ \Eprint
  {http://arxiv.org/abs/1710.00885} {arXiv:1710.00885 [astro-ph.CO]}
  \BibitemShut {NoStop}%
\bibitem [{\citenamefont {{Jarvis}}\ \emph {et~al.}(2016)\citenamefont
  {{Jarvis}}, \citenamefont {{Sheldon}},\ and\ \citenamefont
  {{Zuntz}}}]{jarvis2016}%
  \BibitemOpen
  \bibfield  {author} {\bibinfo {author} {\bibfnamefont {M.}~\bibnamefont
  {{Jarvis}}}, \bibinfo {author} {\bibfnamefont {E.}~\bibnamefont {{Sheldon}}},
  \ and\ \bibinfo {author} {\bibfnamefont {J.}~\bibnamefont {{Zuntz}}},\ }\href
  {\doibase 10.1093/mnras/stw990} {\bibfield  {journal} {\bibinfo  {journal}
  {\mnras}\ }\textbf {\bibinfo {volume} {460}},\ \bibinfo {pages} {2245}
  (\bibinfo {year} {2016})},\ \Eprint {http://arxiv.org/abs/1507.05603}
  {arXiv:1507.05603 [astro-ph.IM]} \BibitemShut {NoStop}%
\bibitem [{\citenamefont {{Bosch}}(2018)}]{bosch2017}%
  \BibitemOpen
  \bibfield  {author} {\bibinfo {author} {\bibfnamefont {J.}~\bibnamefont
  {{Bosch}}},\ }\href {\doibase 10.1093/pasj/psx080} {\bibfield  {journal}
  {\bibinfo  {journal} {\pasj}\ }\textbf {\bibinfo {volume} {70}},\ \bibinfo
  {eid} {S5} (\bibinfo {year} {2018})},\ \Eprint
  {http://arxiv.org/abs/1705.06766} {arXiv:1705.06766 [astro-ph.IM]}
  \BibitemShut {NoStop}%
\bibitem [{\citenamefont {{Samuroff}}\ \emph {et~al.}(2017)\citenamefont
  {{Samuroff}}, \citenamefont {{Troxel}}, \citenamefont {{Bridle}},
  \citenamefont {{Zuntz}}, \citenamefont {{MacCrann}}, \citenamefont
  {{Krause}}, \citenamefont {{Eifler}},\ and\ \citenamefont
  {{Kirk}}}]{samuroff17}%
  \BibitemOpen
  \bibfield  {author} {\bibinfo {author} {\bibfnamefont {S.}~\bibnamefont
  {{Samuroff}}}, \bibinfo {author} {\bibfnamefont {M.~A.}\ \bibnamefont
  {{Troxel}}}, \bibinfo {author} {\bibfnamefont {S.~L.}\ \bibnamefont
  {{Bridle}}}, \bibinfo {author} {\bibfnamefont {J.}~\bibnamefont {{Zuntz}}},
  \bibinfo {author} {\bibfnamefont {N.}~\bibnamefont {{MacCrann}}}, \bibinfo
  {author} {\bibfnamefont {E.}~\bibnamefont {{Krause}}}, \bibinfo {author}
  {\bibfnamefont {T.}~\bibnamefont {{Eifler}}}, \ and\ \bibinfo {author}
  {\bibfnamefont {D.}~\bibnamefont {{Kirk}}},\ }\href {\doibase
  10.1093/mnrasl/slw201} {\bibfield  {journal} {\bibinfo  {journal} {\mnras}\
  }\textbf {\bibinfo {volume} {465}},\ \bibinfo {pages} {L20} (\bibinfo {year}
  {2017})},\ \Eprint {http://arxiv.org/abs/1607.07910} {arXiv:1607.07910
  [astro-ph.CO]} \BibitemShut {NoStop}%
\bibitem [{\citenamefont {{Kannawadi}}\ \emph {et~al.}(2019)\citenamefont
  {{Kannawadi}}, \citenamefont {{Hoekstra}}, \citenamefont {{Miller}},
  \citenamefont {{Viola}}, \citenamefont {{Fenech Conti}}, \citenamefont
  {{Herbonnet}}, \citenamefont {{Erben}}, \citenamefont {{Heymans}},
  \citenamefont {{Hildebrandt}}, \citenamefont {{Kuijken}}, \citenamefont
  {{Vakili}},\ and\ \citenamefont {{Wright}}}]{kannawadi19}%
  \BibitemOpen
  \bibfield  {author} {\bibinfo {author} {\bibfnamefont {A.}~\bibnamefont
  {{Kannawadi}}}, \bibinfo {author} {\bibfnamefont {H.}~\bibnamefont
  {{Hoekstra}}}, \bibinfo {author} {\bibfnamefont {L.}~\bibnamefont
  {{Miller}}}, \bibinfo {author} {\bibfnamefont {M.}~\bibnamefont {{Viola}}},
  \bibinfo {author} {\bibfnamefont {I.}~\bibnamefont {{Fenech Conti}}},
  \bibinfo {author} {\bibfnamefont {R.}~\bibnamefont {{Herbonnet}}}, \bibinfo
  {author} {\bibfnamefont {T.}~\bibnamefont {{Erben}}}, \bibinfo {author}
  {\bibfnamefont {C.}~\bibnamefont {{Heymans}}}, \bibinfo {author}
  {\bibfnamefont {H.}~\bibnamefont {{Hildebrandt}}}, \bibinfo {author}
  {\bibfnamefont {K.}~\bibnamefont {{Kuijken}}}, \bibinfo {author}
  {\bibfnamefont {M.}~\bibnamefont {{Vakili}}}, \ and\ \bibinfo {author}
  {\bibfnamefont {A.~H.}\ \bibnamefont {{Wright}}},\ }\href {\doibase
  10.1051/0004-6361/201834819} {\bibfield  {journal} {\bibinfo  {journal}
  {\aap}\ }\textbf {\bibinfo {volume} {624}},\ \bibinfo {eid} {A92} (\bibinfo
  {year} {2019})},\ \Eprint {http://arxiv.org/abs/1812.03983} {arXiv:1812.03983
  [astro-ph.CO]} \BibitemShut {NoStop}%
\bibitem [{\citenamefont {MacCrann}\ \emph {et~al.}(2020)\citenamefont
  {MacCrann} \emph {et~al.}}]{y3-imagesims}%
  \BibitemOpen
  \bibfield  {author} {\bibinfo {author} {\bibfnamefont {N.}~\bibnamefont
  {MacCrann}} \emph {et~al.} (\bibinfo {collaboration} {DES}),\ }\href@noop {}
  {\bibfield  {journal} {\bibinfo  {journal} {Submitted to MNRAS}\ } (\bibinfo
  {year} {2020})},\ \Eprint {http://arxiv.org/abs/2012.08567} {arXiv:2012.08567
  [astro-ph.CO]} \BibitemShut {NoStop}%
\bibitem [{\citenamefont {{Croft}}\ and\ \citenamefont
  {{Metzler}}(2000)}]{croft2000}%
  \BibitemOpen
  \bibfield  {author} {\bibinfo {author} {\bibfnamefont {R.~A.~C.}\
  \bibnamefont {{Croft}}}\ and\ \bibinfo {author} {\bibfnamefont {C.~A.}\
  \bibnamefont {{Metzler}}},\ }\href {\doibase 10.1086/317856} {\bibfield
  {journal} {\bibinfo  {journal} {\apj}\ }\textbf {\bibinfo {volume} {545}},\
  \bibinfo {pages} {561} (\bibinfo {year} {2000})},\ \Eprint
  {http://arxiv.org/abs/astro-ph/0005384} {arXiv:astro-ph/0005384 [astro-ph]}
  \BibitemShut {NoStop}%
\bibitem [{\citenamefont {{Heavens}}\ \emph {et~al.}(2000)\citenamefont
  {{Heavens}}, \citenamefont {{Refregier}},\ and\ \citenamefont
  {{Heymans}}}]{heavens2000}%
  \BibitemOpen
  \bibfield  {author} {\bibinfo {author} {\bibfnamefont {A.}~\bibnamefont
  {{Heavens}}}, \bibinfo {author} {\bibfnamefont {A.}~\bibnamefont
  {{Refregier}}}, \ and\ \bibinfo {author} {\bibfnamefont {C.}~\bibnamefont
  {{Heymans}}},\ }\href {\doibase 10.1046/j.1365-8711.2000.03907.x} {\bibfield
  {journal} {\bibinfo  {journal} {\mnras}\ }\textbf {\bibinfo {volume} {319}},\
  \bibinfo {pages} {649} (\bibinfo {year} {2000})},\ \Eprint
  {http://arxiv.org/abs/astro-ph/0005269} {arXiv:astro-ph/0005269 [astro-ph]}
  \BibitemShut {NoStop}%
\bibitem [{\citenamefont {{Troxel}}\ and\ \citenamefont
  {{Ishak}}(2015)}]{troxel15}%
  \BibitemOpen
  \bibfield  {author} {\bibinfo {author} {\bibfnamefont {M.~A.}\ \bibnamefont
  {{Troxel}}}\ and\ \bibinfo {author} {\bibfnamefont {M.}~\bibnamefont
  {{Ishak}}},\ }\href {\doibase 10.1016/j.physrep.2014.11.001} {\bibfield
  {journal} {\bibinfo  {journal} {\physrep}\ }\textbf {\bibinfo {volume}
  {558}},\ \bibinfo {pages} {1} (\bibinfo {year} {2015})},\ \Eprint
  {http://arxiv.org/abs/1407.6990} {arXiv:1407.6990 [astro-ph.CO]} \BibitemShut
  {NoStop}%
\bibitem [{\citenamefont {{Chisari}}\ \emph {et~al.}(2018)\citenamefont
  {{Chisari}}, \citenamefont {{Richardson}}, \citenamefont {{Devriendt}},
  \citenamefont {{Dubois}}, \citenamefont {{Schneider}}, \citenamefont {{Le
  Brun}}, \citenamefont {{Beckmann}}, \citenamefont {{Peirani}}, \citenamefont
  {{Slyz}},\ and\ \citenamefont {{Pichon}}}]{chisari}%
  \BibitemOpen
  \bibfield  {author} {\bibinfo {author} {\bibfnamefont {N.~E.}\ \bibnamefont
  {{Chisari}}}, \bibinfo {author} {\bibfnamefont {M.~L.~A.}\ \bibnamefont
  {{Richardson}}}, \bibinfo {author} {\bibfnamefont {J.}~\bibnamefont
  {{Devriendt}}}, \bibinfo {author} {\bibfnamefont {Y.}~\bibnamefont
  {{Dubois}}}, \bibinfo {author} {\bibfnamefont {A.}~\bibnamefont
  {{Schneider}}}, \bibinfo {author} {\bibfnamefont {A.~M.~C.}\ \bibnamefont
  {{Le Brun}}}, \bibinfo {author} {\bibfnamefont {R.~S.}\ \bibnamefont
  {{Beckmann}}}, \bibinfo {author} {\bibfnamefont {S.}~\bibnamefont
  {{Peirani}}}, \bibinfo {author} {\bibfnamefont {A.}~\bibnamefont {{Slyz}}}, \
  and\ \bibinfo {author} {\bibfnamefont {C.}~\bibnamefont {{Pichon}}},\ }\href
  {\doibase 10.1093/mnras/sty2093} {\bibfield  {journal} {\bibinfo  {journal}
  {\mnras}\ }\textbf {\bibinfo {volume} {480}},\ \bibinfo {pages} {3962}
  (\bibinfo {year} {2018})},\ \Eprint {http://arxiv.org/abs/1801.08559}
  {arXiv:1801.08559 [astro-ph.CO]} \BibitemShut {NoStop}%
\bibitem [{\citenamefont {{Wechsler}}\ and\ \citenamefont
  {{Tinker}}(2018)}]{wechsler2018}%
  \BibitemOpen
  \bibfield  {author} {\bibinfo {author} {\bibfnamefont {R.~H.}\ \bibnamefont
  {{Wechsler}}}\ and\ \bibinfo {author} {\bibfnamefont {J.~L.}\ \bibnamefont
  {{Tinker}}},\ }\href {\doibase 10.1146/annurev-astro-081817-051756}
  {\bibfield  {journal} {\bibinfo  {journal} {\araa}\ }\textbf {\bibinfo
  {volume} {56}},\ \bibinfo {pages} {435} (\bibinfo {year} {2018})},\ \Eprint
  {http://arxiv.org/abs/1804.03097} {arXiv:1804.03097 [astro-ph.GA]}
  \BibitemShut {NoStop}%
\bibitem [{\citenamefont {{Krause}}\ and\ \citenamefont
  {{Eifler}}(2017)}]{Krause17}%
  \BibitemOpen
  \bibfield  {author} {\bibinfo {author} {\bibfnamefont {E.}~\bibnamefont
  {{Krause}}}\ and\ \bibinfo {author} {\bibfnamefont {T.}~\bibnamefont
  {{Eifler}}},\ }\href {\doibase 10.1093/mnras/stx1261} {\bibfield  {journal}
  {\bibinfo  {journal} {\mnras}\ }\textbf {\bibinfo {volume} {470}},\ \bibinfo
  {pages} {2100} (\bibinfo {year} {2017})},\ \Eprint
  {http://arxiv.org/abs/1601.05779} {arXiv:1601.05779 [astro-ph.CO]}
  \BibitemShut {NoStop}%
\bibitem [{\citenamefont {{Efstathiou}}\ and\ \citenamefont
  {{Lemos}}(2018)}]{Efstathiou18}%
  \BibitemOpen
  \bibfield  {author} {\bibinfo {author} {\bibfnamefont {G.}~\bibnamefont
  {{Efstathiou}}}\ and\ \bibinfo {author} {\bibfnamefont {P.}~\bibnamefont
  {{Lemos}}},\ }\href {\doibase 10.1093/mnras/sty099} {\bibfield  {journal}
  {\bibinfo  {journal} {\mnras}\ }\textbf {\bibinfo {volume} {476}},\ \bibinfo
  {pages} {151} (\bibinfo {year} {2018})},\ \Eprint
  {http://arxiv.org/abs/1707.00483} {arXiv:1707.00483 [astro-ph.CO]}
  \BibitemShut {NoStop}%
\bibitem [{\citenamefont {{Prat}}\ \emph {et~al.}(2021)\citenamefont {{Prat}}
  \emph {et~al.}}]{y3-gglensing}%
  \BibitemOpen
  \bibfield  {author} {\bibinfo {author} {\bibfnamefont {J.}~\bibnamefont
  {{Prat}}} \emph {et~al.},\ }\href@noop {} {\bibfield  {journal} {\bibinfo
  {journal} {To be submitted to PRD}\ } (\bibinfo {year} {2021})}\BibitemShut
  {NoStop}%
\bibitem [{\citenamefont {{Rodr\'iguez-Monroy}}\ \emph
  {et~al.}(2021)\citenamefont {{Rodr\'iguez-Monroy}} \emph
  {et~al.}}]{y3-galaxyclustering}%
  \BibitemOpen
  \bibfield  {author} {\bibinfo {author} {\bibfnamefont {M.}~\bibnamefont
  {{Rodr\'iguez-Monroy}}} \emph {et~al.},\ }\href@noop {} {\bibfield  {journal}
  {\bibinfo  {journal} {To be submitted to MNRAS}\ } (\bibinfo {year}
  {2021})}\BibitemShut {NoStop}%
\bibitem [{\citenamefont {{Porredon}}\ \emph {et~al.}(2021)\citenamefont
  {{Porredon}} \emph {et~al.}}]{y3-2x2ptaltlensresults}%
  \BibitemOpen
  \bibfield  {author} {\bibinfo {author} {\bibfnamefont {A.}~\bibnamefont
  {{Porredon}}} \emph {et~al.},\ }\href@noop {} {\bibfield  {journal} {\bibinfo
   {journal} {To be submitted to PRD}\ } (\bibinfo {year} {2021})}\BibitemShut
  {NoStop}%
\bibitem [{\citenamefont {{Pandey}}\ \emph {et~al.}(2021)\citenamefont
  {{Pandey}} \emph {et~al.}}]{y3-2x2ptbiasmodelling}%
  \BibitemOpen
  \bibfield  {author} {\bibinfo {author} {\bibfnamefont {S.}~\bibnamefont
  {{Pandey}}} \emph {et~al.},\ }\href@noop {} {\bibfield  {journal} {\bibinfo
  {journal} {To be submitted to MNRAS}\ } (\bibinfo {year} {2021})}\BibitemShut
  {NoStop}%
\bibitem [{\citenamefont {{Elvin-Poole}}\ \emph {et~al.}(2021)\citenamefont
  {{Elvin-Poole}}, \citenamefont {{MacCrann}} \emph
  {et~al.}}]{y3-2x2ptmagnification}%
  \BibitemOpen
  \bibfield  {author} {\bibinfo {author} {\bibfnamefont {J.}~\bibnamefont
  {{Elvin-Poole}}}, \bibinfo {author} {\bibfnamefont {N.}~\bibnamefont
  {{MacCrann}}},  \emph {et~al.},\ }\href@noop {} {\bibfield  {journal}
  {\bibinfo  {journal} {To be submitted to MNRAS}\ } (\bibinfo {year}
  {2021})}\BibitemShut {NoStop}%
\bibitem [{\citenamefont {{DES Collaboration}}(2021)}]{y3-3x2ptkp}%
  \BibitemOpen
  \bibfield  {author} {\bibinfo {author} {\bibnamefont {{DES Collaboration}}},\
  }\href@noop {} {\bibfield  {journal} {\bibinfo  {journal} {To be submitted to
  PRD}\ } (\bibinfo {year} {2021})}\BibitemShut {NoStop}%
\bibitem [{\citenamefont {Sevilla-Noarbe}\ \emph {et~al.}(2020)\citenamefont
  {Sevilla-Noarbe} \emph {et~al.}}]{y3-gold}%
  \BibitemOpen
  \bibfield  {author} {\bibinfo {author} {\bibfnamefont {I.}~\bibnamefont
  {Sevilla-Noarbe}} \emph {et~al.} (\bibinfo {collaboration} {DES}),\
  }\href@noop {} {\bibfield  {journal} {\bibinfo  {journal} {Submitted to
  ApJS}\ } (\bibinfo {year} {2020})},\ \Eprint
  {http://arxiv.org/abs/2011.03407} {arXiv:2011.03407 [astro-ph.CO]}
  \BibitemShut {NoStop}%
\bibitem [{\citenamefont {Hartley}\ \emph {et~al.}(2020)\citenamefont
  {Hartley}, \citenamefont {Choi} \emph {et~al.}}]{y3-deepfields}%
  \BibitemOpen
  \bibfield  {author} {\bibinfo {author} {\bibfnamefont {W.~G.}\ \bibnamefont
  {Hartley}}, \bibinfo {author} {\bibfnamefont {A.}~\bibnamefont {Choi}},
  \emph {et~al.} (\bibinfo {collaboration} {DES}),\ }\href@noop {} {\bibfield
  {journal} {\bibinfo  {journal} {Submitted to MNRAS}\ } (\bibinfo {year}
  {2020})},\ \Eprint {http://arxiv.org/abs/2012.12824} {arXiv:2012.12824
  [astro-ph.CO]} \BibitemShut {NoStop}%
\bibitem [{\citenamefont {Everett}\ \emph {et~al.}(2020)\citenamefont {Everett}
  \emph {et~al.}}]{y3-balrog}%
  \BibitemOpen
  \bibfield  {author} {\bibinfo {author} {\bibfnamefont {S.}~\bibnamefont
  {Everett}} \emph {et~al.} (\bibinfo {collaboration} {DES}),\ }\href@noop {}
  {\bibfield  {journal} {\bibinfo  {journal} {Submitted to ApJS}\ } (\bibinfo
  {year} {2020})},\ \Eprint {http://arxiv.org/abs/2012.12825} {arXiv:2012.12825
  [astro-ph.CO]} \BibitemShut {NoStop}%
\bibitem [{\citenamefont {{Buchs}}\ \emph {et~al.}(2019)\citenamefont
  {{Buchs}}, \citenamefont {{Davis}}, \citenamefont {{Gruen}}, \citenamefont
  {{DeRose}},\ and\ \citenamefont {{DES Collaboration}}}]{buchs19}%
  \BibitemOpen
  \bibfield  {author} {\bibinfo {author} {\bibfnamefont {R.}~\bibnamefont
  {{Buchs}}}, \bibinfo {author} {\bibfnamefont {C.}~\bibnamefont {{Davis}}},
  \bibinfo {author} {\bibfnamefont {D.}~\bibnamefont {{Gruen}}}, \bibinfo
  {author} {\bibfnamefont {J.}~\bibnamefont {{DeRose}}}, \ and\ \bibinfo
  {author} {\bibnamefont {{DES Collaboration}}},\ }\href {\doibase
  10.1093/mnras/stz2162} {\bibfield  {journal} {\bibinfo  {journal} {\mnras}\
  }\textbf {\bibinfo {volume} {489}},\ \bibinfo {pages} {820} (\bibinfo {year}
  {2019})},\ \Eprint {http://arxiv.org/abs/1901.05005} {arXiv:1901.05005
  [astro-ph.CO]} \BibitemShut {NoStop}%
\bibitem [{\citenamefont {Porredon}\ \emph {et~al.}(2021)\citenamefont
  {Porredon} \emph {et~al.}}]{y3-2x2maglimforecast}%
  \BibitemOpen
  \bibfield  {author} {\bibinfo {author} {\bibfnamefont {A.}~\bibnamefont
  {Porredon}} \emph {et~al.} (\bibinfo {collaboration} {DES}),\ }\href
  {\doibase 10.1103/PhysRevD.103.043503} {\bibfield  {journal} {\bibinfo
  {journal} {Phys. Rev. D}\ }\textbf {\bibinfo {volume} {103}},\ \bibinfo
  {pages} {043503} (\bibinfo {year} {2021})},\ \Eprint
  {http://arxiv.org/abs/2011.03411} {arXiv:2011.03411 [astro-ph.CO]}
  \BibitemShut {NoStop}%
\bibitem [{\citenamefont {{Cordero}}\ \emph {et~al.}(2021)\citenamefont
  {{Cordero}}, \citenamefont {{Harrison}} \emph {et~al.}}]{y3-hyperrank}%
  \BibitemOpen
  \bibfield  {author} {\bibinfo {author} {\bibfnamefont {J.~P.}\ \bibnamefont
  {{Cordero}}}, \bibinfo {author} {\bibfnamefont {I.}~\bibnamefont
  {{Harrison}}},  \emph {et~al.},\ }\href@noop {} {\bibfield  {journal}
  {\bibinfo  {journal} {arXiv e-prints}\ ,\ \bibinfo {eid} {arXiv:2109.09636}}
  (\bibinfo {year} {2021})},\ \Eprint {http://arxiv.org/abs/2109.09636}
  {arXiv:2109.09636 [astro-ph.CO]} \BibitemShut {NoStop}%
\bibitem [{\citenamefont {{Krause}}\ \emph {et~al.}(2021)\citenamefont
  {{Krause}} \emph {et~al.}}]{y3-generalmethods}%
  \BibitemOpen
  \bibfield  {author} {\bibinfo {author} {\bibfnamefont {E.}~\bibnamefont
  {{Krause}}} \emph {et~al.},\ }\href@noop {} {\bibfield  {journal} {\bibinfo
  {journal} {To be submitted to PRD}\ } (\bibinfo {year} {2021})}\BibitemShut
  {NoStop}%
\bibitem [{\citenamefont {Friedrich}\ \emph {et~al.}(2020)\citenamefont
  {Friedrich} \emph {et~al.}}]{y3-covariances}%
  \BibitemOpen
  \bibfield  {author} {\bibinfo {author} {\bibfnamefont {O.}~\bibnamefont
  {Friedrich}} \emph {et~al.} (\bibinfo {collaboration} {DES}),\ }\href@noop {}
  {\bibfield  {journal} {\bibinfo  {journal} {Submitted to MNRAS}\ } (\bibinfo
  {year} {2020})},\ \Eprint {http://arxiv.org/abs/2012.08568} {arXiv:2012.08568
  [astro-ph.CO]} \BibitemShut {NoStop}%
\bibitem [{\citenamefont {{DeRose}}\ \emph
  {et~al.}(2021{\natexlab{a}})\citenamefont {{DeRose}} \emph
  {et~al.}}]{y3-simvalidation}%
  \BibitemOpen
  \bibfield  {author} {\bibinfo {author} {\bibfnamefont {J.}~\bibnamefont
  {{DeRose}}} \emph {et~al.},\ }\href@noop {} {\bibfield  {journal} {\bibinfo
  {journal} {To be submitted to MNRAS}\ } (\bibinfo {year}
  {2021}{\natexlab{a}})}\BibitemShut {NoStop}%
\bibitem [{\citenamefont {Doux}\ \emph {et~al.}(2020)\citenamefont {Doux} \emph
  {et~al.}}]{y3-inttensions}%
  \BibitemOpen
  \bibfield  {author} {\bibinfo {author} {\bibfnamefont {C.}~\bibnamefont
  {Doux}} \emph {et~al.} (\bibinfo {collaboration} {DES}),\ }\href {\doibase
  10.1093/mnras/stab526} {\bibfield  {journal} {\bibinfo  {journal} {Mon. Not.
  Roy. Astron. Soc.}\ } (\bibinfo {year} {2020}),\ 10.1093/mnras/stab526},\
  \Eprint {http://arxiv.org/abs/2011.03410} {arXiv:2011.03410 [astro-ph.CO]}
  \BibitemShut {NoStop}%
\bibitem [{\citenamefont {Lemos}\ \emph {et~al.}(2020)\citenamefont {Lemos}
  \emph {et~al.}}]{y3-tensions}%
  \BibitemOpen
  \bibfield  {author} {\bibinfo {author} {\bibfnamefont {M.}~\bibnamefont
  {Lemos}, \bibfnamefont {P.~Raveri}} \emph {et~al.} (\bibinfo {collaboration}
  {DES}),\ }\href@noop {} {\bibfield  {journal} {\bibinfo  {journal} {Submitted
  to MNRAS}\ } (\bibinfo {year} {2020})},\ \Eprint
  {http://arxiv.org/abs/2012.09554} {arXiv:2012.09554 [astro-ph.CO]}
  \BibitemShut {NoStop}%
\bibitem [{\citenamefont {{Flaugher}}\ \emph {et~al.}(2015)\citenamefont
  {{Flaugher}}, \citenamefont {{Diehl}}, \citenamefont {{Honscheid}},\ and\
  \citenamefont {{DES Collaboration}}}]{flaugher15}%
  \BibitemOpen
  \bibfield  {author} {\bibinfo {author} {\bibfnamefont {B.}~\bibnamefont
  {{Flaugher}}}, \bibinfo {author} {\bibfnamefont {H.~T.}\ \bibnamefont
  {{Diehl}}}, \bibinfo {author} {\bibfnamefont {K.}~\bibnamefont
  {{Honscheid}}}, \ and\ \bibinfo {author} {\bibnamefont {{DES
  Collaboration}}},\ }\href {\doibase 10.1088/0004-6256/150/5/150} {\bibfield
  {journal} {\bibinfo  {journal} {\aj}\ }\textbf {\bibinfo {volume} {150}},\
  \bibinfo {eid} {150} (\bibinfo {year} {2015})},\ \Eprint
  {http://arxiv.org/abs/1504.02900} {arXiv:1504.02900 [astro-ph.IM]}
  \BibitemShut {NoStop}%
\bibitem [{\citenamefont {{Dark Energy Survey
  Collaboration}}(2016{\natexlab{b}})}]{des-overview}%
  \BibitemOpen
  \bibfield  {author} {\bibinfo {author} {\bibnamefont {{Dark Energy Survey
  Collaboration}}},\ }\href {\doibase 10.1093/mnras/stw641} {\bibfield
  {journal} {\bibinfo  {journal} {\mnras}\ }\textbf {\bibinfo {volume} {460}},\
  \bibinfo {pages} {1270} (\bibinfo {year} {2016}{\natexlab{b}})},\ \Eprint
  {http://arxiv.org/abs/1601.00329} {arXiv:1601.00329 [astro-ph.CO]}
  \BibitemShut {NoStop}%
\bibitem [{\citenamefont {{Amara}}\ and\ \citenamefont
  {{R{\'e}fr{\'e}gier}}(2007)}]{amara07}%
  \BibitemOpen
  \bibfield  {author} {\bibinfo {author} {\bibfnamefont {A.}~\bibnamefont
  {{Amara}}}\ and\ \bibinfo {author} {\bibfnamefont {A.}~\bibnamefont
  {{R{\'e}fr{\'e}gier}}},\ }\href {\doibase 10.1111/j.1365-2966.2007.12271.x}
  {\bibfield  {journal} {\bibinfo  {journal} {\mnras}\ }\textbf {\bibinfo
  {volume} {381}},\ \bibinfo {pages} {1018} (\bibinfo {year} {2007})},\ \Eprint
  {http://arxiv.org/abs/astro-ph/0610127} {arXiv:astro-ph/0610127 [astro-ph]}
  \BibitemShut {NoStop}%
\bibitem [{\citenamefont {{van Uitert}}\ and\ \citenamefont
  {{Schneider}}(2016)}]{vanuitert16}%
  \BibitemOpen
  \bibfield  {author} {\bibinfo {author} {\bibfnamefont {E.}~\bibnamefont {{van
  Uitert}}}\ and\ \bibinfo {author} {\bibfnamefont {P.}~\bibnamefont
  {{Schneider}}},\ }\href {\doibase 10.1051/0004-6361/201628846} {\bibfield
  {journal} {\bibinfo  {journal} {\aap}\ }\textbf {\bibinfo {volume} {595}},\
  \bibinfo {eid} {A93} (\bibinfo {year} {2016})},\ \Eprint
  {http://arxiv.org/abs/1605.01056} {arXiv:1605.01056 [astro-ph.CO]}
  \BibitemShut {NoStop}%
\bibitem [{\citenamefont {{Kitching}}\ \emph {et~al.}(2021)\citenamefont
  {{Kitching}}, \citenamefont {{Deshpande}},\ and\ \citenamefont
  {{Taylor}}}]{kitching21}%
  \BibitemOpen
  \bibfield  {author} {\bibinfo {author} {\bibfnamefont {T.~D.}\ \bibnamefont
  {{Kitching}}}, \bibinfo {author} {\bibfnamefont {A.~C.}\ \bibnamefont
  {{Deshpande}}}, \ and\ \bibinfo {author} {\bibfnamefont {P.~L.}\ \bibnamefont
  {{Taylor}}},\ }\href@noop {} {\bibfield  {journal} {\bibinfo  {journal}
  {arXiv e-prints}\ ,\ \bibinfo {eid} {arXiv:2110.01275}} (\bibinfo {year}
  {2021})},\ \Eprint {http://arxiv.org/abs/2110.01275} {arXiv:2110.01275
  [astro-ph.CO]} \BibitemShut {NoStop}%
\bibitem [{\citenamefont {{Schneider}}\ \emph {et~al.}(2010)\citenamefont
  {{Schneider}}, \citenamefont {{Eifler}},\ and\ \citenamefont
  {{Krause}}}]{Schneider2010}%
  \BibitemOpen
  \bibfield  {author} {\bibinfo {author} {\bibfnamefont {P.}~\bibnamefont
  {{Schneider}}}, \bibinfo {author} {\bibfnamefont {T.}~\bibnamefont
  {{Eifler}}}, \ and\ \bibinfo {author} {\bibfnamefont {E.}~\bibnamefont
  {{Krause}}},\ }\href {\doibase 10.1051/0004-6361/201014235} {\bibfield
  {journal} {\bibinfo  {journal} {\aap}\ }\textbf {\bibinfo {volume} {520}},\
  \bibinfo {eid} {A116} (\bibinfo {year} {2010})},\ \Eprint
  {http://arxiv.org/abs/1002.2136} {arXiv:1002.2136 [astro-ph.CO]} \BibitemShut
  {NoStop}%
\bibitem [{\citenamefont {De~Vicente}\ \emph {et~al.}(2016)\citenamefont
  {De~Vicente}, \citenamefont {Sánchez},\ and\ \citenamefont
  {Sevilla-Noarbe}}]{dnf}%
  \BibitemOpen
  \bibfield  {author} {\bibinfo {author} {\bibfnamefont {J.}~\bibnamefont
  {De~Vicente}}, \bibinfo {author} {\bibfnamefont {E.}~\bibnamefont
  {Sánchez}}, \ and\ \bibinfo {author} {\bibfnamefont {I.}~\bibnamefont
  {Sevilla-Noarbe}},\ }\href {\doibase 10.1093/mnras/stw857} {\bibfield
  {journal} {\bibinfo  {journal} {Monthly Notices of the Royal Astronomical
  Society}\ }\textbf {\bibinfo {volume} {459}},\ \bibinfo {pages} {3078–3088}
  (\bibinfo {year} {2016})}\BibitemShut {NoStop}%
\bibitem [{\citenamefont {Rozo}\ and\ \citenamefont {Rykoff}(2016)}]{RozoRM}%
  \BibitemOpen
  \bibfield  {author} {\bibinfo {author} {\bibfnamefont {E.}~\bibnamefont
  {Rozo}}\ and\ \bibinfo {author} {\bibfnamefont {E.~S.}\ \bibnamefont
  {Rykoff}},\ }\href {\doibase 10.1093/mnras/stw1281} {\bibfield  {journal}
  {\bibinfo  {journal} {Monthly Notices of the Royal Astronomical Society}\
  }\textbf {\bibinfo {volume} {461}},\ \bibinfo {pages} {1431–1450} (\bibinfo
  {year} {2016})}\BibitemShut {NoStop}%
\bibitem [{\citenamefont {Cawthon}\ \emph {et~al.}(2020)\citenamefont {Cawthon}
  \emph {et~al.}}]{y3-lenswz}%
  \BibitemOpen
  \bibfield  {author} {\bibinfo {author} {\bibfnamefont {R.}~\bibnamefont
  {Cawthon}} \emph {et~al.} (\bibinfo {collaboration} {DES}),\ }\href@noop {}
  {\bibfield  {journal} {\bibinfo  {journal} {Submitted to MNRAS}\ } (\bibinfo
  {year} {2020})},\ \Eprint {http://arxiv.org/abs/2012.12826} {arXiv:2012.12826
  [astro-ph.CO]} \BibitemShut {NoStop}%
\bibitem [{\citenamefont {Rykoff}\ \emph {et~al.}(2014)\citenamefont {Rykoff},
  \citenamefont {Rozo}, \citenamefont {Busha}, \citenamefont {Cunha},
  \citenamefont {Finoguenov}, \citenamefont {Evrard}, \citenamefont {Hao},
  \citenamefont {Koester}, \citenamefont {Leauthaud}, \citenamefont {Nord},\
  and\ \citenamefont {et~al.}}]{rykoffRM}%
  \BibitemOpen
  \bibfield  {author} {\bibinfo {author} {\bibfnamefont {E.~S.}\ \bibnamefont
  {Rykoff}}, \bibinfo {author} {\bibfnamefont {E.}~\bibnamefont {Rozo}},
  \bibinfo {author} {\bibfnamefont {M.~T.}\ \bibnamefont {Busha}}, \bibinfo
  {author} {\bibfnamefont {C.~E.}\ \bibnamefont {Cunha}}, \bibinfo {author}
  {\bibfnamefont {A.}~\bibnamefont {Finoguenov}}, \bibinfo {author}
  {\bibfnamefont {A.}~\bibnamefont {Evrard}}, \bibinfo {author} {\bibfnamefont
  {J.}~\bibnamefont {Hao}}, \bibinfo {author} {\bibfnamefont {B.~P.}\
  \bibnamefont {Koester}}, \bibinfo {author} {\bibfnamefont {A.}~\bibnamefont
  {Leauthaud}}, \bibinfo {author} {\bibfnamefont {B.}~\bibnamefont {Nord}}, \
  and\ \bibinfo {author} {\bibnamefont {et~al.}},\ }\href {\doibase
  10.1088/0004-637x/785/2/104} {\bibfield  {journal} {\bibinfo  {journal} {The
  Astrophysical Journal}\ }\textbf {\bibinfo {volume} {785}},\ \bibinfo {pages}
  {104} (\bibinfo {year} {2014})}\BibitemShut {NoStop}%
\bibitem [{\citenamefont {{Huterer}}\ \emph {et~al.}(2006)\citenamefont
  {{Huterer}}, \citenamefont {{Takada}}, \citenamefont {{Bernstein}},\ and\
  \citenamefont {{Jain}}}]{Huterer06}%
  \BibitemOpen
  \bibfield  {author} {\bibinfo {author} {\bibfnamefont {D.}~\bibnamefont
  {{Huterer}}}, \bibinfo {author} {\bibfnamefont {M.}~\bibnamefont {{Takada}}},
  \bibinfo {author} {\bibfnamefont {G.}~\bibnamefont {{Bernstein}}}, \ and\
  \bibinfo {author} {\bibfnamefont {B.}~\bibnamefont {{Jain}}},\ }\href
  {\doibase 10.1111/j.1365-2966.2005.09782.x} {\bibfield  {journal} {\bibinfo
  {journal} {\mnras}\ }\textbf {\bibinfo {volume} {366}},\ \bibinfo {pages}
  {101} (\bibinfo {year} {2006})},\ \Eprint
  {http://arxiv.org/abs/astro-ph/0506030} {arXiv:astro-ph/0506030 [astro-ph]}
  \BibitemShut {NoStop}%
\bibitem [{\citenamefont {{Van Waerbeke}}\ \emph {et~al.}(2006)\citenamefont
  {{Van Waerbeke}}, \citenamefont {{White}}, \citenamefont {{Hoekstra}},\ and\
  \citenamefont {{Heymans}}}]{vanWaerbeke06}%
  \BibitemOpen
  \bibfield  {author} {\bibinfo {author} {\bibfnamefont {L.}~\bibnamefont {{Van
  Waerbeke}}}, \bibinfo {author} {\bibfnamefont {M.}~\bibnamefont {{White}}},
  \bibinfo {author} {\bibfnamefont {H.}~\bibnamefont {{Hoekstra}}}, \ and\
  \bibinfo {author} {\bibfnamefont {C.}~\bibnamefont {{Heymans}}},\ }\href
  {\doibase 10.1016/j.astropartphys.2006.05.008} {\bibfield  {journal}
  {\bibinfo  {journal} {Astroparticle Physics}\ }\textbf {\bibinfo {volume}
  {26}},\ \bibinfo {pages} {91} (\bibinfo {year} {2006})},\ \Eprint
  {http://arxiv.org/abs/astro-ph/0603696} {arXiv:astro-ph/0603696 [astro-ph]}
  \BibitemShut {NoStop}%
\bibitem [{\citenamefont {{Bonnett}}\ \emph {et~al.}(2016)\citenamefont
  {{Bonnett}}, \citenamefont {{Troxel}}, \citenamefont {{Hartley}},\ and\
  \citenamefont {{Dark Energy Survey Collaboration}}}]{Bonnett16}%
  \BibitemOpen
  \bibfield  {author} {\bibinfo {author} {\bibfnamefont {C.}~\bibnamefont
  {{Bonnett}}}, \bibinfo {author} {\bibfnamefont {M.~A.}\ \bibnamefont
  {{Troxel}}}, \bibinfo {author} {\bibfnamefont {W.}~\bibnamefont {{Hartley}}},
  \ and\ \bibinfo {author} {\bibnamefont {{Dark Energy Survey
  Collaboration}}},\ }\href {\doibase 10.1103/PhysRevD.94.042005} {\bibfield
  {journal} {\bibinfo  {journal} {\prd}\ }\textbf {\bibinfo {volume} {94}},\
  \bibinfo {eid} {042005} (\bibinfo {year} {2016})},\ \Eprint
  {http://arxiv.org/abs/1507.05909} {arXiv:1507.05909 [astro-ph.CO]}
  \BibitemShut {NoStop}%
\bibitem [{\citenamefont {{Masters}}\ \emph {et~al.}(2015)\citenamefont
  {{Masters}}, \citenamefont {{Capak}}, \citenamefont {{Stern}},\ and\
  \citenamefont {{Ilbert}}}]{Masters15}%
  \BibitemOpen
  \bibfield  {author} {\bibinfo {author} {\bibfnamefont {D.}~\bibnamefont
  {{Masters}}}, \bibinfo {author} {\bibfnamefont {P.}~\bibnamefont {{Capak}}},
  \bibinfo {author} {\bibfnamefont {D.}~\bibnamefont {{Stern}}}, \ and\
  \bibinfo {author} {\bibfnamefont {O.}~\bibnamefont {{Ilbert}}},\ }\href
  {\doibase 10.1088/0004-637X/813/1/53} {\bibfield  {journal} {\bibinfo
  {journal} {\apj}\ }\textbf {\bibinfo {volume} {813}},\ \bibinfo {eid} {53}
  (\bibinfo {year} {2015})},\ \Eprint {http://arxiv.org/abs/1509.03318}
  {arXiv:1509.03318 [astro-ph.CO]} \BibitemShut {NoStop}%
\bibitem [{\citenamefont {{S{\'a}nchez}}\ and\ \citenamefont
  {{Bernstein}}(2019)}]{Sanchez19}%
  \BibitemOpen
  \bibfield  {author} {\bibinfo {author} {\bibfnamefont {C.}~\bibnamefont
  {{S{\'a}nchez}}}\ and\ \bibinfo {author} {\bibfnamefont {G.~M.}\ \bibnamefont
  {{Bernstein}}},\ }\href {\doibase 10.1093/mnras/sty3222} {\bibfield
  {journal} {\bibinfo  {journal} {\mnras}\ }\textbf {\bibinfo {volume} {483}},\
  \bibinfo {pages} {2801} (\bibinfo {year} {2019})},\ \Eprint
  {http://arxiv.org/abs/1807.11873} {arXiv:1807.11873 [astro-ph.CO]}
  \BibitemShut {NoStop}%
\bibitem [{\citenamefont {{Ross}}\ \emph {et~al.}(2020)\citenamefont {{Ross}},
  \citenamefont {{Bautista}},\ and\ \citenamefont {{Tojeiro}}}]{Ross20}%
  \BibitemOpen
  \bibfield  {author} {\bibinfo {author} {\bibfnamefont {A.~J.}\ \bibnamefont
  {{Ross}}}, \bibinfo {author} {\bibfnamefont {J.}~\bibnamefont {{Bautista}}},
  \ and\ \bibinfo {author} {\bibfnamefont {R.}~\bibnamefont {{Tojeiro}}},\
  }\href {\doibase 10.1093/mnras/staa2416} {\bibfield  {journal} {\bibinfo
  {journal} {\mnras}\ } (\bibinfo {year} {2020}),\ 10.1093/mnras/staa2416},\
  \Eprint {http://arxiv.org/abs/2007.09000} {arXiv:2007.09000 [astro-ph.CO]}
  \BibitemShut {NoStop}%
\bibitem [{\citenamefont {Jain}\ and\ \citenamefont
  {Taylor}(2003)}]{JainTaylor}%
  \BibitemOpen
  \bibfield  {author} {\bibinfo {author} {\bibfnamefont {B.}~\bibnamefont
  {Jain}}\ and\ \bibinfo {author} {\bibfnamefont {A.}~\bibnamefont {Taylor}},\
  }\href {\doibase 10.1103/physrevlett.91.141302} {\bibfield  {journal}
  {\bibinfo  {journal} {Physical Review Letters}\ }\textbf {\bibinfo {volume}
  {91}} (\bibinfo {year} {2003}),\ 10.1103/physrevlett.91.141302}\BibitemShut
  {NoStop}%
\bibitem [{\citenamefont {{Schneider}}(2016)}]{schneider2016}%
  \BibitemOpen
  \bibfield  {author} {\bibinfo {author} {\bibfnamefont {P.}~\bibnamefont
  {{Schneider}}},\ }\href {\doibase 10.1051/0004-6361/201628506} {\bibfield
  {journal} {\bibinfo  {journal} {\aap}\ }\textbf {\bibinfo {volume} {592}},\
  \bibinfo {eid} {L6} (\bibinfo {year} {2016})},\ \Eprint
  {http://arxiv.org/abs/1603.04226} {arXiv:1603.04226 [astro-ph.CO]}
  \BibitemShut {NoStop}%
\bibitem [{\citenamefont {Mandelbaum}\ \emph {et~al.}(2005)\citenamefont
  {Mandelbaum}, \citenamefont {Hirata}, \citenamefont {Seljak}, \citenamefont
  {Guzik}, \citenamefont {Padmanabhan}, \citenamefont {Blake}, \citenamefont
  {Blanton}, \citenamefont {Lupton},\ and\ \citenamefont
  {Brinkmann}}]{mandelbaum2005}%
  \BibitemOpen
  \bibfield  {author} {\bibinfo {author} {\bibfnamefont {R.}~\bibnamefont
  {Mandelbaum}}, \bibinfo {author} {\bibfnamefont {C.~M.}\ \bibnamefont
  {Hirata}}, \bibinfo {author} {\bibfnamefont {U.}~\bibnamefont {Seljak}},
  \bibinfo {author} {\bibfnamefont {J.}~\bibnamefont {Guzik}}, \bibinfo
  {author} {\bibfnamefont {N.}~\bibnamefont {Padmanabhan}}, \bibinfo {author}
  {\bibfnamefont {C.}~\bibnamefont {Blake}}, \bibinfo {author} {\bibfnamefont
  {M.~R.}\ \bibnamefont {Blanton}}, \bibinfo {author} {\bibfnamefont
  {R.}~\bibnamefont {Lupton}}, \ and\ \bibinfo {author} {\bibfnamefont
  {J.}~\bibnamefont {Brinkmann}},\ }\href {\doibase
  10.1111/j.1365-2966.2005.09282.x} {\bibfield  {journal} {\bibinfo  {journal}
  {Monthly Notices of the Royal Astronomical Society}\ }\textbf {\bibinfo
  {volume} {361}},\ \bibinfo {pages} {1287–1322} (\bibinfo {year}
  {2005})}\BibitemShut {NoStop}%
\bibitem [{\citenamefont {{Hoekstra}}\ \emph {et~al.}(2006)\citenamefont
  {{Hoekstra}}, \citenamefont {{Mellier}}, \citenamefont {{van Waerbeke}},
  \citenamefont {{Semboloni}}, \citenamefont {{Fu}}, \citenamefont {{Hudson}},
  \citenamefont {{Parker}}, \citenamefont {{Tereno}},\ and\ \citenamefont
  {{Benabed}}}]{hoekstra2005}%
  \BibitemOpen
  \bibfield  {author} {\bibinfo {author} {\bibfnamefont {H.}~\bibnamefont
  {{Hoekstra}}}, \bibinfo {author} {\bibfnamefont {Y.}~\bibnamefont
  {{Mellier}}}, \bibinfo {author} {\bibfnamefont {L.}~\bibnamefont {{van
  Waerbeke}}}, \bibinfo {author} {\bibfnamefont {E.}~\bibnamefont
  {{Semboloni}}}, \bibinfo {author} {\bibfnamefont {L.}~\bibnamefont {{Fu}}},
  \bibinfo {author} {\bibfnamefont {M.~J.}\ \bibnamefont {{Hudson}}}, \bibinfo
  {author} {\bibfnamefont {L.~C.}\ \bibnamefont {{Parker}}}, \bibinfo {author}
  {\bibfnamefont {I.}~\bibnamefont {{Tereno}}}, \ and\ \bibinfo {author}
  {\bibfnamefont {K.}~\bibnamefont {{Benabed}}},\ }\href {\doibase
  10.1086/503249} {\bibfield  {journal} {\bibinfo  {journal} {\apj}\ }\textbf
  {\bibinfo {volume} {647}},\ \bibinfo {pages} {116} (\bibinfo {year}
  {2006})},\ \Eprint {http://arxiv.org/abs/astro-ph/0511089}
  {arXiv:astro-ph/0511089 [astro-ph]} \BibitemShut {NoStop}%
\bibitem [{\citenamefont {{Heymans}}\ \emph {et~al.}(2012)\citenamefont
  {{Heymans}}, \citenamefont {{Van Waerbeke}},\ and\ \citenamefont
  {{Miller}}}]{Heymans2012}%
  \BibitemOpen
  \bibfield  {author} {\bibinfo {author} {\bibfnamefont {C.}~\bibnamefont
  {{Heymans}}}, \bibinfo {author} {\bibfnamefont {L.}~\bibnamefont {{Van
  Waerbeke}}}, \ and\ \bibinfo {author} {\bibnamefont {{Miller}}},\ }\href
  {\doibase 10.1111/j.1365-2966.2012.21952.x} {\bibfield  {journal} {\bibinfo
  {journal} {\mnras}\ }\textbf {\bibinfo {volume} {427}},\ \bibinfo {pages}
  {146} (\bibinfo {year} {2012})},\ \Eprint {http://arxiv.org/abs/1210.0032}
  {arXiv:1210.0032} \BibitemShut {NoStop}%
\bibitem [{\citenamefont {{Lilly}}\ \emph {et~al.}(2009)\citenamefont
  {{Lilly}}, \citenamefont {{Le Brun}}, \citenamefont {{Maier}},\ and\
  \citenamefont {{Mainieri}}}]{lilly09}%
  \BibitemOpen
  \bibfield  {author} {\bibinfo {author} {\bibfnamefont {S.~J.}\ \bibnamefont
  {{Lilly}}}, \bibinfo {author} {\bibfnamefont {V.}~\bibnamefont {{Le Brun}}},
  \bibinfo {author} {\bibfnamefont {C.}~\bibnamefont {{Maier}}}, \ and\
  \bibinfo {author} {\bibfnamefont {V.}~\bibnamefont {{Mainieri}}},\ }\href
  {\doibase 10.1088/0067-0049/184/2/218} {\bibfield  {journal} {\bibinfo
  {journal} {\apjs}\ }\textbf {\bibinfo {volume} {184}},\ \bibinfo {pages}
  {218} (\bibinfo {year} {2009})}\BibitemShut {NoStop}%
\bibitem [{\citenamefont {{Masters}}\ \emph {et~al.}(2017)\citenamefont
  {{Masters}}, \citenamefont {{Stern}}, \citenamefont {{Cohen}}, \citenamefont
  {{Capak}}, \citenamefont {{Rhodes}}, \citenamefont {{Castander}},\ and\
  \citenamefont {{Paltani}}}]{Masters17}%
  \BibitemOpen
  \bibfield  {author} {\bibinfo {author} {\bibfnamefont {D.~C.}\ \bibnamefont
  {{Masters}}}, \bibinfo {author} {\bibfnamefont {D.~K.}\ \bibnamefont
  {{Stern}}}, \bibinfo {author} {\bibfnamefont {J.~G.}\ \bibnamefont
  {{Cohen}}}, \bibinfo {author} {\bibfnamefont {P.~L.}\ \bibnamefont
  {{Capak}}}, \bibinfo {author} {\bibfnamefont {J.~D.}\ \bibnamefont
  {{Rhodes}}}, \bibinfo {author} {\bibfnamefont {F.~J.}\ \bibnamefont
  {{Castander}}}, \ and\ \bibinfo {author} {\bibfnamefont {S.}~\bibnamefont
  {{Paltani}}},\ }\href {\doibase 10.3847/1538-4357/aa6f08} {\bibfield
  {journal} {\bibinfo  {journal} {\apj}\ }\textbf {\bibinfo {volume} {841}},\
  \bibinfo {eid} {111} (\bibinfo {year} {2017})},\ \Eprint
  {http://arxiv.org/abs/1704.06665} {arXiv:1704.06665 [astro-ph.CO]}
  \BibitemShut {NoStop}%
\bibitem [{\citenamefont {{Masters}}\ \emph {et~al.}(2019)\citenamefont
  {{Masters}}, \citenamefont {{Stern}}, \citenamefont {{Cohen}}, \citenamefont
  {{Capak}}, \citenamefont {{Stanford}}, \citenamefont {{Hernitschek}},
  \citenamefont {{Galametz}}, \citenamefont {{Davidzon}}, \citenamefont
  {{Rhodes}}, \citenamefont {{Sand ers}}, \citenamefont {{Mobasher}},
  \citenamefont {{Castander}}, \citenamefont {{Pruett}},\ and\ \citenamefont
  {{Fotopoulou}}}]{Masters19}%
  \BibitemOpen
  \bibfield  {author} {\bibinfo {author} {\bibfnamefont {D.~C.}\ \bibnamefont
  {{Masters}}}, \bibinfo {author} {\bibfnamefont {D.~K.}\ \bibnamefont
  {{Stern}}}, \bibinfo {author} {\bibfnamefont {J.~G.}\ \bibnamefont
  {{Cohen}}}, \bibinfo {author} {\bibfnamefont {P.~L.}\ \bibnamefont
  {{Capak}}}, \bibinfo {author} {\bibfnamefont {S.~A.}\ \bibnamefont
  {{Stanford}}}, \bibinfo {author} {\bibfnamefont {N.}~\bibnamefont
  {{Hernitschek}}}, \bibinfo {author} {\bibfnamefont {A.}~\bibnamefont
  {{Galametz}}}, \bibinfo {author} {\bibfnamefont {I.}~\bibnamefont
  {{Davidzon}}}, \bibinfo {author} {\bibfnamefont {J.~D.}\ \bibnamefont
  {{Rhodes}}}, \bibinfo {author} {\bibfnamefont {D.}~\bibnamefont {{Sand
  ers}}}, \bibinfo {author} {\bibfnamefont {B.}~\bibnamefont {{Mobasher}}},
  \bibinfo {author} {\bibfnamefont {F.}~\bibnamefont {{Castander}}}, \bibinfo
  {author} {\bibfnamefont {K.}~\bibnamefont {{Pruett}}}, \ and\ \bibinfo
  {author} {\bibfnamefont {S.}~\bibnamefont {{Fotopoulou}}},\ }\href {\doibase
  10.3847/1538-4357/ab184d} {\bibfield  {journal} {\bibinfo  {journal} {\apj}\
  }\textbf {\bibinfo {volume} {877}},\ \bibinfo {eid} {81} (\bibinfo {year}
  {2019})},\ \Eprint {http://arxiv.org/abs/1904.06394} {arXiv:1904.06394
  [astro-ph.GA]} \BibitemShut {NoStop}%
\bibitem [{\citenamefont {{Le F{\`e}vre}}(2013)}]{lefevre13}%
  \BibitemOpen
  \bibfield  {author} {\bibinfo {author} {\bibfnamefont {O.}~\bibnamefont {{Le
  F{\`e}vre}}},\ }\href {\doibase 10.1051/0004-6361/201322179} {\bibfield
  {journal} {\bibinfo  {journal} {\aap}\ }\textbf {\bibinfo {volume} {559}},\
  \bibinfo {eid} {A14} (\bibinfo {year} {2013})},\ \Eprint
  {http://arxiv.org/abs/1307.0545} {arXiv:1307.0545 [astro-ph.CO]} \BibitemShut
  {NoStop}%
\bibitem [{\citenamefont {{Guzzo}}\ and\ \citenamefont
  {{Scodeggio}}(2014)}]{guzzo14}%
  \BibitemOpen
  \bibfield  {author} {\bibinfo {author} {\bibfnamefont {L.}~\bibnamefont
  {{Guzzo}}}\ and\ \bibinfo {author} {\bibfnamefont {M.}~\bibnamefont
  {{Scodeggio}}},\ }\href {\doibase 10.1051/0004-6361/201321489} {\bibfield
  {journal} {\bibinfo  {journal} {\aap}\ }\textbf {\bibinfo {volume} {566}},\
  \bibinfo {eid} {A108} (\bibinfo {year} {2014})},\ \Eprint
  {http://arxiv.org/abs/1303.2623} {arXiv:1303.2623 [astro-ph.CO]} \BibitemShut
  {NoStop}%
\bibitem [{\citenamefont {{Garilli}}\ and\ \citenamefont
  {{Guzzo}}(2014)}]{garilli14}%
  \BibitemOpen
  \bibfield  {author} {\bibinfo {author} {\bibfnamefont {B.}~\bibnamefont
  {{Garilli}}}\ and\ \bibinfo {author} {\bibfnamefont {L.}~\bibnamefont
  {{Guzzo}}},\ }\href {\doibase 10.1051/0004-6361/201322790} {\bibfield
  {journal} {\bibinfo  {journal} {\aap}\ }\textbf {\bibinfo {volume} {562}},\
  \bibinfo {eid} {A23} (\bibinfo {year} {2014})},\ \Eprint
  {http://arxiv.org/abs/1310.1008} {arXiv:1310.1008 [astro-ph.CO]} \BibitemShut
  {NoStop}%
\bibitem [{\citenamefont {{Scodeggio}}\ \emph {et~al.}(2018)\citenamefont
  {{Scodeggio}}, \citenamefont {{Guzzo}},\ and\ \citenamefont
  {{Garilli}}}]{scodeggio18}%
  \BibitemOpen
  \bibfield  {author} {\bibinfo {author} {\bibfnamefont {M.}~\bibnamefont
  {{Scodeggio}}}, \bibinfo {author} {\bibfnamefont {L.}~\bibnamefont
  {{Guzzo}}}, \ and\ \bibinfo {author} {\bibfnamefont {B.}~\bibnamefont
  {{Garilli}}},\ }\href {\doibase 10.1051/0004-6361/201630114} {\bibfield
  {journal} {\bibinfo  {journal} {\aap}\ }\textbf {\bibinfo {volume} {609}},\
  \bibinfo {eid} {A84} (\bibinfo {year} {2018})},\ \Eprint
  {http://arxiv.org/abs/1611.07048} {arXiv:1611.07048 [astro-ph.GA]}
  \BibitemShut {NoStop}%
\bibitem [{\citenamefont {{Laigle}}\ \emph {et~al.}(2016)\citenamefont
  {{Laigle}}, \citenamefont {{McCracken}},\ and\ \citenamefont
  {{Ilbert}}}]{laigle16}%
  \BibitemOpen
  \bibfield  {author} {\bibinfo {author} {\bibfnamefont {C.}~\bibnamefont
  {{Laigle}}}, \bibinfo {author} {\bibfnamefont {H.~J.}\ \bibnamefont
  {{McCracken}}}, \ and\ \bibinfo {author} {\bibfnamefont {O.}~\bibnamefont
  {{Ilbert}}},\ }\href {\doibase 10.3847/0067-0049/224/2/24} {\bibfield
  {journal} {\bibinfo  {journal} {\apjs}\ }\textbf {\bibinfo {volume} {224}},\
  \bibinfo {eid} {24} (\bibinfo {year} {2016})},\ \Eprint
  {http://arxiv.org/abs/1604.02350} {arXiv:1604.02350 [astro-ph.GA]}
  \BibitemShut {NoStop}%
\bibitem [{\citenamefont {{Alarcon}}\ \emph {et~al.}(2021)\citenamefont
  {{Alarcon}}, \citenamefont {{Gaztanaga}},\ and\ \citenamefont
  {{Eriksen}}}]{Alarcon20}%
  \BibitemOpen
  \bibfield  {author} {\bibinfo {author} {\bibfnamefont {A.}~\bibnamefont
  {{Alarcon}}}, \bibinfo {author} {\bibfnamefont {E.}~\bibnamefont
  {{Gaztanaga}}}, \ and\ \bibinfo {author} {\bibfnamefont {P.}~\bibnamefont
  {{Eriksen}}},\ }\href {\doibase 10.1093/mnras/staa3659} {\bibfield  {journal}
  {\bibinfo  {journal} {\mnras}\ }\textbf {\bibinfo {volume} {501}},\ \bibinfo
  {pages} {6103} (\bibinfo {year} {2021})},\ \Eprint
  {http://arxiv.org/abs/2007.11132} {arXiv:2007.11132 [astro-ph.GA]}
  \BibitemShut {NoStop}%
\bibitem [{\citenamefont {Troxel}\ \emph
  {et~al.}(2018{\natexlab{b}})\citenamefont {Troxel}, \citenamefont {MacCrann},
  \citenamefont {Zuntz}, \citenamefont {Eifler}, \citenamefont {Krause},
  \citenamefont {Dodelson}, \citenamefont {Gruen}, \citenamefont {Blazek},
  \citenamefont {Friedrich}, \citenamefont {Samuroff} \emph
  {et~al.}}]{troxel2018dark}%
  \BibitemOpen
  \bibfield  {author} {\bibinfo {author} {\bibfnamefont {M.~A.}\ \bibnamefont
  {Troxel}}, \bibinfo {author} {\bibfnamefont {N.}~\bibnamefont {MacCrann}},
  \bibinfo {author} {\bibfnamefont {J.}~\bibnamefont {Zuntz}}, \bibinfo
  {author} {\bibfnamefont {T.}~\bibnamefont {Eifler}}, \bibinfo {author}
  {\bibfnamefont {E.}~\bibnamefont {Krause}}, \bibinfo {author} {\bibfnamefont
  {S.}~\bibnamefont {Dodelson}}, \bibinfo {author} {\bibfnamefont
  {D.}~\bibnamefont {Gruen}}, \bibinfo {author} {\bibfnamefont
  {J.}~\bibnamefont {Blazek}}, \bibinfo {author} {\bibfnamefont
  {O.}~\bibnamefont {Friedrich}}, \bibinfo {author} {\bibfnamefont
  {S.}~\bibnamefont {Samuroff}},  \emph {et~al.},\ }\href@noop {} {\bibfield
  {journal} {\bibinfo  {journal} {Physical Review D}\ }\textbf {\bibinfo
  {volume} {98}},\ \bibinfo {pages} {043528} (\bibinfo {year}
  {2018}{\natexlab{b}})}\BibitemShut {NoStop}%
\bibitem [{\citenamefont {{Leauthaud}}\ \emph {et~al.}(2007)\citenamefont
  {{Leauthaud}}, \citenamefont {{Massey}}, \citenamefont {{Kneib}},\ and\
  \citenamefont {{Rhodes}}}]{leauthaudHST}%
  \BibitemOpen
  \bibfield  {author} {\bibinfo {author} {\bibfnamefont {A.}~\bibnamefont
  {{Leauthaud}}}, \bibinfo {author} {\bibfnamefont {R.}~\bibnamefont
  {{Massey}}}, \bibinfo {author} {\bibfnamefont {J.-P.}\ \bibnamefont
  {{Kneib}}}, \ and\ \bibinfo {author} {\bibfnamefont {J.}~\bibnamefont
  {{Rhodes}}},\ }\href {\doibase 10.1086/516598} {\bibfield  {journal}
  {\bibinfo  {journal} {\apjs}\ }\textbf {\bibinfo {volume} {172}},\ \bibinfo
  {pages} {219} (\bibinfo {year} {2007})},\ \Eprint
  {http://arxiv.org/abs/astro-ph/0702359} {arXiv:astro-ph/0702359 [astro-ph]}
  \BibitemShut {NoStop}%
\bibitem [{\citenamefont {{Wechsler}}\ \emph {et~al.}(2021)\citenamefont
  {{Wechsler}}, \citenamefont {{DeRose}}, \citenamefont {{Busha}},
  \citenamefont {{Becker}}, \citenamefont {{Rykoff}},\ and\ \citenamefont
  {{Evrard}}}]{addgals}%
  \BibitemOpen
  \bibfield  {author} {\bibinfo {author} {\bibfnamefont {R.~H.}\ \bibnamefont
  {{Wechsler}}}, \bibinfo {author} {\bibfnamefont {J.}~\bibnamefont
  {{DeRose}}}, \bibinfo {author} {\bibfnamefont {M.~T.}\ \bibnamefont
  {{Busha}}}, \bibinfo {author} {\bibfnamefont {M.~R.}\ \bibnamefont
  {{Becker}}}, \bibinfo {author} {\bibfnamefont {E.}~\bibnamefont {{Rykoff}}},
  \ and\ \bibinfo {author} {\bibfnamefont {A.}~\bibnamefont {{Evrard}}},\
  }\href@noop {} {\bibfield  {journal} {\bibinfo  {journal} {arXiv e-prints}\
  ,\ \bibinfo {eid} {arXiv:2105.12105}} (\bibinfo {year} {2021})},\ \Eprint
  {http://arxiv.org/abs/2105.12105} {arXiv:2105.12105 [astro-ph.CO]}
  \BibitemShut {NoStop}%
\bibitem [{\citenamefont {{DeRose}}\ \emph
  {et~al.}(2021{\natexlab{b}})\citenamefont {{DeRose}}, \citenamefont
  {{Becker}},\ and\ \citenamefont {{Wechsler}}}]{DeRose2021}%
  \BibitemOpen
  \bibfield  {author} {\bibinfo {author} {\bibfnamefont {J.}~\bibnamefont
  {{DeRose}}}, \bibinfo {author} {\bibfnamefont {M.~R.}\ \bibnamefont
  {{Becker}}}, \ and\ \bibinfo {author} {\bibfnamefont {R.~H.}\ \bibnamefont
  {{Wechsler}}},\ }\href@noop {} {\bibfield  {journal} {\bibinfo  {journal}
  {arXiv e-prints}\ ,\ \bibinfo {eid} {arXiv:2105.12104}} (\bibinfo {year}
  {2021}{\natexlab{b}})},\ \Eprint {http://arxiv.org/abs/2105.12104}
  {arXiv:2105.12104 [astro-ph.CO]} \BibitemShut {NoStop}%
\bibitem [{\citenamefont {{Springel}}(2005)}]{Springel2005}%
  \BibitemOpen
  \bibfield  {author} {\bibinfo {author} {\bibfnamefont {V.}~\bibnamefont
  {{Springel}}},\ }\href {\doibase 10.1111/j.1365-2966.2005.09655.x} {\bibfield
   {journal} {\bibinfo  {journal} {\mnras}\ }\textbf {\bibinfo {volume}
  {364}},\ \bibinfo {pages} {1105} (\bibinfo {year} {2005})},\ \Eprint
  {http://arxiv.org/abs/astro-ph/0505010} {arXiv:astro-ph/0505010 [astro-ph]}
  \BibitemShut {NoStop}%
\bibitem [{\citenamefont {{Crocce}}\ \emph {et~al.}(2012)\citenamefont
  {{Crocce}}, \citenamefont {{Pueblas}},\ and\ \citenamefont
  {{Scoccimarro}}}]{2LPTIC}%
  \BibitemOpen
  \bibfield  {author} {\bibinfo {author} {\bibfnamefont {M.}~\bibnamefont
  {{Crocce}}}, \bibinfo {author} {\bibfnamefont {S.}~\bibnamefont {{Pueblas}}},
  \ and\ \bibinfo {author} {\bibfnamefont {R.}~\bibnamefont {{Scoccimarro}}},\
  }\href@noop {} {\enquote {\bibinfo {title} {{2LPTIC: 2nd-order Lagrangian
  Perturbation Theory Initial Conditions}},}\ } (\bibinfo {year} {2012}),\
  \Eprint {http://arxiv.org/abs/1201.005} {ascl:1201.005} \BibitemShut
  {NoStop}%
\bibitem [{\citenamefont {{Becker}}(2013)}]{Becker2013}%
  \BibitemOpen
  \bibfield  {author} {\bibinfo {author} {\bibfnamefont {M.~R.}\ \bibnamefont
  {{Becker}}},\ }\emph {\bibinfo {title} {{CALCLENS: Weak lensing
  simulations}}},\ \href@noop {} {Ph.D. thesis},\ \bibinfo  {school} {The
  University of Chicago} (\bibinfo {year} {2013})\BibitemShut {NoStop}%
\bibitem [{\citenamefont {{G{\'o}rski}}\ \emph
  {et~al.}(2005{\natexlab{a}})\citenamefont {{G{\'o}rski}}, \citenamefont
  {{Hivon}}, \citenamefont {{Banday}}, \citenamefont {{Wandelt}}, \citenamefont
  {{Hansen}}, \citenamefont {{Reinecke}},\ and\ \citenamefont
  {{Bartelmann}}}]{healpix}%
  \BibitemOpen
  \bibfield  {author} {\bibinfo {author} {\bibfnamefont {K.~M.}\ \bibnamefont
  {{G{\'o}rski}}}, \bibinfo {author} {\bibfnamefont {E.}~\bibnamefont
  {{Hivon}}}, \bibinfo {author} {\bibfnamefont {A.~J.}\ \bibnamefont
  {{Banday}}}, \bibinfo {author} {\bibfnamefont {B.~D.}\ \bibnamefont
  {{Wandelt}}}, \bibinfo {author} {\bibfnamefont {F.~K.}\ \bibnamefont
  {{Hansen}}}, \bibinfo {author} {\bibfnamefont {M.}~\bibnamefont
  {{Reinecke}}}, \ and\ \bibinfo {author} {\bibfnamefont {M.}~\bibnamefont
  {{Bartelmann}}},\ }\href {\doibase 10.1086/427976} {\bibfield  {journal}
  {\bibinfo  {journal} {\apj}\ }\textbf {\bibinfo {volume} {622}},\ \bibinfo
  {pages} {759} (\bibinfo {year} {2005}{\natexlab{a}})},\ \Eprint
  {http://arxiv.org/abs/astro-ph/0409513} {arXiv:astro-ph/0409513 [astro-ph]}
  \BibitemShut {NoStop}%
\bibitem [{\citenamefont {{DeRose}}\ \emph {et~al.}(2019)\citenamefont
  {{DeRose}}, \citenamefont {{Wechsler}},\ and\ \citenamefont
  {{Becker}}}]{DeRose2019}%
  \BibitemOpen
  \bibfield  {author} {\bibinfo {author} {\bibfnamefont {J.}~\bibnamefont
  {{DeRose}}}, \bibinfo {author} {\bibfnamefont {R.~H.}\ \bibnamefont
  {{Wechsler}}}, \ and\ \bibinfo {author} {\bibfnamefont {M.~R.}\ \bibnamefont
  {{Becker}}},\ }\href@noop {} {\bibfield  {journal} {\bibinfo  {journal}
  {arXiv e-prints}\ ,\ \bibinfo {eid} {arXiv:1901.02401}} (\bibinfo {year}
  {2019})},\ \Eprint {http://arxiv.org/abs/1901.02401} {arXiv:1901.02401
  [astro-ph.CO]} \BibitemShut {NoStop}%
\bibitem [{\citenamefont {{Jarvis}}\ \emph {et~al.}(2004)\citenamefont
  {{Jarvis}}, \citenamefont {{Bernstein}},\ and\ \citenamefont
  {{Jain}}}]{TreeCorr}%
  \BibitemOpen
  \bibfield  {author} {\bibinfo {author} {\bibfnamefont {M.}~\bibnamefont
  {{Jarvis}}}, \bibinfo {author} {\bibfnamefont {G.}~\bibnamefont
  {{Bernstein}}}, \ and\ \bibinfo {author} {\bibfnamefont {B.}~\bibnamefont
  {{Jain}}},\ }\href {\doibase 10.1111/j.1365-2966.2004.07926.x} {\bibfield
  {journal} {\bibinfo  {journal} {\mnras}\ }\textbf {\bibinfo {volume} {352}},\
  \bibinfo {pages} {338} (\bibinfo {year} {2004})},\ \Eprint
  {http://arxiv.org/abs/astro-ph/0307393} {arXiv:astro-ph/0307393 [astro-ph]}
  \BibitemShut {NoStop}%
\bibitem [{\citenamefont {{Fang}}\ \emph {et~al.}(2020)\citenamefont {{Fang}},
  \citenamefont {{Eifler}},\ and\ \citenamefont {{Krause}}}]{fang20b}%
  \BibitemOpen
  \bibfield  {author} {\bibinfo {author} {\bibfnamefont {X.}~\bibnamefont
  {{Fang}}}, \bibinfo {author} {\bibfnamefont {T.}~\bibnamefont {{Eifler}}}, \
  and\ \bibinfo {author} {\bibfnamefont {E.}~\bibnamefont {{Krause}}},\ }\href
  {\doibase 10.1093/mnras/staa1726} {\bibfield  {journal} {\bibinfo  {journal}
  {\mnras}\ }\textbf {\bibinfo {volume} {497}},\ \bibinfo {pages} {2699}
  (\bibinfo {year} {2020})},\ \Eprint {http://arxiv.org/abs/2004.04833}
  {arXiv:2004.04833 [astro-ph.CO]} \BibitemShut {NoStop}%
\bibitem [{\citenamefont {Hildebrandt}\ \emph {et~al.}(2016)\citenamefont
  {Hildebrandt}, \citenamefont {Choi}, \citenamefont {Heymans}, \citenamefont
  {Blake}, \citenamefont {Erben}, \citenamefont {Miller}, \citenamefont
  {Nakajima}, \citenamefont {van Waerbeke}, \citenamefont {Viola},
  \citenamefont {Buddendiek},\ and\ \citenamefont {et~al.}}]{rcslens}%
  \BibitemOpen
  \bibfield  {author} {\bibinfo {author} {\bibfnamefont {H.}~\bibnamefont
  {Hildebrandt}}, \bibinfo {author} {\bibfnamefont {A.}~\bibnamefont {Choi}},
  \bibinfo {author} {\bibfnamefont {C.}~\bibnamefont {Heymans}}, \bibinfo
  {author} {\bibfnamefont {C.}~\bibnamefont {Blake}}, \bibinfo {author}
  {\bibfnamefont {T.}~\bibnamefont {Erben}}, \bibinfo {author} {\bibfnamefont
  {L.}~\bibnamefont {Miller}}, \bibinfo {author} {\bibfnamefont
  {R.}~\bibnamefont {Nakajima}}, \bibinfo {author} {\bibfnamefont
  {L.}~\bibnamefont {van Waerbeke}}, \bibinfo {author} {\bibfnamefont
  {M.}~\bibnamefont {Viola}}, \bibinfo {author} {\bibfnamefont
  {A.}~\bibnamefont {Buddendiek}}, \ and\ \bibinfo {author} {\bibnamefont
  {et~al.}},\ }\href {\doibase 10.1093/mnras/stw2013} {\bibfield  {journal}
  {\bibinfo  {journal} {Monthly Notices of the Royal Astronomical Society}\
  }\textbf {\bibinfo {volume} {463}},\ \bibinfo {pages} {635–654} (\bibinfo
  {year} {2016})}\BibitemShut {NoStop}%
\bibitem [{\citenamefont {{Zuntz}}\ \emph {et~al.}(2018)\citenamefont
  {{Zuntz}}, \citenamefont {{Sheldon}}, \citenamefont {{Samuroff}},
  \citenamefont {{Troxel}}, \citenamefont {{Jarvis}}, \citenamefont
  {{MacCrann}}, \citenamefont {{Gruen}},\ and\ \citenamefont {{DES
  Collaboration}}}]{y1shearcat}%
  \BibitemOpen
  \bibfield  {author} {\bibinfo {author} {\bibfnamefont {J.}~\bibnamefont
  {{Zuntz}}}, \bibinfo {author} {\bibfnamefont {E.}~\bibnamefont {{Sheldon}}},
  \bibinfo {author} {\bibfnamefont {S.}~\bibnamefont {{Samuroff}}}, \bibinfo
  {author} {\bibfnamefont {M.~A.}\ \bibnamefont {{Troxel}}}, \bibinfo {author}
  {\bibfnamefont {M.}~\bibnamefont {{Jarvis}}}, \bibinfo {author}
  {\bibfnamefont {N.}~\bibnamefont {{MacCrann}}}, \bibinfo {author}
  {\bibfnamefont {D.}~\bibnamefont {{Gruen}}}, \ and\ \bibinfo {author}
  {\bibnamefont {{DES Collaboration}}},\ }\href {\doibase
  10.1093/mnras/sty2219} {\bibfield  {journal} {\bibinfo  {journal} {\mnras}\
  }\textbf {\bibinfo {volume} {481}},\ \bibinfo {pages} {1149} (\bibinfo {year}
  {2018})},\ \Eprint {http://arxiv.org/abs/1708.01533} {arXiv:1708.01533
  [astro-ph.CO]} \BibitemShut {NoStop}%
\bibitem [{\citenamefont {Muir}\ \emph {et~al.}(2020)\citenamefont {Muir} \emph
  {et~al.}}]{y3-blinding}%
  \BibitemOpen
  \bibfield  {author} {\bibinfo {author} {\bibfnamefont {J.}~\bibnamefont
  {Muir}} \emph {et~al.} (\bibinfo {collaboration} {DES}),\ }\href {\doibase
  10.1093/mnras/staa965} {\bibfield  {journal} {\bibinfo  {journal} {Mon. Not.
  Roy. Astron. Soc.}\ }\textbf {\bibinfo {volume} {494}},\ \bibinfo {pages}
  {4454} (\bibinfo {year} {2020})},\ \Eprint {http://arxiv.org/abs/1911.05929}
  {arXiv:1911.05929 [astro-ph.CO]} \BibitemShut {NoStop}%
\bibitem [{\citenamefont {Olive}(2016)}]{Patrignani}%
  \BibitemOpen
  \bibfield  {author} {\bibinfo {author} {\bibfnamefont {K.}~\bibnamefont
  {Olive}},\ }\href {\doibase 10.1088/1674-1137/40/10/100001} {\bibfield
  {journal} {\bibinfo  {journal} {Chinese Physics C}\ }\textbf {\bibinfo
  {volume} {40}},\ \bibinfo {pages} {100001} (\bibinfo {year}
  {2016})}\BibitemShut {NoStop}%
\bibitem [{\citenamefont {{Stebbins}}(1996)}]{Stebbins}%
  \BibitemOpen
  \bibfield  {author} {\bibinfo {author} {\bibfnamefont {A.}~\bibnamefont
  {{Stebbins}}},\ }\href@noop {} {\bibfield  {journal} {\bibinfo  {journal}
  {arXiv e-prints}\ ,\ \bibinfo {eid} {astro-ph/9609149}} (\bibinfo {year}
  {1996})},\ \Eprint {http://arxiv.org/abs/astro-ph/9609149}
  {arXiv:astro-ph/9609149 [astro-ph]} \BibitemShut {NoStop}%
\bibitem [{\citenamefont {{Limber}}(1953)}]{Limber53}%
  \BibitemOpen
  \bibfield  {author} {\bibinfo {author} {\bibfnamefont {D.~N.}\ \bibnamefont
  {{Limber}}},\ }\href {\doibase 10.1086/145672} {\bibfield  {journal}
  {\bibinfo  {journal} {\apj}\ }\textbf {\bibinfo {volume} {117}},\ \bibinfo
  {pages} {134} (\bibinfo {year} {1953})}\BibitemShut {NoStop}%
\bibitem [{\citenamefont {LoVerde}\ and\ \citenamefont
  {Afshordi}(2008)}]{Limber_LoVerde2008}%
  \BibitemOpen
  \bibfield  {author} {\bibinfo {author} {\bibfnamefont {M.}~\bibnamefont
  {LoVerde}}\ and\ \bibinfo {author} {\bibfnamefont {N.}~\bibnamefont
  {Afshordi}},\ }\href {\doibase 10.1103/PhysRevD.78.123506} {\bibfield
  {journal} {\bibinfo  {journal} {Phys. Rev. D}\ }\textbf {\bibinfo {volume}
  {78}},\ \bibinfo {pages} {123506} (\bibinfo {year} {2008})}\BibitemShut
  {NoStop}%
\bibitem [{\citenamefont {{Dodelson}}\ \emph {et~al.}(2006)\citenamefont
  {{Dodelson}}, \citenamefont {{Shapiro}},\ and\ \citenamefont
  {{White}}}]{dodelson2006}%
  \BibitemOpen
  \bibfield  {author} {\bibinfo {author} {\bibfnamefont {S.}~\bibnamefont
  {{Dodelson}}}, \bibinfo {author} {\bibfnamefont {C.}~\bibnamefont
  {{Shapiro}}}, \ and\ \bibinfo {author} {\bibfnamefont {M.}~\bibnamefont
  {{White}}},\ }\href {\doibase 10.1103/PhysRevD.73.023009} {\bibfield
  {journal} {\bibinfo  {journal} {\prd}\ }\textbf {\bibinfo {volume} {73}},\
  \bibinfo {eid} {023009} (\bibinfo {year} {2006})},\ \Eprint
  {http://arxiv.org/abs/astro-ph/0508296} {arXiv:astro-ph/0508296 [astro-ph]}
  \BibitemShut {NoStop}%
\bibitem [{\citenamefont {{Shapiro}}(2009)}]{shapiro2009}%
  \BibitemOpen
  \bibfield  {author} {\bibinfo {author} {\bibfnamefont {C.}~\bibnamefont
  {{Shapiro}}},\ }\href {\doibase 10.1088/0004-637X/696/1/775} {\bibfield
  {journal} {\bibinfo  {journal} {\apj}\ }\textbf {\bibinfo {volume} {696}},\
  \bibinfo {pages} {775} (\bibinfo {year} {2009})},\ \Eprint
  {http://arxiv.org/abs/0812.0769} {arXiv:0812.0769 [astro-ph]} \BibitemShut
  {NoStop}%
\bibitem [{\citenamefont {{Schneider}}\ \emph {et~al.}(2002)\citenamefont
  {{Schneider}}, \citenamefont {{van Waerbeke}}, \citenamefont {{Kilbinger}},\
  and\ \citenamefont {{Mellier}}}]{Schneider2002}%
  \BibitemOpen
  \bibfield  {author} {\bibinfo {author} {\bibfnamefont {P.}~\bibnamefont
  {{Schneider}}}, \bibinfo {author} {\bibfnamefont {L.}~\bibnamefont {{van
  Waerbeke}}}, \bibinfo {author} {\bibfnamefont {M.}~\bibnamefont
  {{Kilbinger}}}, \ and\ \bibinfo {author} {\bibfnamefont {Y.}~\bibnamefont
  {{Mellier}}},\ }\href {\doibase 10.1051/0004-6361:20021341} {\bibfield
  {journal} {\bibinfo  {journal} {\aap}\ }\textbf {\bibinfo {volume} {396}},\
  \bibinfo {pages} {1} (\bibinfo {year} {2002})},\ \Eprint
  {http://arxiv.org/abs/astro-ph/0206182} {arXiv:astro-ph/0206182 [astro-ph]}
  \BibitemShut {NoStop}%
\bibitem [{\citenamefont {{Schmidt}}\ \emph {et~al.}(2009)\citenamefont
  {{Schmidt}}, \citenamefont {{Rozo}}, \citenamefont {{Dodelson}},
  \citenamefont {{Hui}},\ and\ \citenamefont {{Sheldon}}}]{Schmidt2009}%
  \BibitemOpen
  \bibfield  {author} {\bibinfo {author} {\bibfnamefont {F.}~\bibnamefont
  {{Schmidt}}}, \bibinfo {author} {\bibfnamefont {E.}~\bibnamefont {{Rozo}}},
  \bibinfo {author} {\bibfnamefont {S.}~\bibnamefont {{Dodelson}}}, \bibinfo
  {author} {\bibfnamefont {L.}~\bibnamefont {{Hui}}}, \ and\ \bibinfo {author}
  {\bibfnamefont {E.}~\bibnamefont {{Sheldon}}},\ }\href {\doibase
  10.1088/0004-637X/702/1/593} {\bibfield  {journal} {\bibinfo  {journal}
  {\apj}\ }\textbf {\bibinfo {volume} {702}},\ \bibinfo {pages} {593} (\bibinfo
  {year} {2009})},\ \Eprint {http://arxiv.org/abs/0904.4703} {arXiv:0904.4703
  [astro-ph.CO]} \BibitemShut {NoStop}%
\bibitem [{\citenamefont {{Seljak}}(1996)}]{seljak96}%
  \BibitemOpen
  \bibfield  {author} {\bibinfo {author} {\bibfnamefont {U.}~\bibnamefont
  {{Seljak}}},\ }\href {\doibase 10.1086/177218} {\bibfield  {journal}
  {\bibinfo  {journal} {\apj}\ }\textbf {\bibinfo {volume} {463}},\ \bibinfo
  {pages} {1} (\bibinfo {year} {1996})},\ \Eprint
  {http://arxiv.org/abs/astro-ph/9505109} {arXiv:astro-ph/9505109 [astro-ph]}
  \BibitemShut {NoStop}%
\bibitem [{\citenamefont {{Dodelson}}\ \emph {et~al.}(2008)\citenamefont
  {{Dodelson}}, \citenamefont {{Schmidt}},\ and\ \citenamefont
  {{Vallinotto}}}]{dodelson08}%
  \BibitemOpen
  \bibfield  {author} {\bibinfo {author} {\bibfnamefont {S.}~\bibnamefont
  {{Dodelson}}}, \bibinfo {author} {\bibfnamefont {F.}~\bibnamefont
  {{Schmidt}}}, \ and\ \bibinfo {author} {\bibfnamefont {A.}~\bibnamefont
  {{Vallinotto}}},\ }\href {\doibase 10.1103/PhysRevD.78.043508} {\bibfield
  {journal} {\bibinfo  {journal} {\prd}\ }\textbf {\bibinfo {volume} {78}},\
  \bibinfo {eid} {043508} (\bibinfo {year} {2008})},\ \Eprint
  {http://arxiv.org/abs/0806.0331} {arXiv:0806.0331 [astro-ph]} \BibitemShut
  {NoStop}%
\bibitem [{\citenamefont {{Hirata}}\ and\ \citenamefont
  {{Seljak}}(2004)}]{hirata04}%
  \BibitemOpen
  \bibfield  {author} {\bibinfo {author} {\bibfnamefont {C.~M.}\ \bibnamefont
  {{Hirata}}}\ and\ \bibinfo {author} {\bibfnamefont {U.}~\bibnamefont
  {{Seljak}}},\ }\href {\doibase 10.1103/PhysRevD.70.063526} {\bibfield
  {journal} {\bibinfo  {journal} {\prd}\ }\textbf {\bibinfo {volume} {70}},\
  \bibinfo {eid} {063526} (\bibinfo {year} {2004})},\ \Eprint
  {http://arxiv.org/abs/astro-ph/0406275} {arXiv:astro-ph/0406275 [astro-ph]}
  \BibitemShut {NoStop}%
\bibitem [{\citenamefont {{Kiessling}}\ \emph {et~al.}(2015)\citenamefont
  {{Kiessling}}, \citenamefont {{Cacciato}}, \citenamefont {{Joachimi}},
  \citenamefont {{Kirk}}, \citenamefont {{Kitching}}, \citenamefont
  {{Leonard}}, \citenamefont {{Mandelbaum}}, \citenamefont {{Sch{\"a}fer}},
  \citenamefont {{Sif{\'o}n}}, \citenamefont {{Brown}},\ and\ \citenamefont
  {{Rassat}}}]{kiessling15}%
  \BibitemOpen
  \bibfield  {author} {\bibinfo {author} {\bibfnamefont {A.}~\bibnamefont
  {{Kiessling}}}, \bibinfo {author} {\bibfnamefont {M.}~\bibnamefont
  {{Cacciato}}}, \bibinfo {author} {\bibfnamefont {B.}~\bibnamefont
  {{Joachimi}}}, \bibinfo {author} {\bibfnamefont {D.}~\bibnamefont {{Kirk}}},
  \bibinfo {author} {\bibfnamefont {T.~D.}\ \bibnamefont {{Kitching}}},
  \bibinfo {author} {\bibfnamefont {A.}~\bibnamefont {{Leonard}}}, \bibinfo
  {author} {\bibfnamefont {R.}~\bibnamefont {{Mandelbaum}}}, \bibinfo {author}
  {\bibfnamefont {B.~M.}\ \bibnamefont {{Sch{\"a}fer}}}, \bibinfo {author}
  {\bibfnamefont {C.}~\bibnamefont {{Sif{\'o}n}}}, \bibinfo {author}
  {\bibfnamefont {M.~L.}\ \bibnamefont {{Brown}}}, \ and\ \bibinfo {author}
  {\bibfnamefont {A.}~\bibnamefont {{Rassat}}},\ }\href {\doibase
  10.1007/s11214-015-0203-6} {\bibfield  {journal} {\bibinfo  {journal} {\ssr}\
  }\textbf {\bibinfo {volume} {193}},\ \bibinfo {pages} {67} (\bibinfo {year}
  {2015})},\ \Eprint {http://arxiv.org/abs/1504.05546} {arXiv:1504.05546
  [astro-ph.GA]} \BibitemShut {NoStop}%
\bibitem [{\citenamefont {{Blazek}}\ \emph {et~al.}(2019)\citenamefont
  {{Blazek}}, \citenamefont {{MacCrann}}, \citenamefont {{Troxel}},\ and\
  \citenamefont {{Fang}}}]{blazek19}%
  \BibitemOpen
  \bibfield  {author} {\bibinfo {author} {\bibfnamefont {J.~A.}\ \bibnamefont
  {{Blazek}}}, \bibinfo {author} {\bibfnamefont {N.}~\bibnamefont
  {{MacCrann}}}, \bibinfo {author} {\bibfnamefont {M.~A.}\ \bibnamefont
  {{Troxel}}}, \ and\ \bibinfo {author} {\bibfnamefont {X.}~\bibnamefont
  {{Fang}}},\ }\href {\doibase 10.1103/PhysRevD.100.103506} {\bibfield
  {journal} {\bibinfo  {journal} {\prd}\ }\textbf {\bibinfo {volume} {100}},\
  \bibinfo {eid} {103506} (\bibinfo {year} {2019})},\ \Eprint
  {http://arxiv.org/abs/1708.09247} {arXiv:1708.09247 [astro-ph.CO]}
  \BibitemShut {NoStop}%
\bibitem [{\citenamefont {{Mackey}}\ \emph {et~al.}(2002)\citenamefont
  {{Mackey}}, \citenamefont {{White}},\ and\ \citenamefont
  {{Kamionkowski}}}]{mackey02}%
  \BibitemOpen
  \bibfield  {author} {\bibinfo {author} {\bibfnamefont {J.}~\bibnamefont
  {{Mackey}}}, \bibinfo {author} {\bibfnamefont {M.}~\bibnamefont {{White}}}, \
  and\ \bibinfo {author} {\bibfnamefont {M.}~\bibnamefont {{Kamionkowski}}},\
  }\href {\doibase 10.1046/j.1365-8711.2002.05337.x} {\bibfield  {journal}
  {\bibinfo  {journal} {\mnras}\ }\textbf {\bibinfo {volume} {332}},\ \bibinfo
  {pages} {788} (\bibinfo {year} {2002})},\ \Eprint
  {http://arxiv.org/abs/astro-ph/0106364} {arXiv:astro-ph/0106364 [astro-ph]}
  \BibitemShut {NoStop}%
\bibitem [{\citenamefont {{Codis}}\ \emph {et~al.}(2015)\citenamefont
  {{Codis}}, \citenamefont {{Pichon}},\ and\ \citenamefont
  {{Pogosyan}}}]{codis15}%
  \BibitemOpen
  \bibfield  {author} {\bibinfo {author} {\bibfnamefont {S.}~\bibnamefont
  {{Codis}}}, \bibinfo {author} {\bibfnamefont {C.}~\bibnamefont {{Pichon}}}, \
  and\ \bibinfo {author} {\bibfnamefont {D.}~\bibnamefont {{Pogosyan}}},\
  }\href {\doibase 10.1093/mnras/stv1570} {\bibfield  {journal} {\bibinfo
  {journal} {\mnras}\ }\textbf {\bibinfo {volume} {452}},\ \bibinfo {pages}
  {3369} (\bibinfo {year} {2015})},\ \Eprint {http://arxiv.org/abs/1504.06073}
  {arXiv:1504.06073 [astro-ph.CO]} \BibitemShut {NoStop}%
\bibitem [{\citenamefont {{Blazek}}\ \emph {et~al.}(2015)\citenamefont
  {{Blazek}}, \citenamefont {{Vlah}},\ and\ \citenamefont
  {{Seljak}}}]{blazek15}%
  \BibitemOpen
  \bibfield  {author} {\bibinfo {author} {\bibfnamefont {J.}~\bibnamefont
  {{Blazek}}}, \bibinfo {author} {\bibfnamefont {Z.}~\bibnamefont {{Vlah}}}, \
  and\ \bibinfo {author} {\bibfnamefont {U.}~\bibnamefont {{Seljak}}},\ }\href
  {\doibase 10.1088/1475-7516/2015/08/015} {\bibfield  {journal} {\bibinfo
  {journal} {\jcap}\ }\textbf {\bibinfo {volume} {2015}},\ \bibinfo {eid} {015}
  (\bibinfo {year} {2015})},\ \Eprint {http://arxiv.org/abs/1504.02510}
  {arXiv:1504.02510 [astro-ph.CO]} \BibitemShut {NoStop}%
\bibitem [{\citenamefont {Brown}\ \emph {et~al.}(2002)\citenamefont {Brown},
  \citenamefont {Taylor}, \citenamefont {Hambly},\ and\ \citenamefont
  {Dye}}]{brown02}%
  \BibitemOpen
  \bibfield  {author} {\bibinfo {author} {\bibfnamefont {M.~L.}\ \bibnamefont
  {Brown}}, \bibinfo {author} {\bibfnamefont {A.~N.}\ \bibnamefont {Taylor}},
  \bibinfo {author} {\bibfnamefont {N.~C.}\ \bibnamefont {Hambly}}, \ and\
  \bibinfo {author} {\bibfnamefont {S.}~\bibnamefont {Dye}},\ }\href {\doibase
  10.1046/j.1365-8711.2002.05354.x} {\bibfield  {journal} {\bibinfo  {journal}
  {Monthly Notices of the Royal Astronomical Society}\ }\textbf {\bibinfo
  {volume} {333}},\ \bibinfo {pages} {501} (\bibinfo {year} {2002})},\ \Eprint
  {http://arxiv.org/abs/https://academic.oup.com/mnras/article-pdf/333/3/501/3215948/333-3-501.pdf}
  {https://academic.oup.com/mnras/article-pdf/333/3/501/3215948/333-3-501.pdf}
  \BibitemShut {NoStop}%
\bibitem [{\citenamefont {{van Daalen}}\ \emph {et~al.}(2014)\citenamefont
  {{van Daalen}}, \citenamefont {{Schaye}}, \citenamefont {{McCarthy}},
  \citenamefont {{Booth}},\ and\ \citenamefont {{Dalla
  Vecchia}}}]{vandaalen2014}%
  \BibitemOpen
  \bibfield  {author} {\bibinfo {author} {\bibfnamefont {M.~P.}\ \bibnamefont
  {{van Daalen}}}, \bibinfo {author} {\bibfnamefont {J.}~\bibnamefont
  {{Schaye}}}, \bibinfo {author} {\bibfnamefont {I.~G.}\ \bibnamefont
  {{McCarthy}}}, \bibinfo {author} {\bibfnamefont {C.~M.}\ \bibnamefont
  {{Booth}}}, \ and\ \bibinfo {author} {\bibfnamefont {C.}~\bibnamefont {{Dalla
  Vecchia}}},\ }\href {\doibase 10.1093/mnras/stu482} {\bibfield  {journal}
  {\bibinfo  {journal} {\mnras}\ }\textbf {\bibinfo {volume} {440}},\ \bibinfo
  {pages} {2997} (\bibinfo {year} {2014})},\ \Eprint
  {http://arxiv.org/abs/1310.7571} {arXiv:1310.7571 [astro-ph.CO]} \BibitemShut
  {NoStop}%
\bibitem [{\citenamefont {Semboloni}\ \emph {et~al.}(2013)\citenamefont
  {Semboloni}, \citenamefont {Hoekstra},\ and\ \citenamefont
  {Schaye}}]{semboloni2013effect}%
  \BibitemOpen
  \bibfield  {author} {\bibinfo {author} {\bibfnamefont {E.}~\bibnamefont
  {Semboloni}}, \bibinfo {author} {\bibfnamefont {H.}~\bibnamefont {Hoekstra}},
  \ and\ \bibinfo {author} {\bibfnamefont {J.}~\bibnamefont {Schaye}},\
  }\href@noop {} {\bibfield  {journal} {\bibinfo  {journal} {Monthly Notices of
  the Royal Astronomical Society}\ }\textbf {\bibinfo {volume} {434}},\
  \bibinfo {pages} {148} (\bibinfo {year} {2013})}\BibitemShut {NoStop}%
\bibitem [{\citenamefont {Harnois-D{\'e}raps}\ \emph
  {et~al.}(2015)\citenamefont {Harnois-D{\'e}raps}, \citenamefont {van
  Waerbeke}, \citenamefont {Viola},\ and\ \citenamefont
  {Heymans}}]{harnois2015baryons}%
  \BibitemOpen
  \bibfield  {author} {\bibinfo {author} {\bibfnamefont {J.}~\bibnamefont
  {Harnois-D{\'e}raps}}, \bibinfo {author} {\bibfnamefont {L.}~\bibnamefont
  {van Waerbeke}}, \bibinfo {author} {\bibfnamefont {M.}~\bibnamefont {Viola}},
  \ and\ \bibinfo {author} {\bibfnamefont {C.}~\bibnamefont {Heymans}},\
  }\href@noop {} {\bibfield  {journal} {\bibinfo  {journal} {Monthly Notices of
  the Royal Astronomical Society}\ }\textbf {\bibinfo {volume} {450}},\
  \bibinfo {pages} {1212} (\bibinfo {year} {2015})}\BibitemShut {NoStop}%
\bibitem [{\citenamefont {Mead}\ \emph {et~al.}(2015)\citenamefont {Mead},
  \citenamefont {Peacock}, \citenamefont {Heymans}, \citenamefont {Joudaki},\
  and\ \citenamefont {Heavens}}]{mead2015}%
  \BibitemOpen
  \bibfield  {author} {\bibinfo {author} {\bibfnamefont {A.}~\bibnamefont
  {Mead}}, \bibinfo {author} {\bibfnamefont {J.}~\bibnamefont {Peacock}},
  \bibinfo {author} {\bibfnamefont {C.}~\bibnamefont {Heymans}}, \bibinfo
  {author} {\bibfnamefont {S.}~\bibnamefont {Joudaki}}, \ and\ \bibinfo
  {author} {\bibfnamefont {A.}~\bibnamefont {Heavens}},\ }\href@noop {}
  {\bibfield  {journal} {\bibinfo  {journal} {Monthly Notices of the Royal
  Astronomical Society}\ }\textbf {\bibinfo {volume} {454}},\ \bibinfo {pages}
  {1958} (\bibinfo {year} {2015})}\BibitemShut {NoStop}%
\bibitem [{\citenamefont {Schneider}\ \emph {et~al.}(2019)\citenamefont
  {Schneider}, \citenamefont {Teyssier}, \citenamefont {Stadel}, \citenamefont
  {Chisari}, \citenamefont {Le~Brun}, \citenamefont {Amara},\ and\
  \citenamefont {Refregier}}]{schneider2019quantifying}%
  \BibitemOpen
  \bibfield  {author} {\bibinfo {author} {\bibfnamefont {A.}~\bibnamefont
  {Schneider}}, \bibinfo {author} {\bibfnamefont {R.}~\bibnamefont {Teyssier}},
  \bibinfo {author} {\bibfnamefont {J.}~\bibnamefont {Stadel}}, \bibinfo
  {author} {\bibfnamefont {N.~E.}\ \bibnamefont {Chisari}}, \bibinfo {author}
  {\bibfnamefont {A.~M.}\ \bibnamefont {Le~Brun}}, \bibinfo {author}
  {\bibfnamefont {A.}~\bibnamefont {Amara}}, \ and\ \bibinfo {author}
  {\bibfnamefont {A.}~\bibnamefont {Refregier}},\ }\href@noop {} {\bibfield
  {journal} {\bibinfo  {journal} {Journal of Cosmology and Astroparticle
  Physics}\ }\textbf {\bibinfo {volume} {2019}},\ \bibinfo {pages} {020}
  (\bibinfo {year} {2019})}\BibitemShut {NoStop}%
\bibitem [{\citenamefont {Huang}\ \emph {et~al.}(2020)\citenamefont {Huang},
  \citenamefont {Eifler}, \citenamefont {Mandelbaum}, \citenamefont
  {Bernstein}, \citenamefont {Chen}, \citenamefont {Choi}, \citenamefont
  {Garc{\'\i}a-Bellido}, \citenamefont {Huterer}, \citenamefont {Krause},
  \citenamefont {Rozo} \emph {et~al.}}]{huang2020dark}%
  \BibitemOpen
  \bibfield  {author} {\bibinfo {author} {\bibfnamefont {H.-J.}\ \bibnamefont
  {Huang}}, \bibinfo {author} {\bibfnamefont {T.}~\bibnamefont {Eifler}},
  \bibinfo {author} {\bibfnamefont {R.}~\bibnamefont {Mandelbaum}}, \bibinfo
  {author} {\bibfnamefont {G.~M.}\ \bibnamefont {Bernstein}}, \bibinfo {author}
  {\bibfnamefont {A.}~\bibnamefont {Chen}}, \bibinfo {author} {\bibfnamefont
  {A.}~\bibnamefont {Choi}}, \bibinfo {author} {\bibfnamefont {J.}~\bibnamefont
  {Garc{\'\i}a-Bellido}}, \bibinfo {author} {\bibfnamefont {D.}~\bibnamefont
  {Huterer}}, \bibinfo {author} {\bibfnamefont {E.}~\bibnamefont {Krause}},
  \bibinfo {author} {\bibfnamefont {E.}~\bibnamefont {Rozo}},  \emph {et~al.},\
  }\href@noop {} {\bibfield  {journal} {\bibinfo  {journal} {arXiv preprint
  arXiv:2007.15026}\ } (\bibinfo {year} {2020})}\BibitemShut {NoStop}%
\bibitem [{\citenamefont {McAlpine}\ \emph {et~al.}(2016)\citenamefont
  {McAlpine}, \citenamefont {Helly}, \citenamefont {Schaller}, \citenamefont
  {Trayford}, \citenamefont {Qu}, \citenamefont {Furlong}, \citenamefont
  {Bower}, \citenamefont {Crain}, \citenamefont {Schaye}, \citenamefont
  {Theuns} \emph {et~al.}}]{mcalpine}%
  \BibitemOpen
  \bibfield  {author} {\bibinfo {author} {\bibfnamefont {S.}~\bibnamefont
  {McAlpine}}, \bibinfo {author} {\bibfnamefont {J.~C.}\ \bibnamefont {Helly}},
  \bibinfo {author} {\bibfnamefont {M.}~\bibnamefont {Schaller}}, \bibinfo
  {author} {\bibfnamefont {J.~W.}\ \bibnamefont {Trayford}}, \bibinfo {author}
  {\bibfnamefont {Y.}~\bibnamefont {Qu}}, \bibinfo {author} {\bibfnamefont
  {M.}~\bibnamefont {Furlong}}, \bibinfo {author} {\bibfnamefont {R.~G.}\
  \bibnamefont {Bower}}, \bibinfo {author} {\bibfnamefont {R.~A.}\ \bibnamefont
  {Crain}}, \bibinfo {author} {\bibfnamefont {J.}~\bibnamefont {Schaye}},
  \bibinfo {author} {\bibfnamefont {T.}~\bibnamefont {Theuns}},  \emph
  {et~al.},\ }\href@noop {} {\bibfield  {journal} {\bibinfo  {journal}
  {Astronomy and Computing}\ }\textbf {\bibinfo {volume} {15}},\ \bibinfo
  {pages} {72} (\bibinfo {year} {2016})}\BibitemShut {NoStop}%
\bibitem [{\citenamefont {{van Daalen}}\ \emph {et~al.}(2011)\citenamefont
  {{van Daalen}}, \citenamefont {{Schaye}}, \citenamefont {{Booth}},\ and\
  \citenamefont {{Dalla Vecchia}}}]{vanDaalen11}%
  \BibitemOpen
  \bibfield  {author} {\bibinfo {author} {\bibfnamefont {M.~P.}\ \bibnamefont
  {{van Daalen}}}, \bibinfo {author} {\bibfnamefont {J.}~\bibnamefont
  {{Schaye}}}, \bibinfo {author} {\bibfnamefont {C.~M.}\ \bibnamefont
  {{Booth}}}, \ and\ \bibinfo {author} {\bibfnamefont {C.}~\bibnamefont {{Dalla
  Vecchia}}},\ }\href {\doibase 10.1111/j.1365-2966.2011.18981.x} {\bibfield
  {journal} {\bibinfo  {journal} {\mnras}\ }\textbf {\bibinfo {volume} {415}},\
  \bibinfo {pages} {3649} (\bibinfo {year} {2011})},\ \Eprint
  {http://arxiv.org/abs/1104.1174} {arXiv:1104.1174 [astro-ph.CO]} \BibitemShut
  {NoStop}%
\bibitem [{\citenamefont {{Smith}}\ \emph {et~al.}(2003)\citenamefont
  {{Smith}}, \citenamefont {{Peacock}}, \citenamefont {{Jenkins}},
  \citenamefont {{White}}, \citenamefont {{Frenk}}, \citenamefont {{Pearce}},
  \citenamefont {{Thomas}}, \citenamefont {{Efstathiou}},\ and\ \citenamefont
  {{Couchman}}}]{smith2003}%
  \BibitemOpen
  \bibfield  {author} {\bibinfo {author} {\bibfnamefont {R.~E.}\ \bibnamefont
  {{Smith}}}, \bibinfo {author} {\bibfnamefont {J.~A.}\ \bibnamefont
  {{Peacock}}}, \bibinfo {author} {\bibfnamefont {A.}~\bibnamefont
  {{Jenkins}}}, \bibinfo {author} {\bibfnamefont {S.~D.~M.}\ \bibnamefont
  {{White}}}, \bibinfo {author} {\bibfnamefont {C.~S.}\ \bibnamefont
  {{Frenk}}}, \bibinfo {author} {\bibfnamefont {F.~R.}\ \bibnamefont
  {{Pearce}}}, \bibinfo {author} {\bibfnamefont {P.~A.}\ \bibnamefont
  {{Thomas}}}, \bibinfo {author} {\bibfnamefont {G.}~\bibnamefont
  {{Efstathiou}}}, \ and\ \bibinfo {author} {\bibfnamefont {H.~M.~P.}\
  \bibnamefont {{Couchman}}},\ }\href {\doibase
  10.1046/j.1365-8711.2003.06503.x} {\bibfield  {journal} {\bibinfo  {journal}
  {\mnras}\ }\textbf {\bibinfo {volume} {341}},\ \bibinfo {pages} {1311}
  (\bibinfo {year} {2003})},\ \Eprint {http://arxiv.org/abs/astro-ph/0207664}
  {arXiv:astro-ph/0207664 [astro-ph]} \BibitemShut {NoStop}%
\bibitem [{\citenamefont {{Takahashi}}\ \emph {et~al.}(2012)\citenamefont
  {{Takahashi}}, \citenamefont {{Sato}}, \citenamefont {{Nishimichi}},
  \citenamefont {{Taruya}},\ and\ \citenamefont {{Oguri}}}]{takahashi2012}%
  \BibitemOpen
  \bibfield  {author} {\bibinfo {author} {\bibfnamefont {R.}~\bibnamefont
  {{Takahashi}}}, \bibinfo {author} {\bibfnamefont {M.}~\bibnamefont {{Sato}}},
  \bibinfo {author} {\bibfnamefont {T.}~\bibnamefont {{Nishimichi}}}, \bibinfo
  {author} {\bibfnamefont {A.}~\bibnamefont {{Taruya}}}, \ and\ \bibinfo
  {author} {\bibfnamefont {M.}~\bibnamefont {{Oguri}}},\ }\href {\doibase
  10.1088/0004-637X/761/2/152} {\bibfield  {journal} {\bibinfo  {journal}
  {\apj}\ }\textbf {\bibinfo {volume} {761}},\ \bibinfo {eid} {152} (\bibinfo
  {year} {2012})},\ \Eprint {http://arxiv.org/abs/1208.2701} {arXiv:1208.2701
  [astro-ph.CO]} \BibitemShut {NoStop}%
\bibitem [{\citenamefont {{Choi}}\ \emph {et~al.}(2016)\citenamefont {{Choi}},
  \citenamefont {{Heymans}}, \citenamefont {{Blake}}, \citenamefont
  {{Hildebrandt}}, \citenamefont {{Duncan}}, \citenamefont {{Erben}},
  \citenamefont {{Nakajima}}, \citenamefont {{Van Waerbeke}},\ and\
  \citenamefont {{Viola}}}]{choideltaz}%
  \BibitemOpen
  \bibfield  {author} {\bibinfo {author} {\bibfnamefont {A.}~\bibnamefont
  {{Choi}}}, \bibinfo {author} {\bibfnamefont {C.}~\bibnamefont {{Heymans}}},
  \bibinfo {author} {\bibfnamefont {C.}~\bibnamefont {{Blake}}}, \bibinfo
  {author} {\bibfnamefont {H.}~\bibnamefont {{Hildebrandt}}}, \bibinfo {author}
  {\bibfnamefont {C.~A.~J.}\ \bibnamefont {{Duncan}}}, \bibinfo {author}
  {\bibfnamefont {T.}~\bibnamefont {{Erben}}}, \bibinfo {author} {\bibfnamefont
  {R.}~\bibnamefont {{Nakajima}}}, \bibinfo {author} {\bibfnamefont
  {L.}~\bibnamefont {{Van Waerbeke}}}, \ and\ \bibinfo {author} {\bibfnamefont
  {M.}~\bibnamefont {{Viola}}},\ }\href {\doibase 10.1093/mnras/stw2241}
  {\bibfield  {journal} {\bibinfo  {journal} {\mnras}\ }\textbf {\bibinfo
  {volume} {463}},\ \bibinfo {pages} {3737} (\bibinfo {year} {2016})},\ \Eprint
  {http://arxiv.org/abs/1512.03626} {arXiv:1512.03626 [astro-ph.CO]}
  \BibitemShut {NoStop}%
\bibitem [{\citenamefont {{Joudaki}}\ \emph {et~al.}(2017)\citenamefont
  {{Joudaki}}, \citenamefont {{Blake}}, \citenamefont {{Heymans}},
  \citenamefont {{Choi}}, \citenamefont {{Harnois-Deraps}}, \citenamefont
  {{Hildebrandt}}, \citenamefont {{Joachimi}}, \citenamefont {{Johnson}},
  \citenamefont {{Mead}}, \citenamefont {{Parkinson}}, \citenamefont
  {{Viola}},\ and\ \citenamefont {{van Waerbeke}}}]{joudakicfht}%
  \BibitemOpen
  \bibfield  {author} {\bibinfo {author} {\bibfnamefont {S.}~\bibnamefont
  {{Joudaki}}}, \bibinfo {author} {\bibfnamefont {C.}~\bibnamefont {{Blake}}},
  \bibinfo {author} {\bibfnamefont {C.}~\bibnamefont {{Heymans}}}, \bibinfo
  {author} {\bibfnamefont {A.}~\bibnamefont {{Choi}}}, \bibinfo {author}
  {\bibfnamefont {J.}~\bibnamefont {{Harnois-Deraps}}}, \bibinfo {author}
  {\bibfnamefont {H.}~\bibnamefont {{Hildebrandt}}}, \bibinfo {author}
  {\bibfnamefont {B.}~\bibnamefont {{Joachimi}}}, \bibinfo {author}
  {\bibfnamefont {A.}~\bibnamefont {{Johnson}}}, \bibinfo {author}
  {\bibfnamefont {A.}~\bibnamefont {{Mead}}}, \bibinfo {author} {\bibfnamefont
  {D.}~\bibnamefont {{Parkinson}}}, \bibinfo {author} {\bibfnamefont
  {M.}~\bibnamefont {{Viola}}}, \ and\ \bibinfo {author} {\bibfnamefont
  {L.}~\bibnamefont {{van Waerbeke}}},\ }\href {\doibase 10.1093/mnras/stw2665}
  {\bibfield  {journal} {\bibinfo  {journal} {\mnras}\ }\textbf {\bibinfo
  {volume} {465}},\ \bibinfo {pages} {2033} (\bibinfo {year} {2017})},\ \Eprint
  {http://arxiv.org/abs/1601.05786} {arXiv:1601.05786 [astro-ph.CO]}
  \BibitemShut {NoStop}%
\bibitem [{\citenamefont {{Handley}}\ \emph {et~al.}(2015)\citenamefont
  {{Handley}}, \citenamefont {{Hobson}},\ and\ \citenamefont
  {{Lasenby}}}]{Handley15}%
  \BibitemOpen
  \bibfield  {author} {\bibinfo {author} {\bibfnamefont {W.~J.}\ \bibnamefont
  {{Handley}}}, \bibinfo {author} {\bibfnamefont {M.~P.}\ \bibnamefont
  {{Hobson}}}, \ and\ \bibinfo {author} {\bibfnamefont {A.~N.}\ \bibnamefont
  {{Lasenby}}},\ }\href {\doibase 10.1093/mnras/stv1911} {\bibfield  {journal}
  {\bibinfo  {journal} {\mnras}\ }\textbf {\bibinfo {volume} {453}},\ \bibinfo
  {pages} {4384} (\bibinfo {year} {2015})},\ \Eprint
  {http://arxiv.org/abs/1506.00171} {arXiv:1506.00171 [astro-ph.IM]}
  \BibitemShut {NoStop}%
\bibitem [{\citenamefont {{Zuntz}}\ \emph {et~al.}(2015)\citenamefont
  {{Zuntz}}, \citenamefont {{Paterno}}, \citenamefont {{Jennings}},
  \citenamefont {{Rudd}}, \citenamefont {{Manzotti}}, \citenamefont
  {{Dodelson}}, \citenamefont {{Bridle}}, \citenamefont {{Sehrish}},\ and\
  \citenamefont {{Kowalkowski}}}]{zuntz15}%
  \BibitemOpen
  \bibfield  {author} {\bibinfo {author} {\bibfnamefont {J.}~\bibnamefont
  {{Zuntz}}}, \bibinfo {author} {\bibfnamefont {M.}~\bibnamefont {{Paterno}}},
  \bibinfo {author} {\bibfnamefont {E.}~\bibnamefont {{Jennings}}}, \bibinfo
  {author} {\bibfnamefont {D.}~\bibnamefont {{Rudd}}}, \bibinfo {author}
  {\bibfnamefont {A.}~\bibnamefont {{Manzotti}}}, \bibinfo {author}
  {\bibfnamefont {S.}~\bibnamefont {{Dodelson}}}, \bibinfo {author}
  {\bibfnamefont {S.}~\bibnamefont {{Bridle}}}, \bibinfo {author}
  {\bibfnamefont {S.}~\bibnamefont {{Sehrish}}}, \ and\ \bibinfo {author}
  {\bibfnamefont {J.}~\bibnamefont {{Kowalkowski}}},\ }\href {\doibase
  10.1016/j.ascom.2015.05.005} {\bibfield  {journal} {\bibinfo  {journal}
  {Astronomy and Computing}\ }\textbf {\bibinfo {volume} {12}},\ \bibinfo
  {pages} {45} (\bibinfo {year} {2015})},\ \Eprint
  {http://arxiv.org/abs/1409.3409} {arXiv:1409.3409 [astro-ph.CO]} \BibitemShut
  {NoStop}%
\bibitem [{\citenamefont {Howlett}\ \emph {et~al.}(2012)\citenamefont
  {Howlett}, \citenamefont {Lewis}, \citenamefont {Hall},\ and\ \citenamefont
  {Challinor}}]{Howlett2012}%
  \BibitemOpen
  \bibfield  {author} {\bibinfo {author} {\bibfnamefont {C.}~\bibnamefont
  {Howlett}}, \bibinfo {author} {\bibfnamefont {A.}~\bibnamefont {Lewis}},
  \bibinfo {author} {\bibfnamefont {A.}~\bibnamefont {Hall}}, \ and\ \bibinfo
  {author} {\bibfnamefont {A.}~\bibnamefont {Challinor}},\ }\href {\doibase
  10.1088/1475-7516/2012/04/027} {\bibfield  {journal} {\bibinfo  {journal}
  {Journal of Cosmology and Astroparticle Physics}\ }\textbf {\bibinfo {volume}
  {2012}},\ \bibinfo {pages} {027–027} (\bibinfo {year} {2012})}\BibitemShut
  {NoStop}%
\bibitem [{\citenamefont {Lewis}\ \emph {et~al.}(2000)\citenamefont {Lewis},
  \citenamefont {Challinor},\ and\ \citenamefont {Lasenby}}]{CAMB}%
  \BibitemOpen
  \bibfield  {author} {\bibinfo {author} {\bibfnamefont {A.}~\bibnamefont
  {Lewis}}, \bibinfo {author} {\bibfnamefont {A.}~\bibnamefont {Challinor}}, \
  and\ \bibinfo {author} {\bibfnamefont {A.}~\bibnamefont {Lasenby}},\ }\href
  {\doibase 10.1086/309179} {\bibfield  {journal} {\bibinfo  {journal} {\apj}\
  }\textbf {\bibinfo {volume} {538}},\ \bibinfo {pages} {473} (\bibinfo {year}
  {2000})},\ \Eprint {http://arxiv.org/abs/astro-ph/9911177}
  {arXiv:astro-ph/9911177 [astro-ph]} \BibitemShut {NoStop}%
\bibitem [{\citenamefont {{Lemos}}\ \emph {et~al.}(2020)\citenamefont
  {{Lemos}}, \citenamefont {{Raveri}}, \citenamefont {{Campos}}, \citenamefont
  {{Park}}, \citenamefont {{Chang}}, \citenamefont {{Weaverdyck}},
  \citenamefont {{Huterer}},\ and\ \citenamefont {{Liddle}}}]{Lemos2020}%
  \BibitemOpen
  \bibfield  {author} {\bibinfo {author} {\bibfnamefont {P.}~\bibnamefont
  {{Lemos}}}, \bibinfo {author} {\bibfnamefont {M.}~\bibnamefont {{Raveri}}},
  \bibinfo {author} {\bibfnamefont {A.}~\bibnamefont {{Campos}}}, \bibinfo
  {author} {\bibfnamefont {Y.}~\bibnamefont {{Park}}}, \bibinfo {author}
  {\bibfnamefont {C.}~\bibnamefont {{Chang}}}, \bibinfo {author} {\bibfnamefont
  {N.}~\bibnamefont {{Weaverdyck}}}, \bibinfo {author} {\bibfnamefont
  {D.}~\bibnamefont {{Huterer}}}, \ and\ \bibinfo {author} {\bibfnamefont
  {A.~R.}\ \bibnamefont {{Liddle}}},\ }\href@noop {} {\bibfield  {journal}
  {\bibinfo  {journal} {arXiv e-prints}\ ,\ \bibinfo {eid} {arXiv:2012.09554}}
  (\bibinfo {year} {2020})},\ \Eprint {http://arxiv.org/abs/2012.09554}
  {arXiv:2012.09554 [astro-ph.CO]} \BibitemShut {NoStop}%
\bibitem [{\citenamefont {{Chang}}\ \emph {et~al.}(2019)\citenamefont
  {{Chang}}, \citenamefont {{Wang}},\ and\ \citenamefont {{LSST Dark Energy
  Science Collaboration}}}]{chang2018}%
  \BibitemOpen
  \bibfield  {author} {\bibinfo {author} {\bibfnamefont {C.}~\bibnamefont
  {{Chang}}}, \bibinfo {author} {\bibfnamefont {M.}~\bibnamefont {{Wang}}}, \
  and\ \bibinfo {author} {\bibnamefont {{LSST Dark Energy Science
  Collaboration}}},\ }\href {\doibase 10.1093/mnras/sty2902} {\bibfield
  {journal} {\bibinfo  {journal} {\mnras}\ }\textbf {\bibinfo {volume} {482}},\
  \bibinfo {pages} {3696} (\bibinfo {year} {2019})},\ \Eprint
  {http://arxiv.org/abs/1808.07335} {arXiv:1808.07335 [astro-ph.CO]}
  \BibitemShut {NoStop}%
\bibitem [{\citenamefont {{Huterer}}\ and\ \citenamefont
  {{Turner}}(2001)}]{huterer01}%
  \BibitemOpen
  \bibfield  {author} {\bibinfo {author} {\bibfnamefont {D.}~\bibnamefont
  {{Huterer}}}\ and\ \bibinfo {author} {\bibfnamefont {M.~S.}\ \bibnamefont
  {{Turner}}},\ }\href {\doibase 10.1103/PhysRevD.64.123527} {\bibfield
  {journal} {\bibinfo  {journal} {\prd}\ }\textbf {\bibinfo {volume} {64}},\
  \bibinfo {pages} {123527} (\bibinfo {year} {2001})},\ \Eprint
  {http://arxiv.org/abs/astro-ph/0012510} {arXiv:astro-ph/0012510 [astro-ph]}
  \BibitemShut {NoStop}%
\bibitem [{\citenamefont {Wang}(2008)}]{wang08}%
  \BibitemOpen
  \bibfield  {author} {\bibinfo {author} {\bibfnamefont {Y.}~\bibnamefont
  {Wang}},\ }\href {\doibase 10.1103/physrevd.77.123525} {\bibfield  {journal}
  {\bibinfo  {journal} {Physical Review D}\ }\textbf {\bibinfo {volume} {77}}
  (\bibinfo {year} {2008}),\ 10.1103/physrevd.77.123525}\BibitemShut {NoStop}%
\bibitem [{\citenamefont {{Albrecht}}\ \emph {et~al.}(2006)\citenamefont
  {{Albrecht}}, \citenamefont {{Bernstein}}, \citenamefont {{Cahn}},
  \citenamefont {{Freedman}}, \citenamefont {{Hewitt}}, \citenamefont {{Hu}},
  \citenamefont {{Huth}}, \citenamefont {{Kamionkowski}}, \citenamefont
  {{Kolb}}, \citenamefont {{Knox}}, \citenamefont {{Mather}}, \citenamefont
  {{Staggs}},\ and\ \citenamefont {{Suntzeff}}}]{fom}%
  \BibitemOpen
  \bibfield  {author} {\bibinfo {author} {\bibfnamefont {A.}~\bibnamefont
  {{Albrecht}}}, \bibinfo {author} {\bibfnamefont {G.}~\bibnamefont
  {{Bernstein}}}, \bibinfo {author} {\bibfnamefont {R.}~\bibnamefont {{Cahn}}},
  \bibinfo {author} {\bibfnamefont {W.~L.}\ \bibnamefont {{Freedman}}},
  \bibinfo {author} {\bibfnamefont {J.}~\bibnamefont {{Hewitt}}}, \bibinfo
  {author} {\bibfnamefont {W.}~\bibnamefont {{Hu}}}, \bibinfo {author}
  {\bibfnamefont {J.}~\bibnamefont {{Huth}}}, \bibinfo {author} {\bibfnamefont
  {M.}~\bibnamefont {{Kamionkowski}}}, \bibinfo {author} {\bibfnamefont
  {E.~W.}\ \bibnamefont {{Kolb}}}, \bibinfo {author} {\bibfnamefont
  {L.}~\bibnamefont {{Knox}}}, \bibinfo {author} {\bibfnamefont {J.~C.}\
  \bibnamefont {{Mather}}}, \bibinfo {author} {\bibfnamefont {S.}~\bibnamefont
  {{Staggs}}}, \ and\ \bibinfo {author} {\bibfnamefont {N.~B.}\ \bibnamefont
  {{Suntzeff}}},\ }\href@noop {} {\bibfield  {journal} {\bibinfo  {journal}
  {arXiv e-prints}\ ,\ \bibinfo {eid} {astro-ph/0609591}} (\bibinfo {year}
  {2006})},\ \Eprint {http://arxiv.org/abs/astro-ph/0609591}
  {arXiv:astro-ph/0609591 [astro-ph]} \BibitemShut {NoStop}%
\bibitem [{\citenamefont {{Raveri}}\ and\ \citenamefont
  {{Hu}}(2019)}]{raverihu}%
  \BibitemOpen
  \bibfield  {author} {\bibinfo {author} {\bibfnamefont {M.}~\bibnamefont
  {{Raveri}}}\ and\ \bibinfo {author} {\bibfnamefont {W.}~\bibnamefont
  {{Hu}}},\ }\href {\doibase 10.1103/PhysRevD.99.043506} {\bibfield  {journal}
  {\bibinfo  {journal} {\prd}\ }\textbf {\bibinfo {volume} {99}},\ \bibinfo
  {eid} {043506} (\bibinfo {year} {2019})},\ \Eprint
  {http://arxiv.org/abs/1806.04649} {arXiv:1806.04649 [astro-ph.CO]}
  \BibitemShut {NoStop}%
\bibitem [{\citenamefont {{Handley}}\ and\ \citenamefont
  {{Lemos}}(2019)}]{Handley:2019wlz}%
  \BibitemOpen
  \bibfield  {author} {\bibinfo {author} {\bibfnamefont {W.}~\bibnamefont
  {{Handley}}}\ and\ \bibinfo {author} {\bibfnamefont {P.}~\bibnamefont
  {{Lemos}}},\ }\href@noop {} {\bibfield  {journal} {\bibinfo  {journal} {arXiv
  e-prints}\ ,\ \bibinfo {eid} {arXiv:1903.06682}} (\bibinfo {year} {2019})},\
  \Eprint {http://arxiv.org/abs/1903.06682} {arXiv:1903.06682 [astro-ph.CO]}
  \BibitemShut {NoStop}%
\bibitem [{\citenamefont {{Joachimi}}\ and\ \citenamefont
  {{Bridle}}(2010)}]{joachimi10}%
  \BibitemOpen
  \bibfield  {author} {\bibinfo {author} {\bibfnamefont {B.}~\bibnamefont
  {{Joachimi}}}\ and\ \bibinfo {author} {\bibfnamefont {S.~L.}\ \bibnamefont
  {{Bridle}}},\ }\href {\doibase 10.1051/0004-6361/200913657} {\bibfield
  {journal} {\bibinfo  {journal} {\aap}\ }\textbf {\bibinfo {volume} {523}},\
  \bibinfo {eid} {A1} (\bibinfo {year} {2010})},\ \Eprint
  {http://arxiv.org/abs/0911.2454} {arXiv:0911.2454 [astro-ph.CO]} \BibitemShut
  {NoStop}%
\bibitem [{\citenamefont {{Samuroff}}\ \emph {et~al.}(2019)\citenamefont
  {{Samuroff}}, \citenamefont {{Blazek}}, \citenamefont {{Troxel}},
  \citenamefont {{MacCrann}}, \citenamefont {{Krause}},\ and\ \citenamefont
  {{DES Collaboration}}}]{samuroff18}%
  \BibitemOpen
  \bibfield  {author} {\bibinfo {author} {\bibfnamefont {S.}~\bibnamefont
  {{Samuroff}}}, \bibinfo {author} {\bibfnamefont {J.}~\bibnamefont
  {{Blazek}}}, \bibinfo {author} {\bibfnamefont {M.~A.}\ \bibnamefont
  {{Troxel}}}, \bibinfo {author} {\bibfnamefont {N.}~\bibnamefont
  {{MacCrann}}}, \bibinfo {author} {\bibfnamefont {E.}~\bibnamefont
  {{Krause}}}, \ and\ \bibinfo {author} {\bibnamefont {{DES Collaboration}}},\
  }\href {\doibase 10.1093/mnras/stz2197} {\bibfield  {journal} {\bibinfo
  {journal} {\mnras}\ }\textbf {\bibinfo {volume} {489}},\ \bibinfo {pages}
  {5453} (\bibinfo {year} {2019})},\ \Eprint {http://arxiv.org/abs/1811.06989}
  {arXiv:1811.06989 [astro-ph.CO]} \BibitemShut {NoStop}%
\bibitem [{\citenamefont {{van den Busch}}\ \emph {et~al.}(2020)\citenamefont
  {{van den Busch}}, \citenamefont {{Hildebrandt}}, \citenamefont {{Wright}},
  \citenamefont {{Morrison}}, \citenamefont {{Blake}}, \citenamefont
  {{Joachimi}}, \citenamefont {{Erben}}, \citenamefont {{Heymans}},
  \citenamefont {{Kuijken}},\ and\ \citenamefont {{Taylor}}}]{vandenbusch2020}%
  \BibitemOpen
  \bibfield  {author} {\bibinfo {author} {\bibfnamefont {J.~L.}\ \bibnamefont
  {{van den Busch}}}, \bibinfo {author} {\bibfnamefont {H.}~\bibnamefont
  {{Hildebrandt}}}, \bibinfo {author} {\bibfnamefont {A.~H.}\ \bibnamefont
  {{Wright}}}, \bibinfo {author} {\bibfnamefont {C.~B.}\ \bibnamefont
  {{Morrison}}}, \bibinfo {author} {\bibfnamefont {C.}~\bibnamefont {{Blake}}},
  \bibinfo {author} {\bibfnamefont {B.}~\bibnamefont {{Joachimi}}}, \bibinfo
  {author} {\bibfnamefont {T.}~\bibnamefont {{Erben}}}, \bibinfo {author}
  {\bibfnamefont {C.}~\bibnamefont {{Heymans}}}, \bibinfo {author}
  {\bibfnamefont {K.}~\bibnamefont {{Kuijken}}}, \ and\ \bibinfo {author}
  {\bibfnamefont {E.~N.}\ \bibnamefont {{Taylor}}},\ }\href {\doibase
  10.1051/0004-6361/202038835} {\bibfield  {journal} {\bibinfo  {journal}
  {\aap}\ }\textbf {\bibinfo {volume} {642}},\ \bibinfo {eid} {A200} (\bibinfo
  {year} {2020})},\ \Eprint {http://arxiv.org/abs/2007.01846} {arXiv:2007.01846
  [astro-ph.CO]} \BibitemShut {NoStop}%
\bibitem [{\citenamefont {{Asgari}}\ \emph {et~al.}(2020)\citenamefont
  {{Asgari}}, \citenamefont {{Tr{\"o}ster}},\ and\ \citenamefont
  {{Heymans}}}]{asgari_baryons}%
  \BibitemOpen
  \bibfield  {author} {\bibinfo {author} {\bibfnamefont {M.}~\bibnamefont
  {{Asgari}}}, \bibinfo {author} {\bibfnamefont {T.}~\bibnamefont
  {{Tr{\"o}ster}}}, \ and\ \bibinfo {author} {\bibfnamefont {C.}~\bibnamefont
  {{Heymans}}},\ }\href {\doibase 10.1051/0004-6361/201936512} {\bibfield
  {journal} {\bibinfo  {journal} {\aap}\ }\textbf {\bibinfo {volume} {634}},\
  \bibinfo {eid} {A127} (\bibinfo {year} {2020})},\ \Eprint
  {http://arxiv.org/abs/1910.05336} {arXiv:1910.05336 [astro-ph.CO]}
  \BibitemShut {NoStop}%
\bibitem [{\citenamefont {{MacCrann}}\ and\ \citenamefont {{DES
  Collaboration}}(2017)}]{Maccrann2017}%
  \BibitemOpen
  \bibfield  {author} {\bibinfo {author} {\bibfnamefont {N.}~\bibnamefont
  {{MacCrann}}}\ and\ \bibinfo {author} {\bibnamefont {{DES Collaboration}}},\
  }\href {\doibase 10.1093/mnras/stw2849} {\bibfield  {journal} {\bibinfo
  {journal} {\mnras}\ }\textbf {\bibinfo {volume} {465}},\ \bibinfo {pages}
  {2567} (\bibinfo {year} {2017})},\ \Eprint {http://arxiv.org/abs/1608.01838}
  {arXiv:1608.01838 [astro-ph.CO]} \BibitemShut {NoStop}%
\bibitem [{\citenamefont {{Astropy Collaboration}}\ \emph
  {et~al.}(2013)\citenamefont {{Astropy Collaboration}}, \citenamefont
  {{Robitaille}}, \citenamefont {{Tollerud}}, \citenamefont {{Greenfield}},
  \citenamefont {{Droettboom}}, \citenamefont {{Bray}}, \citenamefont
  {{Aldcroft}}, \citenamefont {{Davis}}, \citenamefont {{Ginsburg}},
  \citenamefont {{Price-Whelan}}, \citenamefont {{Kerzendorf}}, \citenamefont
  {{Conley}}, \citenamefont {{Crighton}}, \citenamefont {{Barbary}},
  \citenamefont {{Muna}}, \citenamefont {{Ferguson}}, \citenamefont
  {{Grollier}}, \citenamefont {{Parikh}}, \citenamefont {{Nair}}, \citenamefont
  {{Unther}}, \citenamefont {{Deil}}, \citenamefont {{Woillez}}, \citenamefont
  {{Conseil}}, \citenamefont {{Kramer}}, \citenamefont {{Turner}},
  \citenamefont {{Singer}}, \citenamefont {{Fox}}, \citenamefont {{Weaver}},
  \citenamefont {{Zabalza}}, \citenamefont {{Edwards}}, \citenamefont {{Azalee
  Bostroem}}, \citenamefont {{Burke}}, \citenamefont {{Casey}}, \citenamefont
  {{Crawford}}, \citenamefont {{Dencheva}}, \citenamefont {{Ely}},
  \citenamefont {{Jenness}}, \citenamefont {{Labrie}}, \citenamefont {{Lim}},
  \citenamefont {{Pierfederici}}, \citenamefont {{Pontzen}}, \citenamefont
  {{Ptak}}, \citenamefont {{Refsdal}}, \citenamefont {{Servillat}},\ and\
  \citenamefont {{Streicher}}}]{astropy:2013}%
  \BibitemOpen
  \bibfield  {author} {\bibinfo {author} {\bibnamefont {{Astropy
  Collaboration}}}, \bibinfo {author} {\bibfnamefont {T.~P.}\ \bibnamefont
  {{Robitaille}}}, \bibinfo {author} {\bibfnamefont {E.~J.}\ \bibnamefont
  {{Tollerud}}}, \bibinfo {author} {\bibfnamefont {P.}~\bibnamefont
  {{Greenfield}}}, \bibinfo {author} {\bibfnamefont {M.}~\bibnamefont
  {{Droettboom}}}, \bibinfo {author} {\bibfnamefont {E.}~\bibnamefont
  {{Bray}}}, \bibinfo {author} {\bibfnamefont {T.}~\bibnamefont {{Aldcroft}}},
  \bibinfo {author} {\bibfnamefont {M.}~\bibnamefont {{Davis}}}, \bibinfo
  {author} {\bibfnamefont {A.}~\bibnamefont {{Ginsburg}}}, \bibinfo {author}
  {\bibfnamefont {A.~M.}\ \bibnamefont {{Price-Whelan}}}, \bibinfo {author}
  {\bibfnamefont {W.~E.}\ \bibnamefont {{Kerzendorf}}}, \bibinfo {author}
  {\bibfnamefont {A.}~\bibnamefont {{Conley}}}, \bibinfo {author}
  {\bibfnamefont {N.}~\bibnamefont {{Crighton}}}, \bibinfo {author}
  {\bibfnamefont {K.}~\bibnamefont {{Barbary}}}, \bibinfo {author}
  {\bibfnamefont {D.}~\bibnamefont {{Muna}}}, \bibinfo {author} {\bibfnamefont
  {H.}~\bibnamefont {{Ferguson}}}, \bibinfo {author} {\bibfnamefont
  {F.}~\bibnamefont {{Grollier}}}, \bibinfo {author} {\bibfnamefont {M.~M.}\
  \bibnamefont {{Parikh}}}, \bibinfo {author} {\bibfnamefont {P.~H.}\
  \bibnamefont {{Nair}}}, \bibinfo {author} {\bibfnamefont {H.~M.}\
  \bibnamefont {{Unther}}}, \bibinfo {author} {\bibfnamefont {C.}~\bibnamefont
  {{Deil}}}, \bibinfo {author} {\bibfnamefont {J.}~\bibnamefont {{Woillez}}},
  \bibinfo {author} {\bibfnamefont {S.}~\bibnamefont {{Conseil}}}, \bibinfo
  {author} {\bibfnamefont {R.}~\bibnamefont {{Kramer}}}, \bibinfo {author}
  {\bibfnamefont {J.~E.~H.}\ \bibnamefont {{Turner}}}, \bibinfo {author}
  {\bibfnamefont {L.}~\bibnamefont {{Singer}}}, \bibinfo {author}
  {\bibfnamefont {R.}~\bibnamefont {{Fox}}}, \bibinfo {author} {\bibfnamefont
  {B.~A.}\ \bibnamefont {{Weaver}}}, \bibinfo {author} {\bibfnamefont
  {V.}~\bibnamefont {{Zabalza}}}, \bibinfo {author} {\bibfnamefont {Z.~I.}\
  \bibnamefont {{Edwards}}}, \bibinfo {author} {\bibfnamefont {K.}~\bibnamefont
  {{Azalee Bostroem}}}, \bibinfo {author} {\bibfnamefont {D.~J.}\ \bibnamefont
  {{Burke}}}, \bibinfo {author} {\bibfnamefont {A.~R.}\ \bibnamefont
  {{Casey}}}, \bibinfo {author} {\bibfnamefont {S.~M.}\ \bibnamefont
  {{Crawford}}}, \bibinfo {author} {\bibfnamefont {N.}~\bibnamefont
  {{Dencheva}}}, \bibinfo {author} {\bibfnamefont {J.}~\bibnamefont {{Ely}}},
  \bibinfo {author} {\bibfnamefont {T.}~\bibnamefont {{Jenness}}}, \bibinfo
  {author} {\bibfnamefont {K.}~\bibnamefont {{Labrie}}}, \bibinfo {author}
  {\bibfnamefont {P.~L.}\ \bibnamefont {{Lim}}}, \bibinfo {author}
  {\bibfnamefont {F.}~\bibnamefont {{Pierfederici}}}, \bibinfo {author}
  {\bibfnamefont {A.}~\bibnamefont {{Pontzen}}}, \bibinfo {author}
  {\bibfnamefont {A.}~\bibnamefont {{Ptak}}}, \bibinfo {author} {\bibfnamefont
  {B.}~\bibnamefont {{Refsdal}}}, \bibinfo {author} {\bibfnamefont
  {M.}~\bibnamefont {{Servillat}}}, \ and\ \bibinfo {author} {\bibfnamefont
  {O.}~\bibnamefont {{Streicher}}},\ }\href {\doibase
  10.1051/0004-6361/201322068} {\bibfield  {journal} {\bibinfo  {journal}
  {\aap}\ }\textbf {\bibinfo {volume} {558}},\ \bibinfo {eid} {A33} (\bibinfo
  {year} {2013})},\ \Eprint {http://arxiv.org/abs/1307.6212} {arXiv:1307.6212
  [astro-ph.IM]} \BibitemShut {NoStop}%
\bibitem [{\citenamefont {{Astropy Collaboration}}\ and\ \citenamefont
  {{Astropy Contributors}}(2018)}]{astropy:2018}%
  \BibitemOpen
  \bibfield  {author} {\bibinfo {author} {\bibnamefont {{Astropy
  Collaboration}}}\ and\ \bibinfo {author} {\bibnamefont {{Astropy
  Contributors}}},\ }\href {\doibase 10.3847/1538-3881/aabc4f} {\bibfield
  {journal} {\bibinfo  {journal} {\aj}\ }\textbf {\bibinfo {volume} {156}},\
  \bibinfo {eid} {123} (\bibinfo {year} {2018})},\ \Eprint
  {http://arxiv.org/abs/1801.02634} {arXiv:1801.02634 [astro-ph.IM]}
  \BibitemShut {NoStop}%
\bibitem [{\citenamefont {Lewis}(2019)}]{getdist}%
  \BibitemOpen
  \bibfield  {author} {\bibinfo {author} {\bibfnamefont {A.}~\bibnamefont
  {Lewis}},\ }\href {https://getdist.readthedocs.io} {\  (\bibinfo {year}
  {2019})},\ \Eprint {http://arxiv.org/abs/1910.13970} {arXiv:1910.13970
  [astro-ph.IM]} \BibitemShut {NoStop}%
\bibitem [{\citenamefont {{Hinton}}(2016)}]{Hinton16}%
  \BibitemOpen
  \bibfield  {author} {\bibinfo {author} {\bibfnamefont {S.~R.}\ \bibnamefont
  {{Hinton}}},\ }\href {\doibase 10.21105/joss.00045} {\bibfield  {journal}
  {\bibinfo  {journal} {The Journal of Open Source Software}\ }\textbf
  {\bibinfo {volume} {1}},\ \bibinfo {eid} {00045} (\bibinfo {year}
  {2016})}\BibitemShut {NoStop}%
\bibitem [{\citenamefont {Hunter}(2007)}]{matplotlib}%
  \BibitemOpen
  \bibfield  {author} {\bibinfo {author} {\bibfnamefont {J.~D.}\ \bibnamefont
  {Hunter}},\ }\href {\doibase 10.1109/MCSE.2007.55} {\bibfield  {journal}
  {\bibinfo  {journal} {Computing in Science \& Engineering}\ }\textbf
  {\bibinfo {volume} {9}},\ \bibinfo {pages} {90} (\bibinfo {year}
  {2007})}\BibitemShut {NoStop}%
\bibitem [{\citenamefont {{Paulin-Henriksson}}\ \emph
  {et~al.}(2008)\citenamefont {{Paulin-Henriksson}}, \citenamefont {{Amara}},
  \citenamefont {{Voigt}}, \citenamefont {{Refregier}},\ and\ \citenamefont
  {{Bridle}}}]{Paulin-Henriksson2008}%
  \BibitemOpen
  \bibfield  {author} {\bibinfo {author} {\bibfnamefont {S.}~\bibnamefont
  {{Paulin-Henriksson}}}, \bibinfo {author} {\bibfnamefont {A.}~\bibnamefont
  {{Amara}}}, \bibinfo {author} {\bibfnamefont {L.}~\bibnamefont {{Voigt}}},
  \bibinfo {author} {\bibfnamefont {A.}~\bibnamefont {{Refregier}}}, \ and\
  \bibinfo {author} {\bibfnamefont {S.~L.}\ \bibnamefont {{Bridle}}},\ }\href
  {\doibase 10.1051/0004-6361:20079150} {\bibfield  {journal} {\bibinfo
  {journal} {\aap}\ }\textbf {\bibinfo {volume} {484}},\ \bibinfo {pages} {67}
  (\bibinfo {year} {2008})},\ \Eprint {http://arxiv.org/abs/0711.4886}
  {arXiv:0711.4886 [astro-ph]} \BibitemShut {NoStop}%
\bibitem [{\citenamefont {{Rowe}}(2010)}]{rowe2010}%
  \BibitemOpen
  \bibfield  {author} {\bibinfo {author} {\bibfnamefont {B.}~\bibnamefont
  {{Rowe}}},\ }\href {\doibase 10.1111/j.1365-2966.2010.16277.x} {\bibfield
  {journal} {\bibinfo  {journal} {\mnras}\ }\textbf {\bibinfo {volume} {404}},\
  \bibinfo {pages} {350} (\bibinfo {year} {2010})},\ \Eprint
  {http://arxiv.org/abs/0904.3056} {arXiv:0904.3056 [astro-ph.CO]} \BibitemShut
  {NoStop}%
\bibitem [{\citenamefont {{Hikage}}\ \emph {et~al.}(2011)\citenamefont
  {{Hikage}}, \citenamefont {{Takada}}, \citenamefont {{Hamana}},\ and\
  \citenamefont {{Spergel}}}]{hikage2011}%
  \BibitemOpen
  \bibfield  {author} {\bibinfo {author} {\bibfnamefont {C.}~\bibnamefont
  {{Hikage}}}, \bibinfo {author} {\bibfnamefont {M.}~\bibnamefont {{Takada}}},
  \bibinfo {author} {\bibfnamefont {T.}~\bibnamefont {{Hamana}}}, \ and\
  \bibinfo {author} {\bibfnamefont {D.}~\bibnamefont {{Spergel}}},\ }\href
  {\doibase 10.1111/j.1365-2966.2010.17886.x} {\bibfield  {journal} {\bibinfo
  {journal} {\mnras}\ }\textbf {\bibinfo {volume} {412}},\ \bibinfo {pages}
  {65} (\bibinfo {year} {2011})},\ \Eprint {http://arxiv.org/abs/1004.3542}
  {arXiv:1004.3542 [astro-ph.CO]} \BibitemShut {NoStop}%
\bibitem [{\citenamefont {{Asgari}}\ and\ \citenamefont
  {{Heymans}}(2019)}]{asgari19}%
  \BibitemOpen
  \bibfield  {author} {\bibinfo {author} {\bibfnamefont {M.}~\bibnamefont
  {{Asgari}}}\ and\ \bibinfo {author} {\bibfnamefont {C.}~\bibnamefont
  {{Heymans}}},\ }\href {\doibase 10.1051/0004-6361/201834379} {\bibfield
  {journal} {\bibinfo  {journal} {\aap}\ }\textbf {\bibinfo {volume} {624}},\
  \bibinfo {eid} {A134} (\bibinfo {year} {2019})},\ \Eprint
  {http://arxiv.org/abs/1810.02353} {arXiv:1810.02353 [astro-ph.CO]}
  \BibitemShut {NoStop}%
\bibitem [{\citenamefont {{Asgari}}\ \emph {et~al.}(2019)\citenamefont
  {{Asgari}}, \citenamefont {{Heymans}}, \citenamefont {{Hildebrandt}},
  \citenamefont {{Miller}}, \citenamefont {{Schneider}}, \citenamefont
  {{Amon}}, \citenamefont {{Choi}}, \citenamefont {{Erben}}, \citenamefont
  {{Georgiou}}, \citenamefont {{Harnois-Deraps}},\ and\ \citenamefont
  {{Kuijken}}}]{2019A&A...624A.134A}%
  \BibitemOpen
  \bibfield  {author} {\bibinfo {author} {\bibfnamefont {M.}~\bibnamefont
  {{Asgari}}}, \bibinfo {author} {\bibfnamefont {C.}~\bibnamefont {{Heymans}}},
  \bibinfo {author} {\bibfnamefont {H.}~\bibnamefont {{Hildebrandt}}}, \bibinfo
  {author} {\bibfnamefont {L.}~\bibnamefont {{Miller}}}, \bibinfo {author}
  {\bibfnamefont {P.}~\bibnamefont {{Schneider}}}, \bibinfo {author}
  {\bibfnamefont {A.}~\bibnamefont {{Amon}}}, \bibinfo {author} {\bibfnamefont
  {A.}~\bibnamefont {{Choi}}}, \bibinfo {author} {\bibfnamefont
  {T.}~\bibnamefont {{Erben}}}, \bibinfo {author} {\bibfnamefont
  {C.}~\bibnamefont {{Georgiou}}}, \bibinfo {author} {\bibfnamefont
  {J.}~\bibnamefont {{Harnois-Deraps}}}, \ and\ \bibinfo {author}
  {\bibfnamefont {K.}~\bibnamefont {{Kuijken}}},\ }\href {\doibase
  10.1051/0004-6361/201834379} {\bibfield  {journal} {\bibinfo  {journal}
  {\aap}\ }\textbf {\bibinfo {volume} {624}},\ \bibinfo {eid} {A134} (\bibinfo
  {year} {2019})},\ \Eprint {http://arxiv.org/abs/1810.02353} {arXiv:1810.02353
  [astro-ph.CO]} \BibitemShut {NoStop}%
\bibitem [{\citenamefont {{G{\'o}rski}}\ \emph
  {et~al.}(2005{\natexlab{b}})\citenamefont {{G{\'o}rski}}, \citenamefont
  {{Hivon}}, \citenamefont {{Banday}}, \citenamefont {{Wand elt}},
  \citenamefont {{Hansen}}, \citenamefont {{Reinecke}},\ and\ \citenamefont
  {{Bartelmann}}}]{Gorski05}%
  \BibitemOpen
  \bibfield  {author} {\bibinfo {author} {\bibfnamefont {K.~M.}\ \bibnamefont
  {{G{\'o}rski}}}, \bibinfo {author} {\bibfnamefont {E.}~\bibnamefont
  {{Hivon}}}, \bibinfo {author} {\bibfnamefont {A.~J.}\ \bibnamefont
  {{Banday}}}, \bibinfo {author} {\bibfnamefont {B.~D.}\ \bibnamefont {{Wand
  elt}}}, \bibinfo {author} {\bibfnamefont {F.~K.}\ \bibnamefont {{Hansen}}},
  \bibinfo {author} {\bibfnamefont {M.}~\bibnamefont {{Reinecke}}}, \ and\
  \bibinfo {author} {\bibfnamefont {M.}~\bibnamefont {{Bartelmann}}},\ }\href
  {\doibase 10.1086/427976} {\bibfield  {journal} {\bibinfo  {journal} {\apj}\
  }\textbf {\bibinfo {volume} {622}},\ \bibinfo {pages} {759} (\bibinfo {year}
  {2005}{\natexlab{b}})},\ \Eprint {http://arxiv.org/abs/astro-ph/0409513}
  {arXiv:astro-ph/0409513 [astro-ph]} \BibitemShut {NoStop}%
\bibitem [{\citenamefont {{Alonso}}\ \emph {et~al.}(2019)\citenamefont
  {{Alonso}}, \citenamefont {{Sanchez}}, \citenamefont {{Slosar}},\ and\
  \citenamefont {{LSST Dark Energy Science Collaboration}}}]{Alonso19}%
  \BibitemOpen
  \bibfield  {author} {\bibinfo {author} {\bibfnamefont {D.}~\bibnamefont
  {{Alonso}}}, \bibinfo {author} {\bibfnamefont {J.}~\bibnamefont {{Sanchez}}},
  \bibinfo {author} {\bibfnamefont {A.}~\bibnamefont {{Slosar}}}, \ and\
  \bibinfo {author} {\bibnamefont {{LSST Dark Energy Science Collaboration}}},\
  }\href {\doibase 10.1093/mnras/stz093} {\bibfield  {journal} {\bibinfo
  {journal} {\mnras}\ }\textbf {\bibinfo {volume} {484}},\ \bibinfo {pages}
  {4127} (\bibinfo {year} {2019})},\ \Eprint {http://arxiv.org/abs/1809.09603}
  {arXiv:1809.09603 [astro-ph.CO]} \BibitemShut {NoStop}%
\bibitem [{\citenamefont {{Nicola}}\ \emph {et~al.}(2020)\citenamefont
  {{Nicola}}, \citenamefont {{Garc{\'\i}a-Garc{\'\i}a}}, \citenamefont
  {{Alonso}}, \citenamefont {{Dunkley}}, \citenamefont {{Ferreira}},
  \citenamefont {{Slosar}},\ and\ \citenamefont
  {{Spergel}}}]{2020arXiv201009717N}%
  \BibitemOpen
  \bibfield  {author} {\bibinfo {author} {\bibfnamefont {A.}~\bibnamefont
  {{Nicola}}}, \bibinfo {author} {\bibfnamefont {C.}~\bibnamefont
  {{Garc{\'\i}a-Garc{\'\i}a}}}, \bibinfo {author} {\bibfnamefont
  {D.}~\bibnamefont {{Alonso}}}, \bibinfo {author} {\bibfnamefont
  {J.}~\bibnamefont {{Dunkley}}}, \bibinfo {author} {\bibfnamefont {P.~G.}\
  \bibnamefont {{Ferreira}}}, \bibinfo {author} {\bibfnamefont
  {A.}~\bibnamefont {{Slosar}}}, \ and\ \bibinfo {author} {\bibfnamefont
  {D.~N.}\ \bibnamefont {{Spergel}}},\ }\href@noop {} {\bibfield  {journal}
  {\bibinfo  {journal} {arXiv e-prints}\ ,\ \bibinfo {eid} {arXiv:2010.09717}}
  (\bibinfo {year} {2020})},\ \Eprint {http://arxiv.org/abs/2010.09717}
  {arXiv:2010.09717 [astro-ph.CO]} \BibitemShut {NoStop}%
\bibitem [{\citenamefont {{Doux}}\ \emph {et~al.}(2021)\citenamefont {{Doux}},
  \citenamefont {{Chang}},\ and\ \citenamefont {{(DES
  Collaboration)}}}]{2020arXiv201106469D}%
  \BibitemOpen
  \bibfield  {author} {\bibinfo {author} {\bibfnamefont {C.}~\bibnamefont
  {{Doux}}}, \bibinfo {author} {\bibfnamefont {C.}~\bibnamefont {{Chang}}}, \
  and\ \bibinfo {author} {\bibnamefont {{(DES Collaboration)}}},\ }\href
  {\doibase 10.1093/mnras/stab661} {\bibfield  {journal} {\bibinfo  {journal}
  {\mnras}\ }\textbf {\bibinfo {volume} {503}},\ \bibinfo {pages} {3796}
  (\bibinfo {year} {2021})},\ \Eprint {http://arxiv.org/abs/2011.06469}
  {arXiv:2011.06469 [astro-ph.CO]} \BibitemShut {NoStop}%
\bibitem [{\citenamefont {{Joachimi}}\ and\ \citenamefont
  {{Lin}}(2021)}]{joachimi20}%
  \BibitemOpen
  \bibfield  {author} {\bibinfo {author} {\bibfnamefont {B.}~\bibnamefont
  {{Joachimi}}}\ and\ \bibinfo {author} {\bibnamefont {{Lin}}},\ }\href
  {\doibase 10.1051/0004-6361/202038831} {\bibfield  {journal} {\bibinfo
  {journal} {\aap}\ }\textbf {\bibinfo {volume} {646}},\ \bibinfo {eid} {A129}
  (\bibinfo {year} {2021})},\ \Eprint {http://arxiv.org/abs/2007.01844}
  {arXiv:2007.01844 [astro-ph.CO]} \BibitemShut {NoStop}%
\end{thebibliography}%


%

\appendix

\section{Shear systematics}\label{app:add}

\begin{figure*}
    \centering
    \includegraphics[width=\textwidth]{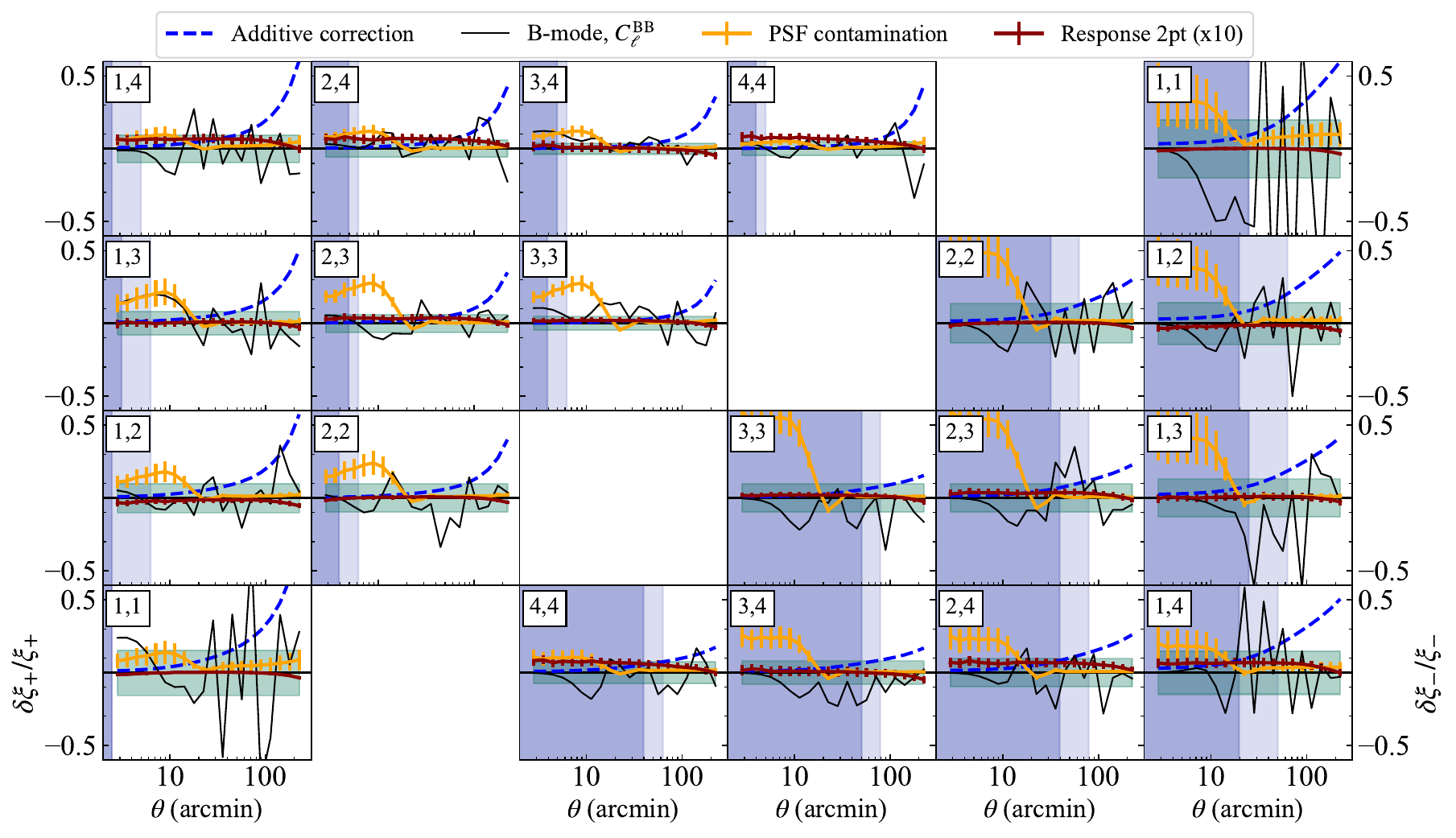}
    \caption{Impact of shear systematics on the $\xi_{\pm}(\theta)$ signal: The fractional impact, $\delta\xi_{\pm}/{\xi}$, of the effects of (i) B-modes (black) (ii) the approximation made in the response correction (red) (iii) the PSF contamination (yellow) and (iv) the additive $c$-correction applied to the shears (blue, dashed). To demonstrate how the fractional biases compare to the uncertainties on a model fit to the data, the green shaded band represents the uncertainty on the amplitude for each bin pair.}
    \label{fig:xitheorytomo}
\end{figure*} 

In addition to multiplicative biases, cosmic shear observables can also be affected by additive systematic errors, potentially arising from issues with the point-spread function modeling or the instrument. The DES Y3 catalog has undergone extensive null testing, as summarised in Section~\ref{sec:data} and presented in \citep{y3-shapecatalog}. Here, we demonstrate the robustness of the tomographic cosmic shear analysis to PSF contamination (see Appendix~\ref{App:PSF}), a spurious B-mode signal (detailed in Appendix~\ref{App:Bmodes}), and a residual additive correction (detailed in Appendix~\ref{App:additivecorr}).

Figure~\ref{fig:xitheorytomo} summarises and compares the predicted impact of these effects on the tomographic cosmic shear measurements. The panels show the fractional impact, $\delta\xi_{\pm}/{\xi}$, of each effect listed in the legend. For comparison, the green shaded region shows the amplitude of the uncertainty or the inverse signal to noise, defined for each bin pair as $\sigma_A = \left( \xi^TC^{-1}\xi \right)^{-0.5}$.
\
\subsection{PSF modeling}\label{App:PSF}

 The observed shape of a galaxy inherits additional contributions due to systematic errors, such as due to PSF misestimation, $\delta \epsilon^{\textrm{sys}}_{\textrm{PSF}}$, and noise, $\delta  \epsilon^{\textrm{noise}}$, as
 \begin{equation}
\label{eq: observed}
\epsilon^{\rm obs}=\gamma +\delta \epsilon^{\textrm{sys}}_{\textrm{PSF}}+\delta  \epsilon^{\textrm{noise}} \, ,
\end{equation} 
neglecting any multiplicative biases (see equation~\ref{eqn:shearbias}). 

It is common to parameterize the uncertainty in modeling the point spread function in terms of $\alpha,\beta$ parameters \cite{Paulin-Henriksson2008}. This can be extended to include a parameter $\eta$ for the PSF size-dependence of the additive error in shear measurement as
\begin{equation}
\label{eq:psf}
\delta \epsilon^{\textrm{sys}}_{\textrm{PSF}}=\alpha \epsilon_{\rm model}+\beta\left(\epsilon_{\rm *}-\epsilon_{\rm model}\right)+\eta\left(\epsilon_{\rm *}\frac{T_{\textrm{\rm *}}-T_{\rm model}}{T_{\rm *}}\right).
\end{equation}
$\epsilon_{\rm model}, T_{\rm model}$ denote the shape and size of the PSF model and $\epsilon_*, T_*$ are measured directly from a field of reserved stars that are not used in the fitting of the PSF model \citep{y3-shapecatalog}. The first term considers linear leakage, which is suppressed by the parameter $\alpha\ll1$. The second and third term are the shape and size residual dependencies, where the parameters $\beta$ and $\eta$ are of order unity in the unweighted moments approximation. 

The uncertainty due to PSF systematics in DES Y3, quantified in terms of the $\rho$-statistics \citep{rowe2010, jarvis2016} are found to have a substantially smaller amplitude compared to DES Y1, owing to improvements detailed in Ref.~\citep{y3-piff}. These statistics are used to fit for $\alpha, \beta, \eta$ per redshift bin, with the best-fit values and their corresponding $\chi^2$ reported in Table~\ref{tab:abe}, with similar amplitudes to DES Y1 \citep{y3-shapecatalog}.  As the inclusion of the $\eta$ parameter does not significantly alter the reduced $\chi^{2}_{\nu}$ model fit, we deduce that the effect of PSF size error is subdominant. However, we preserve this more complete model.

\begin{figure}
\centering
\includegraphics[width=0.45\textwidth]{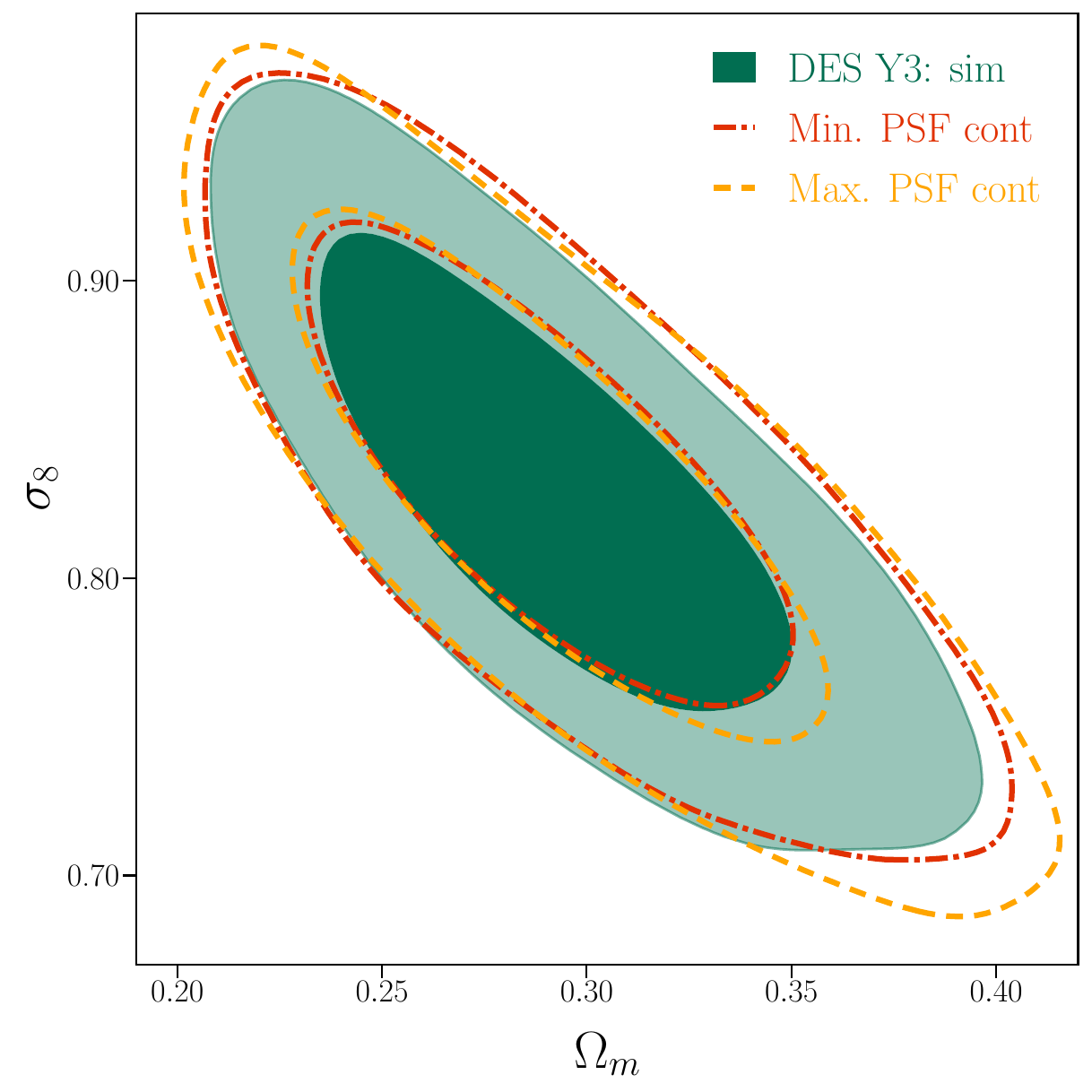}  
\caption{Robustness to PSF modeling systematics: The $\pm 2\sigma$ limits of the expected  additive contamination of PSF residual uncertainty to the true cosmic shear signal,  $\delta \xi^{\textrm{PSF}}$ are propagated to cosmological parameters and shown in red and yellow. For comparison, the uncontaminated simulated data vector is shown in green shaded contours.}
\label{fig:contaminatedpars}
\end{figure}

Here, the impact of PSF modeling uncertainties on cosmological constraints is tested. The best-fit $\alpha, \beta, \eta$ are propagated to determine the expected additive contamination of the true cosmic shear signal 
as $\xi_{+}^{\textrm{obs}}=\xi_{+}+\delta \xi^{\textrm{PSF}}$, where $\delta \xi^{\textrm{PSF}}=\left\langle \delta \epsilon_{\textrm{PSF}}^{\textrm{sys}}\,\delta \epsilon_{\textrm{PSF}}^{\textrm{sys}} \right\rangle $. The fractional PSF contamination with respect to the baseline simulated shear correlation functions , $d\xi/\xi$, is shown in yellow in Figure~\ref{fig:xitheorytomo}. The impact is most significant for the third redshift bin, but in the relevant radial range is at most 30\% of the physical cosmic shear signal and limited to only small-scales of a few bin pairs. The 2$\sigma$ limits of the $\delta \xi^{\textrm{PSF}}$ constraints are used to contaminate a simulated Y3 measurement in order to test the residual impact of the PSF bias on cosmic shear cosmological parameters. The result, shown in Figure~\ref{fig:contaminatedpars} in red and yellow contours, compared to the baseline green, finds the $\Omega_{\rm m}-\sigma_8$ parameters to be insensitive to PSF modeling errors in the Y3 analysis.

\begin{center} 
 \centering 
 \begin{table}
\begin{tabular}{ccccc} \hline 
& Bin 1 &  Bin 2 &  Bin 3 & Bin 4 \\ \hline \rule{0pt}{3ex}
$\alpha$ &  $0.010_{-0.005}^{+0.005}$ & $-0.001_{-0.005}^{+0.005}$ & $-0.004_{-0.005}^{+0.005}$ & $0.014_{-0.006}^{+0.006}$ \\ \rule{0pt}{3ex} 
$\beta$ &  $0.6_{-0.2}^{+0.2}$ &  $1.4_{-0.2}^{+0.2}$ &  $2.5_{-0.2}^{+0.2}$ & $1.3_{-0.3}^{+0.3}$ \\ \rule{0pt}{3ex} 
$\eta$ & $-4.6_{-2.7}^{+2.6}$ & $-4.5_{-2.7}^{+2.7}$ &  $3.0_{-2.7}^{+2.7}$ & $4.2_{-3.0}^{+3.1}$ \\ \rule{0pt}{3ex} 
$\chi^{2}_{\nu}$ &  $1.02_{-0.01}^{+0.02}$ &  $1.43_{-0.01}^{+0.02}$ &  $1.20_{-0.01}^{+0.02}$ & $1.25_{-0.01}^{+0.02}$ \\ \hline
\end{tabular}
 \caption{\label{tab:abe}The values of the parameters $\alpha$,$\beta$ and $\eta$ for each redshift bin, estimated from fits to the mean-substracted $\rho$-statistics, according to equation~\ref{eq:psf}, as well as the reduced goodness-of-fit, $\chi^{2}_{\nu}$ for $\nu=117$ degrees of freedom. }
 \end{table}
\end{center}

\subsection{B-mode contamination}\label{App:Bmodes}

To first order, lensing does not produce B-modes in the shear field for reasons of symmetry. Therefore, a detection of B-modes can indicate either a contamination by observational systematic effects or higher-order lensing or astrophysical effects, such as intrinsic alignments. For a fuller description and discussion of E- and B-mode power spectra, we refer the reader to \citep{y3-generalmethods}. Here, we test for the presence of B-modes in the shape catalog and potential contamination of the two-point functions used in the cosmological analysis. Section 6.4 of \citep{y3-shapecatalog} describes two complementary approaches toward measuring B-modes for the full sample of source galaxies without any redshift binning: the map-based pseudo-$C_\ell$ \citep{hikage2011} and Complete Orthogonal Sets of E/B-Integrals \citep[\textsc{COSEBIs}; ][]{Schneider2010,asgari19}. B-modes computed using both pseudo-$C_\ell$ and \textsc{COSEBIs} methods were shown to be consistent with zero, with $p$-values of {0.06} and {0.87}, respectively. Here, we additionally report tomographic measurements of B-modes using the pseudo-$C_\ell$ approach. We note that a non-detection in harmonic space does not fully rule out contamination, as was shown in \citep{2019A&A...624A.134A}.

For each redshift bin, we build two \code{HEALPix} \citep{Gorski05} maps with resolution ${n_{\rm side}=1024}$ of the cosmic shear signal by computing the weighted average of response-corrected ellipticities of galaxies within each pixel. We then estimate the E- and B-mode power spectra of these maps by the method of pseudo-$C_\ell$ using \code{NaMaster} \citep{Alonso19}, an open-source code that deconvolves the effects of masked regions from the harmonic space coefficients. We use the inverse-variance weight masks given by the weighted count maps and measure auto- and cross-spectra for multipoles in the range $\ell\in[8-2048]$. We turn these into 32 bins evenly spaced on a square-root scale, which spreads signal-to-noise more evenly than linear or logarithmic binning.
The measured auto power spectra receive an additive bias from the shape-noise power spectrum $N_\ell$, which may diverge from the approximation $N_\ell=\sigma_e^2/\bar{n}$ due to masking effects and properties of the pseudo-$C_\ell$ estimator. We therefore employ the analytic formula derived in \citep{2020arXiv201009717N} to estimate the noise power spectrum and subtract it from the initial measurements. 

In order to evaluate the B-mode covariance matrix, we use \num{10000} Gaussian simulations of the shear fields at the baseline cosmology, following the method outlined in \cite{2020arXiv201106469D}. We draw shear maps at resolution ${n_{\rm side}=1024}$, which we then sample at the positions of galaxies in the real data. The intrinsic ellipticities of galaxies are obtained by randomly rotating measured ellipticities. This procedure preserves the shape noise, in terms of galaxy density and ellipticity distribution, and produces a null B-mode signal. For each simulation, we then measure the B-mode power spectra using the data masks. Finally, we compute the sample covariance matrix and use it to form a $\chi^2$ statistic to test the null hypothesis of no B-modes.
We have verified that the number of simulations is sufficient for this statistic to converge.
We report $\chi^2$ statistics for each bin combination as well as the $\chi^2$ for the full data vector, accounting for the full covariance. We obtain $\chi^2=344.0$ for 320 degrees of freedom, consistent with the absence of B-modes.

Finally, we propagate the B-mode contamination to the shear two-point correlation functions, shown in Figure~\ref{fig:xitheorytomo}. 
The two-point correlation functions $\xipm$ are related to angular spectra by expressions of the form (see Eqn.~\ref{eqn:2ptP})
\begin{equation}
    \xipm^{ab}\qty(\theta) = \sum_\ell F_\ell^{\pm}\qty(\theta) \qty(C_\ell^{ab,\rm EE} \pm C_\ell^{ab,\rm BB}),
\end{equation}
where $a$ and $b$ denote redshift bin indices. This allows us to compute an approximate covariance between two-point functions and the bin-averaged B-mode power spectra $C_L^{ab,\rm BB}$, given as
\begin{equation}
\begin{split}
    \mathbf{C}_{\rm B \pm}  & \equiv \expval{\Delta C_L^{ab,\rm BB} \Delta \xipm^{cd}\qty(\theta)} \\ & \approx \sum_{\ell} F_\ell^{\pm}\qty(\theta) \qty[\expval{\Delta C_L^{ab,\rm BB} \Delta C_\ell^{cd,\rm EE}} \pm \expval{\Delta C_L^{ab,\rm BB} \Delta C_\ell^{cd,\rm BB}}] .
\end{split}
\end{equation}
We then approximate $F_\ell^{\pm}\qty(\theta)$ and the signal to be piece-wise constant and use the covariance matrices of E- and B-mode spectra measured from the simulations described above to compute $\mathbf{C}_{\rm B \pm}$. Assuming all data are Gaussian distributed, we finally compute the conditional distribution of the additive bias to the two-point functions given the B-mode measurements, which is Gaussian with mean
\begin{equation}
    \left. \Delta \xipm\qty(\theta) \,\middle|\, \hat{C}_{\ell,\rm data}^{\rm BB} \right. =
    \mathbf{C}_{\rm B \pm} \vdot \mathbf{C}_{\rm BB}^{-1} \vdot \hat{C}_{\ell,\rm data}^{\rm BB},
\end{equation}
where we omit redshift bin indices to imply vectorization.

\begin{figure*}
\centering
\begin{minipage}{.47\textwidth}
  \centering
  \includegraphics[width=.97\textwidth]{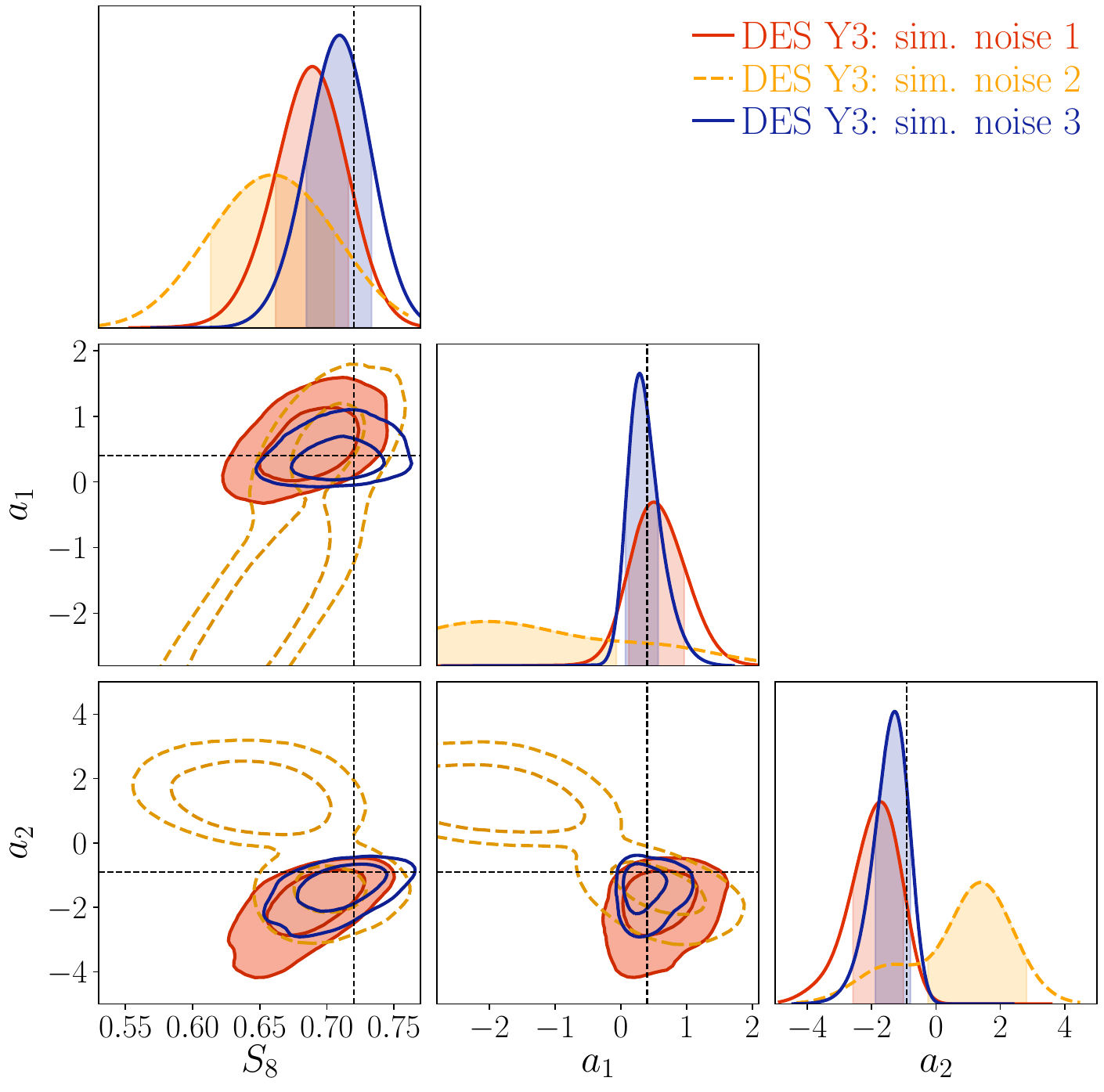}
\end{minipage}
\begin{minipage}{.47\textwidth}
  \centering
  \includegraphics[width=.95\textwidth]{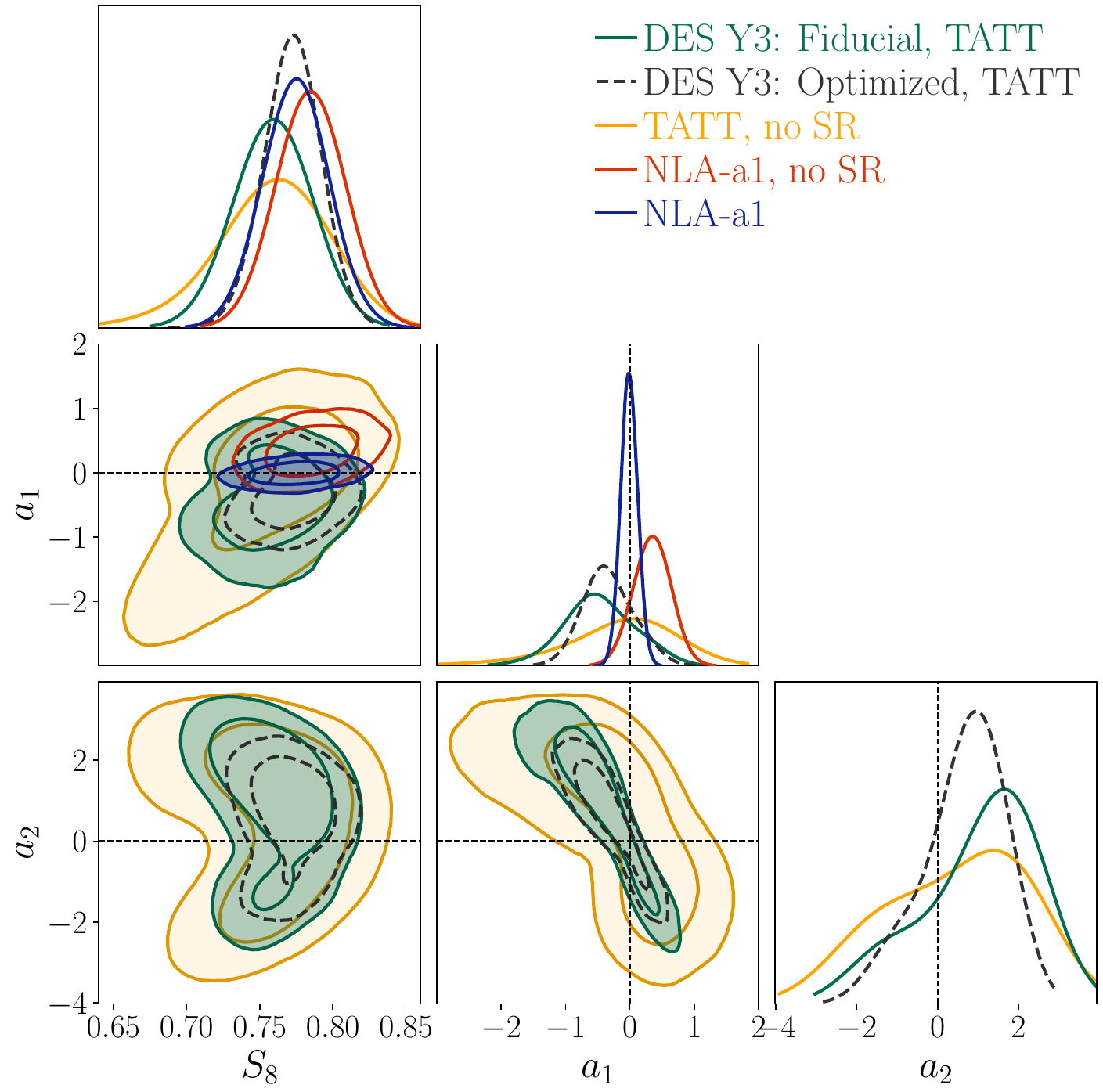}
\end{minipage}
\caption{\label{fig:power} The impact of bimodality on constraining power in $S_8$: The left-hand panel shows analyses of three synthetic Y3-like cosmic shear data vectors with random realizations of noise, determined with the Y3 covariance, one of which presents a bimodality in the $a_1-a_2$ constraints, using with no \textsc{SR} and TATT modeling. The right-hand panel shows posteriors from the data with permutations on the choice of intrinsic alignment model (TATT and NLA-no-$z$) and the inclusion of \textsc{SR}. Compared to the Fiducial analysis (green), the \lcdm-Optimized scales (black), and the redshift-independent NLA model (red), the TATT-no-\textsc{SR} (yellow) case exhibits a degraded $S_8$ constraint. The combination of the conservative Fiducial scale-cuts and TATT model, results in intrinsic alignment parameters that are not well-constrained that are degenerate with, and reduce the precision of the $S_8$ constraint. The inclusion of \textsc{SR} is effective in constraining the intrinsic alignment parameters, and for the TATT case, alleviating the impact of the bimodality. On the other hand, when using a less conservative intrinsic alignment model, which is mildly preferred by our data \citep{y3-cosmicshear2}, while the $a_1$ constraint is substantially improved, the $S_8$ precision is negligibly impacted by including \textsc{SR}.}
\end{figure*}

\subsection{Additive shear}\label{App:additivecorr}

While we conclude that the propagation of PSF model uncertainty and the B-mode signal is below the statistical uncertainty, the catalog gives an unidentified source of mean shear or additive bias (see equation~\ref{eqn:shearbias}).  This is too large in some redshift bins to be consistent with shape noise and cosmic variance. The low-level contribution from the constant additive ellipticity bias is not explicitly marginalized over, but corrected as a global constant per redshift bin and ellipticity component, $c=\langle e_{1,2}^i\rangle$ reported in Table~\ref{tab:datastats}. The impact of this additive correction on the shear two-point correlation functions is
\begin{equation}
    \xi_{\pm}^{ij}=\xi_{\pm, \rm{true}}^{ij} + (\langle e_{1}^i\rangle \langle e_{1}^j\rangle \pm \langle e_{2}^i\rangle \langle e_{2}^j\rangle) \,,
\end{equation}
and the result of artificially introducing this bias due to mean shear to an un-biased simulated data vector is shown in Figure~\ref{fig:xitheorytomo} as a blue dashed line. 
The residual mean shear,  $c=\langle e_{1,2}^i\rangle$, is subtracted from each galaxy's ellipticity, which minimizes its impact on $\xi_{\pm}$. Note that the residual mean shear may be a result of a systematic effect that generates a scale-dependent additive bias in $\xi_{\pm}$, such that simply subtracting the mean will not correct for this scale dependence.

The approximation assumes that the galaxies shapes are not correlated in the absence of lensing, and the mean response is relatively homogeneous across the footprint. The impact of this simplification in the computation of the shear response is shown in red in Figure~\ref{fig:xitheorytomo} to be negligible at the current level of precision.

\section{Constraining power \& bimodality in noisy data}\label{app:power}

In this section, the constraining power of the cosmic shear posteriors is investigated in more detail. In particular, the occurance of bimodality in the intrinsic alignment parameters is investigated for various analysis choices, including the intrinsic alignment model and the inclusion/exclusion of \textsc{SR} data.

Prior to un-blinding, the potential for bimodality in the $a_1-a_2$ space was explored using a simulated data vector, created with the fiducial analysis pipeline. We found that it is possible for particular noise realisations to present such a feature, which, given the degeneracy with $S_8$, substantially alters the precision of that parameter. Ten realisations of noise were added, consistent with the Y3 covariance, and cosmological inference was performed on each, with the baseline TATT model and no \textsc{SR}. 
The resulting parameter constraints from three of these datavectors are shown in the left-hand panel of Figure~\ref{fig:power}. One of the ten realisations, `noise 2' (yellow contour), was found to exhibit bimodality in the $a_1-a_2$ plane. The intrinsic alignment parameters are degenerate with $S_8$. This particular realisation presents significantly degraded cosmological $S_8$ constraint that is scattered toward a lower value of $S_8$, with lower signal-to-noise.

The Y3 data also exhibits a bimodal posterior in the $a_1-a_2$ plane that is less pronounced. The right-hand panel of Figure~\ref{fig:power} shows the Fiducial Y3 cosmic shear result in green filled contours and the \lcdm-Optimized analysis, in black, with significantly improved constraints in the $S_8, a_1, a_2$ parameters. This bimodality and the degraded $S_8$ constraint are consistent with findings of Ref.~\citep{asgari20}, who use the NLA-$a_1$ as their fiducial choice. Similarly, they report that when opting for the more conservative NLA choice, they observe bimodality in these astrophysical parameters, with degraded cosmological posteriors. Therefore, we interpret this feature of a doubly-peaked posterior as an internal degeneracy of the intrinsic alignment model that is reduced or eliminated as statistical power increases.

Given the the degeneracy in $S_8-a_1/a_2$, we find reduced bimodality and significantly improved $S_8$ constraints with either the inclusion of small-scale measurements, such as the \lcdm-Optimized choice,  the inclusion of orthogonal \textsc{SR} instrinsic alignment information, or a more  aggressive intrinsic alignment model, such as NLA-$a_1$. Alongside the Fiducial and Optimized constraints, Figure~\ref{fig:power} shows a
variant of the analysis without \textsc{SR} (yellow). We see a substantial improvement in the precision of the $a_1$, and therefore $S_8$,  compared to the case with \textsc{SR} (green). Moreover, a case that uses a simpler intrinsic alignment model results in substantially tighter $S_8$ posteriors, with a $\times\sim1.5$ improvement from TATT (yellow) to NLA-$a_1$ (red). These improvements can be attributed to the elimination of negative $a_1$ space and the upper and lower bounds of the $a_2$ space. 

We test the impact of \textsc{SR} in the case of the data-preferred NLA-$a_1$ model, shown in Figure~\ref{fig:power} (red to blue). Note that our companion paper \citep{y3-cosmicshear2}, performs a detailed model comparison that finds the data favours a less conservative intrinsic alignment model than TATT (see their Table 3). We find that the inclusion of \textsc{SR} has little effect on the $S_8$ constraint, although it provides significantly tighter constraints on the intrinsic alignment parameters, which are not degenerate with $S_8$ to begin with.

\section{Internal consistency}\label{App:IC}

\begin{table}[h]
    \centering
    \begin{tabular}{l c c c c }
        \hline 
        PPD test \,\, & Calibrated $\tilde{p}$-values & $\xi_+$ & $\xi_-$ \\ 
         \hline 
         \\
        \textit{Goodness-of-fit tests} \\ 
        Cosmic shear & 0.268 & 0.252 & 0.422 \\
        $\xi_+$ & 0.234 & 0.234 & --  \\
        $\xi_-$ & 0.382 & -- & 0.382  \\
        \hline 
        \\
        \textit{Data splits} \\  
        Bin 1 \vs no bin 1 & 0.357 & 0.196 & 0.759  \\
        Bin 2 \vs no bin 2 & 0.394 & 0.547 & 0.132  \\
        Bin 3 \vs no bin 3 & 0.014 & 0.041 & 0.070  \\
        Bin 4 \vs no bin 4 & 0.427 & 0.376 & 0.614  \\
        Low-z \vs High-z   & 0.993 & 0.992 & 0.974  \\
        High-z \vs Low-z   & 0.207 & 0.324 & 0.282  \\
        Large \vs small scales & 0.660 & 0.646 & 0.441 \\ 
        Small \vs large scales & 0.083 & 0.068 & 0.332  \\
        $\xi_+$ \vs $\xi_-$ & 0.601 & 0.601 & -- \\
        $\xi_-$ \vs $\xi_+$ & 0.422 & -- & 0.422 \\
        \hline 
    \end{tabular}
    \caption{\label{tab:IC}Summary of internal consistency test $\tilde{p}$-values. The `PPD test' column specifies the details of the comparison. For consistency tests, `A vs. B' indicates a comparison of observations for data A with PPD realizations for data A derived from data B. The second column shows the calibrated $\tilde{p}$-value for each test, obtained by comparison of PPD tests on simulated data. The third and fourth column show the calibrated $\tilde{p}$-values when PPD metrics are restricted to the $\xi_+$ and $\xi_-$ components respectively. These tests correspond to the parameter constraints shown in Section~\ref{sec:IC}. All internal consistency tests pass the pre-defined (arbitrary) threshold of 0.01. }
\end{table}

Table~\ref{tab:IC} summarizes the calibrated $\tilde{p}$-values for the full data vector as well as each subset considered in Section~\ref{sec:IC} (see Ref.~\cite{y3-inttensions} for details of the methodology). As stated previously, the threshold for consistency is chosen to be 0.01. There are no obvious discrepancies between the PPD realizations of cosmic shear and the actual data. We compute a $\tilde{p}$-value for the goodness-of-fit of cosmic shear of $p = 0.268$, indicating no evidence for tension between the measurements and PPD realizations. 

When splitting the data into subsets, all overall $\tilde{p}$-values are above 0.01, indicating no sign of tensions between redshift bins, angular scales or the two statistics that describe the DES Y3 cosmic shear data. These findings are consistent with the agreement in cosmological parameters measured by the subsets of the data shown in Section~\ref{sec:IC}, validating an internally consistent analysis.

\begin{figure}
    \centering
    \includegraphics[width=\columnwidth]{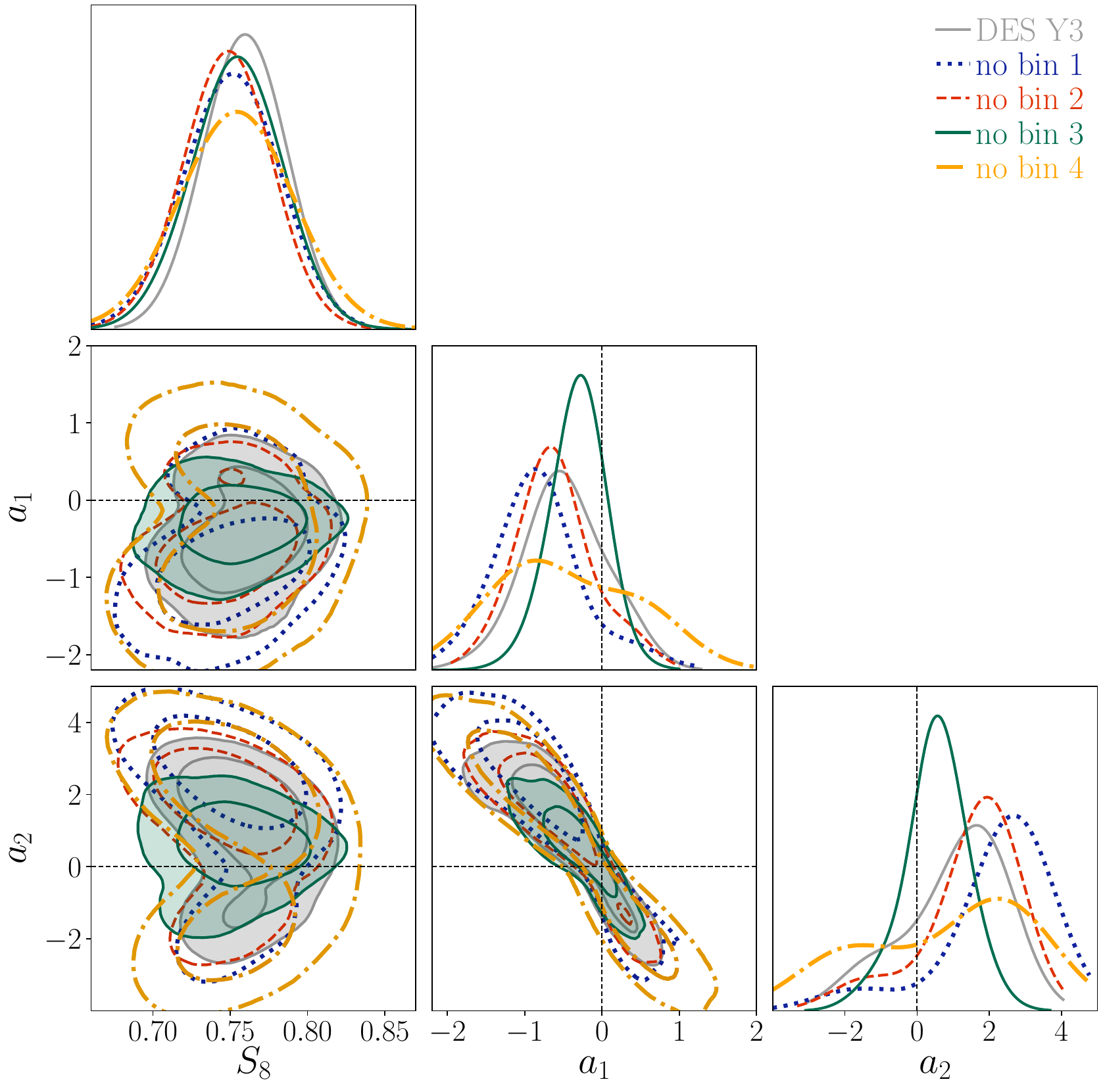}
    \caption{\label{fig:ICfull} Intrinsic alignment consistency across redshift bins: The posteriors for the $S_8$ and intrinsic alignment parameters, $a_1$ and $a_2$, as correlations involving each redshift bin are removed. For comparison, the Fiducial result using all redshift bins is shown in the grey shaded contours. The 68 and 95\% constraints are shown and a zero-line is marked for reference. }
\end{figure}

The modeling of intrinsic alignments is degenerate with the $\Omega_{\rm m}-S_8$ parameters, and systematic errors in the data can be absorbed by the intrinsic alignment model, impacting the cosmological posteriors. Thus, disentangling the true cosmic shear signal from intrinsic alignments requires accurate knowledge of the redshift distributions and their errors. As such, we demonstrate consistency of the intrinsic alignment constraints with redshift \citep{Efstathiou18}. Figure~\ref{fig:ICfull} demonstrates the stability of the $S_8$ and intrinsic alignment $a_1-a_2$ solution across redshift for the DES data: the posteriors across the five parameters are consistent for each subset of the data within $1\sigma$. The first point to note is that the $S_8$ parameter is stable in the removal of any of these subsets of the data, and unsurprisingly, Bin 4 holds the most cosmological constraining power. Next, the intrinsic alignment amplitude, $a_1$, is stable to the removal of photometric redshift bins, within 1$\sigma$. Moreover, all five intrinsic alignment posteriors, $[a_1, a_2, \eta_1, \eta_2, b_{\rm ta}]$, a subset of which are shown in Figure~\ref{fig:ICfull}, are consistent with the intrinsic alignment solution from the full dataset. Bins 1 and 4 carry a higher weight in fixing the amplitudes, $a_1$ and $a_2$.

\section{Robust redshift calibration in the presence of intrinsic alignment}\label{App:IAnz}

\begin{figure*}

\begin{minipage}[c]{0.48\linewidth}
\includegraphics[width=\linewidth]{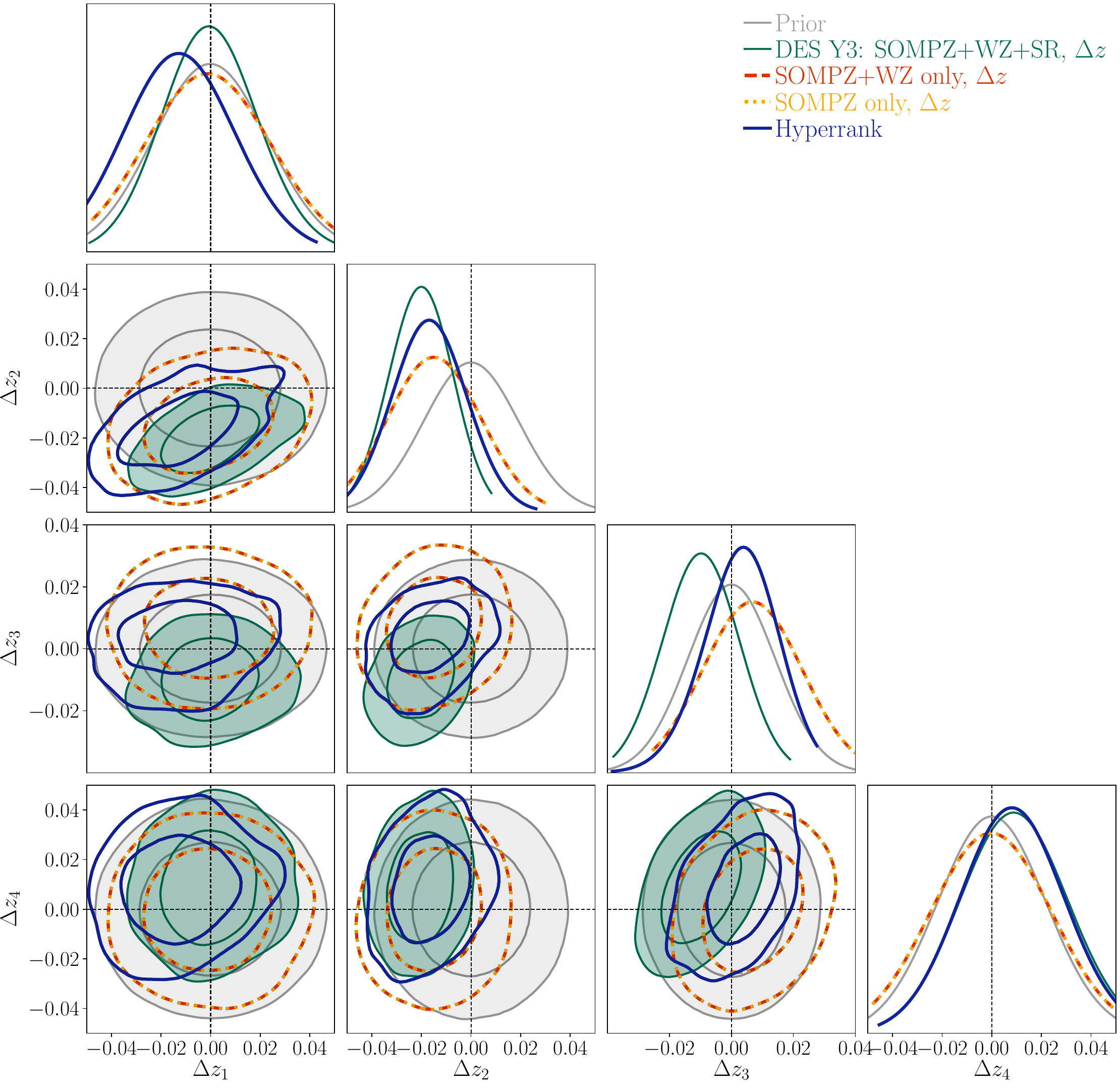}
\caption{\label{fig:deltaz} Stability of redshift uncertainty posteriors, $\Delta z$, from the cosmological-inference chains, with the varying redshift methods and modeling. The filled grey contours correspond to the priors, listed in Table~\ref{tab:priors}. The Fiducial analysis (green) is compared to decreasing complexity in methodology, removing the \textsc{SR}, `\textsc{SOMPZ+WZ}' (red), as well as the clustering redshift likelihood, `\textsc{SOMPZ} only' (yellow), as well as the translation of \textsc{Hyperrank} parameters to $\Delta z$ (blue), when the uncertainty in the full shape of the $n(z)$ is accounted for.  While the approximation of the uncertainty as a shift in the mean does not impact $S_8$ at the precision of this analysis, a substantial shift in the mean of the $n(z)$ can compensate for a change in shape, as seen in bin 3.}
\end{minipage}%
\hfill
\begin{minipage}[c]{0.48\linewidth}
\includegraphics[width=\linewidth]{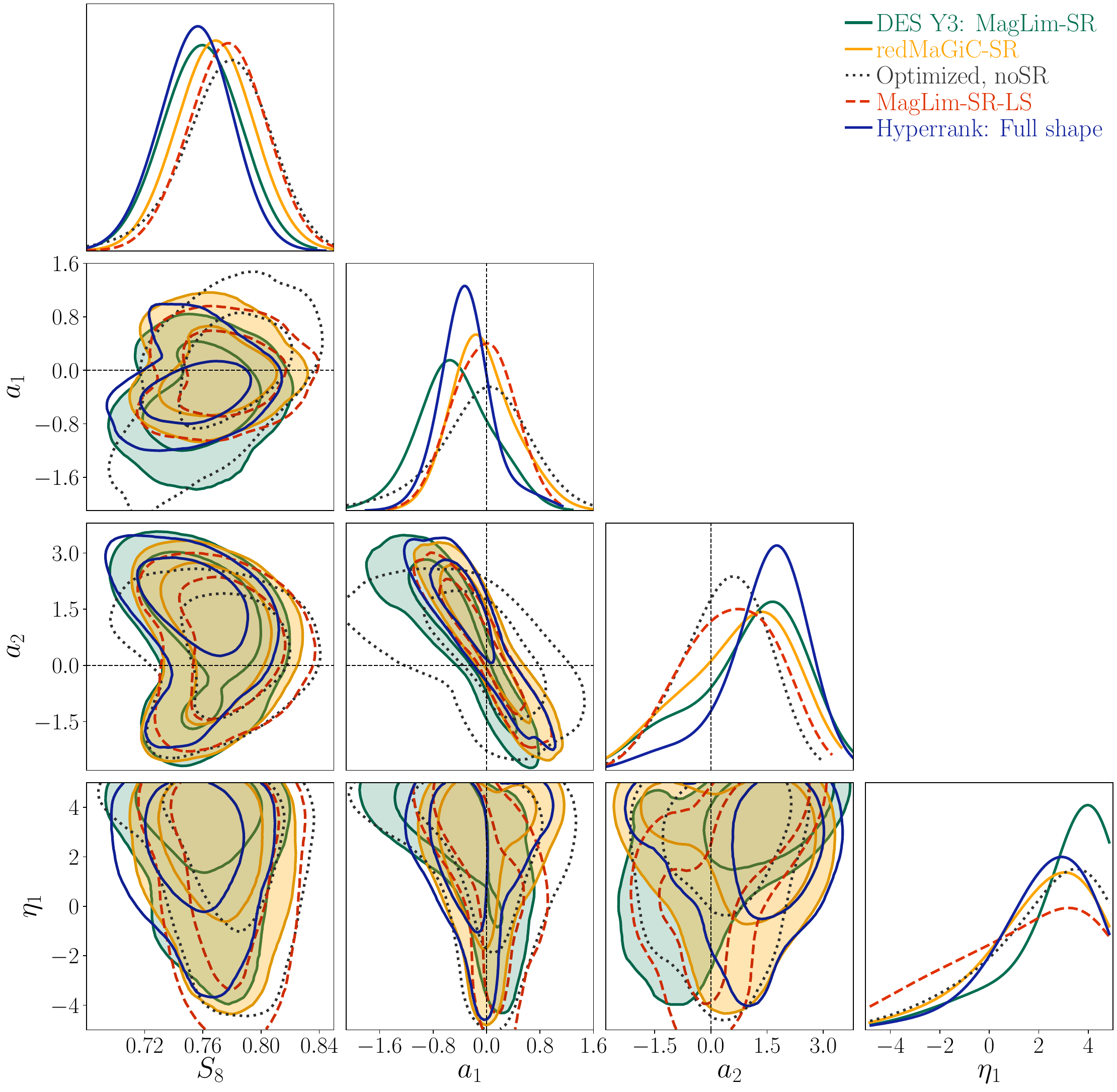}
\caption{\label{fig:nzIA}Stability of $S_8$, the intrinsic alignment model parameters, $a_1$, $a_2$, $\eta_1$, when including additional information: The filled green contours show the Fiducial analysis, which includes \textsc{SR} measured with the \textsc{MagLim} sample, consistent with one that includes \textsc{SR} measured with the \textsc{redMaGiC} lens sample (yellow). Using the large-scale-only \textsc{MagLim} \textsc{SR} (red),  tests the robustness of the modeling along with the result incorporating the \lcdm-Optimized cosmic shear without any \textsc{SR} (black). In addition, the Fiducial \textsc{Hyperrank} result (blue) modulates the shape of the $n(z)$ and finds more constraining $a_1$-$a_2$ posteriors that are consistent.}
\end{minipage}

\end{figure*}

The redshift uncertainty parameters, $\Delta z$, are shown in Figure~\ref{fig:deltaz} with varying redshift methodologies: removing the SR method (red), removing the SR and WZ methods (yellow) and using \textsc{Hyperrank} (blue). The corresponding constraints on the $\Omega_{\rm {m}}-S_8$ parameters are shown in Figure~\ref{fig:zuncertainty}. We find that these nuisance parameters are most constrained by the addition of the \textsc{SR}, but that each tier of the methodology gives consistent posteriors. When accounting for the full-shape uncertainty in the $n(z)$ with \textsc{Hyperrank}, we find that the need for shifts in $\Delta z$, such as the case for Bin 3, are alleviated. This suggests that an uncertainty in the shape of the $n(z)$ is compensated for by a more substantial shift in the mean of the distribution. While the approximation of the uncertainty as a shift in the mean does not impact the cosmological parameters at the precision of this analysis, this demonstrates that shifts in the mean have the potential to be misleading.

Uncalibrated redshift error can result in an incorrectly inferred intrinsic alignment signal or be absorbed by unconstrained model nuisance parameters in the likelihood analysis, as suggested in \citep{wright20, joachimi20}. To test that such an effect is not occurring in the DES Y3 analysis, we include a detailed account of the full uncertainty in the $n(z)$ estimates \citep{y3-sompz}. Section~\ref{sec:robustredshift} shows that approximating uncertainty in the $n(z)$ to a shift in the mean leads to a consistent set of cosmological parameters for the Y3 analysis. In this Appendix, we add to this a demonstration that the intrinsic alignment parameter constraints are fully consistent when marginalizing over the full shape uncertainty of $n(z)$ with \textsc{Hyperrank}, instead of the approximation of the mean shift (green). Interestingly, Figure~\ref{fig:nzIA} shows the $a_1, a_2$ posteriors in the \textsc{Hyperrank} analysis are significantly tighter, though consistent. This demonstrates that shifts in the mean have the potential to be misleading and could bias the measurement of intrinsic alignment parameters, a possibility we now explore. We find no evidence that the nuisance model parameters absorb residual observational systematics in the analysis.

In this Appendix, the robustness of the results of the analysis to choices in the \textsc{SR} is also investigated. More specifically, Figure~\ref{fig:nzIA} shows the constraints on the $S_8$ parameters, the intrinsic alignment amplitudes, $a_1$ and $a_2$, and their redshift dependence, $\eta_1$, when the alternative lens sample, \textsc{redMaGiC}, is used in the measurement (yellow) instead of the Fiducial \textsc{MagLim} (green). In addition, a \lcdm-Optimized case that includes smaller-scale measurements and neglects \textsc{SR} is shown in black. While this method has been validated in a simulated framework \citep{y3-shearratio}, here, the stability of the data constraints is demonstrated: The $S_8$ and intrinsic alignment posteriors are consistent. In all variations tested, the $a_1$ amplitude is consistent with zero, with the Fiducial \textsc{MagLim} result finding the most negative value. The `\lcdm-Optimized, no SR' case, that does not depend on either lens sample, or the \textsc{SR} method is centred on $a_1=0$.

Furthermore, Figure~\ref{fig:nzIA} shows the consistency of the constraints when large-scale (LS) \textsc{SR} are used (red). The fiducial \textsc{SR} measurements use small angular scales to compute the ratios, which can be subject to systematics in the modeling of non-linear galaxy bias, baryonic effects, and intrinsic alignments. The \textsc{LS-SR} measurements are assumed to be independent of those from the small scales. For this consistency check, we assume any cross-covariance between the \textsc{LS-SR} and the cosmic shear measurements is negligible. The results are found to be consistent, demonstrating the robustness to these modeling effects, with the \textsc{LS-SR} preferring a slightly higher value of $S_8$, and $a_1$ and $a_2$ values closer to zero. Overall, the cosmic shear $S_8$ and intrinsic alignment posteriors are robust to the inclusion of \textsc{SR}, as well as the lens sample and scales used in that method.

\section{The unblinding process}\label{app:unblinding}

After unblinding the cosmological parameters, two updates to the analysis were made that marginally impacted the results of this  work: a standard, planned update to the covariance matrix to the `3x2pt' best-fit parameters, as well as a change of the fiducial lens sample, from \textsc{redMaGiC} to \textsc{Maglim}, which impacts the \textsc{SR} method (see Ref.~\citep{y3-3x2ptkp} for a discussion). While the use of either of the two \textsc{SR} give consistent cosmological constraints, it does shift the $S_8$ parameter by 0.3$\sigma$. This is primarily attributed to a shift in the intrinsic alignment $a_1-a_2$ space, as demonstrated in Figure~\ref{fig:nzIA} (the difference between the green and yellow contours) and discussed in more detail in Appendix~\ref{App:IAnz}.

\section{Cosmological parameters}\label{app:as}

\begin{figure*}
    \centering
    \includegraphics[width=1.01\textwidth]{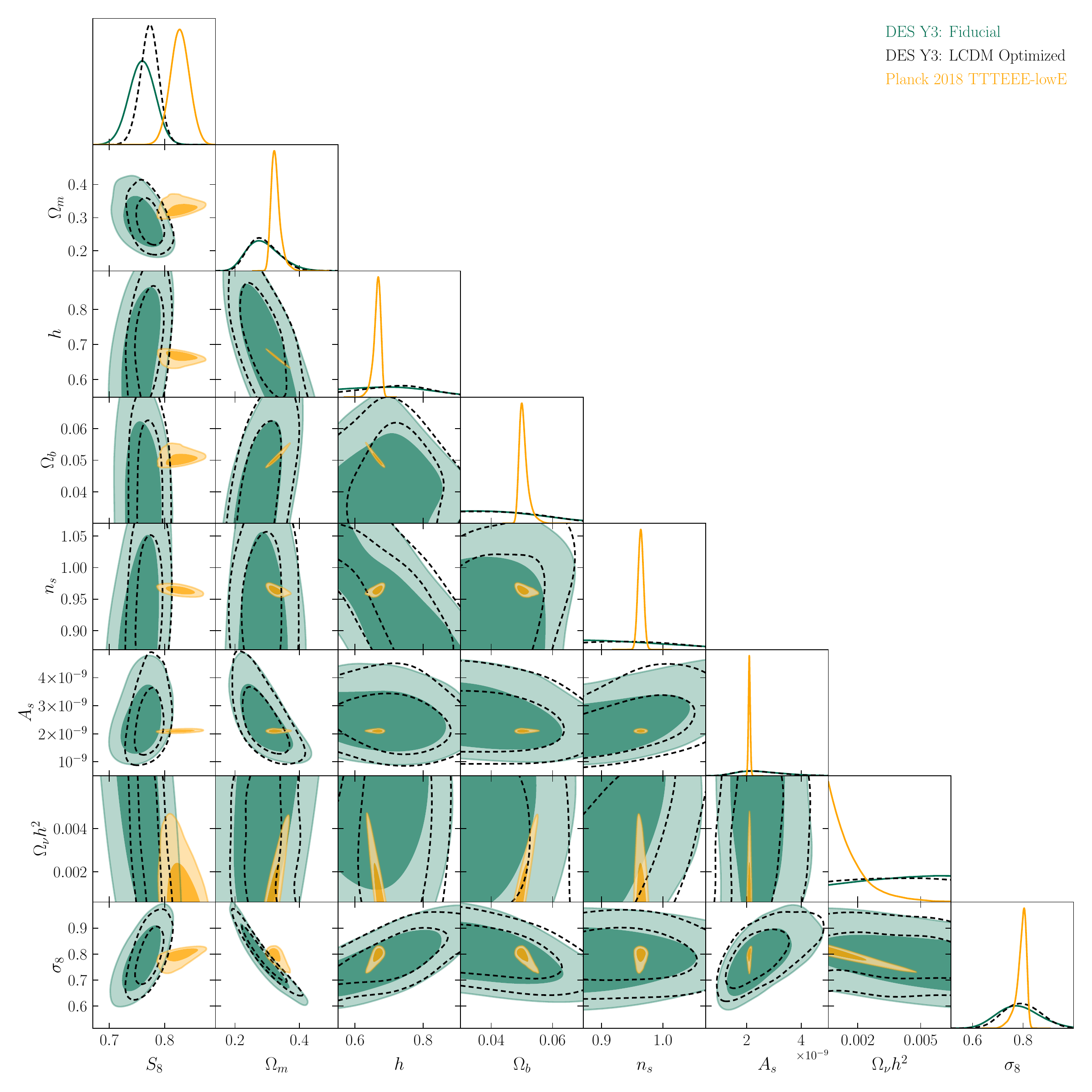}
    \caption{\label{fig:cosmo} The posteriors for a subset of the \lcdm cosmological parameters for the Fiducial and \lcdm-Optimized analyses.}
    
\end{figure*}

The 1D posteriors for the full \lcdm cosmological parameter space is shown in Figure~\ref{fig:cosmo}. While the DES constraint on $A_{\rm s}$ is weak, it is interesting that there is no evidence for a discrepancy with \textit{Planck} in this parameter. We find no significant constraints beyond the prior imposed on the parameters $\Omega_{\rm b}$, $H_0$, $n_{\rm s}$, and $\Omega_{\nu}h^2$. The priors for all six parameters are listed in Table~\ref{tab:priors}.

\label{lastpage}
\end{document}